\documentclass[11pt,3p,doublespacing]{elsarticle}
\journal{.}
\usepackage{times}
\emergencystretch=\maxdimen
\biboptions{sort&compress}
\usepackage{hyperref}
 \hypersetup{
 	plainpages = false, 
	bookmarksopen = true,
	bookmarksnumbered = true,
	breaklinks = true,
	linktocpage,
	colorlinks = true,
	linkcolor = purple,
	urlcolor  = black,
	citecolor = purple,
	}
\usepackage{multicol}
\usepackage{multirow}
\usepackage{imakeidx}
\makeindex
\usepackage{nomencl}
\makenomenclature
\usepackage[xindy]{glossaries}
\usepackage{graphics}
\usepackage{graphicx}
\usepackage{epsf,psfrag}
\usepackage{epstopdf}
\usepackage[para]{threeparttable}
\usepackage{subfigure,subcaption}
\usepackage{epsfig}
\usepackage{pdflscape}
\usepackage{rotating}
\usepackage{lscape}
\usepackage{float}
\usepackage{stfloats}
\usepackage{textgreek}
\usepackage{longtable}
\usepackage{lscape}
\usepackage{capt-of}
\usepackage{comment}
\usepackage{tablefootnote}
\usepackage{footnote}
\usepackage{caption}
\usepackage[autopunct=true]{csquotes}
\usepackage{rotating}
\usepackage{txfonts}
\usepackage[normalem]{ulem}
\usepackage{resizegather}
\usepackage{siunitx}
\usepackage{setspace}
\usepackage{color,soul}
\usepackage[usenames,dvipsnames]{xcolor}
\usepackage{amsfonts,amsmath,amssymb}
\usepackage{latexsym,array}
\usepackage{dashrule}
\usepackage{textcomp}
\usepackage{standalone}
\usepackage{tikz,tikzscale}
\usetikzlibrary{shapes.geometric}
\usetikzlibrary{scopes}
\usetikzlibrary {arrows.meta} 
\usetikzlibrary{arrows,calc,patterns}

\newcommand{\RomanNumeralCaps}[1]
\linenumbers
\makesavenoteenv{tabular}
\makesavenoteenv{table}
\usepackage[left,displaymath, mathlines]{lineno}
\DeclareMathAlphabet{\mathpzc}{OT1}{pzc}{m}{it}
\def\fig{Fig.~}
\def\figs{Figs.~}
\def\eqn{Eq.~}
\def\eqns{Eqs.~}
\def\tab{Table~}
\def\tabs{Tables~}

\def\tsc#1{\csdef{#1}{\textsc{\lowercase{#1}}\xspace}}
\tsc{WGM}
\tsc{QE}
\tsc{EP}
\tsc{PMS}
\tsc{BEC}
\tsc{DE}
\usepackage{ltablex,threeparttablex, longtable}
\usepackage{array, multirow, tabularx, booktabs}

\newcolumntype{L}{>{\raggedright\arraybackslash}X} 
\newcolumntype{C}{>{\centering\arraybackslash}X} 
\usepackage{makecell}
\graphicspath{{../}{Figures/}{Figures/Validation/}}
\DeclareGraphicsExtensions{.png, .PNG,.pdf,.PDF,.jpg,.JPG,.eps,.EPS}
\begin{document}
%
%
\setcounter{page}{1}
\begin{frontmatter} 
%
%
%
\title{Computational insights into the unconfined flow of a non-Newtonian power-law fluid past counter-rotating circular cylinders}
\author[labela]{Lekhraj Malviya} 
\author[labela]{Ram Prakash Bharti\corref{cor3}}
\emailauthor{rpbharti@iitr.ac.in}{RPB}
\author[labela]{Abhishek Kumar Lal} 
\address[labela]{Complex Fluid Dynamics and Microfluidics (CFDM) Lab, Department of Chemical Engineering, Indian Institute of Technology Roorkee, Roorkee - 247667, Uttarakhand, India}
%
\cortext[cor3]{Corresponding author}
\begin{abstract}
	\fontsize{11}{16pt}\selectfont
	\noindent
	This study numerically examines the steady unconfined laminar flow of incompressible non-Newtonian power-law fluids past a pair of side-by-side counter-rotating circular cylinders using the finite element method. The cylinders simultaneously rotate at equal angular speeds in opposite directions, with the upper cylinder (UC) rotating clockwise and the lower cylinder (LC) counterclockwise. The numerical simulations are performed over wide parameter ranges: power-law index ($0.2 \le n \le 1.8$), rotational rate ($0 \le \alpha \le 2$), gap ratio ($0.2 \le G \le 1$), and Reynolds number ($1 \le Re \le 40$). The detailed influence of these flow parameters on key flow characteristics, including streamlines, surface pressure, centerline velocity, and individual as well as total drag and lift coefficients, is systematically analyzed. For stationary cylinders ($\alpha = 0$), a twin-vortex structure forms in the wake, which progressively vanishes with increasing rotation ($\alpha = 2$). The surface pressure coefficient attains its maximum values for shear-thickening fluids ($n > 1$) at higher rotational speeds, irrespective of the gap ratio. The total drag coefficient ($C_D$) decreases with increasing $\alpha$ and $n$, whereas the lift coefficient ($C_L$) exhibits a more intricate dependence on the flow parameters. The results provide fundamental insights into the interplay between rotation, non-Newtonian rheology, and geometric confinement in determining the hydrodynamic behavior of rotating cylinder systems.
%
\end{abstract}
\begin{keyword}
\fontsize{11}{16pt}\selectfont
Rotating cylinders\sep Gap spacing\sep Drag and lift coefficient\sep Non-Newtonian fluids\sep Power-law index
\end{keyword}
\end{frontmatter}
%
\section{Background}\label{sec:intro}
%
\noindent
The flow over cylinders has been the classical problem in fluid mechanics over the century \citep{Streeter1961,Zdravkovich1977, Zdravkovich1997,Zdravkovich2003,Zapryanov1999,Clift2005,Michaelides2006,Johnson2016,Hishikar2022,Chhabra2023} due to the simplicity of the flow configuration, yet providing enormous flow features essential to a variety of practical industrial applications, including paper and pulp industries, heat exchanger, cooling tower arrays, nuclear reactors (fuel and control rods), chimney stacks, bridge piers, and offshore structures. The cylinders of different cross-sectional (e.g., circular and non-circular) shapes and sizes (equal or unequal diameter, length) can be arranged in various configurations (single, tandem, side-by-side, triangular, rectangular, and staggered arrangements) to represent the wide-ranging flow problems  of industrial importance.  
Furthermore, the industrial applications involving cylindrical structures use a wide variety of fluids (e.g., polymer solutions, emulsions, suspensions, high molecular weight substances, paper and pulp suspensions, drilling mud, crude oil, and multiphase mixtures) displaying the non-Newtonian natures such as shear-dependent, viscoelastic, viscoplastic and time-dependent characteristics  \mbox{\citep{Chhabra2008}}. The rheological nature of wide-ranging shear-dependent fluids is typically expressed by the power-law constitutive model.
However, to the best of our knowledge, the unconfined steady flow of incompressible power-law fluids across side-by-side rotating cylinders remains unexplored in the literature, despite signifying numerous practically applications.
Therefore, this article aims to provide a novel contribution by examining the effects of rotational rate and gap ratio on the hydrodynamics of non-Newtonian power-law fluids flowing across two side-by-side rotating cylinders. It is, however, imperative to present the relevant literature to understand better the existing knowledge gaps for flow around circular cylinders and news results presented in this work. 
%
\noindent
\begin{table}[!htb]
	\centering
	\caption{Literature on the flow of Newtonian ($n=1$) and non-Newtonian ($n\neq 1$) fluids across a pair of the side-by-side stationary {($\alpha = 0 $)} and rotating {($\alpha \neq 0 $)} circular cylinders}\label{tab:lit}	
	\resizebox{\linewidth}{!}
	{\renewcommand{\arraystretch}{1.3}
	\begin{tabular}{|c|c|c|c|c|l|l|}
			\hline
			Source                                   & $\alpha$     & $G$        & $Re$           & $n$               & Technique          & Remarks                            \\
			\hline
			\textbf{Present study}                  & {0 -- 2} &{0.2 -- 1} & {1 -- 40}  & {0.2--1.8} & {N (FEM)}        &    Counter-rotating cylinders             \\ \hline			\cite{meneghini2001numerical}                & 0            & 0.5--3         & 100--200       & 1                 & N (FEM)           &       \\
			\cite{kang2003characteristics}               & 0            & <5             & 40--160        & 1                 & N (IBM)                  &              \\
			\cite{xu2003reynolds}                        & 0            & 0.2--0.6       & 150--14300     & 1                 & E (LIF, PIV)     &                  \\
			\cite{kun2007wake}                           & 0            & 0.1--2         & 30--100        & 1                 & N (SEM) & \\
			\cite{mizushima2008stability}                & 0            & 0.3--1         & 20--80         & 1                 & N (VSFM)          &\\			
			\cite{sarvghad2011numerical}                 & 0            & 0.5--3         & 100, 300, $10^4$ & 1                 & N (FVM)                  &              \\
			\cite{vakil2011two}                          & 0            & 0.1--30        & 1--20          & 1                 & N (FVM)                               & \\
			\cite{bao2013flow}                           & 0            & 0.2--3         & 100            & 1                 & N (FEM)                            &\\
			\cite{supradeepan2014characterisation}       & 0            & 0.1--7         & 100            & 1                 & N (FVM)            &                    \\
			\cite{carini2015secondary}                   & 0            & 0.7, 1.8       & 50--70         & 1                 & N (DNS)                         &       \\
			\cite{thapa2015three}                        & 0            & 0.5--3.5       & 500            & 1                 & N (FEM)                             & \\
			\cite{singha2016numerical}                   & 0            & 0.2--4         & 20--160        & 1                 & N (FVM)                       &         \\
			\cite{qiu2017vortex}                    & 0            & 0.2--1.7       & 3900           & 1                 & N (LES, FVM)                   &   Unequal sized cylinders; 3-D turbulent flow    \\
			\cite{adeeb2018flow}                         & 0            & 0.5--4         & 100            & 1                 & N (LBM)                         &    \\
			\cite{sanyal2022impact}                      & 0            & 0.5--3         & 5--40          & 1                 & N (FVM)                         &       \\\hline
			\cite{Yoon2007,yoon2009flow,yoon2010laminar} & $\le 2$      & 0.2--3         & 100            & 1                 & N (IBM)         & Counter-rotating cylinders      \\
			\cite{guo2009flow}                           & 0--4         & 0.11           & 425--1130      & 1                 & N (LBM); E (PIV)    &     \\
			\cite{xiao2009vortex}                        & 2            & 0.7--1.5       & 160--200       & 1                 & N (LBM)               &               \\
			\cite{nemati2010numerical}                   & 0--2         & 0.2--3         & 100            & 1                 & N (LBM)              &                  \\
			\cite{chan2011vortex}                        & 0--5         & 1--5           & 100--200       & 1                 & N (SDM); E (PIV)      &      \\
			\cite{kumar2011flow}                         & 0--5         & 0.8--6.5       & 100--500       & 1                 & E (PIV)                 &              \\
			\cite{supradeepan2015low}                    & 0.5--1.25    & 0.1--3.5       & 100            & 1                 & N (FVM)                &                \\
			\cite{dou2018effect}                         & 0--3         & 1              & 40--200        & 1                 & N (FVM)        &                        \\
			\cite{darvishyadegari2018analysis}           & 0--4         & 0.5--2         & 200            & 1                 & N (FVM)      &                          \\
			\cite{hassanzadeh2019analysis}               & 0--4         & 0.5--2         & 200            & 1                 & N (FVM)  &                              \\
			\cite{chaitanya2023mixed}                    & $\le 2$      & 0.2--3         & 100            & 1                 & N (FVM)  &  Counter-rotating cylinders                                \\\hline
			\cite{chaitanya2012non}                      & 0            & 0.5--3         & 1--40          & 0.4--1.8          & N (FVM)                      &          \\
			\cite{daniel2013aiding}                      & 0            & 0.5            & 1--40          & 0.2--1.0            & N (FVM)                              &  \\
			\cite{Panda2017}                             & 0            & 0.2--3         & 0.1--100       & 0.2--1.8          & N (FVM)                             &   \\
			\cite{Zhu2024}                    						& 0      & 1.1--1.6         & 100            & 0.8--1.5                 & N (FVM)                    &            \\\hline
			\multicolumn{7}{|p{1.2\linewidth}|}{$\rightarrow$ N (Numerical): FDM  (finite difference method); FEM (finite element method); SEM (spectral element method); FVM (finite volume method); IBM  (immersed boundary method); LBM (lattice Boltzmann method); SDM (spectral difference method); VSFM (vorticity - stream function method);   $\rightarrow$ E (Experimental): LIF (laser-induced fluorescence); PIV (particle image velocimetry);  $\rightarrow$ $Re$ (Reynolds number); $n$ (power-law index); $G$ (gap ratio); $\alpha$ (rotational velocity)}\\\hline
	\end{tabular}
	}
\end{table}

\noindent 
Enormous knowledge framework comprehends various features of the flow over  cylinders subjected to  unconfined (and confined) Newtonian (and non-Newtonian) fluids,  as presented in excellent reviews, articles, and books \citep{Streeter1961,Zdravkovich1977, Zdravkovich1997,Zdravkovich2003,badr1989steady,Kang1999,Zapryanov1999,mittal2003flow,Chhabra2004,Clift2005,Michaelides2006,Johnson2016,Khan2006,Kang2006,Bharti2006,Bharti2007a,Bharti2007b,Senthil2008,Paramane2009,Patnana2009,Sen2009d,Rao2011,Sarkar2011,Thakur2016,suru2021non,Hishikar2022,kumar2022non,Malviya2022,Chhabra2023,Mohanty2024,Zhu2024}. For instance,  \citet{Zdravkovich1977} has categorized the wake regime for laminar flow around a single stationary cylinder as  steady unseparated wakes ($Re\leq 5$), steady separated near-wake ($6\leq Re \leq30$), steady near-wake oscillation ($30\leq Re \leq 60$), and unsteady periodic wakes ($Re\geq 60$). 
Various studies \citep{chhabra2001a,Chhabra2004,Bharti2006,Bharti2007a, Sivakumar2006a, Sivakumar2007, Patil2008a, Patil2008b, Patnana2009, Patnana2010, Nirmalkar2014,Chhabra2023} have extensively investigated unconfined non-Newtonian power-law fluid flow around a stationary cylinder for wide range of the Reynolds number ($Re=1-100$) and power-law index ($n=0.2 - 1.8$). Furthermore, the flow regimes, in particular, the critical Reynolds numbers ($Re_{c}$ and $Re^{c}$) indicating the range of Reynolds number ($Re_{c}\le Re \le Re^{c}$) for symmetrical wake formation around the cylinder, have been characterized numerically for power-law fluid ($n=0.2 - 1.8$) flow across an unconfined \citep{Sivakumar2006a} and a channel-confined \citep{Vishal2021} cylinder. Few studies \citep{Patnana2009,Sabarinath2018,Bailoor2019} have also numerically examined the vortex shedding characteristics of power-law flow over a cylinder. Recently, \citet{Mohanty2024} have explored the roles of the solvent viscosity contribution and Deborah number ($De$) on the creeping flow of  Oldroyd-B viscoelastic fluid over a channel confined circular cylinder.
On the other hand, a few studies have numerically revealed negative drag and lift forces in the non-Newtonian flow characteristics of a rotating single circular cylinder ($0.2 \leq n \leq 1$; $0 \leq \alpha \leq 6$; $0.1 \leq Re \leq 40$) \citep{Panda2010,Panda2011}  and elliptical cylinder ($0.4 \leq n \leq 1.8$; $1 \leq \alpha \leq 4$; $5 \leq Re \leq 40$; $0.1 \leq e \leq 0.7$) \citep{kumar2022non}.
Recently, \citet{Kumar_2018,Kumar_2021} have used experimental (2D-PIV) and numerical (FEM) techniques to explore the friction stir welding (FSW) process wherein the welding tool of cylindrical shape rotates and translates in the direction of welding of the material sheets behaving as non-Newtonian fluids.

\noindent
Furthermore, researchers have extensively explored the hydrodynamics of multiple cylinder configurations (such as tandem, staggered, or side-by-side) submerged in steady cross-flow and revealed complex flow phenomena influenced by the positioning of the cylinders, flow governing parameters and the nature of the fluid \citep{Chhabra2023}. In addition to the flow governing parameters for a single cylinder, the gap spacing ($G=h/D$) between the surface of two cylinders is pivotal and causes significant changes in flow characteristics.
\tab\ref{tab:lit} lists the relevant prior studies on the flow across the pair of the side-by-side arrangements of stationary ($\alpha = 0$) and rotating ($\alpha \neq 0$) cylinders. Further, several studies \citep{Patil2008a, Patil2008b, dwivedi2019flow, darvishyadegari2018convective, darvishyadegari2019heat, rastan2021flow,durga2014cfd, Sanyal2017, adeeb2018flow, patel2018mixed}, not listed in \tab \ref{tab:lit}, have examined the cross-flow across a pair of tandem cylinders spanning over the wide ranges of flow governing parameters ($Re$, $n$, $G$, $\alpha$). These studies (\tab \ref{tab:lit}) have revealed complex flow features as a function of the parameters governing the flow around side-by-side circular cylinders. However, most of the existing studies have dealt with Newtonian ($n=1$) fluid flow around a pair of side-by-side circular cylinders over the wide range of gap ratio ($G$), Reynolds number ($Re$), and rotational rate ($\alpha$). While few studies  \citep{chaitanya2012non,daniel2013aiding,Panda2017,Zhu2024} deal with non-Newtonian ($n\neq 1$) fluid, they are limited to the stationary ($\alpha = 0$) cylinders; conversely, none of the literature has explored the combined influence of non-Newtonian fluids and rotational rate ($n\neq 1$, $\alpha \neq 0$) for the flow around a pair of the side-by-side circular cylinders. 
\noindent
Therefore, this study aims to explore the hydrodynamic characteristics of non-Newtonian power-law fluid flowing across a pair of counter-rotating circular cylinders arranged in an unconfined side-by-side configuration.  The two-dimensional flow governing equations have been solved numerically using the finite element method spanning over the wide range of dimensionless conditions (Reynolds number, $1\leq Re \leq 40$; power-law index, $0.2\leq n \leq 1.8$; gap ratio, $0.2\leq G\leq1$; cylinder rotational rate, $0\leq\alpha\leq2$) to obtain the flow field and subsequent analysis of local and global flow characteristics, including streamline profiles, centerline velocity, pressure coefficients on the cylinder surface, and individual and total force coefficients.
\section{Problem Description}
\begin{figure}[!b]
	\centering
		\includegraphics[width=1\linewidth]{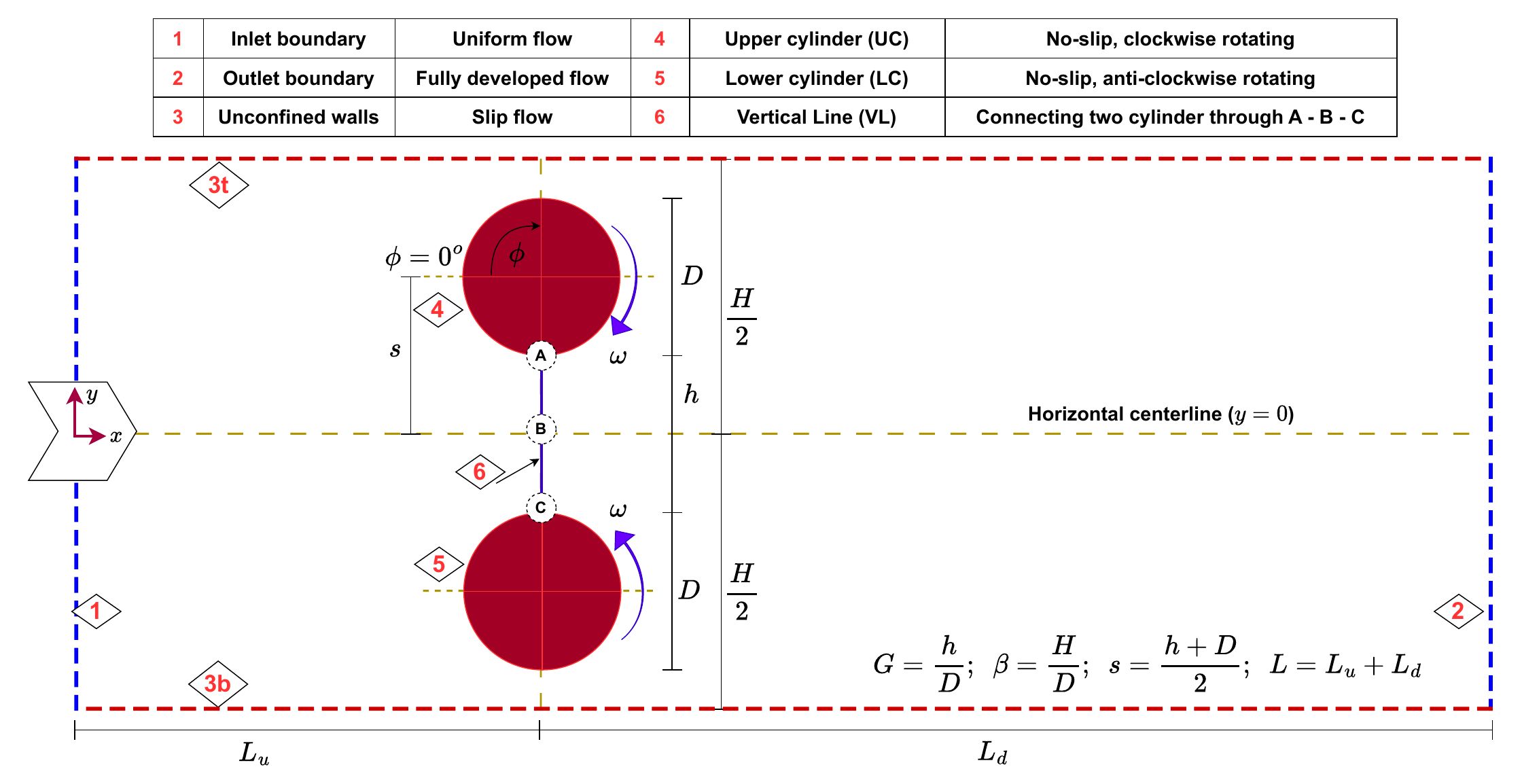} 
	\caption{Schematic representation of an unconfined flow across a pair of side-by-side counter-rotating circular cylinders.}
	\label{fig:1}
\end{figure}

\noindent
Consider the two-dimensional steady incompressible flow across an unconfined pair of counter-rotating side-by-side circular cylinders, as sketched in \fig\ref{fig:1}. The cylinders of equal diameter ($D$) are arranged side by side and positioned vertically symmetric ($y=\pm s$) about the horizontal mid-plane ($y=0$) at a streamwise distance $L_u$ (upstream length) from the inlet and $L_d$ (downstream length) from the outlet, both measured to the center of cylinders. The spacing between the cylinders is $h$ (surface-to-surface) and $2s$ (center-to-center), where $s=(D+h)/2$. The centers ($x_c, y_c$) of the upper cylinder (UC)  and lower cylinder (LC) are located at ($L_u$, $s$) and ($L_u$, $-s$), respectively. 
In addition, the position of any point on the surface of the cylinders is specified by the angle $\phi$, measured clockwise from the mid-plane ($y=\pm s$) through the center the cylinders. Accordingly, the front stagnation point (FSP) and the rear stagnation point (RSP) correspond to $\phi=0$ and $\phi=\pi$, respectively, for both cylinders.
Furthermore, the lateral boundaries are placed symmetrically at $y=\pm H/2$ about the mid-plane. The physical domain therefore spans an axial streamwise length $L \ (=L_u+L_d)$ and a lateral height $H$, respectively. The gap ratio between cylinders is defined as $G = h/D$, and the wall blockage ratio as $\beta = D/H$. To represent the unconfined flow, the top and bottom boundaries are treated as imaginary ($\beta \to \infty$), and the outlet is placed sufficiently far downstream of the cylinders, allowing the flow to be fully developed before exiting.
All length variables are measured in meters.

\noindent
The streaming liquid, with uniform velocity ($U_0$, m/s), approaches the counter-rotating cylinders, where the upper cylinder (UC) rotates clockwise and the lower cylinder (LC) counter-clockwise about their centers at a fixed angular velocity ($\omega$, rad/s). 
The physical properties of the fluid, such as density ($\rho$, kg/m$^3$) and viscosity ($\eta$, Pa$\cdot$s), are assumed independent of temperature and pressure. Furthermore, the viscous behavior of the incompressible fluid is represented using the non-Newtonian power-law model \citep{Bird2002,Chhabra2008} as follows.
\begin{gather}
	\eta=m (I_2/2) ^{[(n-1)/2]} \label{eq7a} 	\\	
	I_2=2\left(\varepsilon^2_{xx} +\varepsilon^2_{yy}+\varepsilon^2_{xy}+\varepsilon^2_{yx}\right) ;	
	\qquad	\varepsilon(U)= (1/2)\left[ (\nabla U)+(\nabla U)^T \right] 	\label{eq7b}\\
	n  < 1: \text{shear-thinning}; \quad n = 1: \text{Newtonian}; \quad  n > 1: \text{shear-thickening}\nonumber
\end{gather}
where $m$ is the flow consistency index (Pa$\cdot$s$^n$), $n$ is the flow behavior index (dimensionless), $\varepsilon$ is the rate-of-strain tensor (s$^{-1}$), and $I_{2}$ is the second invariant of the strain-rate tensor (s$^{-2}$).

\noindent
The physical problem described above is governed by the steady incompressible Navier-Stokes (N-S) equations \citep{Patil2008a,Patil2008b,Panda2017}. In this study, the field variables are non-dimensionalized using the characteristic scales $U_0$, $D$, $\rho U_0^2$, $m(U_0/D)^{n-1}$, and $m(U_0/D)^n$ for velocity, length, pressure, viscosity, and shear stress, respectively. Consequently, the dimensionless parameters derived from the scaling of the field variables and equations are defined as follows.
\begin{align}
	\alpha=\frac{\omega (D/2)}{U_0};	 \qquad Re=\frac{D^{n}U_{0}^{2-n}\rho }{m};  \qquad Re_{\omega}=\alpha^{(2-n)}Re;  
	\label{eq12} 
\end{align}
Here, $\alpha$, $Re$, and $Re_{\omega}$ denote the dimensionless rotational (angular) velocity of cylinders, the Reynolds number for power-law fluids, and the rotational Reynolds number of the cylinders, respectively. Henceforth, all equations and variables are expressed in dimensionless form unless stated otherwise.

\noindent
Subsequently, the governing equations and boundary conditions in dimensionless form are presented as follows.
\begin{gather}
	\nabla \cdot \boldsymbol{U} = 0, \label{eq1} \\
	(\boldsymbol{U} \cdot \nabla) \boldsymbol{U} = - \nabla p +  {Re^{-1}} \, (\nabla \cdot \boldsymbol{\tau}), \label{eq2}
\end{gather}
where, $\boldsymbol{U}$, $p$, and $\boldsymbol{\tau}$ denote the velocity vector (with components $U_x$ and $U_y$ in Cartesian coordinates), the isotropic pressure, and the deviatoric stress tensor, respectively. The stress tensor ($\boldsymbol{\tau}$) for an incompressible fluid is related to the rate-of-strain tensor ($\boldsymbol{\varepsilon}$) as follows \citep{Bird2002,Chhabra2008}.
\begin{gather}
	\boldsymbol{\tau} = 2\eta\boldsymbol{\varepsilon} \qquad\text{where}\qquad  	
	\eta=(I_2/2) ^{(n-1)/2} 
	\label{eq5}
\end{gather}
Here, $\eta$ denotes the viscosity of the power-law fluid (dimensionless form of \eqn\ref{eq7a}), and $\boldsymbol{\varepsilon}$ is the rate-of-strain tensor (dimensionless form of \eqn\ref{eq7b}).

\noindent
The physically relevant boundary conditions (BCs) for the considered flow configuration (see \fig\ref{fig:1}) are specified as follows.
\begin{itemize}
	\item BC 1 (inlet boundary): A uniform velocity field prescribed in the streamwise direction as follows.
	\begin{gather}
		 \boldsymbol{U} = (U_x,\ U_y) = (1,\, 0)
		\label{eq8}
	\end{gather}
	\item 
	BC 2 (outlet boundary): A homogeneous Neumann condition, representing fully developed flow at the outlet, is prescribed as follows.
	\begin{gather}
		\frac{\partial U_{x}}{\partial x}=0; \qquad \frac{\partial U_{y}}{\partial x}=0 ; \qquad\text{and}\qquad p=0	\label{eq9}
	\end{gather}
	\item 
	BC 3 (top and bottom boundaries): To represent the unconfined flow, the top and bottom boundaries are treated as imaginary with no dissipation and are subject to a slip condition, as follows.
	\begin{gather}
		\frac{\partial U_{x}}{\partial y}=0; \qquad\text{and} \qquad U_{y}=0	\label{eq10}
	\end{gather}
	\item 
	BCs 4 and 5 (surface of the cylinders): A no-slip condition is imposed on the surfaces of the solid, impermeable cylinders, which rotate with a fixed rotational rate ($\alpha$) about their axes in the clockwise (upper cylinder) and counter-clockwise (lower cylinder) directions, as follows:
\begin{gather}
	\boldsymbol{U} = (U_x,\ U_y) =
	\begin{cases}
		(-\alpha \sin\phi,\ +\alpha \cos\phi), & \text{for UC}, \\[-4pt]
		(+\alpha \sin\phi,\ -\alpha \cos\phi), & \text{for LC} 
	\end{cases}
	\label{eq11}
\end{gather}
	%
\end{itemize}
The governing equations (\eqns\ref{eq1}--\ref{eq2}), together with the boundary conditions (\eqns\ref{eq8}--\ref{eq11}), describe the flow field in terms of the primitive variables, namely the velocity ($\boldsymbol{U}$) and pressure ($p$). Subsequently, the hydrodynamic characteristics such as streamlines, pressure coefficient, and drag and lift coefficients are evaluated following standard definitions \citep{Patil2008b,Panda2017}. For completeness, the relevant fluid dynamic quantities are defined below.
	\newline 
	The pressure coefficient ($C_p$) over the surface of the cylinders is obtained as follows:
	\begin{gather}
		C_p =\frac{(p-{p_0})}{p_d}; \qquad p_d = \frac{1}{2}(\rho {U_{0}}^2)	\label{eq13} 
	\end{gather}
	where, $p$, $p_0$, and $p_d$ denote the local pressure at an angular position $\phi$ on the cylinder surface and the reference pressure in the far-field region (i.e., at the outlet boundary, BC 2), dynamic pressure of the flow, respectively.
	\newline
	The force coefficients ($C_i $), denoting the dimensionless hydrodynamic forces ($F_i$), are defined as follows. 
	\begin{gather}
	C_D = \frac{F_D}{p_d D} = C_{DP} + C_{DF}; \qquad 
	C_L = \frac{F_L}{p_d D} = C_{LP} + C_{LF}
	\label{eq14}	
	\end{gather}
	where $C_D$ and $C_L$ denote the total drag and lift coefficients, respectively, corresponding to the total drag ($F_D = F_{DP} + F_{DF}$, N/m) and lift ($F_L = F_{LP} + F_{LF}$, N/m) forces per unit length of the cylinder. 
	The pressure contributions to the drag and lift coefficients ($C_{DP}$ and $C_{LP}$) are expressed as follows.
	\begin{gather}
		C_{DP}=\frac{F_{DP}}{p_d D}=\int_{S}C_p n_x \text{d}S; \label{eq15}  \qquad\text{and}\qquad  
		C_{LP}=\frac{F_{LP}}{p_d D}=\int_{S}C_p n_ y \text{d}S 
	\end{gather}
	where $F_{DP}$ and $F_{LP}$ denote the pressure components of the drag and lift forces, respectively, and $S$ represents the surface of the cylinders. 
	Similarly, the viscous components of the drag and lift coefficients ($C_{DF}$ and $C_{LF}$) are expressed as follows.
	\begin{gather}
		C_{DF} = \frac{F_{DF}}{p_d D} 
		= \frac{2^{\,n+1}}{Re} \int_{S} [(\boldsymbol{\tau} \cdot \mathbf{n}_s) \cdot \hat{\mathbf{i}}] \, \text{d}S 
		= \frac{2^{\,n+1}}{Re} \int_{S} (\tau_{xx} n_x + \tau_{xy} n_y) \, \text{d}S, \label{eq16} \\[6pt]
		C_{LF} = \frac{F_{LF}}{p_d D} 
		= \frac{2^{\,n+1}}{Re} \int_{S} [(\boldsymbol{\tau} \cdot \mathbf{n}_s) \cdot \hat{\mathbf{j}}] \, \text{d}S 
		= \frac{2^{\,n+1}}{Re} \int_{S} (\tau_{yx} n_x + \tau_{yy} n_y) \, \text{d}S, \label{eq16L}
	\end{gather}
	where, $F_{DF}$ and $F_{LF}$ denote the frictional components of the drag and lift forces, respectively. The outward unit normal vector to the cylinder surface ($\mathbf{n}_s$) is defined as follows.
	\begin{align}
		\mathbf{n}_s=\frac{x\mathbf{e}_x+y\mathbf{e}_y}{\sqrt{x^2+y^2}}={n}_x \mathbf{e}_x+{n}_y \mathbf{e}_y	\label{eq17} 
	\end{align}
	where $\mathbf{e}_x$,  and $\mathbf{e}_y$ denote the unit vectors along the $x-$ and $y-$directions, respectively.
\newline
The numerical solution of the flow governing equations (\eqns\ref{eq1}--\ref{eq11}) is obtained to analyze the dynamics of local and global flow characteristics as functions of the dimensionless parameters ($n$, $Re$, $\alpha$, $G$). The following section presents the numerical methodology and the selection of parameters used to obtain the final results. It is reiterated that, hereafter, all quantities are expressed in dimensionless form unless otherwise stated.
\section{Numerical approach and parameters}
\noindent
In this study, the flow governing equations (\eqns\ref{eq1} -- \ref{eq11}) are solved numerically using the finite element method (FEM) based the commercial CFD solver COMSOL Multiphysics. The two-dimensional (2D) single-phase (spf) laminar flow module is employed to model the physical system, and the fluid viscosity is defined using a shear-dependent, inelastic, non-Newtonian power-law model.
\newline
The computational domain (\fig\ref{fig:1}) is discretized using a hybrid mesh of uniform triangular elements near the cylinder surfaces and non-uniform quadrilateral elements elsewhere. The mesh density gradually decreases away from the cylinders, and the grid is generated using integrated CAD geometry and mesh generation tools in COMSOL.
%
The mesh near the cylinder surfaces has a uniform spacing of $\delta_r$, with $N_c$ points on the surface of each cylinder. The mesh is gradually stretched from $\delta_r$ near the cylinders to $\delta$ and finally to $\Delta$ in the faraway from cylinders.
\newline
The discretized flow governing equations on the non-uniform grid result in a coupled set of simultaneous  algebraic equations, which are solved using parallel direct linear solver (PARDISO) based on LU matrix factorization, combined with an automatic highly non-linear (Newton) method and a relative convergence criterion of $10^{-6}$. The steady-state solutions yield the flow field ($\mathbf{U}, p$) as functions of the governing parameters ($n$, $\alpha$, $Re$, $G$). All reported local and global flow quantities are rounded to four significant digits. A parametric sweep was performed over $n$ and $\alpha$ sequentially to reduce computational effort.
%
%

\noindent
Furthermore, the existing literature \citep{Chhabra2004, Bharti2006, Bharti2007a, Bharti2007b, Senthil2008, Paramane2009, Patnana2009, Sen2009d, Rao2011, Sarkar2011, Thakur2016, kumar2022non, Mohanty2024, Zhu2024, Patil2008a, Patil2008b, Panda2017, Bharti2022} has recognized that the sizes of the computational domain and mesh are critical numerical parameters for obtaining accurate and reliable results. In the case of unconfined flow, the computational domain is characterized by the upstream length ($L_u$), downstream length ($L_d$), and height ($H$), as illustrated in \fig\ref{fig:1}. In the present study, detailed investigations on domain and mesh independence have been carried out and are presented in \tabs\ref{Tab:dom_Lu-Ld}–\ref{tab:grid1} in \ref{appendix:domain-grid}, to determine suitable values for the computational domain ($L_u$, $L_d$, $H$) and mesh resolution. Based on this comprehensive analysis and our previous experience, a computational domain of $L_u = 120$, $L_d = 300$, and $H = 300$, along with a non-uniform unstructured mesh ($G_2$ consisting of $N_c=135$, $\delta_r-\delta-\Delta=0.005-0.1-0.5$; refer to \tab\ref{tab:grid} for details), has been adopted to obtain the final results discussed in the subsequent sections.
\section{Results and Discussion}
\noindent
In this study, numerical simulations have been performed to examine the influence of the flow parameters, such as Reynolds number ($Re$), power-law index ($n$), rotational rate ($\alpha$), and gap ratio ($G$), on the unconfined flow characteristics of a pair of counter-rotating circular cylinders submerged in a power-law fluid medium. The simulations cover a broad range of conditions: $Re = 1$, 5, 10, 20, 40; $0.2\le n \le 1.8$ (steps of 0.2, covering shear-thinning, Newtonian, and shear-thickening fluid behaviors); $0.2\le G \le 1$ (steps of 0.2); and $0\le \alpha \le 2$ (steps of 0.5). Prior to presenting the new findings, the accuracy and reliability of the present modeling and simulation methodology are verified by comparing the results with available data \citep{kang2003characteristics, Panda2017} for limited benchmark cases.
%
%
\begin{table}[!tb]
	\centering
	\caption{Comparison of present and literature values of force coefficients ($C_D$, $C_L$) for a fixed condition ($G=1$, $Re=40$, $n=1$). The $\delta$ denotes the relative deviation between the present and literature results.} 	\label{tab:val}
	\footnotesize
	\renewcommand{\arraystretch}{1.3}
	\begin{tabular}{|c| c| c| c| c|}
		\hline
		Cylinder & {Parameters}	&	Present work	&	\citet{Panda2017} & \citet{kang2003characteristics}	\\
		\hline
		{UC}	& $C_D$	&	1.6836	& 1.7064	&	1.7000 \\
			& ($\delta$ \%)	&		--	& 	(1.35) 	&	(0.97) \\ \cline{2-5}
		& $C_L$	&	0.3591	& 0.3665	&	0.3651 \\
			& ($\delta$ \%)	&		--	& 	(2.06) 	&	(1.67) \\		
		\hline
		{LC}	& $C_D$	&	1.6835	& 1.7066	&	1.7000 \\
			& ($\delta$ \%)	&		--	& 	(1.37) 	&	(0.98)\\ \cline{2-5}		
		& $C_L$	&	$-0.3594$	&	$-0.3658$	&	$-0.3706$ \\	
			& ($\delta$ \%)	&		--	& 	(1.78) 	&	(3.12) \\		
		\hline
	\end{tabular}
\end{table} 
\newline
\tab\ref{tab:val} presents a comparison of the present and literature values of the force coefficients ($C_D$, $C_L$) and their relative deviations ($\delta\ \%$) for a cylinder under the limiting condition ($G=1$, $Re=40$, $n=1$). An analysis of \tab\ref{tab:val}  shows good agreement between the present and literature results, with maximum deviations of 1.35\% for drag ($C_D$) and 3.12\% for lift ($C_L$) coefficients. While the drag ($C_D$) deviations are nearly identical for UC and LC, the lift ($C_L$) deviation is slightly larger for LC.
Similar deviations have been reported in the literature \citep{Chhabra2004, Bharti2006,Bharti2007a,Bharti2007b, Senthil2008,Paramane2009,Patnana2009,Sen2009d,Rao2011,Sarkar2011,Thakur2016,kumar2022non,Mohanty2024,Zhu2024,Patil2008a,Patil2008b,Panda2017,Bharti2022}, attributed to differences in numerical methods (FEM, FVM, FDM, LBM), solvers, and mesh refinement levels. Therefore, the present results are considered reliable and accurate within an acceptable range of $\pm 1 - 2\%$ for design and engineering applications.
%
%
\begin{figure}[htbp]
	\centering
	\subfigure[$G=0.4$, $Re=1$]{\includegraphics[width=0.4\linewidth]{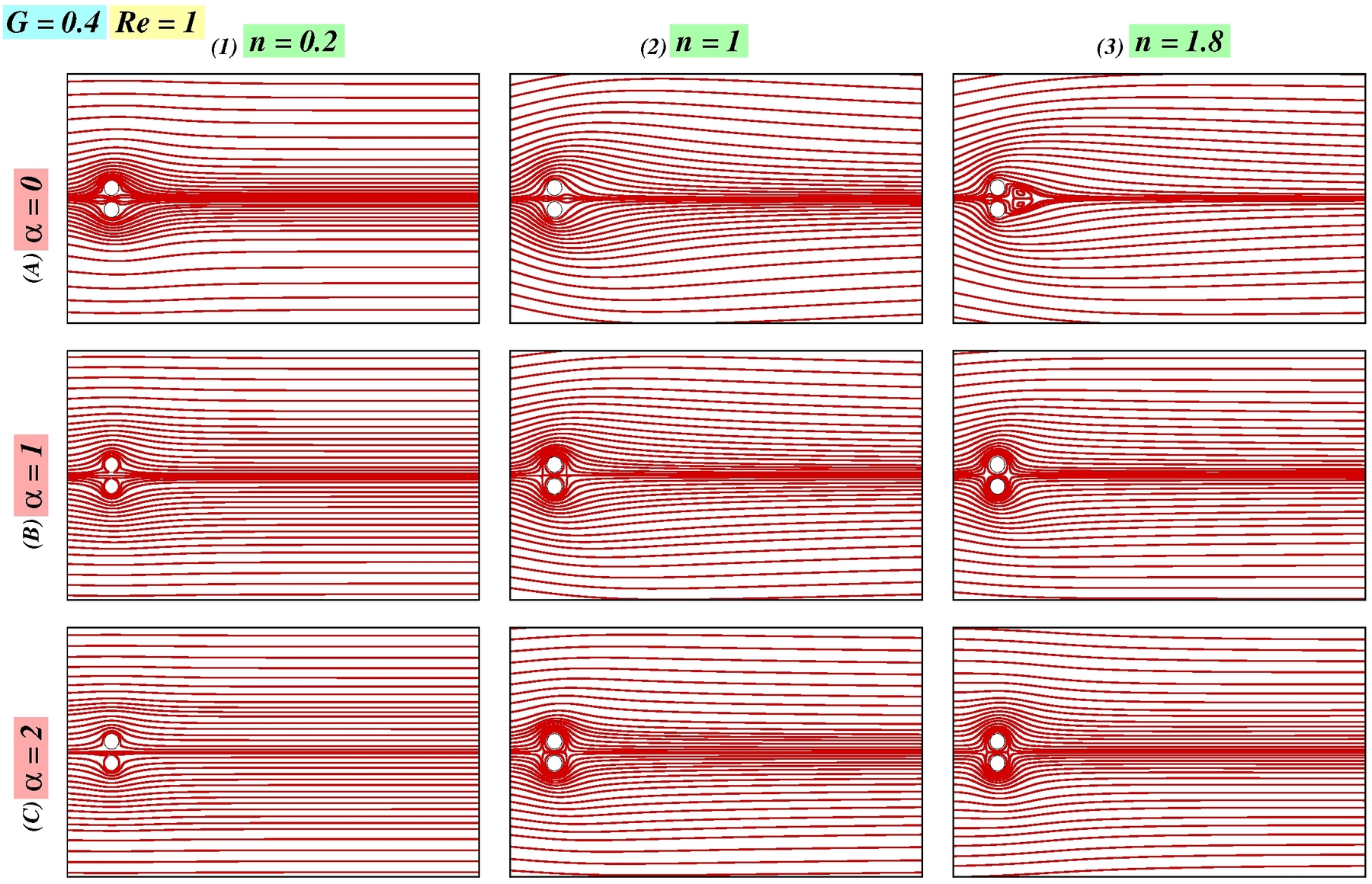}\label{fig:g-0.4}}
	\subfigure[$G=0.4$, $Re=40$]{\includegraphics[width=0.4\linewidth]{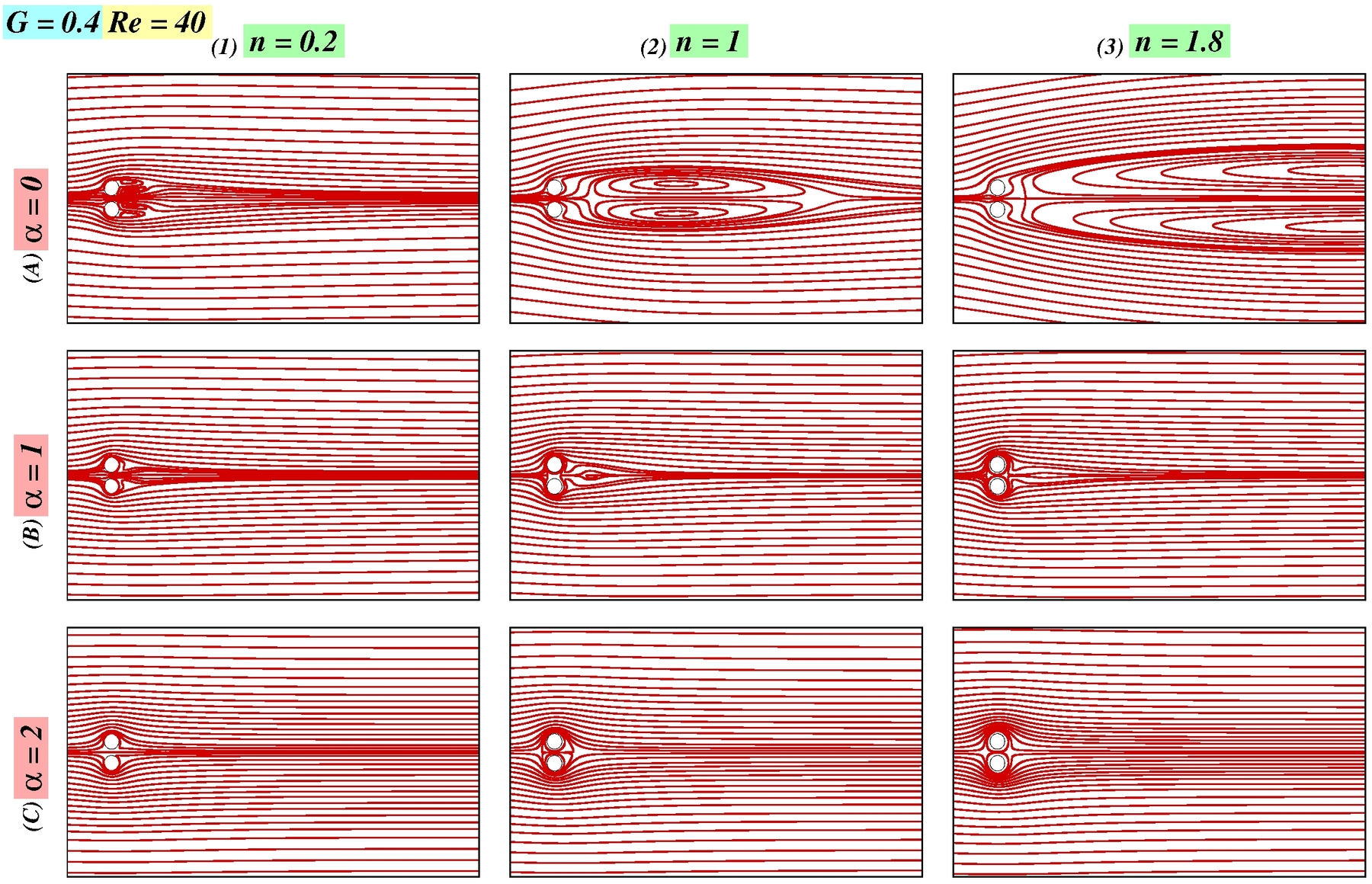}\label{fig:R40g-0.4}}
	\subfigure[$G=0.6$, $Re=1$]{\includegraphics[width=0.4\linewidth]{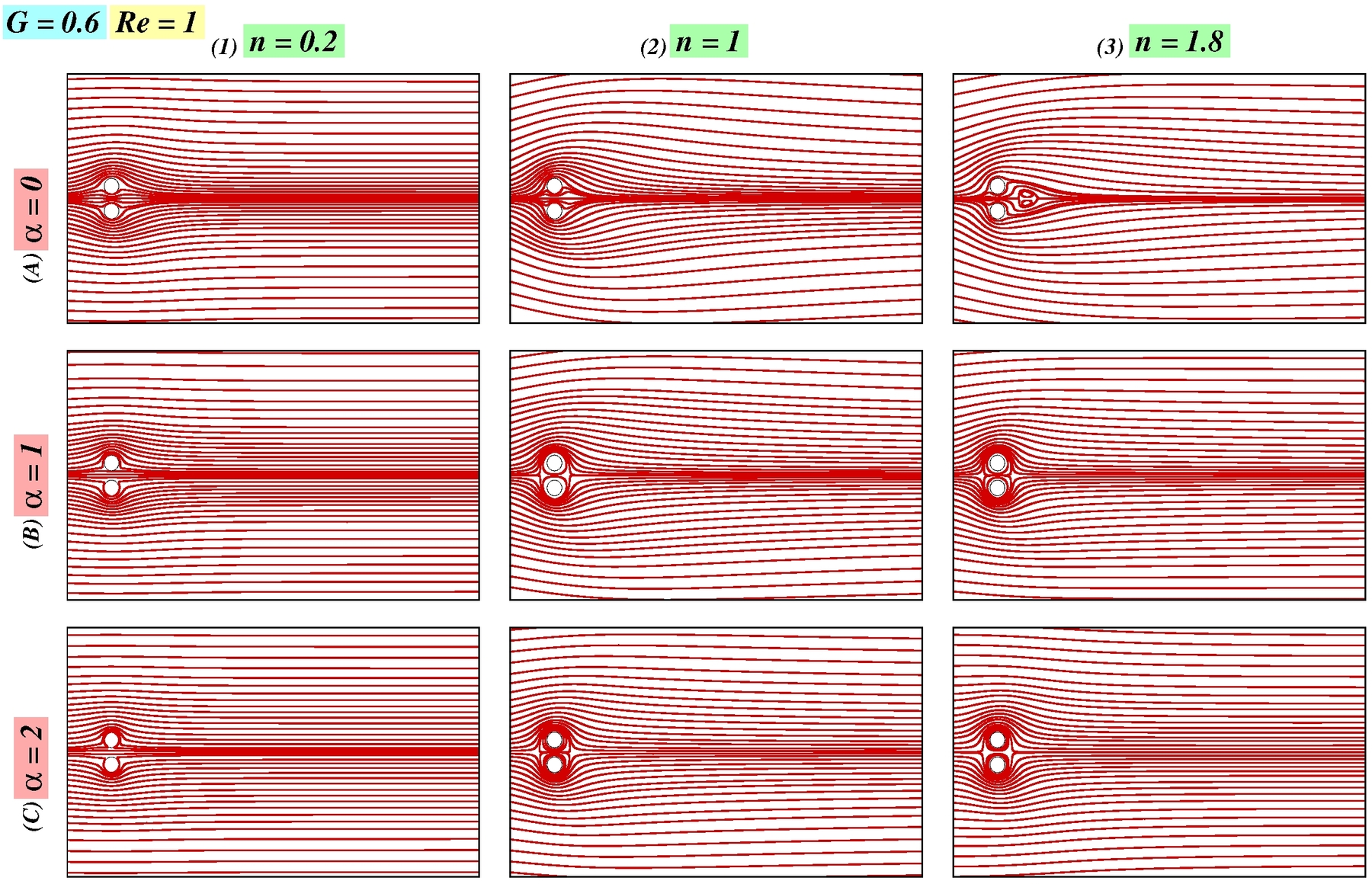}\label{fig:g-0.6}}
	\subfigure[$G=0.6$, $Re=40$]{\includegraphics[width=0.4\linewidth]{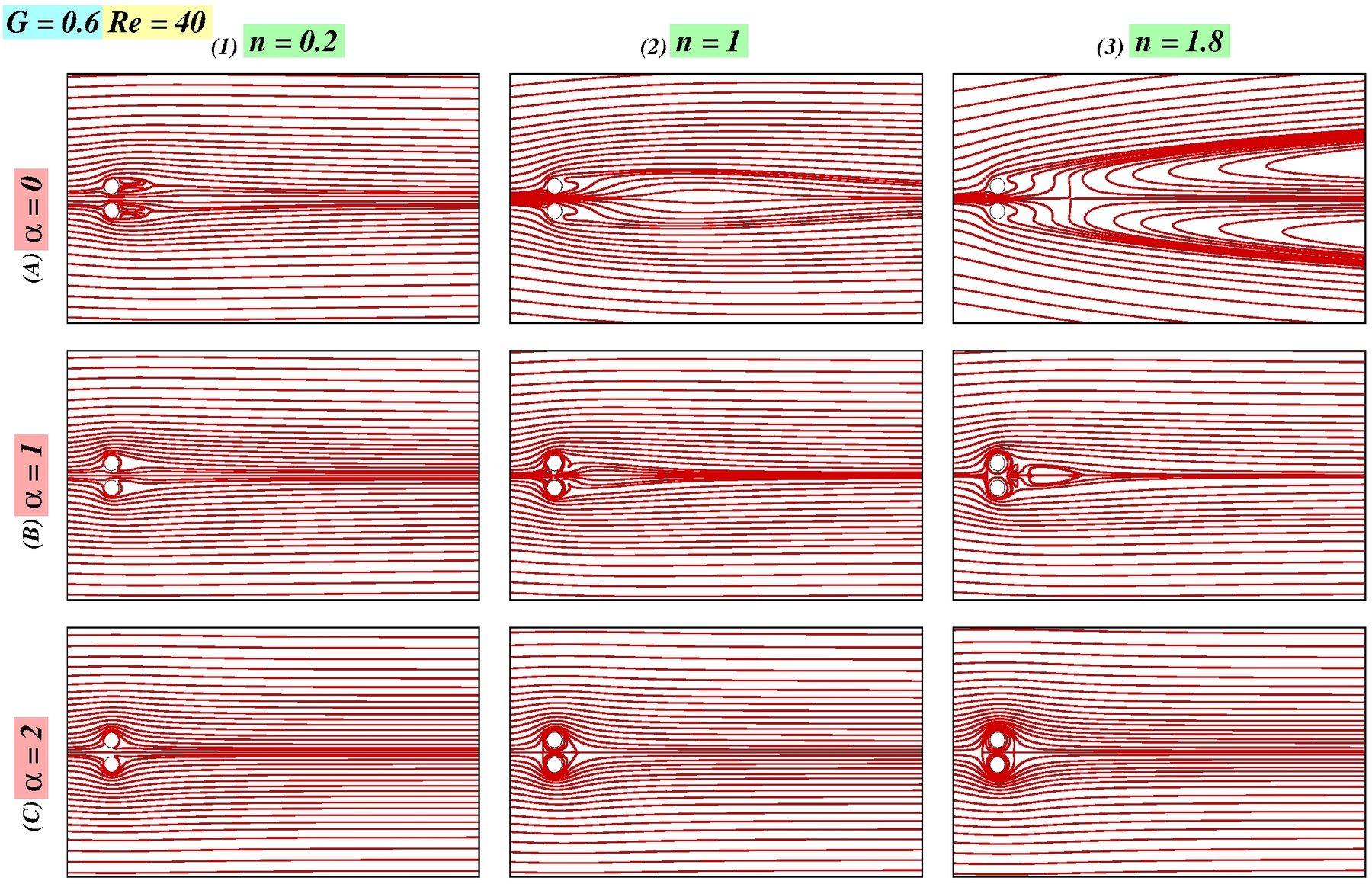}\label{fig:R40g-0.6}}
	\subfigure[$G=0.8$, $Re=1$]{\includegraphics[width=0.4\linewidth]{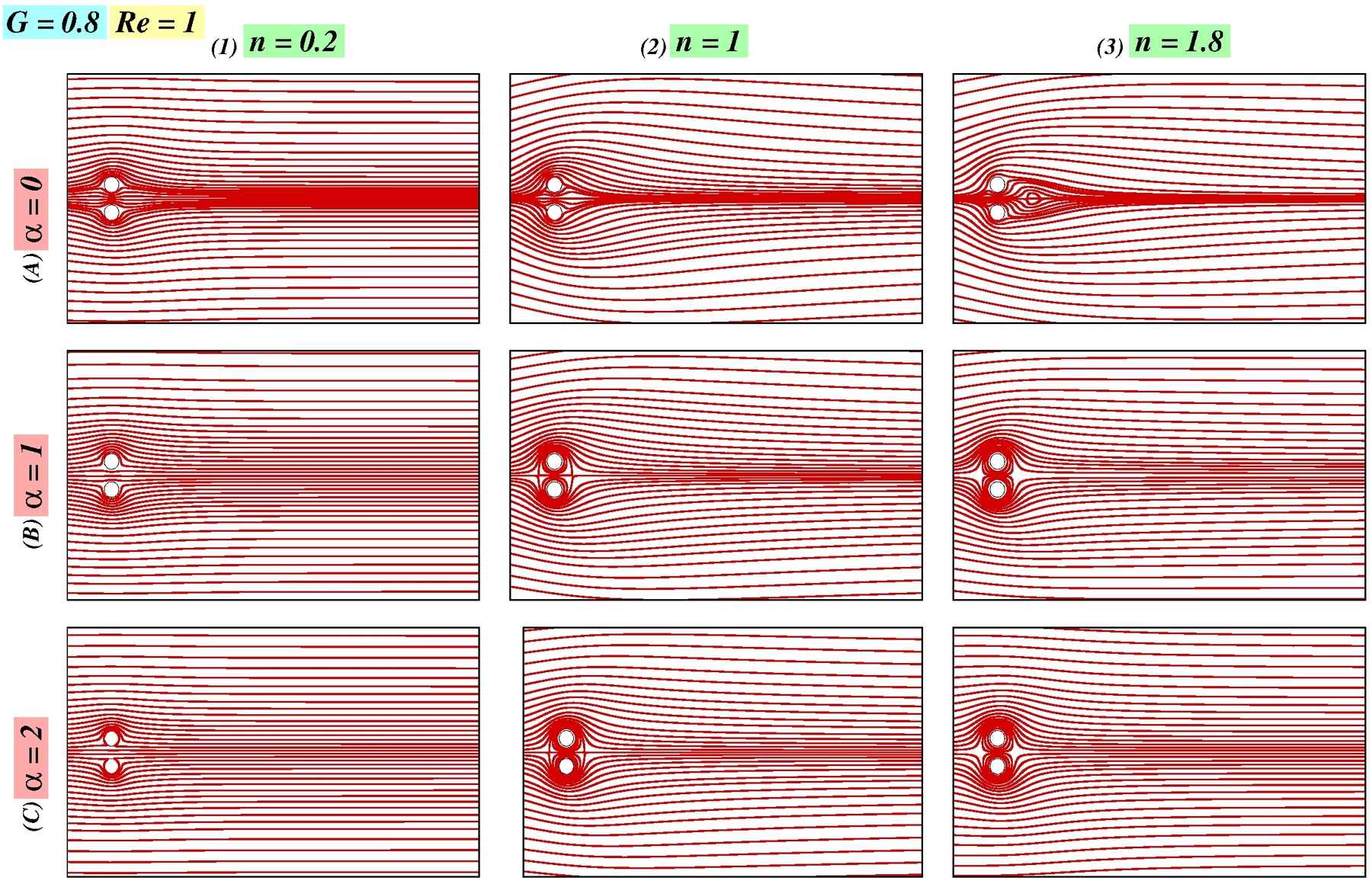}\label{fig:g-0.8}}
	\subfigure[$G=0.8$, $Re=40$]{\includegraphics[width=0.4\linewidth]{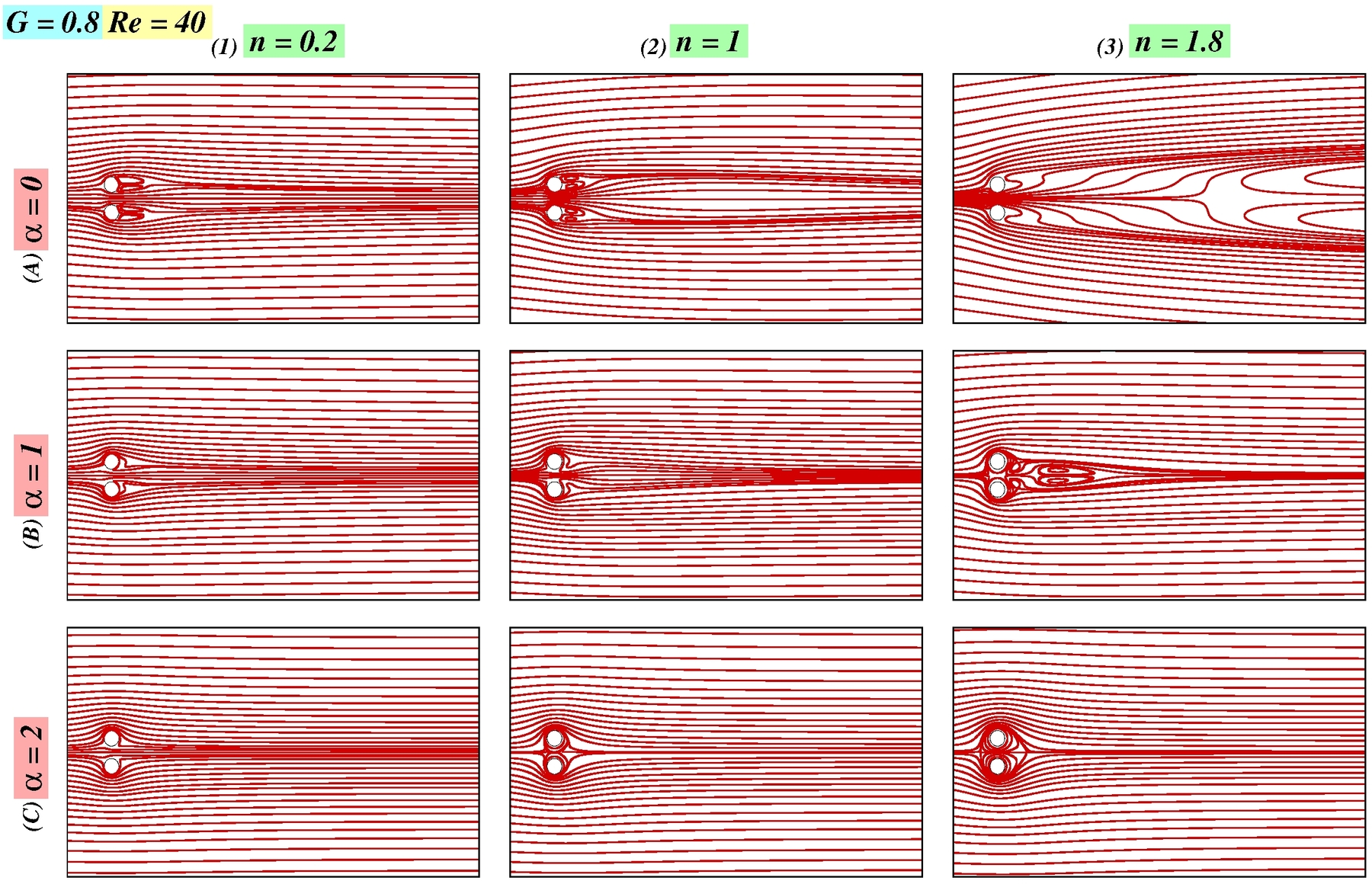}\label{fig:R40g-0.8}}
	\subfigure[$G=1$, $Re=1$]{\includegraphics[width=0.4\linewidth]{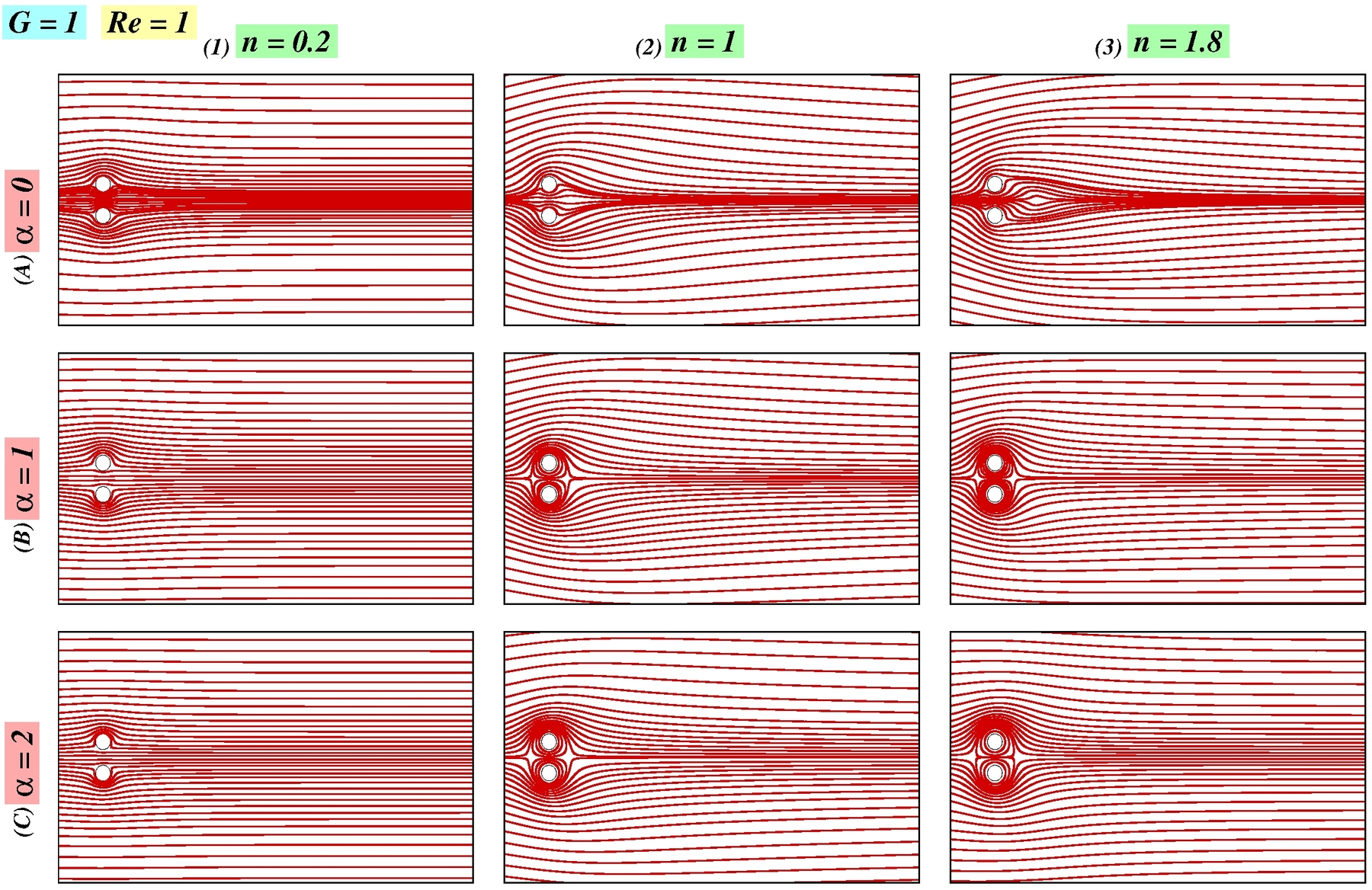}\label{fig:g-1}}
	\subfigure[$G=1$, $Re=40$]{\includegraphics[width=0.4\linewidth]{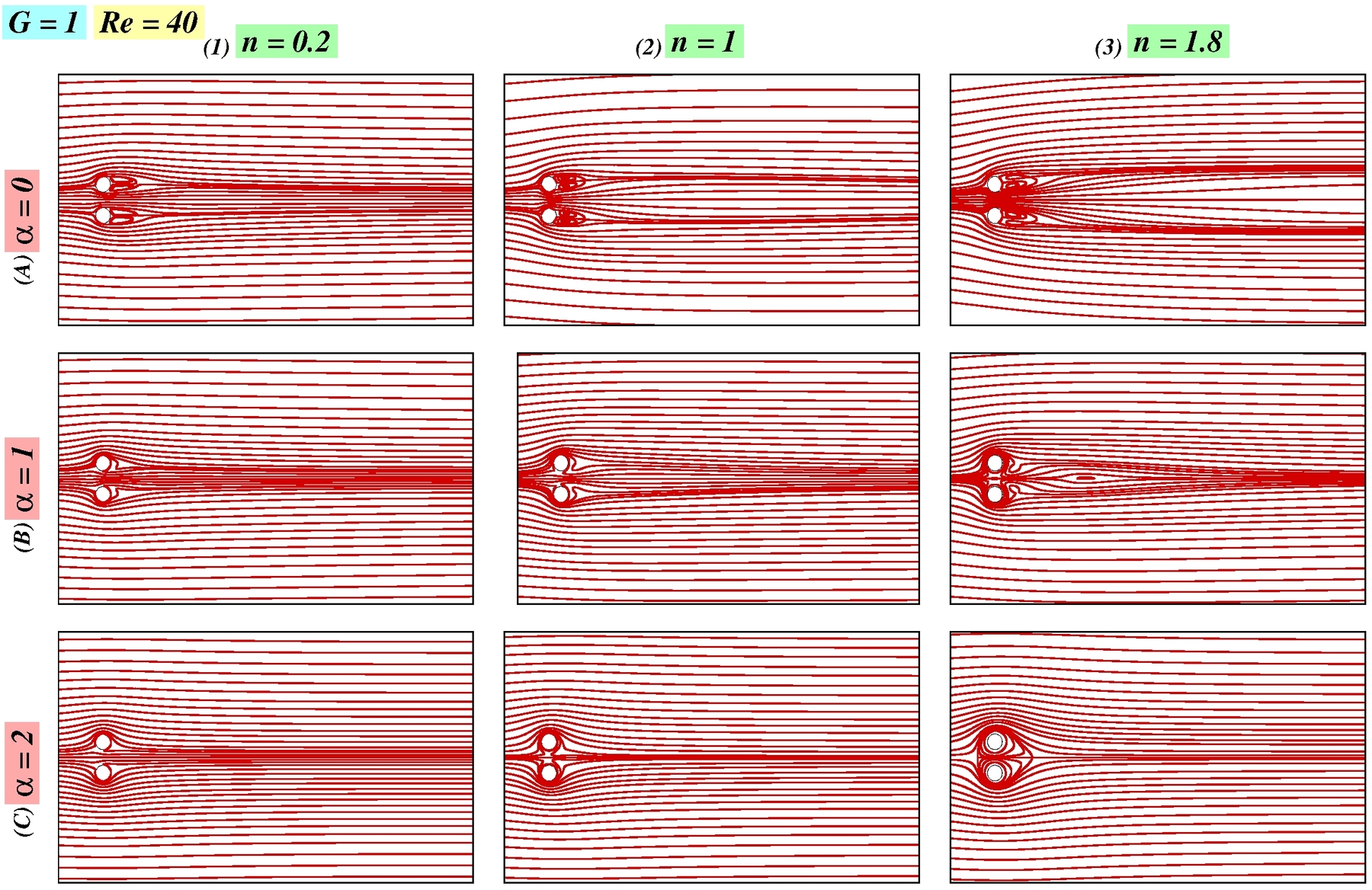}\label{fig:R40g-1}}
	\caption{Influence of gap ratio ($0.4\le G\le 1$), power-law index ($0.2\le n\le 1.8$) and rotational rate ($0\le \alpha\le 2$) on the streamline profiles for the extreme values of Reynolds number ($Re=1$, 40). The full-size images are included in \ref{appendix:streamline}.}
	\label{fig:R-1-40} \label{fig:R-1}\label{fig:R40}
\end{figure}
%
\subsection{Detailed flow kinematics}
\noindent
The detailed insights into the flow kinematics are obtained by examining the streamline patterns, velocity fields, and pressure distributions in the following sections.
\begin{figure}[htbp]
	\centering
	\subfigure[$G=0.2$, $Re=1$] {\includegraphics[width=0.4\linewidth]{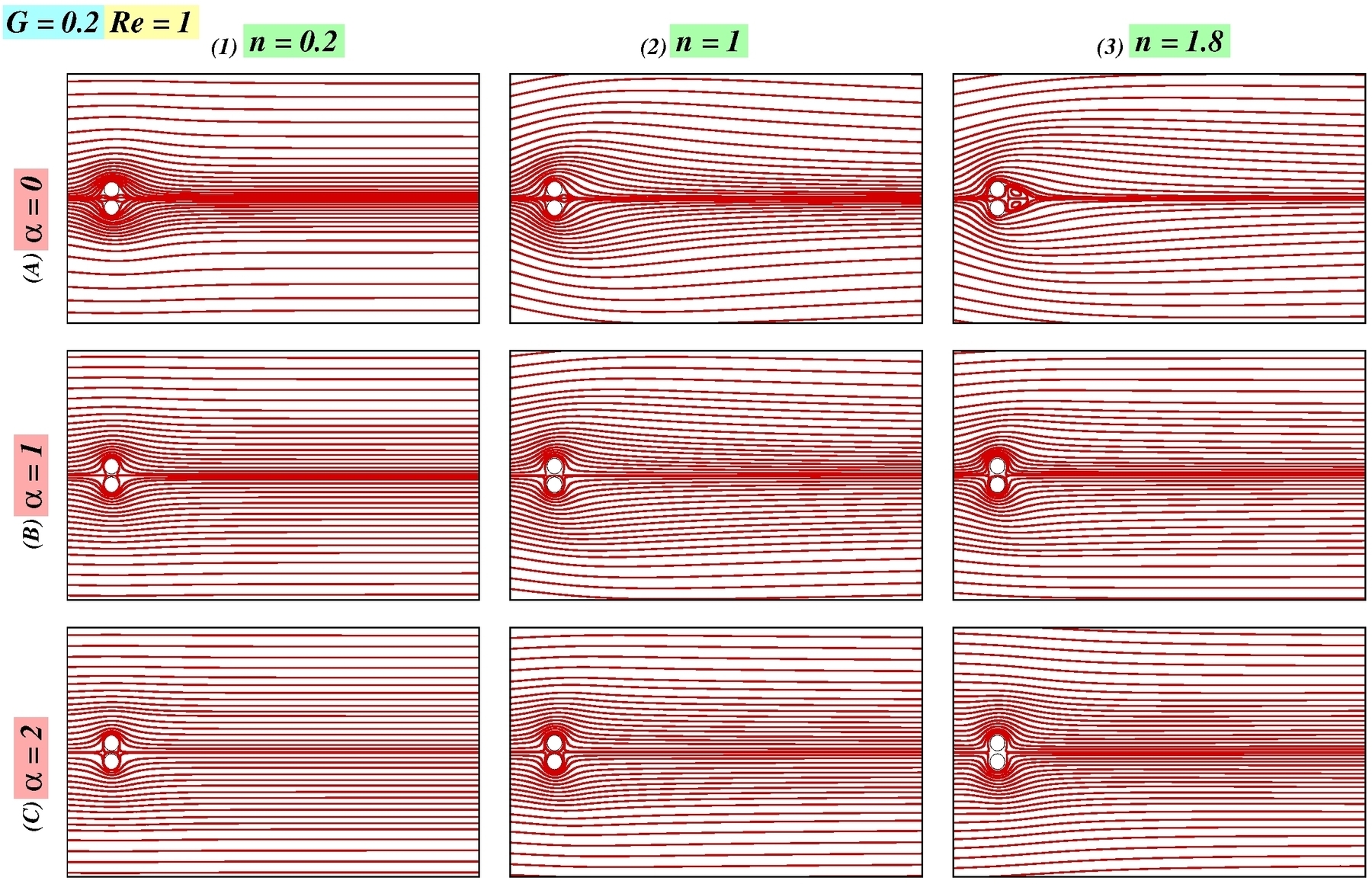}\label{fig:R1g-0.2}}
	\subfigure[$G=1$, $Re=1$] {\includegraphics[width=0.4\linewidth]{streamtraces/G1-Re1.jpg}\label{fig:R1g1}}	
	\subfigure[$G=0.2$, $Re=10$] {\includegraphics[width=0.4\linewidth]{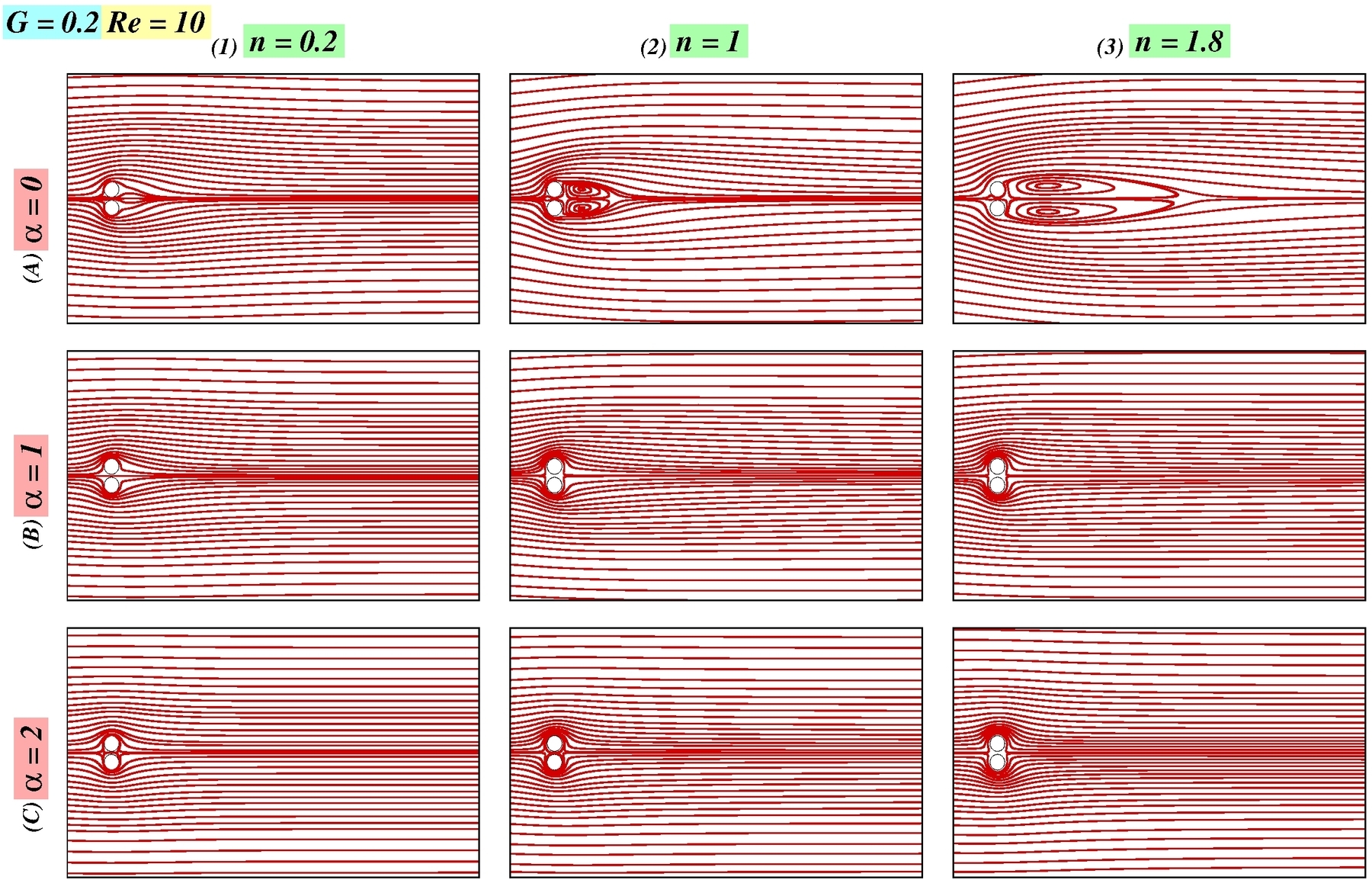}\label{fig:R10g-0.2}}
	\subfigure[$G=1$, $Re=10$] {\includegraphics[width=0.4\linewidth]{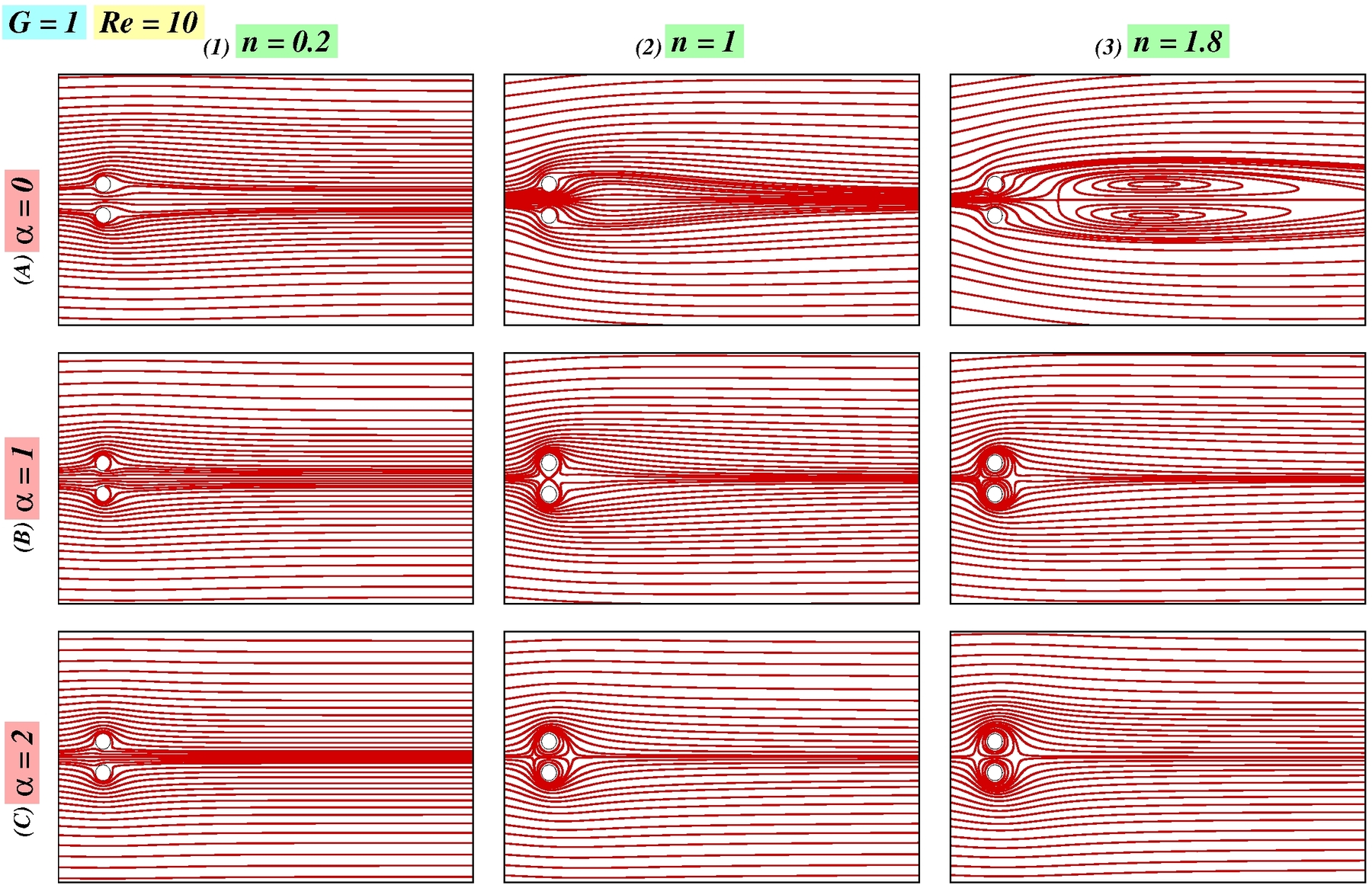}\label{fig:R10g1}}	
	\subfigure[$G=0.2$, $Re=20$] {\includegraphics[width=0.4\linewidth]{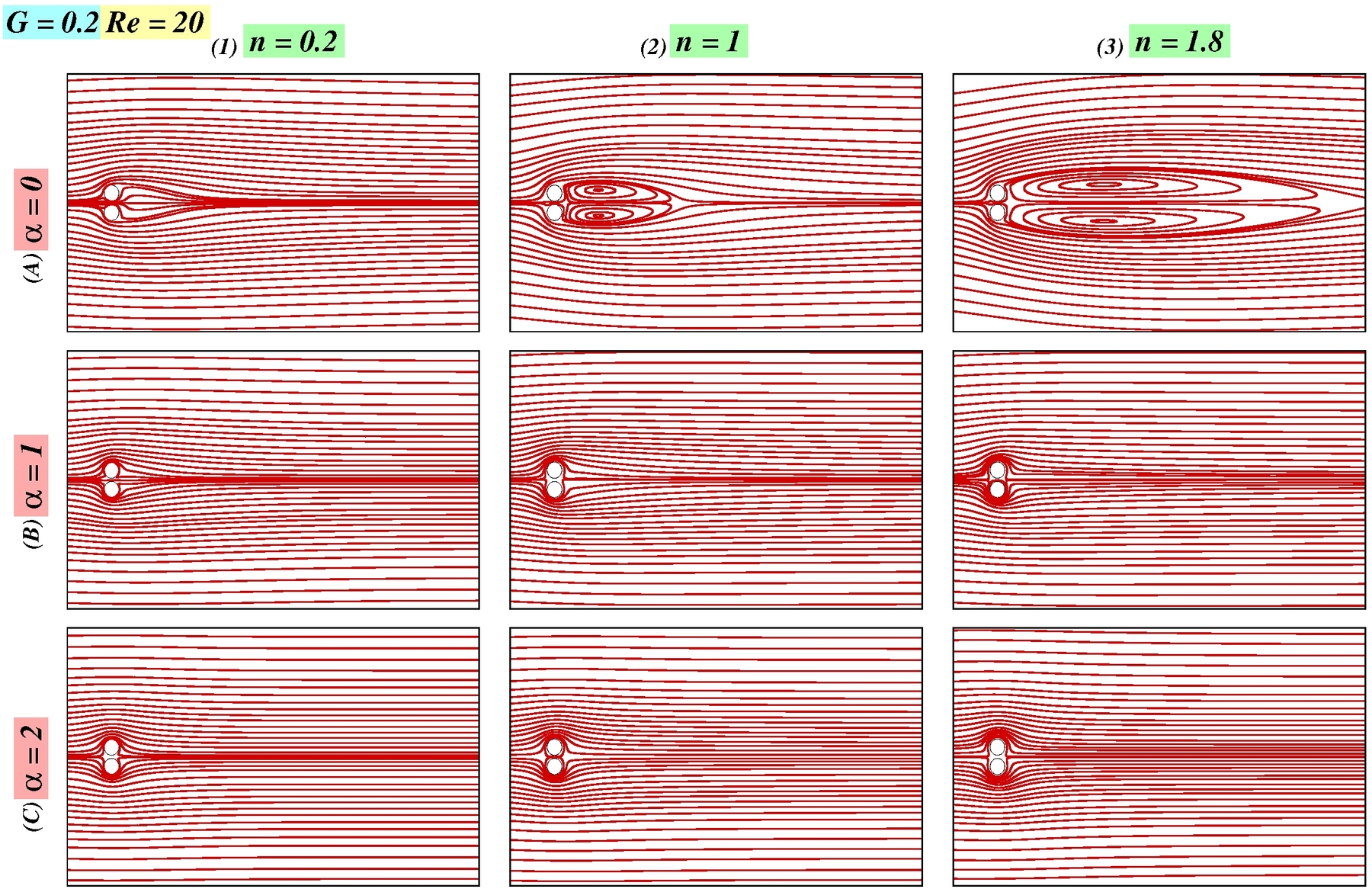}\label{fig:R20g-0.2}}
	\subfigure[$G=1$, $Re=20$] {\includegraphics[width=0.4\linewidth]{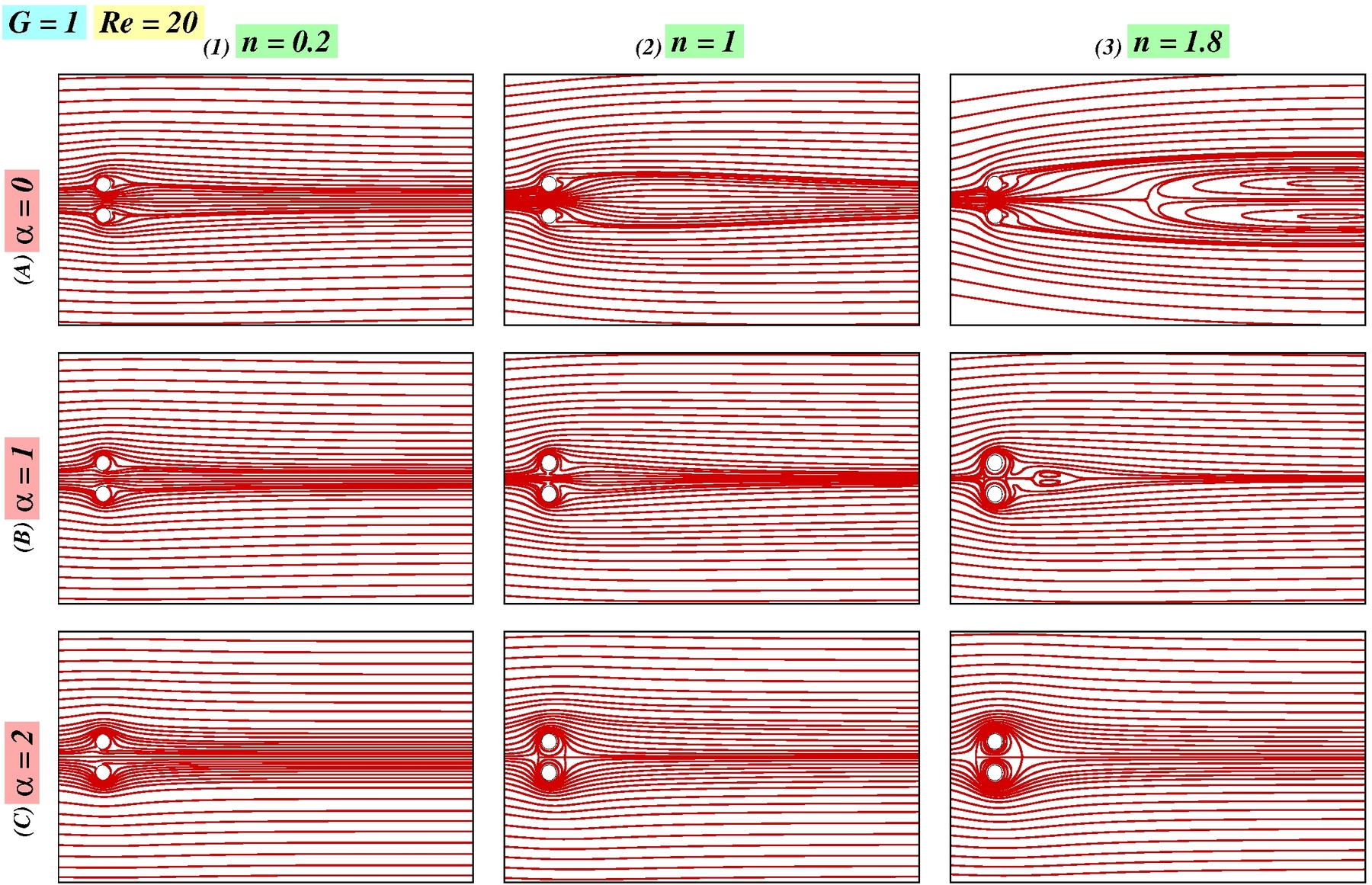}\label{fig:R20g1}}
	\subfigure[$G=0.2$, $Re=40$] {\includegraphics[width=0.4\linewidth]{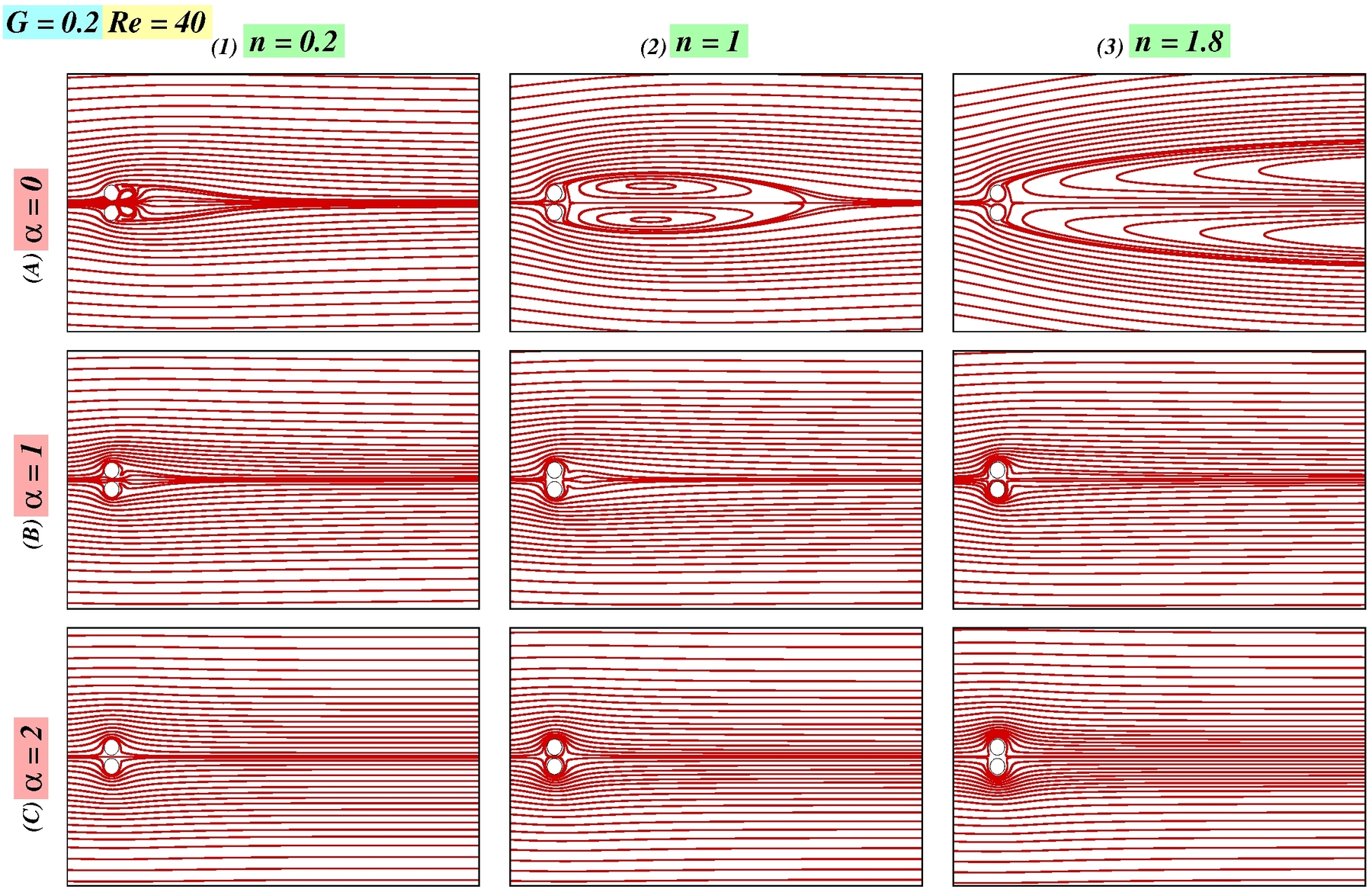}\label{fig:R40g-0.2}}
	\subfigure[$G=1$, $Re=40$] {\includegraphics[width=0.4\linewidth]{streamtraces/G1-Re40.jpg}\label{fig:R40g1}}
	\caption{Influence of Reynolds number ($1\le Re\le 40$), power-law index ($0.2\le n\le 1.8$) and rotational rate ($0\le \alpha\le 2$) on the streamline profiles for the extreme values of gap ratio ($G=0.2$, 1). The full-size images are included in \ref{appendix:streamline}.}
	\label{fig:g02-1}\label{fig:g02}\label{fig:g1}
\end{figure}
%
\subsubsection{Streamline profiles}
%
\noindent
\figs\ref{fig:R-1} and \ref{fig:g1} show the normalized streamline, $\psi = (\psi^{\ast} - \psi_{min}^{\ast})/(\psi_{max}^{\ast} - \psi_{min}^{\ast})$, profiles as a function of the dimensionless parameters ($Re$, $n$, $\alpha$, $G$). Each figure presents uniformly distributed streamlines, with sub-figures organized such that rows correspond to rotational rates (A, B, C for $\alpha = 0, 1, 2$) and columns correspond to power-law indices (1, 2, 3 for $n = 0.2, 1, 1.8$). The streamline profiles for other conditions are shown in  \ref{appendix:streamline}. The streamline profiles are symmetric about the mid-plane for all considered flow conditions ($Re$, $n$, $\alpha$, $G$), owing to the symmetric placement of the counter-rotating side-by-side cylinders (see \fig\ref{fig:1}).

\noindent
For stationary cylinders ($\alpha=0$) at $Re = 1$, shear-thinning and Newtonian fluids ($n \le 1$) exhibit uniform flow without separation for all gaps ($G$) (see panels A1-A2 in \figs \ref{fig:g-0.4}-\ref{fig:g-1}), consistent with previous studies \citep{Daniel2013}. In contrast, shear-thickening fluids ($n > 1$) show flow separation and the onset of wake formation at low gap ratios ($G \le 0.6$) (see panel A3 in \figs \ref{fig:g-0.4} and \ref{fig:g-0.6}). For $G \le 0.6$, $Re \le 20$, and $n \ge 1$, the two stationary cylinders are hydrodynamically isolated and behave similarly to a single cylinder. At larger gaps ($G \ge 0.8$), a small residual bulge forms behind the cylinders due to the increased flow resistance of shear-thickening fluids, causing localized fluid accumulation (see panel A3 in \figs \ref{fig:g-0.8} and \ref{fig:g-1}). The critical phenomena, such as vortex formation and vortex shedding, are suppressed between the cylinders due to their close proximity, an effect referred to as the ``near-wall effect'' \citep{xu2003reynolds}.

\noindent
For intermediate Reynolds number ($Re = 10$ and 20), wake formation occurs for Newtonian and shear-thickening fluids ($n \ge 1$), whereas no wake forms for shear-thinning fluids ($n < 1$) across all gap ratios ($G$) (see \figs \ref{fig:R10g-0.2} and \ref{fig:R20g-0.2}). However, the wake disappears for Newtonian fluids ($n=1$) at larger gaps ($G \ge  0.8$) (see panel A2 in \figs \ref{fig:st9} and \ref{fig:st10}). Broadly, wakes are generated by the shear layers of the upper and lower cylinders. As the gap widens, the interior shear layers interact within the gap. For small gaps, transitions between single and larger wakes increase momentum transfer. At $Re \ge 10$, increasing $G$ from 0.2 to 0.6 produces symmetric single-body vortices downstream for $n \ge 1$, which grow and shift downstream. For larger gaps ($G \ge 0.8$), separated double-body wakes appear behind each cylinder for $n \le 1$ (panel A2 in \figs \ref{fig:R40g-0.8} and \ref{fig:R40g-1}), slightly misaligned with the incoming flow due to gap size and cylinder interactions. Shear-thickening fluids, in contrast, exhibit symmetric vortex pairs behind the cylinders at $G \ge 0.8$ for $10 \le Re \le 20$.

\noindent
At $Re = 40$, wake interference between the stationary cylinders is observed for shear-thinning fluids ($n < 1$) for $G = 0.2$ (panel A1 in \fig\ref{fig:R40g-0.2}). The flow exhibits a distinct ``double-body wake'' structure characterized by two separated vortices behind each cylinder. The interacting shear layers promote the formation of a coherent ``vortex array'' in the wake region. A separated double-body wakes appear for $n < 1$ (at $G \ge 0.4$), for $n = 1$ (at $G \ge 0.8$), and for $n > 1$ (at $G = 1$), consistent with the literature \citep{singha2016numerical}. Furthermore, a transition from double-body to delayed wake formation is observed for $n > 1$ (at $G = 0.6$ and $0.8$) (panel A3 in \figs\ref{fig:R40g-0.6} and \ref{fig:R40g-0.8}).
%

\noindent
When the rotational rate increases from $\alpha = 0$ to 2 at $Re = 1$, the wake phenomenon disappears for shear-thinning fluids ($n < 1$). Instead, a thin fluid layer forms around the cylinders (panels B and C in \figs\ref{fig:g-0.4}-\ref{fig:g-1}), becoming more prominent with higher $n$ and larger gap ratios ($0.2 \le G \le 1$). As $\alpha$ increases, vortex shedding is fully suppressed, leading to a steady, vortex-free flow \citep{yoon2010laminar}. This behavior is observed for $G \le 0.6$ (at $Re \le 20$) and for $G \ge 0.8$ (at $Re \le 10$), where the cylinders are surrounded by increasingly thicker fluid layers with increasing $G$ and $n$. Further, at $Re = 40$ and $\alpha = 1$, a small residual bulge appears for Newtonian fluids at $G \le 0.6$, becoming more pronounced as $n$ increases. For $G \ge 0.8$, a residual bulge forms for $n > 1$, growing in size with both $G$ and $\alpha$. At $G = 1$ and $\alpha = 2$, a faint wake re-emerges behind both cylinders when $n > 1$ (panel C3 in \fig\ref{fig:R40g-1}). In all cases, the flow separation points shift in the direction of rotation due to the induced circumferential flow. This displacement becomes more evident with increasing $Re$ and $G$. Overall, the combined influence of rotation, inertia, and gap causes the separation points to migrate toward the direction of rotation, regardless of $n$ (\fig\ref{fig:g1}).
\begin{figure}[!hbtp]
	\centering
	\subfigure[$G=0.2$]{\includegraphics[width=0.44\linewidth]{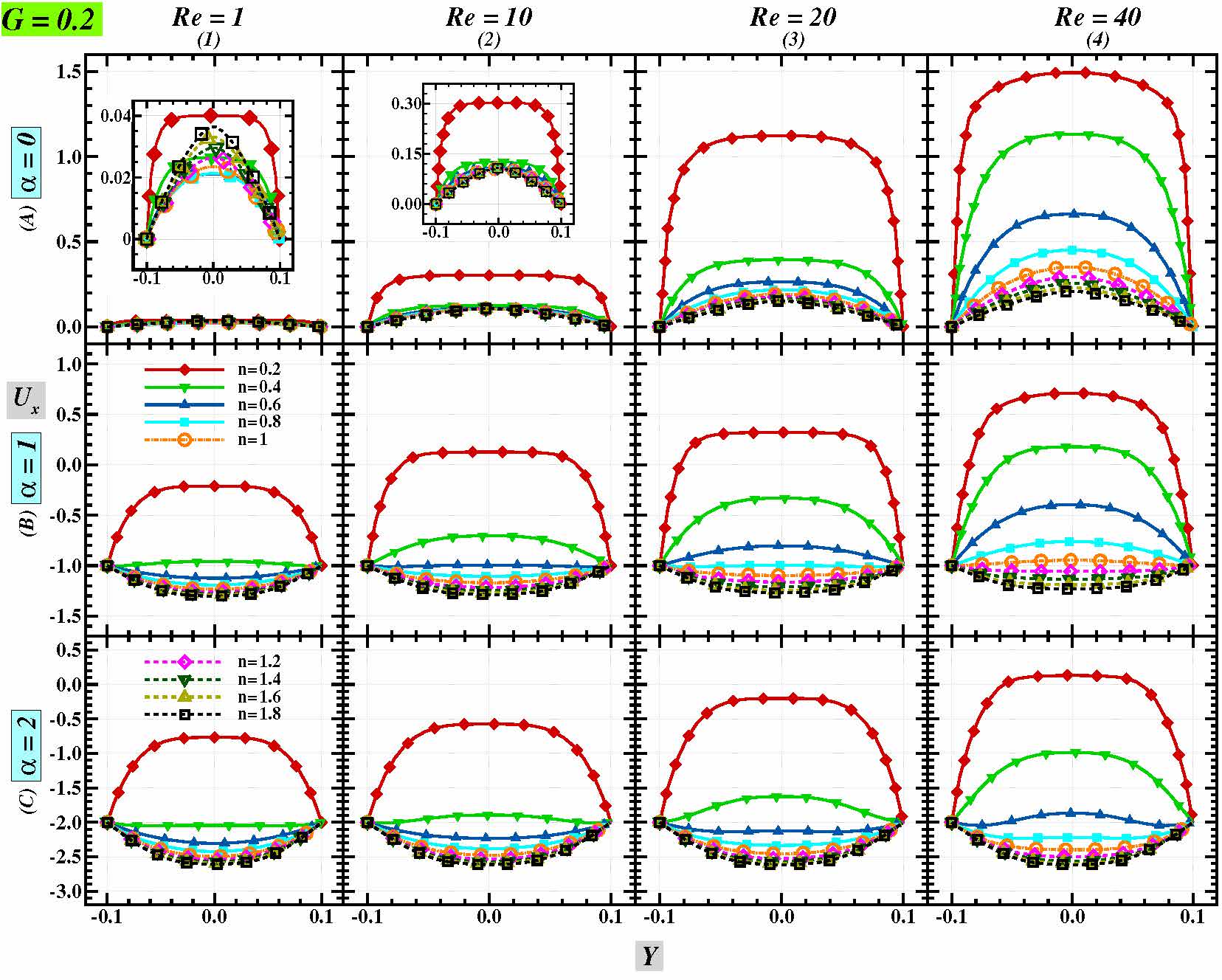}\label{fig:velo-0.2}}
	\subfigure[$G=0.4$]{\includegraphics[width=0.44\linewidth]{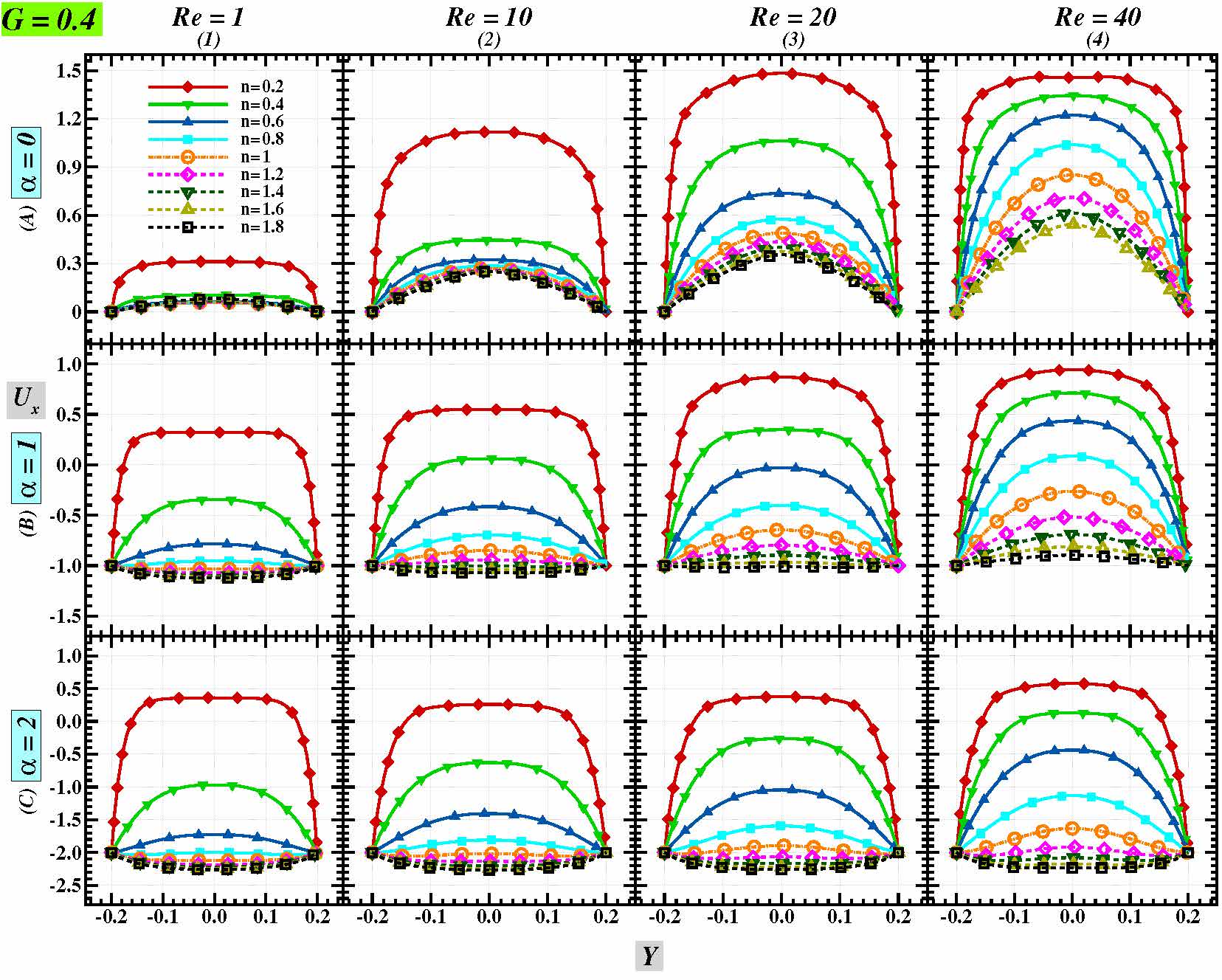}\label{fig:velo-0.4}}
	\subfigure[$G=0.6$]{\includegraphics[width=0.44\linewidth]{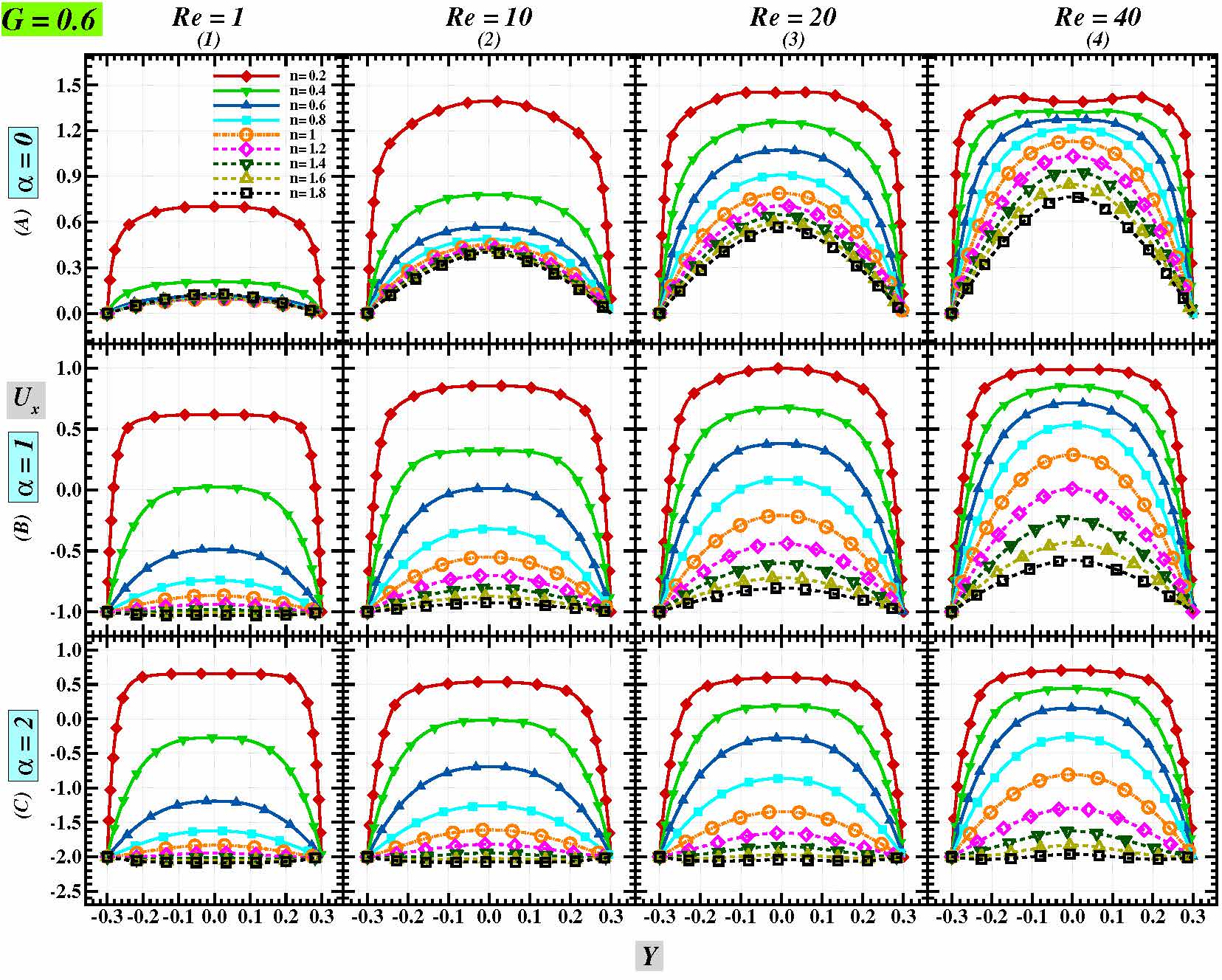}\label{fig:velo-0.6}}
	\subfigure[$G=0.8$]{\includegraphics[width=0.44\linewidth]{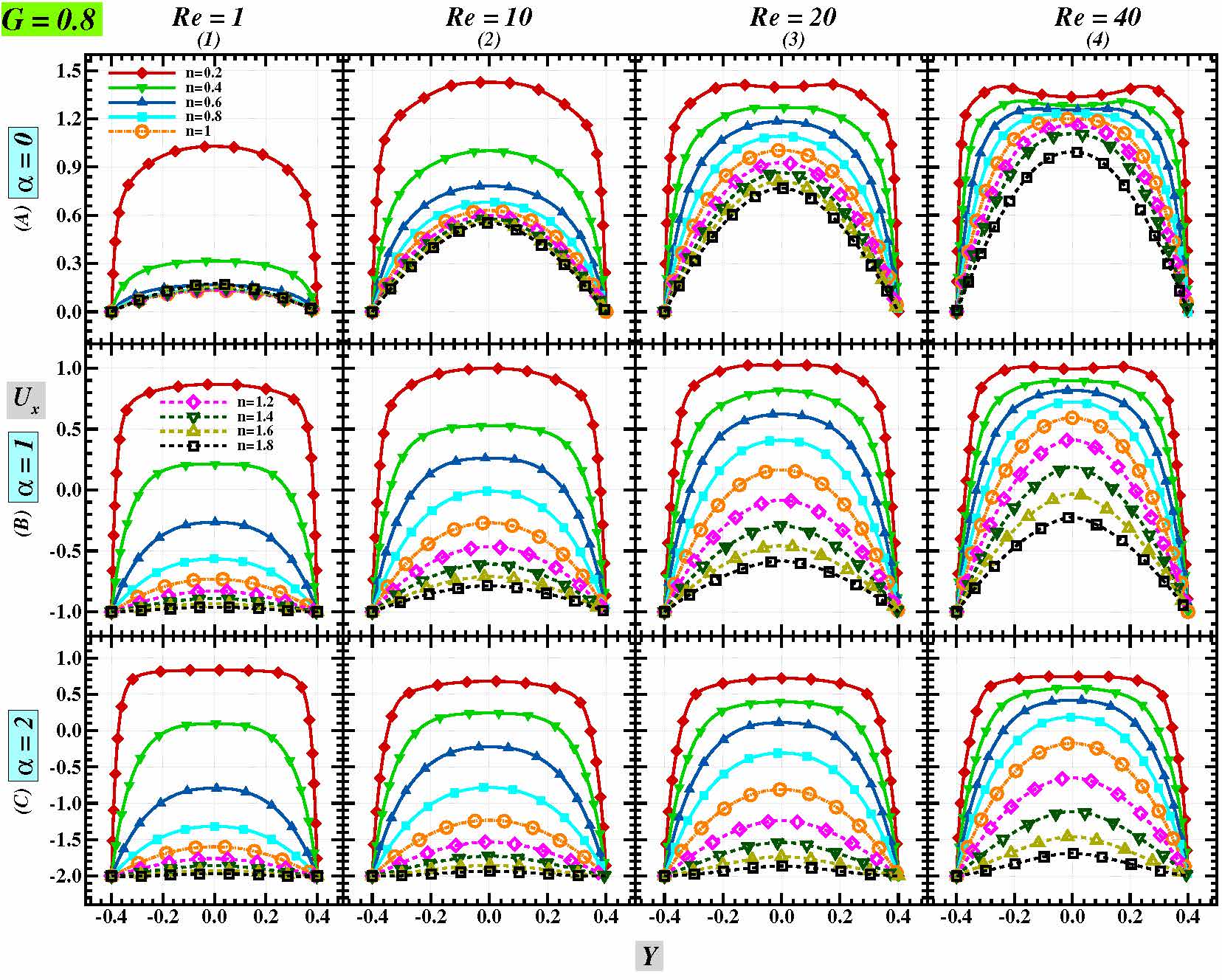}\label{fig:velo-0.8}}
	\subfigure[$G=1$]{\includegraphics[width=0.44\linewidth]{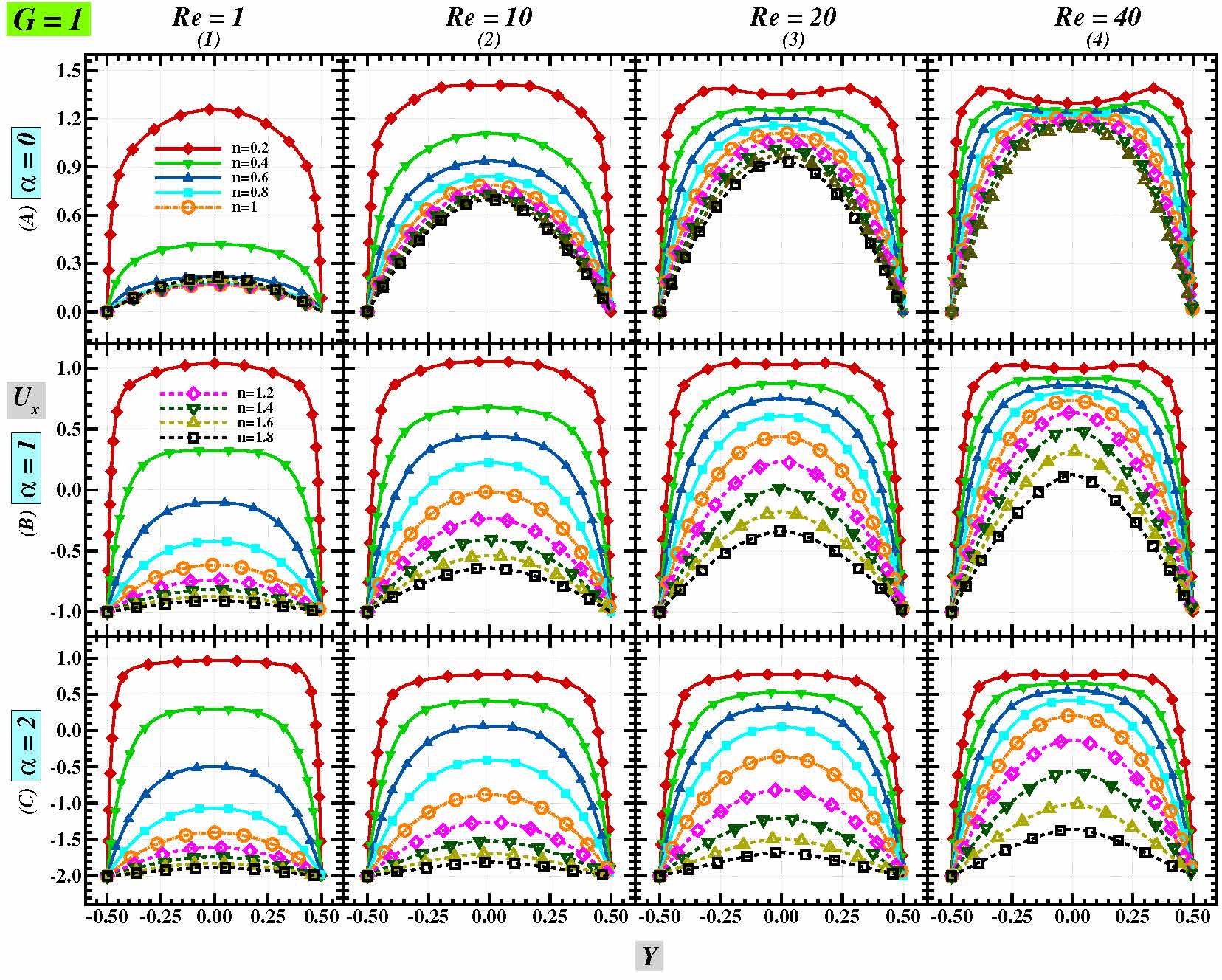}\label{fig:velo-1}}
	\caption{Centerline velocity profiles ($U_x$ at line 6, \fig\ref{fig:1}) for different power-law indices ($n$) at various rotational rates ($\alpha$ in rows) and Reynolds numbers ($Re$ in columns) for gap ratios $G = 0.2-1$. The full-size images are included in \ref{appendix:velocity}.}
	\label{fig:velo}
\end{figure}
%
\subsubsection{Centerline velocity profiles}
\noindent
\fig\ref{fig:velo} 
illustrates the normalized streamwise velocity component ($U_x$) along the centerline ($x = L_u$), measured along line 6 (connecting points A--B--C) in \fig\ref{fig:1}, between the two cylinders. The plots depict the distinct trends in the variation of $U_x$ under different combinations of Reynolds number ($1\le Re\le 40$), power-law index ($0.2\le n\le 1.8$), rotational rate ($0\le \alpha\le 2$), and gap ratio ($0.2\le G\le 1$), where in each panel, $Re$ and $\alpha$ vary along columns and rows, respectively.  In general, the velocity profile exhibits a parabolic shape with maximum velocity at the mid-plane ($y=0$), which becomes flatter (i.e., plug shapped) as $Re$, $G$, or $\alpha$ increase. The profiles remain symmetric about the mid-plane due to the symmetry of the geometrical configuration at all conditions. For low Reynolds numbers ($Re=1$), the flow is dominated by viscous effects, resulting in smaller and more rounded velocity distributions, whereas at higher Reynolds numbers ($Re=40$), the profiles become flatter and exhibit higher velocity magnitudes, indicating stronger inertial effects. As the power-law index ($n$) increases, the velocity magnitude rises, reflecting enhanced momentum transfer in shear-thickening fluids  ($n>1$), while shear-thinning fluids ($n<1$) show suppressed velocities due to higher effective viscosity. The influence of cylinder rotation ($\alpha$) is evident in the downward shift of velocity profiles and the expansion of negative velocity regions near the cylinder surfaces, particularly at higher $Re$, demonstrating the suppression of wake formation and the transition toward steady flow. Furthermore, increasing the gap ratio ($G$) broadens the flow passage and enhances the centerline velocity, as the interference between the cylinders weakens and the pressure drop across the gap decreases. Overall, the combined effects of inertia, non-Newtonian rheology, and cylinder rotation govern the shape and magnitude of the velocity profiles within the cylinders-gap region. Consistent with previous findings \citep{kang2003characteristics, xu2003reynolds, singha2016numerical}, closely spaced cylinders ($G \le  0.2$) exhibit weak interstitial flow characterized by reduced velocity and momentum. As $G$ increases, the pressure in the gap decreases, leading to an enhancement of the streamwise velocity ($U_x$) with increasing $Re$.

\noindent
The velocity ($U_x$) increases with increasing gap ($G$), Reynolds numbers ($Re$), and rotation rates ($\alpha$), indicating enhanced fluid acceleration and reduced inter-cylinder resistance, as shown in \fig\ref{fig:velo}.  However, for lower gap ratio ($G \leq 0.4$), the influence of the power-law index ($n$) becomes more complex due to strong shear interactions and geometric confinement between the cylinders (\figs\ref{fig:velo-0.2} and \ref{fig:velo-0.4}). For all $Re$ at smaller gaps ($G < 0.6$) (\fig\ref{fig:velo-0.2}) and for $Re \leq 10$ at moderate gaps ($G \geq 0.6$), see \figs\ref{fig:velo-0.6}--\ref{fig:velo-1}, the maximum velocity ($U_{\text{x,max}}$) peak appears at the mid-plane (point B in \fig\ref{fig:1}) for all $n$ and $\alpha$, indicating symmetric flow distribution dominated by viscous effects. In contrast, as the gap widens ($G \geq 0.6$) and $Re$ increases ($Re \geq 20$) for slow rotations ($\alpha \leq 1$) (refer panels A3-A4 and B3-B4 in \figs\ref{fig:velo-0.6}--\ref{fig:velo-1}), $U_{\text{x,max}}$ progressively shifts toward the surface of cylinders (i.e., towards points A and C), especially for shear-thinning ($n<1$) fluids. This shift corresponds to the emergence of separated double wakes behind each cylinder, resulting in an inverse parabolic velocity distribution within the cylinders gap region, highlighting the growing dominance of inertial and non-Newtonian effects over viscous confinement  at larger $G$ and $Re$.

\noindent
In shear-thinning fluids  ($n<1$), decreasing viscosity, with increasing shear rate, enhances flow acceleration in the gap between the cylinders, leading to the formation of separated double wakes and a noticeable shift of the maximum axial velocity ($U_\text{x,max}$) from the mid-plane toward the cylinders. As the rotational rate ($\alpha$) increases, thin fluid layers form around the cylinders, which suppress wake formation and yield an inverse parabolic or concave velocity profile for Newtonian and shear-thickening fluids ($n \geq 1$) (see panels B and C in \figs\ref{fig:velo-0.2} - \ref{fig:velo-0.6}). This effect becomes more prominent with increasing gap ($G$) and rotation ($\alpha$). For instance, at $G=0.2$ and $\alpha>0$, concave profiles appear even at $Re=1$ for $n \geq 0.4$, indicating reverse flow near the cylinders. Overall, these results demonstrate that fluid rheology ($n$) and flow parameters ($Re$, $G$, $\alpha$) strongly govern the velocity field characteristics: shear-thinning behavior promotes near-wall acceleration and double wake formation, while increasing $\alpha$ stabilizes the flow by forming circumferential fluid layers. The complex interplay between the flow parameters ($n$, $G$, $Re$, $\alpha$) is clearly illustrated in \fig\ref{fig:velo}, where the shift in peak velocity ($U_\text{x,max}$) towards the cylinders for $G \geq 0.6$, $Re \geq 20$, and $\alpha \leq 1$ highlights the transition from symmetric to wake-dominated flow in non-Newtonian regimes.
%
\subsubsection{Pressure coefficient ($C_p$) profiles}
\noindent
The pressure coefficient ($C_p$) distributions over the surfaces of the cylinders ($\phi=0^\circ$ to $360^\circ$) , presented in \fig\ref{fig:pre} (\figs\ref{fig:pre-up0.2}--\ref{fig:pre-low1} in \ref{appendix:pressure}), display a clear inverse symmetry, where the $C_p$ profiles of upper cylinder (UC) and lower cylinder (LC) mirror each other. This symmetric behavior persists across all examined conditions ($Re$, $G$, $n$, $\alpha$). The polar plots effectively capture $C_p(\phi)$ variation over the surface of cylinders, highlighting the opposing $C_p$ distributions on the upper and lower cylinders resulting from their counter-rotating motion and geometric symmetry.
\noindent
\begin{figure}[!htbp]
	\centering
	\subfigure[$G=0.2$ for Upper cylinder ({UC})]{\includegraphics[width=0.44\linewidth]{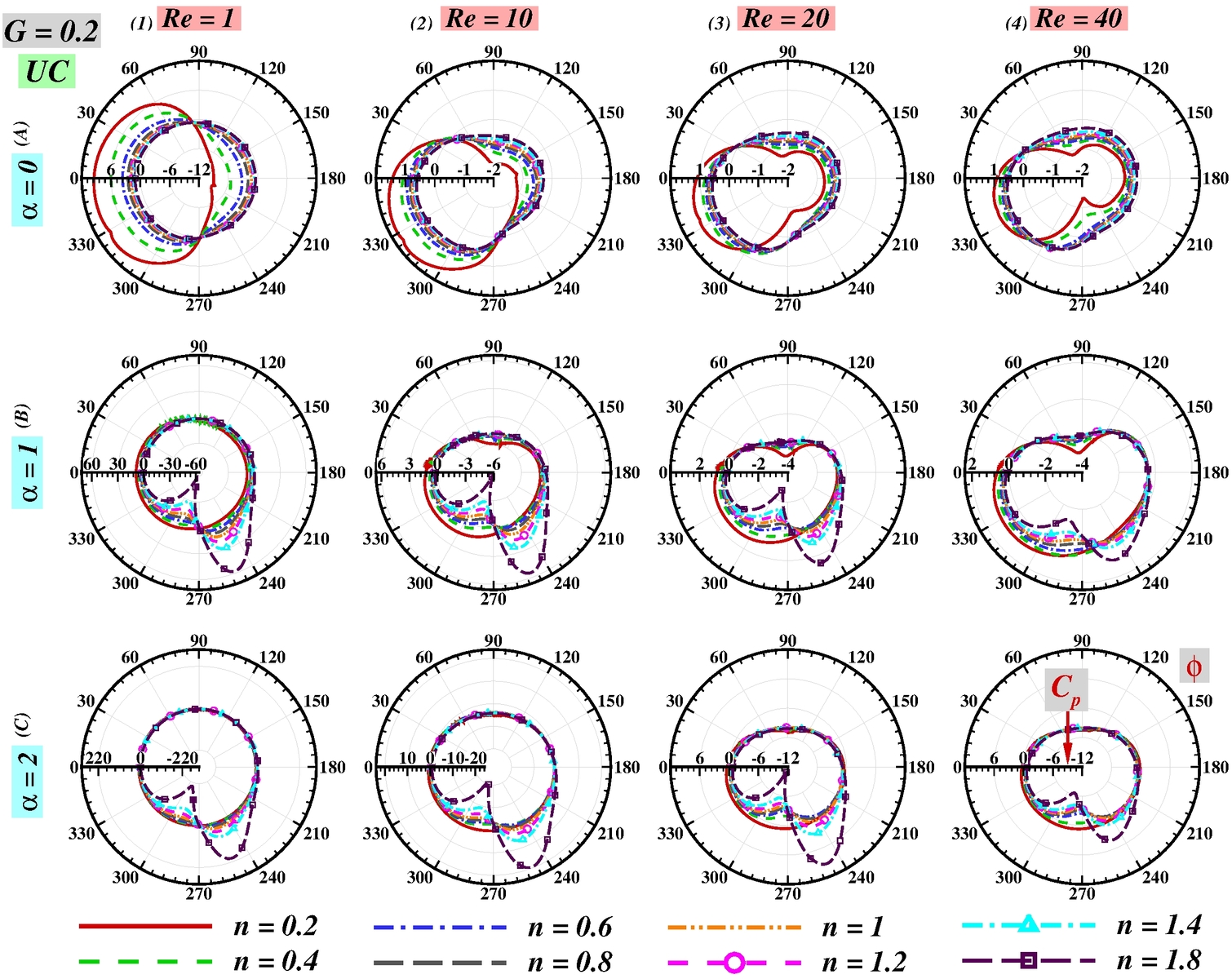}\label{fig:pre-uc0.2}}
	\subfigure[$G=0.2$ for Lower cylinder ({LC})]{\includegraphics[width=0.44\linewidth]{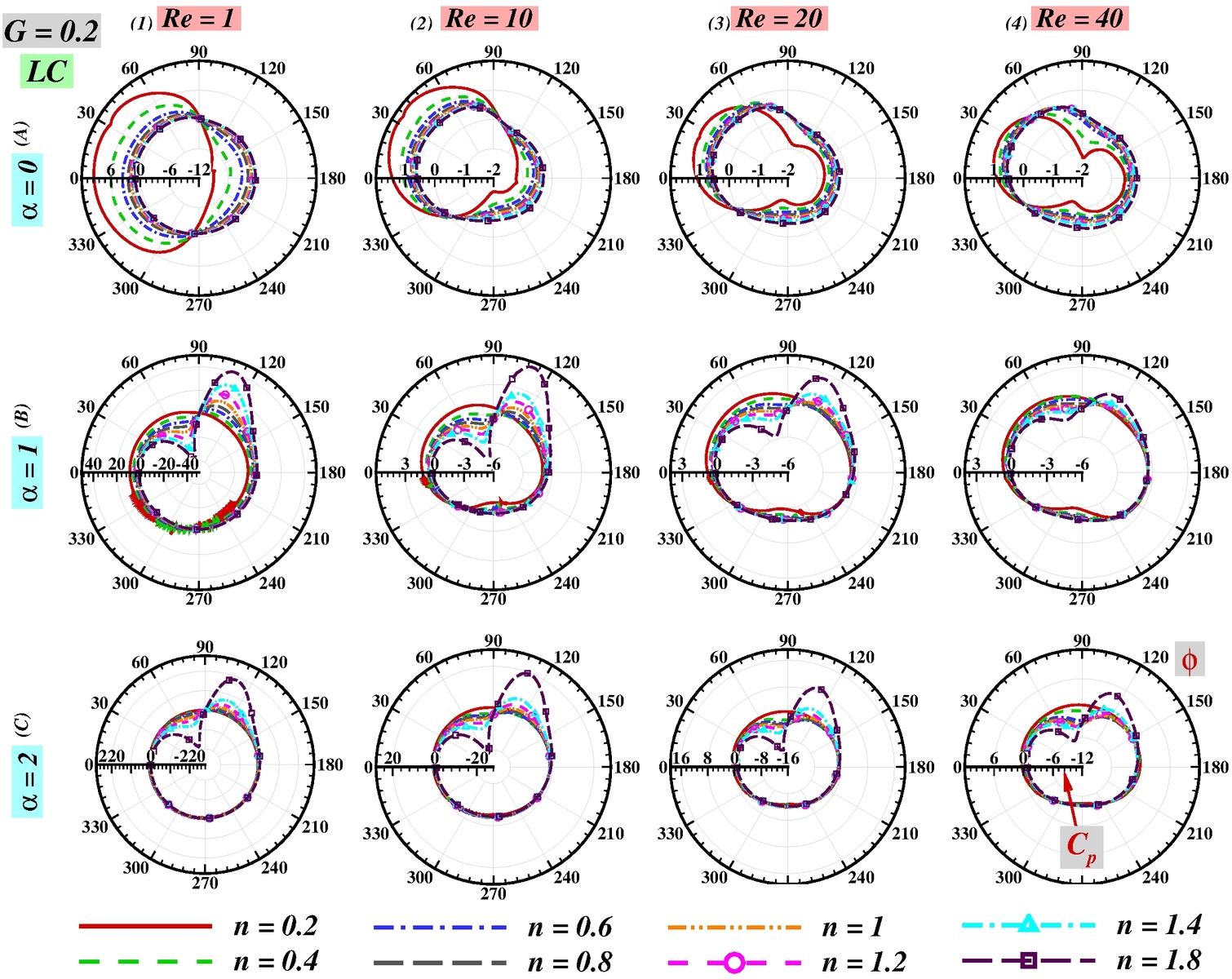}\label{fig:pre-lc0.2}}
	\subfigure[$G=0.6$ for Upper cylinder ({UC})]{\includegraphics[width=0.44\linewidth]{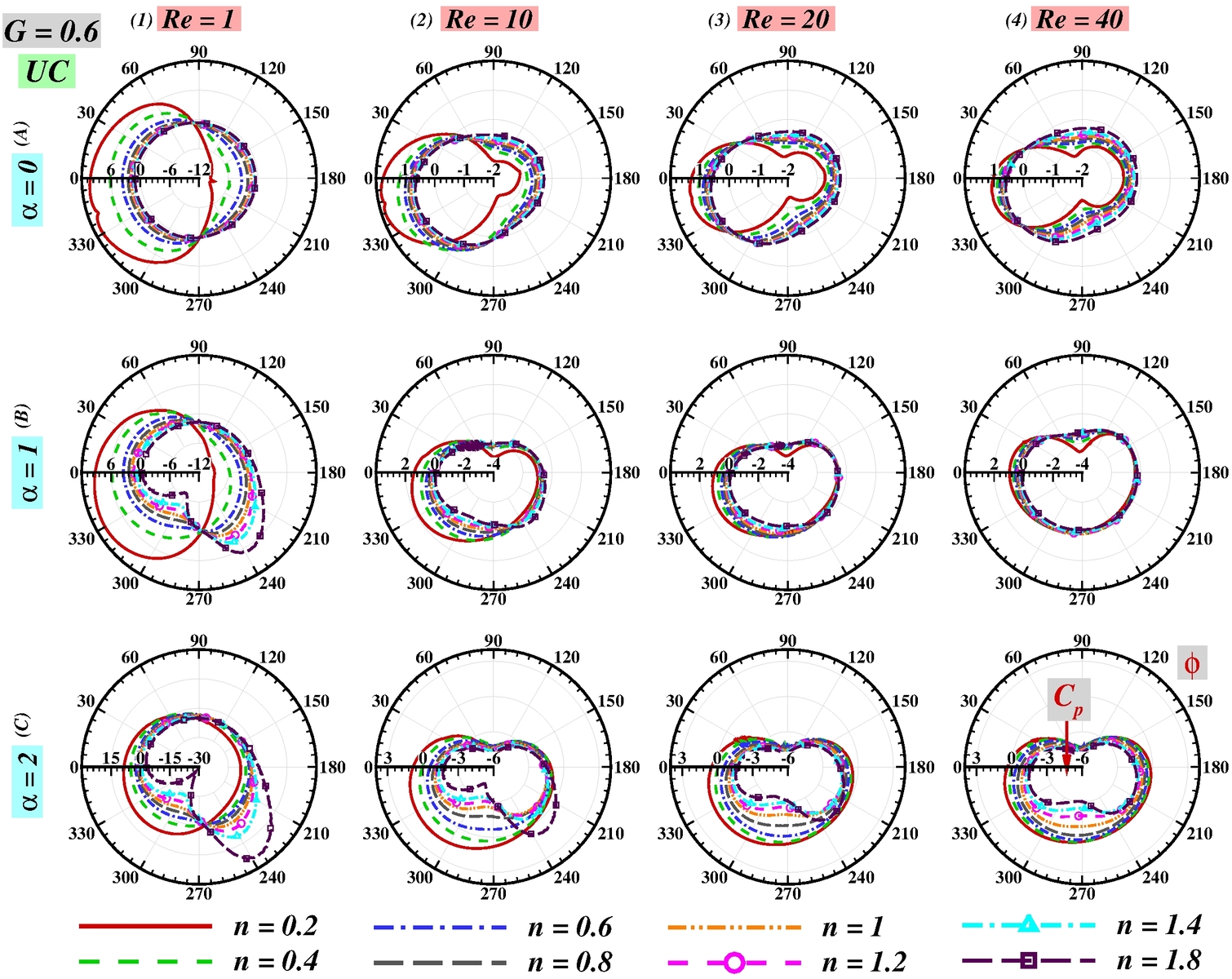}\label{fig:pre-uc0.6}}
	\subfigure[$G=0.6$ for Lower cylinder ({LC})]{\includegraphics[width=0.44\linewidth]{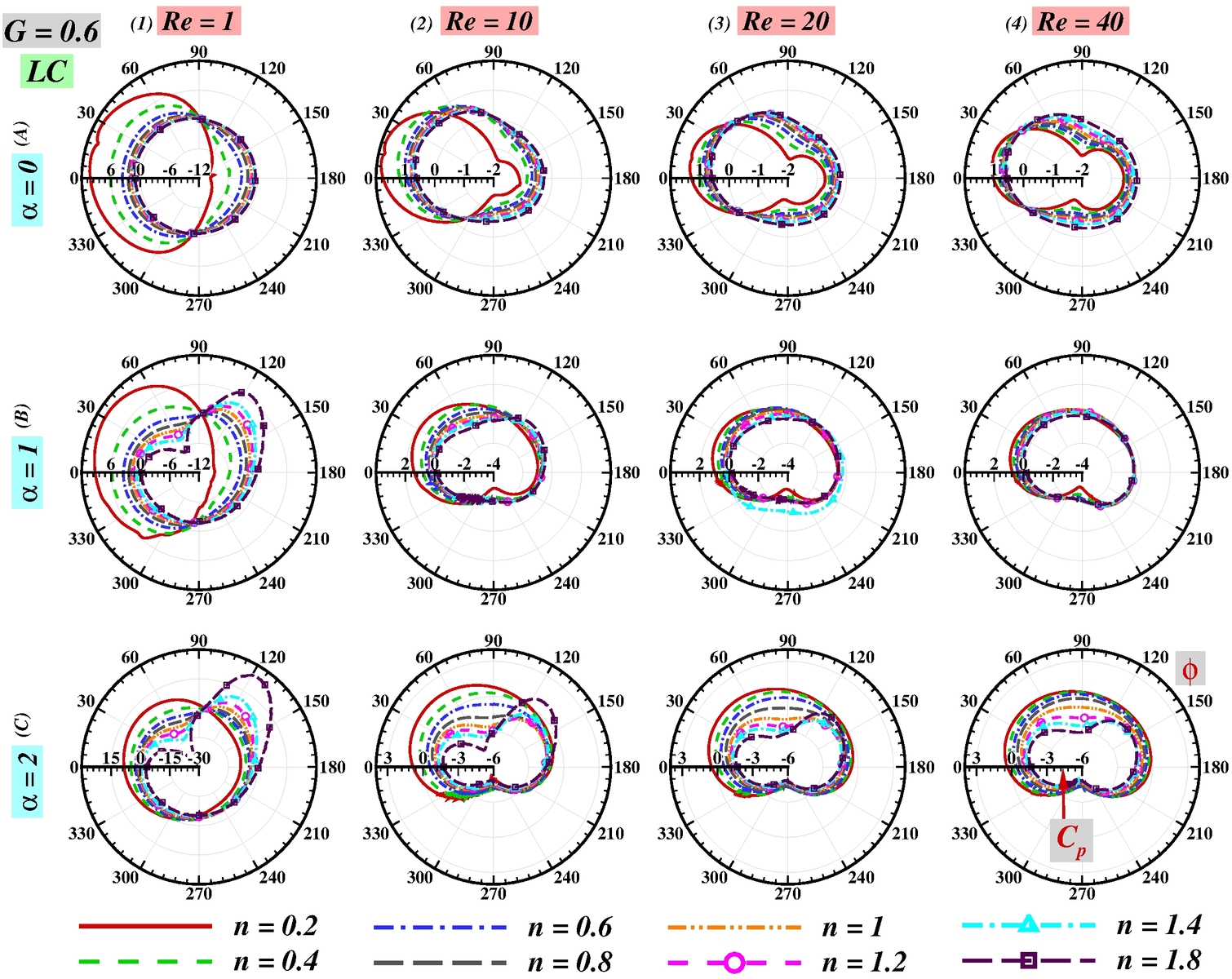}\label{fig:pre-lc0.6}}
	\subfigure[$G=1$ for Upper cylinder ({UC})]{\includegraphics[width=0.44\linewidth]{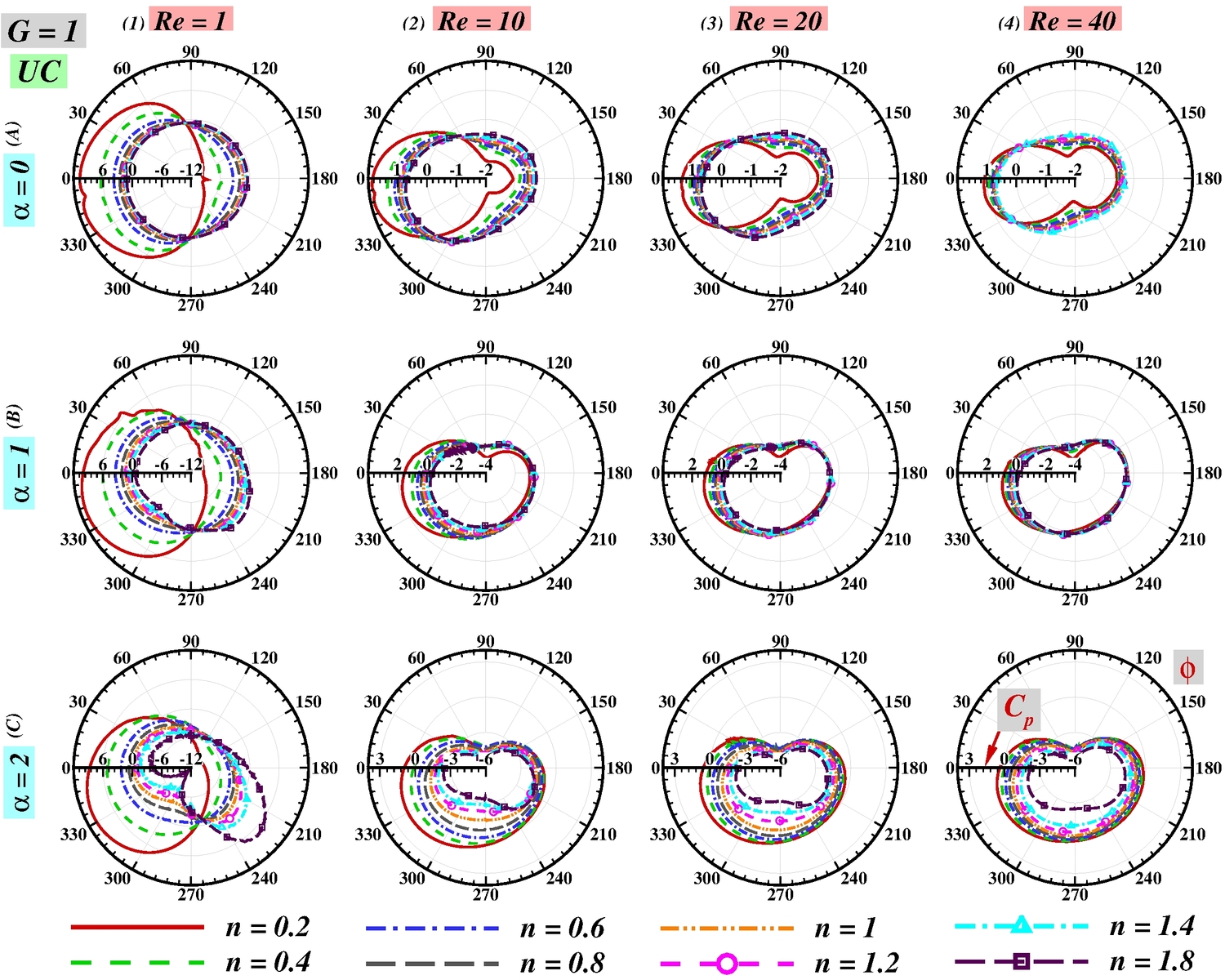}\label{fig:pre-uc1}}
	\subfigure[$G=1$ for Lower cylinder ({LC})]{\includegraphics[width=0.44\linewidth]{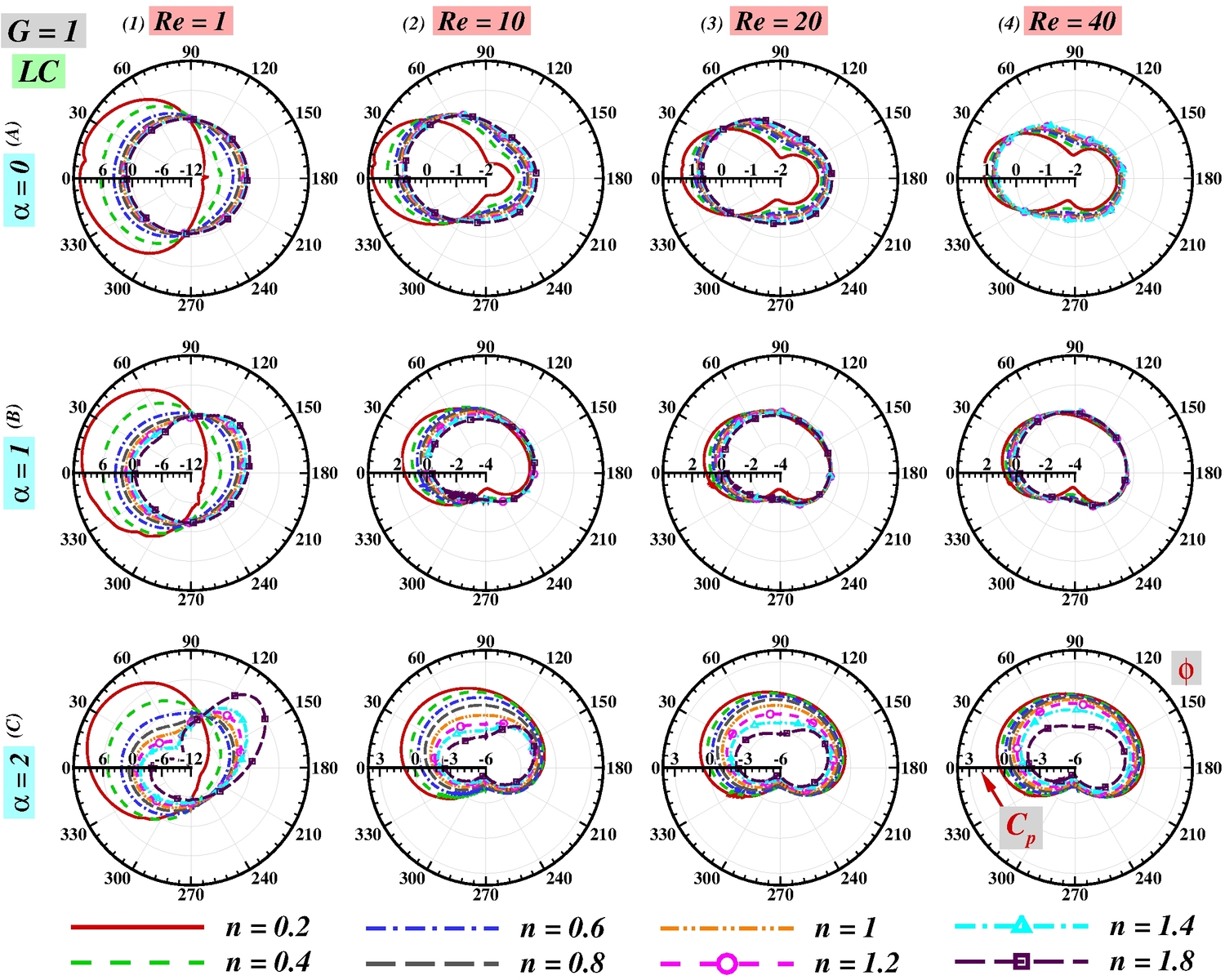}\label{fig:pre-lc1}}
	\caption{Distribution of pressure coefficient on the surface of the upper and lower cylinders ({UC \& LC}) for different $n$ with $\alpha$ (in rows) and $Re$ (in columns) at gap ratio $G=0.2-1$.}
	\label{fig:pre}
\end{figure}
%

\noindent
At $Re=1$, the pressure coefficient ($C_p$) profiles maintain a nearly circular shape for all gaps ($G$) when the power-law index ($n > 0.6$), indicating uniform pressure distribution around both cylinders. As $Re$ increases for stationary cylinders ($\alpha=0$), the polar plots gradually transform into hypercardioid shapes, characterized by a large high-pressure lobe at the front and a smaller one at the rear. The minima in these hypercardioid profiles correspond to regions of flow separation and wake formation behind the cylinders, marking the onset of vortex structures. With increasing rotational rate ($\alpha$), the pressure contours evolve distinctly depending on the rheological behavior of the fluid. For shear-thinning and Newtonian ($n\le 1$) fluids, the plots largely retain their circular symmetry, whereas shear-thickening ($n>1$) fluids develop an inverse one-loop limaçon shape with a slight tilt, reflecting asymmetric pressure distribution caused by rotation-induced shear effects.
Furthermore, at $\alpha=0$, the maximum $C_p$ occurs at the front stagnation point (FSP, $\phi = 0^{\circ}$), shifting inward for smaller $G$ and outward as the gap ($G$) increases. As the cylinders rotate ($\alpha>0$), the maxima of negative $C_p$ shifts towards the outer side at specific angles ($\phi$), and negative $C_p$ values are prevalent across the entire surface of the cylinders. 
When $\alpha > 0$, negative $C_p$ values emerge along the outer cylinder surfaces, indicating recirculation zones. 
The magnitude of these negative $C_p$ values intensify with increasing rotations ($\alpha$), increasing difference in $C_p$ between the inner and outer sides of the cylinders. Similar flow and pressure patterns have also been reported in literature \citep{chaitanya2023mixed,darvishyadegari2018analysis}.

\noindent
This symmetry in the pressure coefficient ($C_p$) distribution results from the counter-rotating motion of the cylinders, producing mirror-like pressure patterns on the upper surface of UC ($0^\circ \leq \phi \leq 180^\circ$) and the lower surface of LC ($180^\circ \leq \phi \leq 360^\circ$). Therefore, subsequent discussions focus on the upper cylinder (UC) for clarity. The angular positions ($\phi$) corresponding to maximum pressure coefficient ($C_{p,\text{max}}$) and minimum pressure coefficient ($C_{p,\text{min}}$) are denoted as $\phi_{h}$ and $\phi_{l}$, respectively. 

\noindent
For stationary cylinders ($\alpha=0$), as the Reynolds number ($Re$) increases from 1 to 40, the surface locations ($\phi$) on the UC, corresponding to the maximum and minimum $C_p$ values, shift noticeably. Specifically, at $G=0.2$, $G=0.6$, and $G=1$, see panel A in \figs\ref{fig:pre-uc0.2}, \ref{fig:pre-uc0.6}, and \ref{fig:pre-uc1}, the angular positions, $\phi_{h}$ and $\phi_{l}$, vary by approximately 5\% and 52\% with the power-law index ($n$), respectively. This substantial shift in $\phi_{l}$ highlights the growing influence of inertial and rheological effects with increasing $Re$ and $n$, which alter the surface pressure distribution and promote asymmetric flow separation around the cylinders. Furthermore, at low $Re$, the minimum pressure zones are shifted anticlockwise from the rear stagnation point (RSP, $\phi = 180^\circ$), indicating that they do not form directly behind the cylinder. This offset enhances the dominance of positive shear stress on one side of the cylinder, inducing a net clockwise torque. As $Re$ increases, the single low-pressure region divides into two distinct minima, and one maxima on $C_p$ curves; both minima shift azimuthally along the cylinder surface. This asymmetrical $C_p$ distribution causes unequal surface areas to experience positive and negative shear stresses. With further increases in $Re$, these minimum $C_p$ zones gradually become more symmetric and their magnitudes converge, reducing the imbalance between positive and negative shear stresses and leading to a more stable and uniform flow pattern around the cylinders.
%
\begin{table}[t!]
	\centering
	\caption{The maximum and minimum values of the pressure coefficients ($C_{p,\text{max}}$ and $C_{p,\text{min}}$) with their angular positions ($\phi_{h}$ and $\phi_{l}$) over the surface of the upper cylinder (UC) as a function of flow parameters ($Re$, $\alpha$, $n$, $G$).}
	\resizebox{\columnwidth}{!}{%
	\renewcommand{\arraystretch}{1.3}
		\begin{tabular}{|c|c|r|r|r|r|r|r|r|r|r|r|r|r|r|r|r|r|}	 %
			\hline
			\multirow{2}[0]{*}{$G$}& \multirow{2}[0]{*}{$n$}      & \multicolumn{4}{c|}{$Re=1$}      & \multicolumn{4}{c|}{$Re=10$}     & \multicolumn{4}{c|}{$Re=20$}     & \multicolumn{4}{c|}{$Re=40$} \\
			\cline{3-18}
			&       & $C_{p,\text{max}}$  & $\phi_\text{h}$ & $C_{p,\text{min}}$  & $\phi_\text{l}$ & $C_{p,\text{max}}$  & $\phi_\text{h}$ & $C_{p,\text{min}}$  & $\phi_\text{l}$ & $C_{p,\text{max}}$  & $\phi_\text{h}$ & $C_{p,\text{min}}$  & $\phi_\text{l}$ & $C_{p,\text{max}}$  & $\phi_\text{h}$ & $C_{p,\text{min}}$  & $\phi_\text{l}$ \\ \hline
			\multicolumn{2}{|c|}{}  & \multicolumn{16}{c|}{$\alpha=0$} \\ \hline
			\multirow{3}[0]{*}{0.2} & 0.2   & 9.680 & 320.7 & -9.020 & 178.7 & 1.7989 & 318.7 & -1.5410 & 95.3  & 1.3514 & 330.0 & 1.2373 & 78.0  & 1.1031 & 336.0 & -1.3204 & 266.0 \\\cline{2-18}
			& 1     & 1.916 & 357.3 & -1.605 & 138.7 & 0.8260 & 338.7 & -0.8025 & 100.7 & 0.7517 & 332.0 & -0.6889 & 89.3  & 0.7174 & 329.3 & -0.5582 & 78.7 \\ \cline{2-18}
			& 1.8   & 1.054 & 351.3 & -0.839 & 120.7 & 0.6812 & 338.7 & -0.5641 & 96.0  & 0.6502 & 334.0 & -0.4588 & 89.3  & 0.6358 & 330.0 & -0.3094 & 79.3 \\
			 \hline
			\multirow{3}[0]{*}{0.4} & 0.2   & 10.347 & 327.3 & -9.516 & 208.0 & 1.8556 & 335.3 & -1.4660 & 95.3  & 1.3612 & 342.0 & -1.3571 & 264.0 & 1.1066 & 343.3 & -1.2817 & 268.0 \\\cline{2-18}
			& 1     & 1.912 & 357.3 & -1.610 & 137.3 & 0.8528 & 337.3 & -0.7952 & 99.3  & 0.7823 & 335.3 & -0.6666 & 90.0  & 0.7411 & 337.3 & -0.5498 & 82.0 \\\cline{2-18}
			& 1.8   & 1.066 & 351.3 & -0.854 & 121.3 & 0.7075 & 337.3 & -0.5482 & 95.3  & 0.6754 & 335.3 & -0.4317 & 90.0  & 0.6881 & 335.3 & -0.4242 & 82.0 \\
			\hline
			\multirow{3}[0]{*}{0.6} & 0.2   & 10.663 & 332.0 & -9.650 & 178.0 & 1.8672 & 341.3 & -1.4207 & 98.7  & 1.3648 & 344.7 & -1.3497 & 264.0 & 1.1093 & 351.3 & -1.2375 & 270.7 \\\cline{2-18}
			& 1     & 1.928 & 352.7 & -1.620 & 139.3 & 0.8865 & 336.0 & -0.7786 & 98.7  & 0.8088 & 339.3 & -0.6482 & 89.3  & 0.7550 & 343.3 & -0.6056 & 82.7 \\\cline{2-18}
			& 1.8   & 1.094 & 348.7 & -0.866 & 118.7 & 0.7393 & 339.3 & -0.5301 & 95.3  & 0.6991 & 336.0 & -0.4155 & 89.3  & 0.6701 & 339.3 & -0.3096 & 82.0 \\
			\hline
			\multirow{3}[0]{*}{0.8} & 0.2   & 10.738 & 338.7 & -9.687 & 172.7 & 1.8739 & 346.0 & -1.4667 & 254.0 & 1.3674 & 348.0 & -1.3237 & 265.3 & 1.1117 & 348.0 & -1.2208 & 270.7 \\\cline{2-18}
			& 1     & 1.957 & 351.3 & -1.632 & 137.3 & 0.9171 & 340.7 & -0.7583 & 100.7 & 0.8263 & 343.3 & -0.6561 & 89.3  & 0.7615 & 346.0 & -0.6614 & 253.3 \\\cline{2-18}
			& 1.8   & 1.129 & 348.0 & -0.876 & 118.7 & 0.7690 & 338.7 & -0.5142 & 94.0  & 0.7181 & 340.7 & -0.4245 & 89.3  & 0.6834 & 343.3 & -0.4046 & 83.3 \\
			\hline
			\multirow{3}[0]{*}{1} & 0.2   & 10.735 & 342.0 & -9.658 & 170.0 & 1.8752 & 348.7 & -1.4935 & 256.0 & 1.3695 & 349.3 & -1.3008 & 266.7 & 1.1126 & 354.7 & -1.2108 & 274.7 \\\cline{2-18}
			& 1     & 1.989 & 351.3 & -1.649 & 138.0 & 0.9405 & 342.0 & -0.7432 & 100.0 & 0.8366 & 346.7 & -0.6797 & 89.3  & 0.7657 & 348.7 & -0.6814 & 255.3 \\\cline{2-18}
			& 1.8   & 1.168 & 346.7 & -0.885 & 118.7 & 0.7918 & 341.3 & -0.5055 & 96.0  & 0.7338 & 344.0 & -0.4578 & 89.3  & 0.7056 & 346.7 & -0.5340 & 248.7 \\
			\hline
			\multicolumn{2}{|c|}{}   &\multicolumn{16}{c|}{$\alpha=1$} \\
			\hline
			\multirow{3}[0]{*}{0.2} & 0.2   & 8.9611 & 351.3 & -8.9999 & 138.0 & 1.6283 & 329.3 & -3.0267 & 104.7 & 1.2683 & 314.0 & -2.8813 & 85.3  & 1.0834 & 322.0 & -2.9155 & 80.7 \\\cline{2-18}
			& 1     & 13.5150 & 252.0 & -13.6770 & 287.3 & 0.9590 & 250.0 & -1.9795 & 90.7  & 0.5337 & 333.3 & -2.0506 & 83.3  & 0.5221 & 326.7 & -2.1204 & 79.3 \\\cline{2-18}
			& 1.8   & 57.5850 & 250.0 & -57.3600 & 290.0 & 5.4734 & 249.3 & -5.5715 & 289.3 & 2.5557 & 248.0 & -2.7350 & 288.0 & 1.1009 & 247.3 & -1.9864 & 85.3 \\		
			\hline
			\multirow{3}[0]{*}{0.4} & 0.2   & 9.0062 & 360.0 & -8.4540 & 163.3 & 1.7852 & 323.3 & -2.9535 & 94.0  & 1.3415 & 327.3 & -2.8490 & 84.7  & 1.0570 & 327.3 & -3.7585 & 84.7 \\\cline{2-18}
			& 1     & 4.2206 & 248.0 & -4.6468 & 292.7 & 0.4603 & 338.7 & -2.0780 & 91.3  & 0.4483 & 327.3 & -2.1434 & 82.7  & 0.1673 & 324.7 & -3.2651 & 82.7 \\\cline{2-18}
			& 1.8   & 15.7560 & 243.3 & -15.6860 & 298.0 & 1.1678 & 241.3 & -1.9415 & 91.3  & 0.3544 & 240.7 & -2.0519 & 80.7  & 0.0424 & 234.7 & -3.2285 & 80.7 \\
			\hline
			\multirow{3}[0]{*}{0.6} & 0.2   & 10.1751 & 330.0 & -8.9016 & 147.3 & 1.8436 & 331.3 & -2.9399 & 94.0  & 1.3494 & 334.0 & -2.8636 & 85.3  & 0.9447 & 357.3 & -2.5709 & 106.0 \\\cline{2-18}
			& 1     & 1.7999 & 245.3 & -2.7724 & 129.3 & 0.4149 & 334.0 & -2.1571 & 90.7  & 0.4820 & 322.7 & -2.1703 & 81.3  & 0.3773 & 341.3 & -1.4369 & 101.3 \\\cline{2-18}
			& 1.8   & 6.9144 & 237.3 & -7.0066 & 303.3 & 0.2078 & 232.7 & -2.0558 & 90.7  & 0.1172 & 334.0 & -2.1432 & 79.3  & 0.1402 & 348.7 & -1.3052 & 102.7 \\
			\hline
			\multirow{3}[0]{*}{0.8} & 0.2   & 10.6179 & 336.0 & -9.2196 & 184.7 & 1.8620 & 335.3 & -2.9373 & 94.7  & 0.5423 & 318.7 & -2.1620 & 80.7  & 1.1067 & 337.3 & -3.0113 & 82.7 \\\cline{2-18}
			& 1     & 1.5071 & 4.0   & -2.7431 & 134.0 & 0.4479 & 326.7 & -2.1853 & 90.7  & 0.5423 & 318.7 & -2.1620 & 80.7  & 0.5832 & 322.7 & -2.1663 & 78.0 \\\cline{2-18}
			& 1.8   & 3.6135 & 230.0 & -3.9070 & 307.3 & -0.0248 & 337.3 & -2.1366 & 90.7  & 0.1142 & 326.7 & -2.2057 & 77.3  & 0.2813 & 315.3 & -2.1617 & 79.3 \\
			\hline
			\multirow{3}[0]{*}{1} & 0.2   & 10.7997 & 340.0 & -9.4567 & 178.7 & 1.8711 & 338.0 & -2.9408 & 95.3  & 1.3609 & 338.7 & -2.8908 & 86.7  & 1.1081 & 338.7 & -3.0150 & 82.7 \\\cline{2-18}
			& 1     & 1.3846 & 0.0   & -2.7156 & 127.3 & 0.5119 & 323.3 & -2.1720 & 85.3  & 0.5765 & 322.7 & -2.1521 & 80.7  & 1.1081 & 338.7 & -3.0150 & 82.7 \\\cline{2-18}
			& 1.8   & 1.9938 & 226.0 & -2.4010 & 309.3 & -0.0290 & 328.7 & -2.1888 & 81.3  & 0.1674 & 318.7 & -2.2350 & 77.3  & 0.3253 & 311.3 & -2.1612 & 79.3 \\		
			\hline
			\multicolumn{2}{|c|}{} & \multicolumn{16}{c|}{$\alpha=2$} \\
			\hline
			\multirow{3}[0]{*}{0.2} & 0.2   & 8.3354 & 344.0 & -9.2446 & 145.3 & 1.3464 & 326.7 & -5.1605 & 91.3  & 1.0562 & 322.7 & -4.7635 & 90.7  & 0.9345 & 320.0 & -4.5247 & 88.7 \\\cline{2-18}
			& 1     & 26.9520 & 252.0 & -29.6460 & 287.3 & 1.6472 & 242.7 & -4.8244 & 280.7 & 0.4386 & 237.3 & -3.9452 & 90.7  & -0.0119 & 329.3 & -4.0290 & 87.3 \\\cline{2-18}
			& 1.8   & 200.8700 & 250.0 & -201.3400 & 290.0 & 19.4750 & 248.7 & -21.0370 & 289.3 & 9.3241 & 248.0 & -11.2210 & 288.0 & 4.2200 & 246.0 & -6.4592 & 286.7 \\
			\hline
			\multirow{3}[0]{*}{0.4} & 0.2   & 8.7925 & 336.7 & -9.1592 & 147.3 & 1.4765 & 320.0 & -5.0783 & 91.3  & 1.1547 & 312.7 & -4.8249 & 90.7  & 0.9835 & 320.0 & -4.5691 & 89.3 \\\cline{2-18}
			& 1     & 8.1040 & 247.3 & -11.7890 & 290.7 & -0.5063 & 339.3 & -4.1972 & 91.3  & -0.4948 & 334.7 & -4.3217 & 86.0  & -0.3880 & 332.0 & -4.4529 & 86.0 \\\cline{2-18}
			& 1.8   & 52.8709 & 242.0 & -54.6624 & 298.7 & 4.3929 & 240.7 & -7.1340 & 295.3 & 1.5121 & 237.3 & -4.5043 & 86.0  & 0.0878 & 230.0 & -4.4813 & 86.0 \\
			\hline
			\multirow{3}[0]{*}{0.6} & 0.2   & 9.7286 & 326.0 & -9.2987 & 147.3 & 1.6119 & 316.0 & -4.8948 & 91.3  & 1.1940 & 318.0 & -4.9037 & 90.0  & 0.9894 & 325.3 & -4.6088 & 89.3 \\\cline{2-18}
			& 1     & 3.2190 & 242.0 & -7.2584 & 295.3 & -0.8267 & 337.3 & -4.4667 & 90.7  & 1.1940 & 318.0 & -4.9037 & 90.0  & -0.3525 & 318.0 & -4.6121 & 84.0 \\\cline{2-18}
			& 1.8   & 24.4520 & 237.3 & -27.0910 & 303.3 & 1.0188 & 229.3 & -4.9490 & 90.7  & -0.3613 & 229.3 & -4.8357 & 90.7  & -0.9700 & 218.7 & -4.8401 & 83.3 \\
			\hline
			\multirow{3}[0]{*}{0.8} & 0.2   & 10.1157 & 328.7 & -9.4493 & 147.3 & 1.6472 & 320.7 & -4.9328 & 92.0  & 1.2007 & 322.7 & -4.9281 & 90.0  & 0.9930 & 327.3 & -4.6478 & 90.7 \\\cline{2-18}
			& 1     & 1.0984 & 238.7 & -5.2951 & 294.0 & -0.9755 & 333.3 & -4.6384 & 90.7  & -0.6508 & 320.7 & -4.7056 & 83.3  & -0.1709 & 306.0 & -4.5965 & 83.3 \\\cline{2-18}
			& 1.8   & 13.0660 & 230.0 & -16.4580 & 307.3 & -0.4922 & 223.3 & -5.0829 & 82.0  & -1.2219 & 218.7 & -5.1096 & 90.7  & -1.4234 & 207.3 & -5.1602 & 81.3 \\
			\hline
			\multirow{3}[0]{*}{1} & 0.2   & 10.2471 & 332.0 & -9.5838 & 148.0 & 1.6630 & 324.0 & -4.9429 & 93.3  & 1.2049 & 325.3 & -4.9006 & 90.7  & 0.9956 & 328.7 & -4.6737 & 89.3 \\\cline{2-18}
			& 1     & -0.0343 & 237.3 & -4.9303 & 120.7 & -0.9612 & 323.3 & -4.7148 & 83.3  & -0.4773 & 311.3 & -4.7258 & 81.3  & -0.0892 & 300.0 & -4.5813 & 83.3 \\\cline{2-18}
			& 1.8   & 7.4525 & 226.0 & -11.2970 & 310.0 & -1.3064 & 218.7 & -5.3396 & 81.3  & -1.6917 & 210.0 & -5.3666 & 82.0  & -1.5865 & 202.0 & -5.4126 & 81.3 \\
			\hline
		\end{tabular}%
	}
	\label{tab:cp_min-max}%
\end{table}%

\noindent
As the rotational rate ($\alpha$) increases from 0 to 2, the pressure coefficient ($C_p$) exhibits a general rise with increasing power-law index ($n$) but decreases with Reynolds number ($Re$) and gap ratio ($G$). Broadly, $C_{p,\text{max}}$ typically occurs between $300^\circ \le \phi_{h} \le 360^\circ$ for shear-thinning ($n < 1$) fluids and $240^\circ \le \phi_{h} \le 270^\circ$ for shear-thickening ($n > 1$) fluids.  For example, when $Re$ increases from 1 to 40 at $\alpha=2$, the angular location ($\phi_{h}$) of maxima varies by approximately 7\%, 31\%, and 2\% for $n=0.2$, 1, and 1.8 at $G=0.2$ (panel C in \fig\ref{fig:pre-uc0.2}); by 0\%, 31\%, and 8\% at $G=0.6$ (panel C in \fig\ref{fig:pre-uc0.6}); and by 1\%, 26\%, and 11\% at $G=1$ (panel C in \fig\ref{fig:pre-uc1}). Similarly, the angular location ($\phi_{l}$) of minima  shifts by approximately 39\%, 70\%, and 1\% at $G=0.2$; 39\%, 72\%, and 73\% at $G=0.6$; and 40\%, 31\%, and 74\% at $G=1$ for $n=0.2$, 1, and 1.8, respectively. These variations indicate that flow acceleration around the rotating cylinders, influenced by fluid rheology and gap spacing, governs the azimuthal position of $C_p$ extrema, with stronger sensitivity observed for Newtonian and shear-thickening regimes.

\noindent
Table \ref{tab:cp_min-max} presents a detailed analysis of the maximum ($C_{p,\text{max}}$) and minimum ($C_{p,\text{min}}$) pressure coefficients, along with their corresponding angular positions ($\phi_{h}$, and $\phi_{l}$) for UC across different flow conditions ($1\le Re\le 40$, $0\le \alpha\le 2$, $0.2\le G\le 1$, $0.2\le n\le 1.8$). Overall, $C_{p,\text{max}}$ decreases with increasing $Re$ and $\alpha$, while $C_{p,\text{min}}$ shows an increasing trend. At $\alpha=0$, $C_{p,\text{max}}$ decreases and $C_{p,\text{min}}$ increases with increasing power-law index ($n$) and gap ratio ($G$). For example, at $G=0.2$ and $\alpha=0$ (see panel A in \fig\ref{fig:pre-uc0.2}), when $Re$ increases from 1 to 40, $C_{p,\text{max}}$ decreases by approximately 89\%, 63\%, and 40\%, while $C_{p,\text{min}}$ increases by 85\%, 65\%, and 63\% for $n=0.2$, 1, and 1.8, respectively.
For $G=0.2$ and $\alpha=2$ (\fig\ref{fig:pre-uc0.2}.C), as $Re$ increases from 1 to 40, the $C_{p,{\text{max}}}$ decreases by $89\%$, $100\%$, and $98\%$, while $C_{p,{\text{min}}}$ increases by $51\%$, $86\%$, and $97\%$ for $n=0.2$, 1, and 1.8, respectively; similarly, for $G=1$ and $\alpha=2$ (\fig\ref{fig:pre-uc1}.C), the $C_{p,{\text{max}}}$ decreases by $90\%$, $160\%$, and $121\%$, whereas $C_{p,{\text{min}}}$ increases by $51\%$, $7\%$, and $74\%$ for $n=0.2$, 1, and 1.8, respectively. These trends highlight the contrasting rheological behaviors of non-Newtonian fluids, i.e., shear-thinning fluids ($n<1$) exhibit reduced viscosity and resistance with increasing shear rate, resulting in lower pressure coefficients, while shear-thickening fluids ($n>1$) show the opposite behavior, with increased viscosity and higher pressure coefficients.
Further, as the power-law index ($n$) increases, the angular positions ($\phi$) progressively shifts in the direction of cylinder rotation. With increasing $Re$ (from 1 to 40), $\phi_{h}$ shows a $5\%$ increase for $n=0.2$, while it decreases by approximately $8\%$ and $6\%$ for $n=1$ and $n=1.8$, respectively.

\noindent
The analysis also indicates that increasing gap ($G$) from 0.2 to 1 results in a systematic increase in $C_{p,{\text{max}}}$ for $\alpha=0$, irrespective of the power-law index ($n$). However, at a higher rotational rate ($\alpha=2$), $C_{p,{\text{max}}}$ increases with $G$ for shear-thinning fluids ($n< 1$) but decreases for Newtonian and shear-thickening ($n > 1$) fluids. For instance, when $G$ increases from 0.2 to 1, $C_{p,{\text{max}}}$ reduces by $11\%$, $4\%$, and $11\%$ at $Re=1$ (see panel A1 in \fig\ref{fig:pre}) and by $1\%$, $6\%$, and $10\%$ at $Re=40$ (see panel A4 in \fig\ref{fig:pre}) for $n=0.2$, 1, and 1.8, respectively. In contrast, at $\alpha=2$, $C_{p,{\text{max}}}$ increases by $23\%$ for $n=0.2$ but decreases by $100\%$ and $96\%$ for $n=1$ and 1.8 at $Re=1$ (see panel C1 in \fig\ref{fig:pre}), following similar trends at $Re=40$ (see panel C4 in \fig\ref{fig:pre}) as increase by $7\%$, $600\%$, and $138\%$ for $n=0.2$, 1, and 1.8, respectively..
For stationary cylinders, the angular position ($\phi_{h}$) shifts by $5\%$ for $n=0.2$ and $-7\%$ for $n=1$ and $n=1.8$ as $G$ increases from 0.2 to 1 with rising $Re$. At $\alpha=2$, $\phi_{h}$ decreases by $7\%$ for $n=0.2$, increases by $31\%$ for $n=1$, and decreases by $2\%$ for $n=1.8$ over the same range of $G$ and $Re$.
The higher $C_p$ values observed at lower gaps and higher rotations, particularly for shear-thickening fluids ($n=1.8$), stem from enhanced flow confinement and intensified fluid-cylinder interactions. The smaller spacing increases local shear and pressure buildup, while rotation amplifies shear-induced viscosity changes, collectively leading to elevated pressure coefficients.

\noindent
In context of the fluid behavior, it is observed that $C_{p,{\text{max}}}$ increases with decreasing $n$ (from Newtonian to shear-thinning, $n\leq1$) and decreases with increasing $n$ (from Newtonian to shear-thickening, $n\geq1$) for a given $Re$ and $G$ in the stationary cylinder case. For example, at $G=0.2$ and $\alpha=0$ (see panel A in \fig\ref{fig:pre-uc0.2}), $C_{p,{\text{max}}}$ increases by 405\% and 54\% for $n\leq1$, while it decreases by 45\% and 11\% for $n\geq1$ at $Re=1$ and 40, respectively. Similarly, at $G=1$ and $\alpha=0$ (panel A in\fig\ref{fig:pre-uc1}), $C_{p,{\text{max}}}$ increases by 440\% and 45\% for $n\leq1$, but decreases by 41\% and 8\% for $n\geq1$. When the cylinders rotate, these trends reverse. At $G=0.2$ and $\alpha=2$ (panel A in \fig\ref{fig:pre-uc0.2}), $C_{p,{\text{max}}}$ decreases by 69\% and increases by 79 folds for $n\leq1$, while it increases by 6 and 355 folds for $n\geq1$ at $Re=1$ and 40, respectively. For a higher gap ($G=1$) and $\alpha=2$ (panel A in \fig\ref{fig:pre-uc1}), $C_{p,{\text{max}}}$ increases with decreasing $n$ but decreases (except at $Re=1$) with increasing $n$, showing increases of 300 and 12 folds for $n\leq1$ and an increase of 218 folds followed by a 17-fold decrease for $n\geq1$. The opposite behavior is observed for $C_{p,{\text{min}}}$.
The physical explanation lies in the shear-dependent viscosity relationship. For $n\leq1$, viscosity decreases with shear rate, reducing flow resistance and increasing velocity, thereby increasing $C_{p,{\text{max}}}$. In contrast, for $n\geq1$, viscosity increases with shear rate, enhancing resistance and lowering $C_{p,{\text{max}}}$. Cylinder rotation further modifies flow separation, vortex formation, and pressure distribution, leading to reversed $C_{p,{\text{max}}}$ trends compared to the stationary case. Additionally, higher gap ($G$) and rotation rate ($\alpha$) influence boundary layer interaction and vortex strength, significantly affecting the overall pressure and shear distributions around the cylinders.
Overall, the combination of flow parameters (Reynolds number, gap ratio, and rotational rate) influences the forces distributions of the cylinders, leading to various patterns and symmetries in the pressure coefficients across their surfaces.
%
\subsection{Macroscopic Characteristics}
%
\noindent
This section examines the influences the flow behavior index ($n$), Reynolds number ($Re$), rotation rate ($\alpha$), and gap ratio ($G$) on the individual and global force (drag and lift) coefficients ($C_{DF}$, $C_{DP}$, $C_{LF}$, $C_{LF}$, $C_{D}$, $C_{L}$), which are key indicators of macroscopic flow behavior.
%
\subsubsection{Components of Drag Coefficient}
\begin{figure}[!bt]
	\centering
		\subfigure[Friction drag coefficient ($C_{DF}$)] {\includegraphics[width=0.9\linewidth]{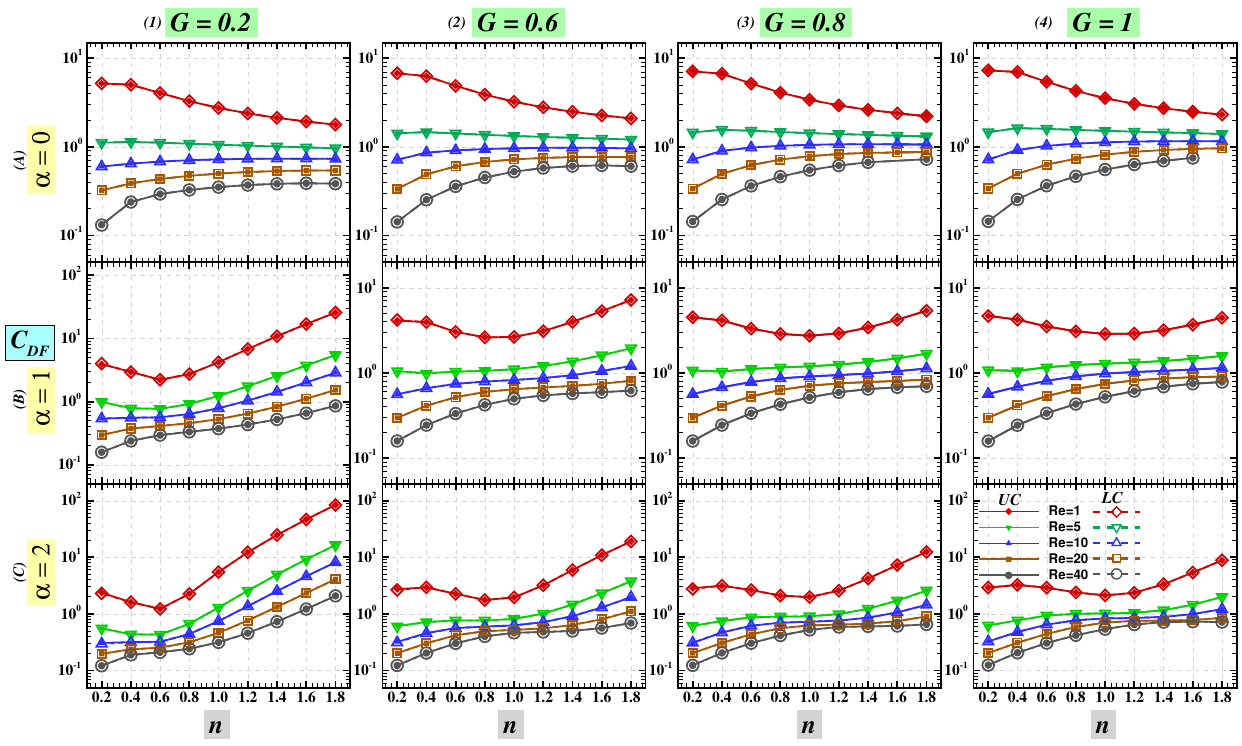}\label{fig:cdf}}
		\subfigure[Pressure drag coefficient ($C_{DP}$)] {\includegraphics[width=0.9\linewidth]{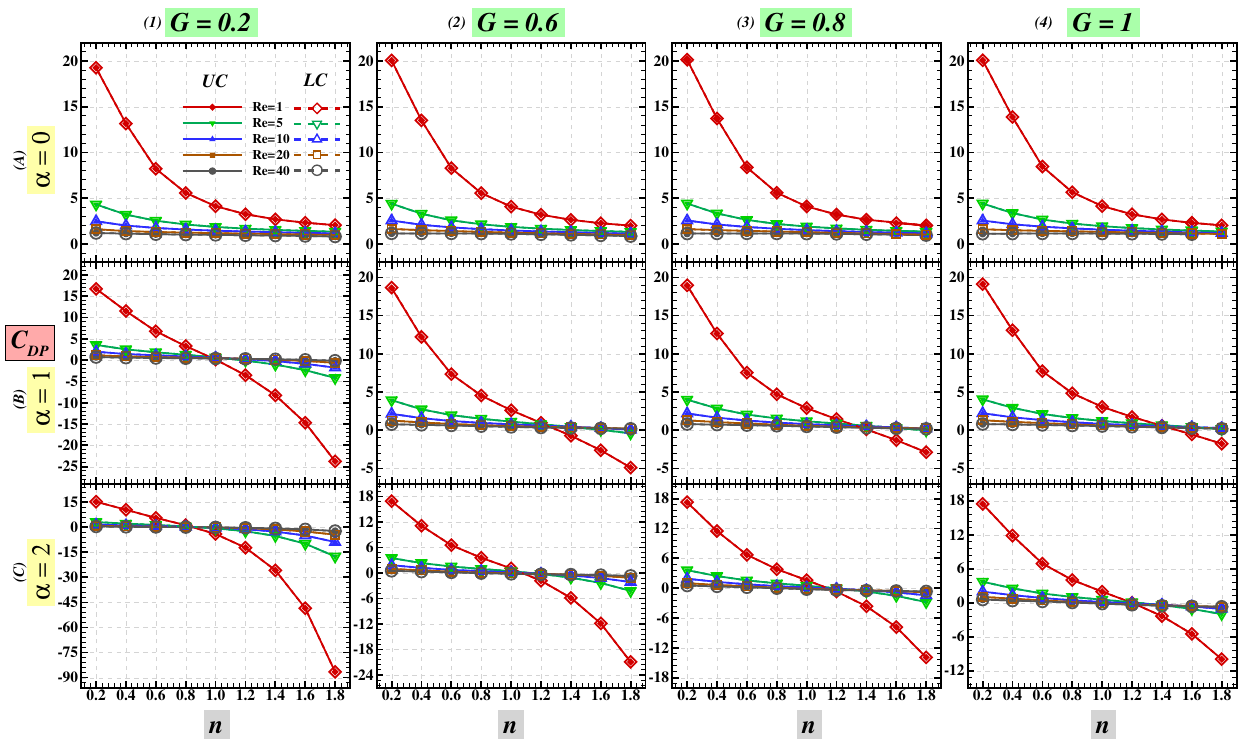}\label{fig:cdp}}
	\caption{Variation of individual drag coefficients ($C_{DP}$ and $C_{DF}$) of the both cylinders ($\rule[0.5ex]{5mm}{.4pt}$ for UC and $\hdashrule[0.5ex]{10mm}{0.4pt}{2mm 1mm}$for LC) with $Re$ and $n$ at different $\alpha$ (in rows) and $G$ (in columns).}\label{fig:cdpcdf}
\end{figure}

\noindent 
\figs\ref{fig:cdpcdf} - \ref{fig:cdfcdp-G} (see also \ref{appendix:Idrag}) illustrate the effects of Reynolds number ($Re$), power-law index ($n$), rotational rate ($\alpha$), and gap ratio ($G$) on the friction and pressure drag coefficients ($C_{DF}$ and $C_{DP}$) for both cylinders, respectively. In these figures, the numerical data for the upper cylinder (UC) are represented by solid lines ($\rule[0.5ex]{5mm}{.4pt}$) with filled markers, while those for the lower cylinder (LC) are shown by dashed lines ($\hdashrule[0.5ex]{10mm}{0.4pt}{1.25mm 1mm}$) with unfilled markers. To avoid redundancy, only the results for the upper cylinder (UC) are discussed, since the upper and lower cylinders exhibit mirror-symmetric behavior.
The comprehensive datasets obtained in this work for the pressure and frictional components of the drag coefficient ($C_{DP}$ and $C_{DF}$) as a function of flow governing parameters ($n$, $Re$, $G$, $\alpha$) are included in \tabs \ref{tab:drag1} to \ref{tab:drag9} in \ref{appendix:dragtable}.
%
%
\subsubsection*{(a) Friction drag coefficient ($C_{DF}$)}
\noindent
The friction drag coefficient ($C_{DF}$) shows a strong dependence on the Reynolds number ($Re$), consistently decreasing with increasing $Re$ across all values of $n$ and $G$; however, distinct behaviors are observed in the low $Re$ ($\leq1$), intermediate $Re$ ($=5$), and high $Re$ ($\geq10$) regimes. The effect of power-law rheology ($n$) on $C_{DF}$ is more pronounced at low and moderate $Re$, which becomes less significant at higher $Re$.
For stationary cylinders ($\alpha=0$) at low $Re$ ($\leq1$), $C_{DF}$ decreases by about 68\% as $n$ increases for all $G$ (see panel A1 in \figs\ref{fig:cdf}). This reduction is attributed to the increase in apparent viscosity with increasing $n$, which enhances resistant to flow and thereby reduces drag, consistent with previous findings \citep{chaitanya2012non,daniel2013aiding}.
However, for rotating cylinders ($\alpha>0$) (see panels B and C in \fig\ref{fig:cdf}) at $Re\leq1$, $C_{DF}$ initially decreases within $0.2 \leq n \leq 0.8$ for $0.2 \leq G \leq 0.6$ and within $0.2 \leq n \leq 1$ for $0.8 \leq G \leq 1$, then increases beyond these ranges. The maximum increase, observed at $\alpha=2$, $Re=1$, and $G=0.2$, is about 34.5 folds as $n$ increases from 0.2 to 1.8 (panel A1 in \fig\ref{fig:cdf}). This occurs because rotation thickens the boundary layer, increasing frictional resistance, initially reducing $C_{DF}$ but later enhancing it as rotational effects dominate.

\noindent
For the intermediate Reynolds number ($Re=5$) at $\alpha=0$, $C_{DF}$ shows a mild increase with $n\leq0.4$, followed by a decrease for $n>0.4$ as $G$ increases from 0.2 to 1. When $\alpha>0$, however, $C_{DF}$ consistently increases with $n$ for all $G \geq 0.4$ (refer panels 2 to 4 in \fig\ref{fig:cdf}). At smaller gaps ($G=0.2$), $C_{DF}$ initially decreases for $n\leq0.6$ and then increases for higher $n$ values (refer panels B1 and C1 in \fig\ref{fig:cdf}), as the flow behavior resembles that around a single cylinder rather than strong cylinder-cylinder interaction, leading to a different trend in the $C_{DF}$ compared to higher $G$ where cylinders interaction is more significant.
At higher $Re$ ($\geq10$) and $\alpha\geq 0$ for all gaps ($G$), $C_{DF}$ increases with $n$, indicating greater frictional resistance with higher viscosity \citep{Chhabra2008}. The highest $C_{DF}$, observed for $G=1$ and $Re=40$ (panel A4 in \fig\ref{fig:cdf}), is about five times larger as $n$ increases from 0.2 to 1.8. As $n$ increases, viscosity and flow resistance increase, enhancing frictional drag. Similarly, increasing $\alpha$ thickens the fluid layer around the cylinders, enhancing viscous forces and $C_{DF}$, though this effect weakens with larger $G$.
%
\begin{figure}[!bt]
	\centering
	\subfigure[Friction drag coefficient ($C_{DF}$)] {\includegraphics[width=0.9\linewidth]{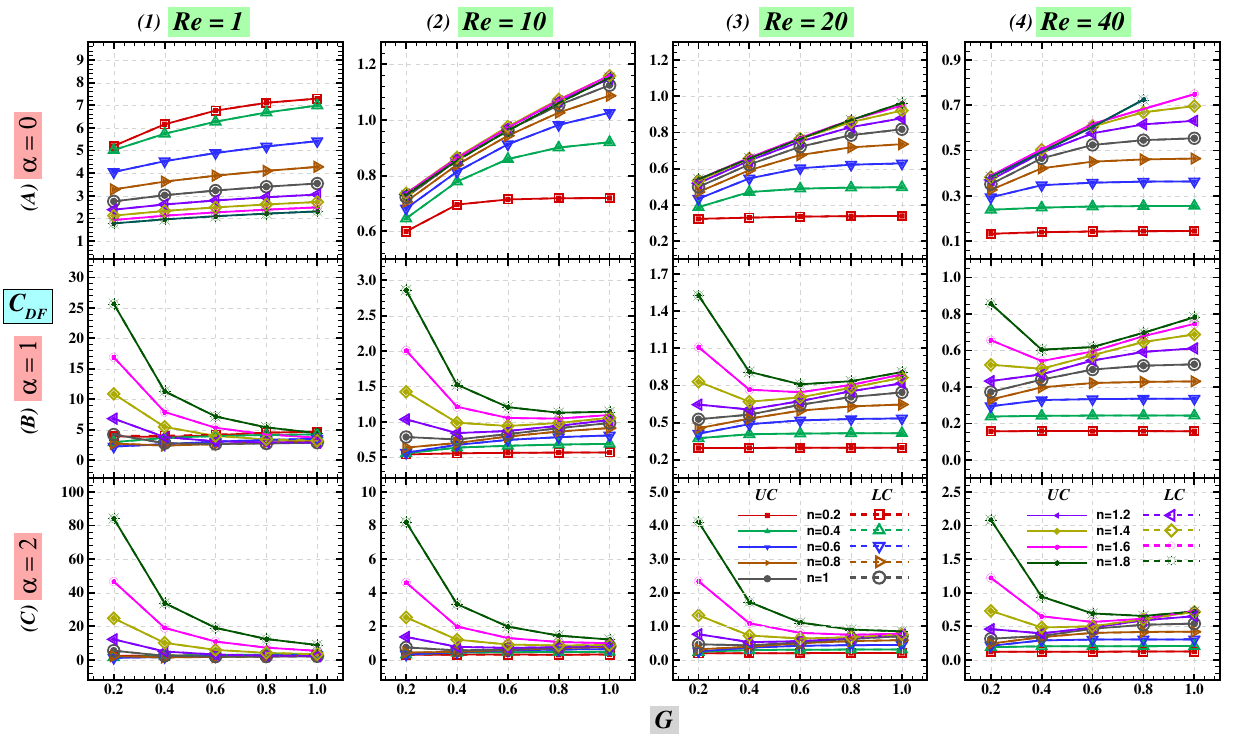}\label{fig:cdf-G}}
	\subfigure[Pressure drag coefficient ($C_{DP}$)] {\includegraphics[width=0.9\linewidth]{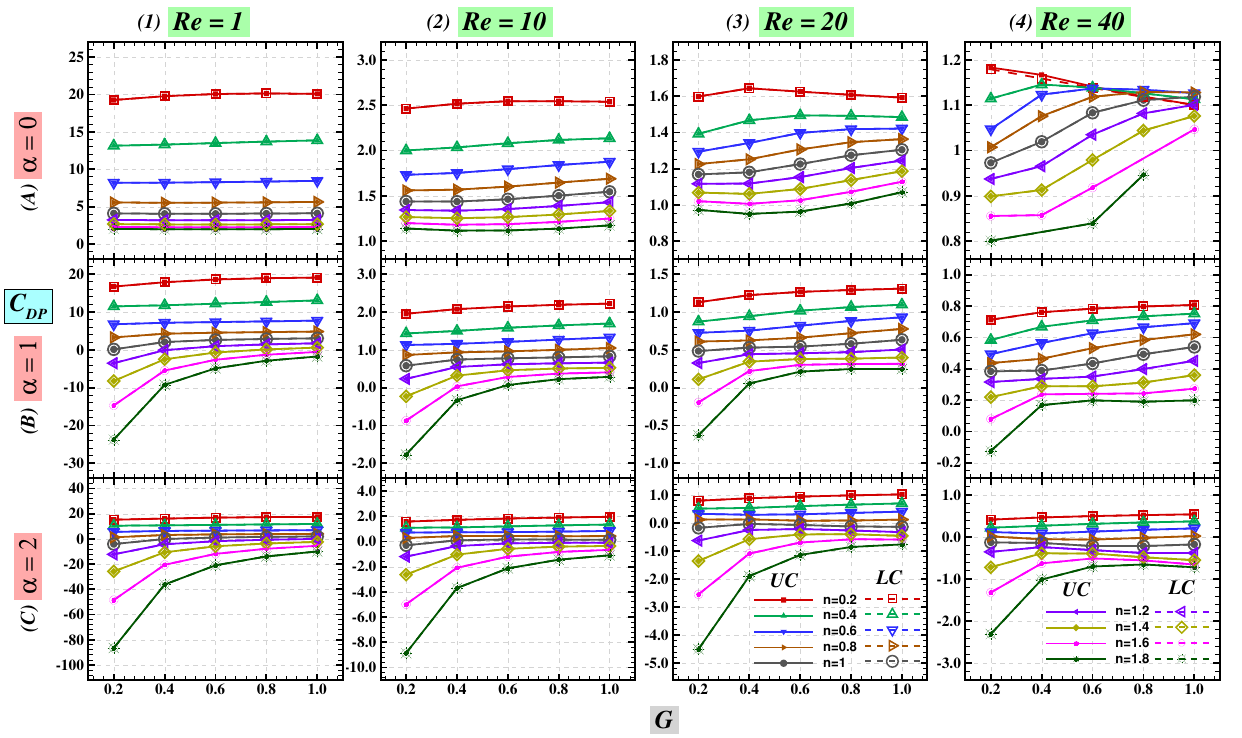}\label{fig:cdp-G}}
	\caption{Variation of individual drag coefficients ($C_{DP}$ and $C_{DF}$) for the both cylinders ($\rule[0.5ex]{5mm}{.4pt}$ for UC and $\hdashrule[0.5ex]{10mm}{0.4pt}{2mm 1mm}$for LC) with $n$ and $G$ at different $\alpha$ (in rows) and $Re$ (in columns).}
	\label{fig:cdfcdp-G}
\end{figure}

\noindent
Regarding the influence of the gap ($G$) between the cylinders on the friction drag coefficient ($C_{DF}$), consistent trends are observed for both stationary and rotating ($\alpha \geq 0$) cylinders. For stationary cylinders ($\alpha=0$), $C_{DF}$ consistently increases with increasing $G$, regardless of $Re$ and $n$. However, the dependence on $n$ shows more complexity, at low Reynolds number ($Re=1$), $C_{DF}$ decreases with increasing $n$, while for $Re>1$, it increases with $n$ (panel A1 in \fig\ref{fig:cdf-G}). Overall, $C_{DF}$ generally increases with increasing $n$. When the cylinders rotate ($\alpha>0$), $C_{DF}$ decreases with $G$ for $Re\leq10$ and $n>1$ (panels 1 and 2 in \fig\ref{fig:cdf-G}); conversely, for shear-thinning fluids ($n\leq1$), it increases with $G$, showing a maximum increase of about 23\% for $Re=1$, $n=0.2$, and $\alpha=2$ (panel C1 in \fig\ref{fig:cdf-G}). This effect, however, weakens as $Re$ increases. For shear-thickening fluids ($n=1.8$), $C_{DF}$ increases with $G$ for $\alpha=0$ at $Re=40$ (panel A4 in \fig\ref{fig:cdf-G}) but decreases by nearly 89\% for $\alpha=2$ at $Re=1$ (panel C4 in \fig\ref{fig:cdf-G}). Additionally, at $\alpha=1$ (for $Re=20-40$) and $\alpha=2$ (for $Re=40$), $C_{DF}$ first decreases and then increases with $G$ for $n>1$ (panels B3 to C4 in \fig\ref{fig:cdf-G}). Overall, larger $G$ tends to reduce friction drag, while the combined effects of $Re$, $n$, and $\alpha$ modulate $C_{DF}$ in complex ways. These interactions significantly affect flow resistance and are crucial for optimizing rotating cylinder systems, particularly in shear-thickening fluids.
\subsubsection*{(b) Pressure drag coefficient ($C_{DP}$)}
\noindent
The pressure drag coefficient ($C_{DP}$) decreases exponentially with increasing $n$ and $Re$, as shown in \fig\ref{fig:cdp}, for all $G$ and $\alpha$. For stationary cylinders ($\alpha=0$) at low Reynolds number ($Re=1$), $C_{DP}$ reduces by about 90\% as $n$ increases from 0.2 to 1.8 across all $G$ (panel A1 in \fig\ref{fig:cdp}); however, for rotating cylinders ($\alpha>0$), $C_{DP}$ exhibits a sign reversal with increasing $n$: it changes from positive to negative at $n=1$ for $G=0.2$ (panels B1 to C1 in \fig\ref{fig:cdp}), at $n=1.2$ for $G=0.6$ (panels B2 to C2 in \fig\ref{fig:cdp}), and at $n=1.4$ for $G\geq0.8$ (panels B3 to C4 in \figs\ref{fig:cdp}). Similar partial transitions occur for $Re=5-10$ with $G\leq 0.6$, and for $Re=5$ with $G\geq 0.8$ at $\alpha=2$. The maximum $C_{DP}$ occurs at $\alpha=2$, $G=0.2$, and $Re=1$, increasing nearly seven-folds as $n$ increases from 0.2 to 1.8 (panel C1 in \fig\ref{fig:cdp}). This behavior arises from the increasing apparent viscosity ($\eta$) with $n$, which enhances flow resistance and alters pressure distribution around the cylinders. The observed `positive-to-negative' transitions in $C_{DP}$ is attributed to shear-thickening layer formation and modified vortex structures induced by cylinder rotation.

\noindent
In contrast, the effect of the gap ($G$) on the pressure drag coefficient ($C_{DP}$) is shown in \fig\ref{fig:cdp-G}, where $C_{DP}$ consistently decreases with increasing $n$ and $Re$ across all $G$ and $\alpha>0$ values. For stationary cylinders ($\alpha=0$) at low Reynolds number ($Re=1$), $C_{DP}$ decreases slightly ($\approx$1\%) in shear-thickening fluids (panel A1 in \fig\ref{fig:cdp-G}), while at $Re=40$, a similar trend occurs only for $n=0.2$, with a larger reduction by about 7\% (panel A4 in \fig\ref{fig:cdp-G}). Interestingly, $C_{DP}$ tends to increase with $G$ (from 0.2 to 1) as $\alpha$ increases, especially in shear-thickening fluids. This effect intensifies at higher rotation rates, indicating stronger fluid-structure interactions. For instance, at $\alpha=1$ and $Re=40$, $C_{DP}$ enhances by about 2.6 times as $G$ increases from 0.2 to 1 for $n=1.8$ (panel B4 in \fig\ref{fig:cdp-G}). Conversely, at $\alpha=2$ and $Re=40$, $C_{DP}$ decreases sharply with $G$, showing up to a 69\% reduction (panel C4 in \fig\ref{fig:cdp-G}).

\noindent
These results underscore the complex interplay among $\alpha$, $G$, $Re$, and $n$ in determining the individual drag coefficients ($C_{DF}$ and $C_{DP}$). Understanding these inter-dependencies is essential for optimizing the performance of rotating cylinder systems, particularly in shear-thickening fluids, where such effects strongly influence overall flow behavior.
%
\subsubsection{Components of Lift Coefficient}
\begin{figure}[!bt]
	\centering
	\subfigure[Friction lift coefficient ($C_{LF}$)] {\includegraphics[width=\linewidth]{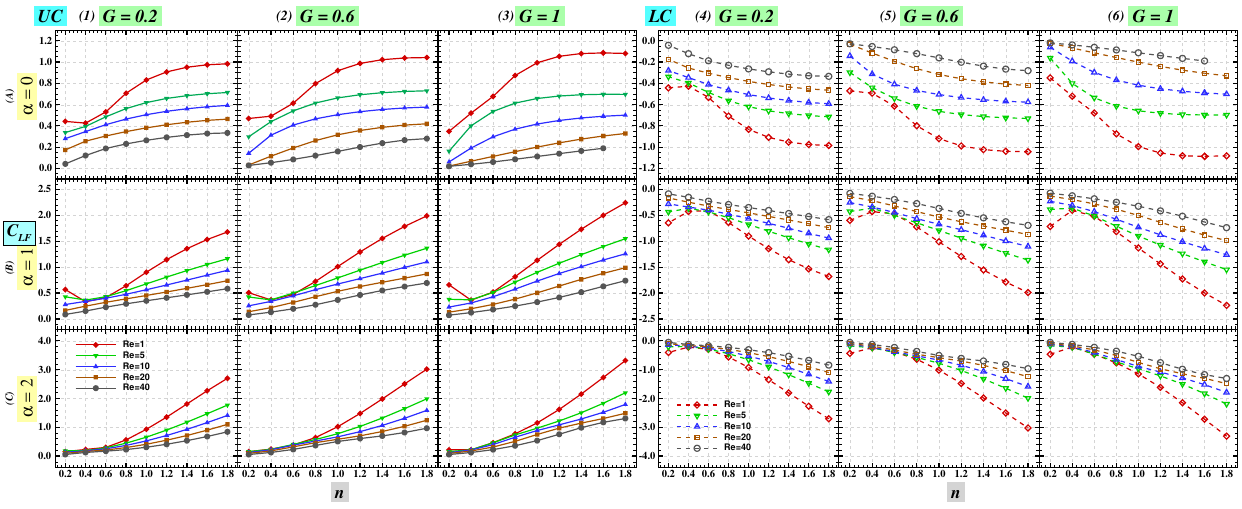}\label{fig:clf}}
	\subfigure[Pressure lift coefficient ($C_{LP}$)] {\includegraphics[width=\linewidth]{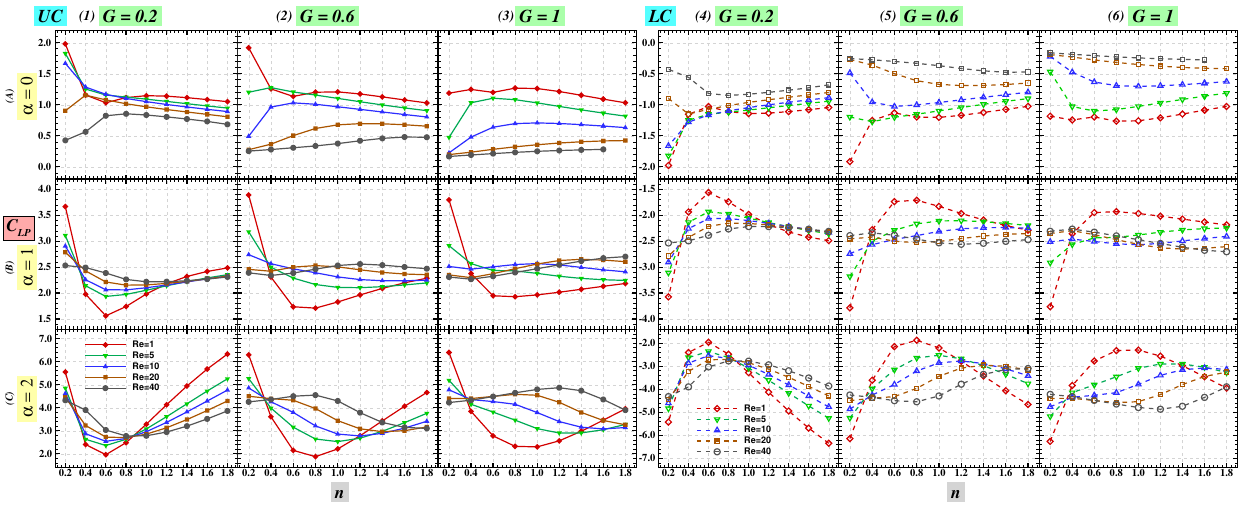}\label{fig:clp}}
	\caption{Variation of individual lift coefficients ($C_{LP}$ and $C_{LF}$) for the both cylinders ($\rule[0.5ex]{5mm}{.4pt}$ for UC and $\hdashrule[0.5ex]{10mm}{0.4pt}{2mm 1mm}$for LC) with $n$ and $Re$ at different $\alpha$ (in rows) and $G$ (in columns).}
	\label{fig:clpf0}
\end{figure}

%
\noindent 
\figs\ref{fig:clpf0} - \ref{fig:clpf_G} (see also \ref{appendix:Ilift}) illustrate the effects of Reynolds number ($Re$), power-law index ($n$), rotational rate ($\alpha$), and gap ratio ($G$) on the friction and pressure lift coefficients ($C_{LF}$ and $C_{LP}$) for both cylinders, respectively. In these figures, the numerical data for the upper cylinder (UC) are represented by solid lines ($\rule[0.5ex]{5mm}{.4pt}$) with filled markers, while those for the lower cylinder (LC) are shown by dashed lines ($\hdashrule[0.5ex]{10mm}{0.4pt}{1.25mm 1mm}$) with unfilled markers.To avoid redundancy, only the results for the upper cylinder (UC) are discussed, since the upper and lower cylinders exhibit mirror-symmetric behavior.
The comprehensive datasets obtained in this work for the pressure and frictional components of the lift coefficient ($C_{LP}$ and $C_{LF}$) as a function of flow governing parameters ($n$, $Re$, $G$, $\alpha$) are included in \tabs \ref{tab:lift1} to \ref{tab:lift9} in \ref{appendix:lifttable}.
%
%
%
\subsubsection*{(a) Friction lift coefficient ($C_{LF}$)}
\noindent
The friction lift coefficient ($C_{LF}$) generally increases with increasing rotational rate ($\alpha\geq0$) for $0.8\leq n\leq1.8$ across all $G$ and $Re$ values, as shown in \fig\ref{fig:clf}. Higher $\alpha$ and $n$ enhance fluid circulation and flow resistance in non-Newtonian fluids, producing stronger lift forces. For $\alpha=0$ at $Re=1$ and $G\leq0.4$ (panel A1 in \fig\ref{fig:clf}), and for $\alpha=1$ at $Re\leq5$ (panel B1 in \fig\ref{fig:clf}), $C_{LF}$ decreases for $n\leq0.4$ due to dominant shear-thinning effects, while for $n > 0.4$, $C_{LF}$ increases as shear-thinning behavior weakens. The increase in apparent viscosity with higher $n$ enhances flow resistance, and thereby, increasing $C_{LF}$. For $\alpha=2$, $C_{LF}$ varies slightly for shear-thinning fluids but significantly for shear-thickening fluids as $Re$ increases from 1 to 40 (panel C1 in \fig\ref{fig:clf}). At $G=0.4$ and $\alpha=2$ with $Re=1$ (refer \fig\ref{fig:clp-clf-g0.4}), $C_{LF}$ enhances nearly 23-fold as $n$ increases from 0.2 to 1.8. Similarly, notable variations occur at $\alpha=2$, $G=1$, and $Re=40$ (panel C3 in \fig\ref{fig:clf}). Overall, shear-thinning fluids exhibit lower lift due to reduced viscosity, while shear-thickening fluids show higher $C_{LF}$ from increased resistance at elevated $Re$ and $\alpha$.
\begin{figure}[!bt]
	\centering
	\subfigure[Friction lift coefficient ($C_{LF}$)]{\includegraphics[width=0.9\linewidth]{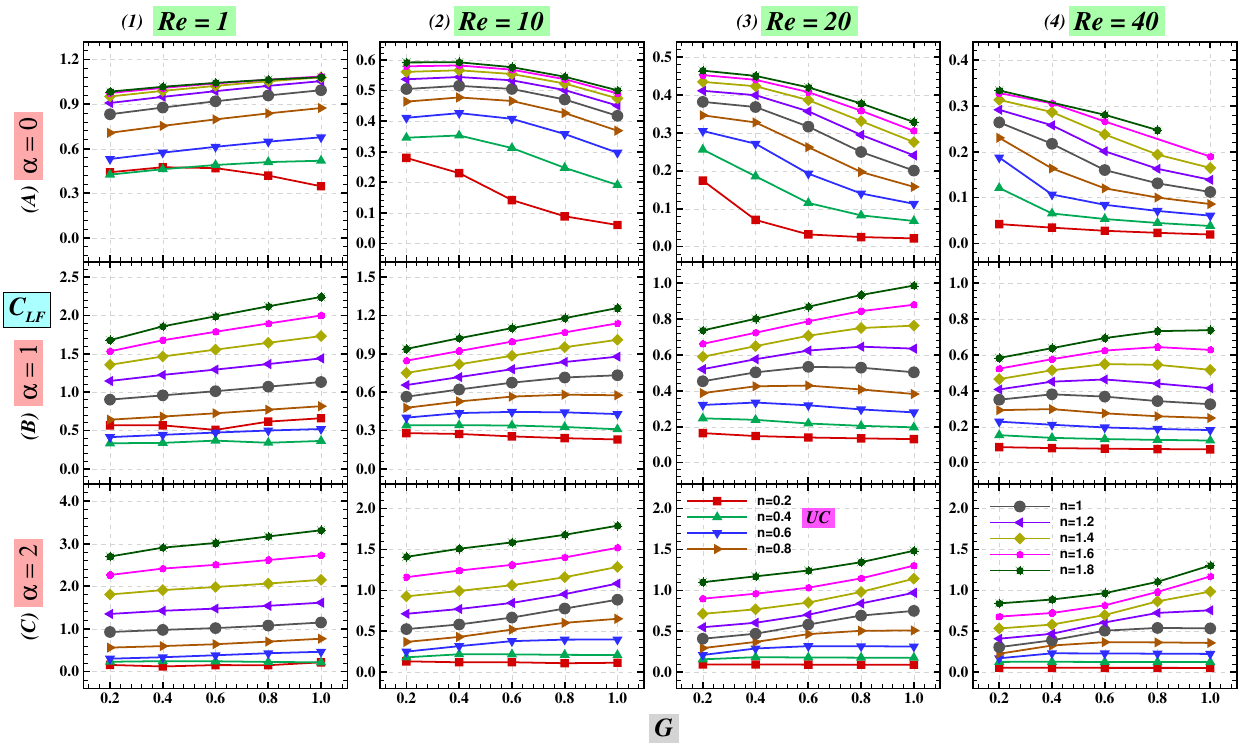}\label{fig:clf_G}}
	\subfigure[Pressure lift coefficient ($C_{LP}$)]{\includegraphics[width=0.9\linewidth]{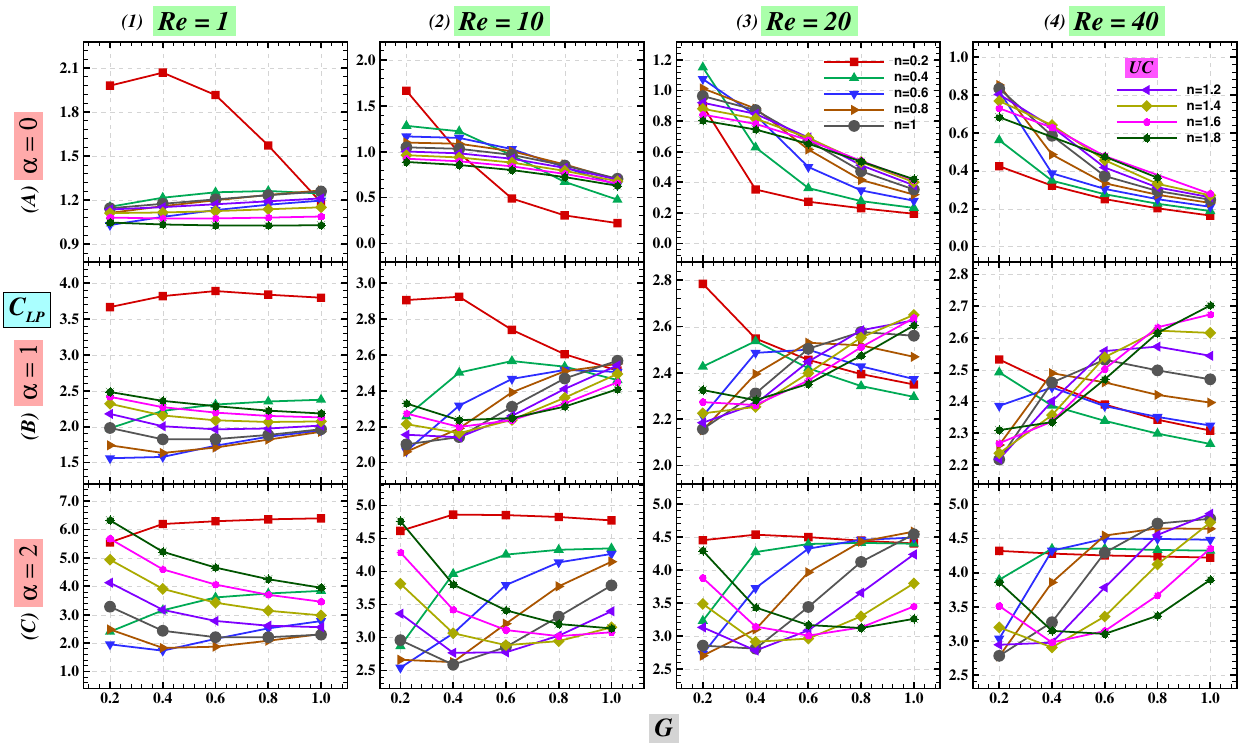}\label{fig:clp_G}}
	\caption{Variation of individual drag coefficients ($C_{DP}$ and $C_{DF}$) for the both cylinders ($\rule[0.5ex]{5mm}{.4pt}$ for UC and $\hdashrule[0.5ex]{10mm}{0.4pt}{2mm 1mm}$for LC) with $n$ and $G$ at different $\alpha$ (in rows) and $Re$ (in columns).}
	\label{fig:clpf_G}
\end{figure}

\noindent
In \fig\ref{fig:clf_G}, the influence of the gap ratio ($G$) on $C_{LF}$ is illustrated, showing its dependence on $n$, $\alpha$, and $Re$. Only the upper cylinder (UC) results are presented due to mirror symmetry; corresponding lower cylinder (LC) results are shown in \fig\ref{fig:clpf_G-LC}. Broadly, $C_{LF}$ decreases with increasing $G$ and $Re$ for all $n$ and $\alpha$, mainly due to enhanced flow stabilization, reduced separation, and thinner boundary layers at higher $G$ and $Re$. For $\alpha=0$, $C_{LF}$ decreases with increasing $G$ for $Re\geq10$ and all $n$, reducing maximally by 88\% at $Re=20$, $n=0.2$ (panel A3 in \fig\ref{fig:clf_G}). At $Re=1$, the reduction is about 21\% for $n=0.2$, while a slight 10\% increase is seen for $n=1.8$ (panel A1 in \fig\ref{fig:clf_G}) due to enhanced flow separation at larger gaps. At $\alpha=1$, $C_{LF}$ increases with increasing $G$ for $Re=1$ across all $n$, with a 34\% maximum increase at $n=1.8$ (panels B1 to B3 in \fig\ref{fig:clf_G}). Similar upward trends occur for $5\leq Re\leq10$ and $n\geq0.6$, and for $Re\geq20$ with $n>1$ (panels B2 to B4 in \fig\ref{fig:clf_G}). At $\alpha=2$, $C_{LF}$ increases with increases $G$ for $Re=1$, reaching a 54\% increase at $n=0.6$ (panel C1 in \fig\ref{fig:clf_G}). For $Re\geq5$ and $n\geq0.4$, the maximum increase is 85\% at $Re=40$, $n=1.2$ (panels C2 to C4 in \fig\ref{fig:clf_G}). Thus, higher rotation enhances lift generation, especially at larger gap ($G$).
\subsubsection*{(b) Pressure lift coefficient ($C_{LP}$)}
\noindent
The pressure lift coefficient ($C_{LP}$) exhibits a complex dependence on the flow parameters ($G$, $Re$, $\alpha$, $n$) in \figs\ref{fig:clp}. For stationary cylinders ($\alpha=0$) and $Re=1$, $C_{LP}$ initially decreases with increasing $n\leq0.6$ (for $G\leq0.8$) (Panels A1 to A2 in \fig\ref{fig:clp} and in \fig\ref{fig:clp-clf-g0.8}), then increases up to $n=1.2$ (for $G\leq0.4$) and $n=1$ (for $0.6\leq G\leq0.8$), followed by another decline. Overall, $C_{LP}$ varies by about 47\% as $n$ increases from 0.2 to 1.8 for $G\leq0.8$. 
For $G=1$, $C_{LP}$ increases with $n\leq0.4$, decreases at $n=0.6$, increases again near $n=1$, and finally reduces at higher $n$ (panel A3 in \fig\ref{fig:clp}). At $Re=5$ and $G\leq0.4$, as well as $Re=10$ and $G=0.2$, $C_{LP}$ decreases with increasing $n$ (panel A1 in \fig\ref{fig:clp}), primarily due to enhanced vortex shedding and stronger flow separation (\figs\ref{fig:R-1} to \ref{fig:g1}), which lower pressure lift ($C_{LP}$). For higher $Re$, $C_{LP}$ shows an interesting non-monotonic trend, i.e., initially enhancing with increasing $n$ (up to a threshold) before declining. This occurs, for instance, at $n\leq0.4$ for $Re=20$ (at $G=0.2$), $Re=10$ (at $G=0.4$), and $Re=5$ (at $G=0.6$); beyond these threshold points, $C_{LP}$ slightly decreases (panel A in \fig\ref{fig:clp}). The increase in $C_{LP}$ with $n$ reflects shear-thickening behavior, where higher $n$ enhances viscosity and lift due to increased flow resistance. The subsequent decline results from competing effects of viscosity, flow inertia, and geometric spacing. When $n$ increases from 0.2 to 1.8, $C_{LP}$ generally rises for $Re=20$ (at $G\geq0.8$) and for $Re=40$ (at $G\geq0.4)$, where larger gap ($G$) weaken cylinder interactions and promote stronger individual vortex formation, leading to elevated pressure lift

\noindent
Further, the pressure lift coefficient ($C_{LP}$) increases with rotations ($\alpha>0$) and remains relatively stable in highly shear-thinning ($n=0.2$) to shear-thickening ($n=1.8$) fluids for all $G$ and $Re$ (panels B to C in \fig\ref{fig:clp}). However, for intermediate $n$ values, $C_{LP}$ decreases with increasing $G$ across all $Re$. The increasing $\alpha$ generates a thin high-velocity fluid layer around the cylinders, enhancing lift through the Magnus effect, which increases $C_{LP}$. Broadly, three distinct trends are observed in  lift coefficient ($C_{LP}$) curves for $\alpha>0$ in \fig\ref{fig:clp} as follows. (i) In the first trend, $C_{LP}$ initially decreases with increasing $n$ up to a critical value, beyond which it rises. For example, when $\alpha\geq1$, $C_{LP}$ decreases for $n<0.6$ at $G=0.2$ and $Re\leq10$, and for $n<0.8$ at $Re\leq40$ (panels B1 to C1 in \fig\ref{fig:clp}). Similar patterns occur at $G=0.6$ and $G=1$ with critical $n$ values shifting higher as $Re$ increases (panels B2 to C3 in \fig\ref{fig:clp}). At $\alpha=1$, $G=0.2$, and $Re=1$, $C_{LP}$ decreases by about 32\% as $n$ increases from 0.2 to 1.8, whereas for $\alpha=2$ it increases by approximately 14\%. These variations arise from changes in fluid resistance and the stability of the thin layer around the cylinders. (ii) The second trend is non-monotonic: $C_{LP}$ decreases initially and then increases, or vice versa. For instance, at $\alpha=1$ and $G\geq0.6$, $C_{LP}$ decreases for $n\leq0.4$, then increases for $0.4\leq n\leq1.2$, and finally decreases again (panels B2 to B3 in \fig\ref{fig:clp})  at $Re\geq20$. This behavior results from competing effects of shear-thickening viscosity and flow separation. (iii) In the third trend, for $\alpha=2$ and $G\geq0.6$, $C_{LP}$ increases with $n<0.8$ at $Re=40$ ($G=0.6$) and $Re=20$ ($G=1$), but decreases beyond these $n$ values (panels C2 to C3 in \fig\ref{fig:clp}). These changes are governed by the interaction between rotation, gap width, and non-Newtonian rheology, which modulate the flow structure, shear rate, and pressure distribution around the cylinders.

\noindent
Furthermore, for stationary ($\alpha=0$) cylinders, $C_{LP}$ generally decreases with increasing $G$ (from 0.2 to 1) for $Re\geq5$ and all values of $n$ (refer \fig\ref{fig:clp_G} for UC and \fig\ref{fig:clpf_G-LC} for LC). However, a distinct trend appears at $Re=1$, i.e., for $n=0.2$ and $1.8$, $C_{LP}$ decreases, whereas for $0.4\leq n\leq1.6$, it increases with $G$ (panel A1 in \fig\ref{fig:clp_G}). This behavior can be attributed to the expanded flow passage and enhanced fluid recirculation in wider gaps, which collectively reduce the net lift on the cylinders. At $\alpha=0$ and $Re=10$, $C_{LP}$ shows a significant reduction of about 87\% as $G$ increases, particularly for the strongly shear-thinning case ($n=0.2$) (panel A2 in \fig\ref{fig:clp_G}). With increasing $\alpha$, the variation of $C_{LP}$ becomes increasingly complex across the range $0.2\leq G\leq1$. The rotational motion alters the vortex-shedding dynamics, and as $G$ increases, the flow transitions from a single attached vortex to separated twin vortices between the cylinders. For shear-thinning fluids, $C_{LP}$ first increases with $G$ and then decreases for higher $n$ values at $Re=1$ ($\alpha=1$) (panel B1 in \fig\ref{fig:clp_G}) and $Re\leq10$ ($\alpha=2$) (panels B1 to B2 in \fig\ref{fig:clp_G}). Similarly, for $Re=5 - 10$ ($\alpha=1$) and $Re\geq20$ ($\alpha=2$) (panels B3 to B4 in \fig\ref{fig:clp_G}), $C_{LP}$ decreases with increasing $G$ for $n=0.2$ before subsequently increasing. At $\alpha=1$, $C_{LP}$ decreases with increasing $G$ by approximately 16\% for $Re=20$ and $n\leq0.4$ (panel B3 in \fig\ref{fig:clp_G}), and by about 9\% for $Re=40$ and $n\leq0.6$ (panel B4 in \fig\ref{fig:clp_G}), followed again by an upward trend. At higher rotation rates ($\alpha=2$), $C_{LP}$ demonstrates contrasting behaviors depending on $Re$ and $n$. For instance, at $Re=1$, $C_{LP}$ decreases markedly by about 40\% as $G$ increases, particularly at $n=1.4$ (panel C1 in \fig\ref{fig:clp_G}). Conversely, for $Re=20$, it increases by nearly 70\% at $n=0.6$ (panel C3 in \fig\ref{fig:clp_G}). Overall, these findings emphasize the intricate coupling between gap geometry, rotational effects, flow inertia, and non-Newtonian rheology in shaping the pressure lift behavior of the cylinders.
\begin{figure}[!hbtp]
	\centering
	\subfigure[Total drag coefficient ($C_D$) for upper and lower cylinders] {\includegraphics[width=0.95\linewidth]{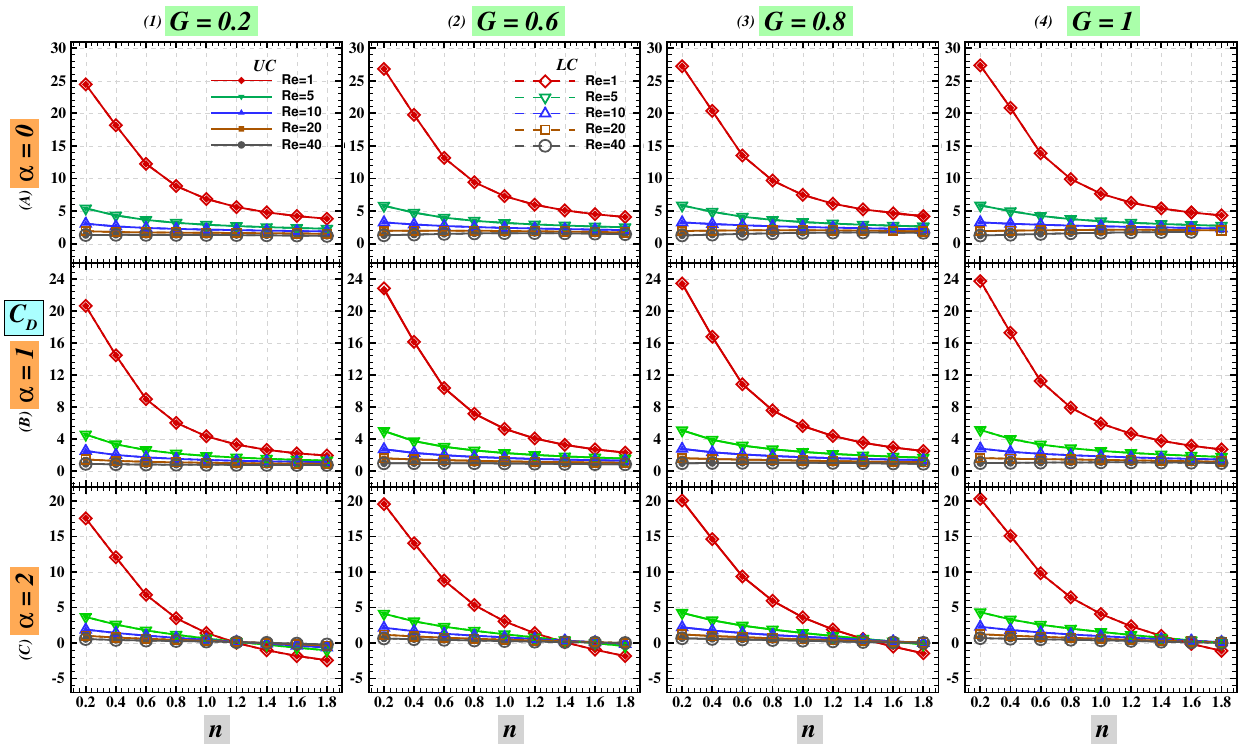}	\label{fig:cd}}
	\subfigure[Total lift coefficient ($C_L$) for upper cylinder] {\includegraphics[width=0.95\linewidth]{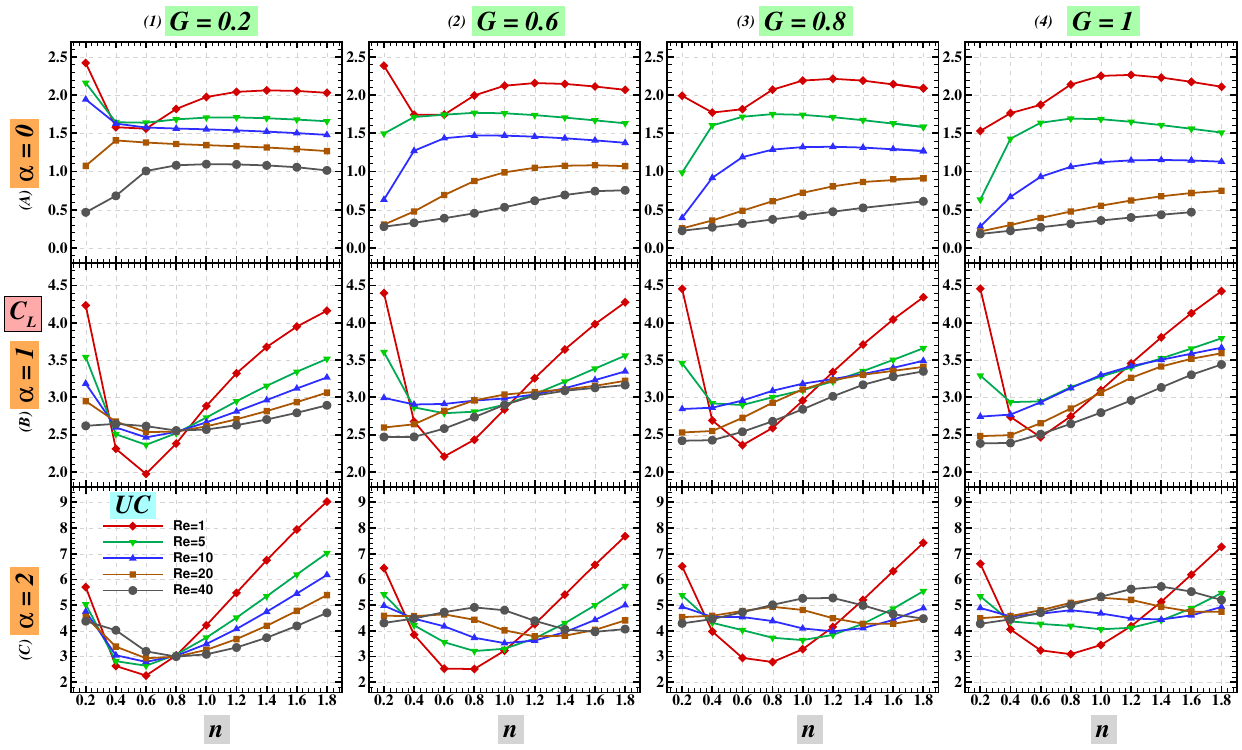}\label{fig:cl}}
	\caption{Variation of total drag and lift coefficient ($C_{D}$ and $C_{L}$) for $Re$ with $n$ at different $\alpha$ (in row) and $G$ (in column).}
	\label{fig:cdclg}
\end{figure}
%
\subsubsection{Total force coefficients over the cylinders}
\noindent
\figs\ref{fig:cdclg} - \ref{fig:cdcl-G-Re1-40} (and \ref{appendix:Tdragnlift}) illustrate the variations of total drag and lift coefficients ($C_D=C_{DP}+C_{DF}$, $C_L=C_{LP}+C_{LF}$) with Reynolds number ($Re$) and power-law index ($n$) for different gap ratios ($G$) and rotational rates ($\alpha$). Solid lines ($\rule[0.5ex]{5mm}{.4pt}$) with filled markers represent the data for the upper cylinder (UC), while dashed lines ($\hdashrule[0.5ex]{12mm}{0.4pt}{1.5mm 1mm}$) with unfilled markers denote for the lower cylinder (LC).
\subsubsection*{(a) Total drag coefficient ($C_D$)}
\noindent
The drag coefficient ($C_D$) for both cylinders broadly exhibit close similarity. The variation of $C_D$ with $n$ (\fig\ref{fig:cd}) closely resembles that of $C_{DP}$ (panel 2 in \fig\ref{fig:cdpcdf}), consistent with earlier studies \citep{chaitanya2012non,daniel2013aiding,Panda2017}. The contribution of $C_{DF}$ shifts the crossover point toward higher $Re$. Moreover, as $n$ increases, lower $C_D$ values are observed at higher $Re$ \citep{Panda2017}. 

\noindent
Generally, for $\alpha\leq1$, $C_D$ decreases with increasing $Re$ and increases with $G$ for all $n$ (panels A and B in \fig\ref{fig:cd}). For $\alpha=0$, $C_D$ decreases exponentially with increasing $n$ across all $G$ for $Re\leq20$, leading to a substantial reduction of about 85\% at $Re=1$ (panel A in \fig\ref{fig:cd}).
However, at $Re=40$, a more complex dependence on $G$ is observed. For example, as $n$ increases from 0.2 to 1.8, $C_D$ decreases by approximately 10\% at $G=0.2$ (panel A1 in \fig\ref{fig:cd}) but increases by 44\% at $G=1$ (panel A4 in \fig\ref{fig:cd}). For intermediate gaps ($0.4\leq G\leq0.8$), $C_D$ first rises up to a certain $n$ and then declines beyond it (panels A2 to A3 in \fig\ref{fig:cd}). Remarkably, at higher rotations ($\alpha=2$), $C_D$ follows a similar trend to lower $\alpha$, but becomes negative ($C_D < 0$) for $n\geq1.4$ when $G\leq0.6$ at all $Re$ (panels C1 to C2 in \fig\ref{fig:cdclg}), and for $n\geq1.6$ at low $Re$ ($\leq5$) when $G\geq0.8$ (panels C3 to C4 in \fig\ref{fig:cdclg}). At sufficiently high rotation speeds ($\alpha=2$), the shear layers can reattach to the cylinder surface, causing wake deflection or even suppression, which results in the formation of bi-stable wakes. This wake asymmetry induces a streamwise thrust, analogous to the Magnus-type propulsion effect, and can consequently yield negative drag ($C_D < 0$). The occurrence of negative drag ($C_D < 0$) is primarily attributed to the pressure distribution around the cylinders rather than viscous shear effects; in such cases, the pressure-induced thrust surpasses the viscous resistance, leading to a net negative drag force. This observation is consistent with earlier studies \citep{meneghini2001numerical,mittal2003flow} conducted for Newtonian fluids ($n=1$), where similar wake suppression and thrust generation phenomena were reported at high rotation rates.
\begin{figure}[!bt]
	\centering
	\subfigure[Total drag coefficient ($C_{D}$) for upper and lower cylinders] {\includegraphics[width=0.95\linewidth]{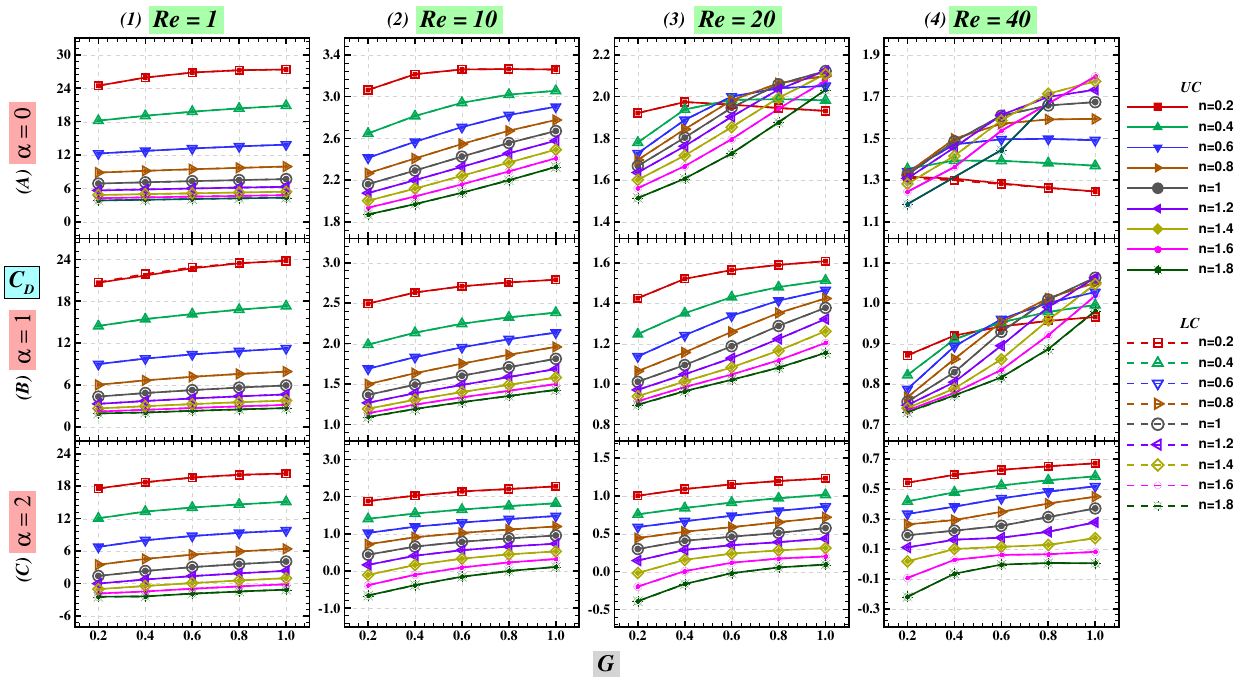}\label{fig:cd_G}}
	\subfigure[Total lift coefficient ($C_{L}$) for upper cylinder] {\includegraphics[width=0.95\linewidth]{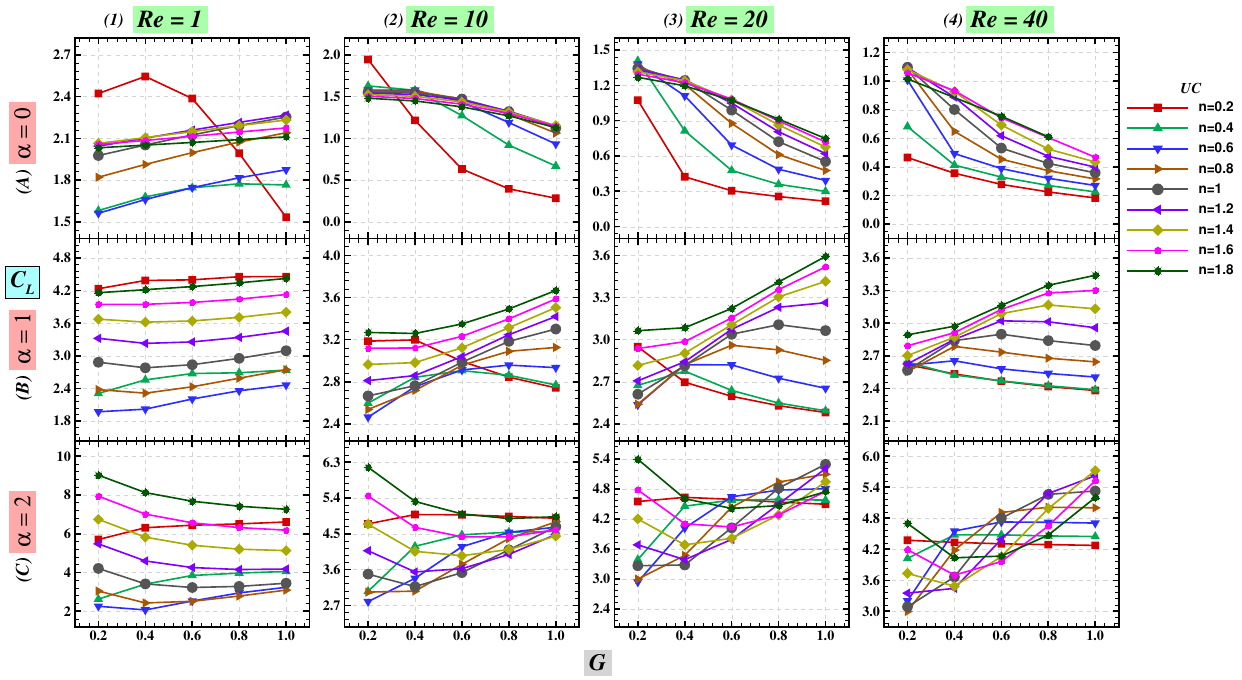}\label{fig:cl_g}}
	\caption{Variation of total drag and lift coefficient ($C_{D}$ and $C_{L}$) for $n$ with $G$ at different $\alpha$ (in rows) and $Re$ (in columns).}
	\label{fig:cdcl-G-Re1-40}
\end{figure}

\noindent
Focusing on the influence of $G$ (\fig\ref{fig:cd_G}), $C_D$ generally increases with $G$ but decreases with increasing other parameters ($n$, $\alpha$, and $Re$), except for $\alpha=0$ (at $Re\geq20$) and $\alpha=1$ (at $Re=40$). For $Re=20$ and $\alpha=0$, $C_D$ initially increases with $G\leq0.4$ for $n\leq0.4$, then decreases for $G>0.4$ (panel A3 in \fig\ref{fig:cd_G}). Similarly, at $Re=40$, $C_D$ decreases with $G$ for strongly shear-thinning ($n\leq0.6$) fluids but increases for weakly shear-thinning to shear-thickening ($n>0.8$) fluids (panel A4 in \fig\ref{fig:cd_G}). Notably, at $\alpha\le 1$ and $G=1$, $C_D$ increases with $n$, particularly for shear-thinning ($n<1$) fluids (panel B4 in \fig\ref{fig:cd_G}). These complex dependencies underscore the sensitivity of drag coefficinet ($C_D$) to flow governing parameters ($G$, $Re$, $\alpha$, and $n$), resulting in diverse hydrodynamic behaviors across flow regimes.
Furthermore, as $\alpha$ increases, $C_D$ consistently decreases with $G$ for all $Re$ and $n$. This is attributed to the larger surface area exposed to the flow at higher $G$, which modifies boundary-layer behavior and drag distribution.
%
\subsubsection*{(b) Total lift coefficient ($C_L$)}
%
\noindent
The total lift coefficient ($C_L$) exhibits opposite signs for the upper and lower cylinders due to symmetry. The variation of $C_L$ with $n$ (\fig\ref{fig:cl} for UC and \fig\ref{fig:cdclg-LC}(a) for LC) resembles that of $C_{LP}$ (\fig\ref{fig:clp}). In general, $C_L$ decreases with increasing $Re$ at both low ($n=0.2$) and high ($n=1.8$) power-law indices, whereas it increases with $\alpha$ for all $G$ and $Re$. Complex transitions are observed at intermediate $n$ due to shifts in rheological behavior from shear-thinning ($n<1$) to Newtonian ($n=1$) and then to shear-thickening ($n>1$).
For stationary cylinders ($\alpha=0$), $C_L$ initially decreases (up to 36\%) with increasing $n\leq0.6$ for $Re\leq5$ and $G\leq0.4$, then increases up to $n=1.4$ up to 32\% (for $G=0.2$) and by  27\% (for $G=0.4$) at $Re=1$, before decreasing again (panels A1 to A2 in \fig\ref{fig:cl}). At $Re=10$, $C_L$ decreases by 24\% with increasing $n$ (from 0.2 to 1.8) for $G=0.2$ (panel A1 in \fig\ref{fig:cl}). Further, $C_L$ increases with $n$ for $0.4\leq G\leq0.6$ at $Re=40$ and for $G\geq0.8$ at $Re\geq20$ (panel A in \fig\ref{fig:cl}). A similar pattern appears at $\alpha=1$, where $C_L$ rises with $n$ for $G=0.4$ at $Re=40$, $G=0.6$ at $Re\geq20$, and $G\geq0.8$ at $Re\geq10$, showing a maximum 44\% increase at $Re=40$ (panel B4 in \fig\ref{fig:cl}).
At lower $Re$ ($\leq10$), viscous effects dominate, and increasing $n$ enhances flow resistance, thus reducing lift ($C_L$). On the other hand, at higher $Re$ ($\geq20$), inertial effects prevail, leading to stronger separation, vortex shedding, and consequently higher lift coefficient ($C_L$).

\noindent
For rotating cylinders ($\alpha\geq1$), $C_L$ often decreases with $n$ up to a threshold (e.g., $n\leq0.6$ for $Re\leq20$ at $G=0.2$), then increases thereafter. The largest decreases ($\sim$ 60\%) and subsequent increases (about 3-fold) occur at $\alpha=2$, $Re=1$ (panel C1 in \fig\ref{fig:cl}). Similar multi-phase complex trends are observed at $\alpha=1$ (for $G=0.4$, $Re=20$) and $\alpha=2$ (for $G=0.4$, $Re=40$) (panels B2 to B3 in \fig\ref{fig:cdcl-g0.4}). For $\alpha=2$, $G=1$, and $Re=10$, $C_L$ decreases by 7\% at $n\leq0.4$, increases by 5\% at $n=0.8$, decreases again by 8\% at $n=1.4$, and rises by 11\% at $n=1.8$ (panel C4 in \fig\ref{fig:cl}). Broadly, the combined effects of higher $Re$, $G$, and $\alpha$ produce a thinner fluid layer around the cylinders, enhancing $C_L$ for shear-thickening ($n>1$) fluids.

\noindent
Regarding the effect of $G$ (\fig\ref{fig:cl_g} for UC and \fig\ref{fig:cdclg-LC}(b) for LC), at $\alpha=0$, $C_L$ decreases with increasing $G$ for $Re\geq5$ and all $n$, and also at $Re=1$, $n=0.2$ (panel A1 in \fig\ref{fig:cl_g}). The largest reduction (about 86\%) occurs at $Re=10$, $n=0.2$ (panel A2 in \fig\ref{fig:cl_g}). Conversely, at $Re=1$ and $n\geq0.4$, $C_L$ increases with $G$.
At $\alpha=1$, $C_L$ increases with $G$ for all $n$ (at $Re=1$) and for $n\geq0.6$ (at $Re\geq5$), reaching a 25\% rise at $Re=1$, $n=0.6$ (panel B1 in \fig\ref{fig:cl_g}). For $n=0.2$, $C_L$ decreases with $G$, the maximum reduction (about 16\%) occurs at $Re=20$ (panel B3 in \fig\ref{fig:cl_g}). Similar declines in $C_L$ occur for $n\leq0.4$ (at $Re=20$) and $n\leq0.6$ (at $Re=40$) (panels B3 to B4 in \fig\ref{fig:cl_g}).
At $\alpha=2$, $C_L$ increases with $G$ for $Re\leq5$ (for $n\leq1$), $Re=10$ (for $n\leq1.2$), $Re=20$ (for $0.4\leq n\leq1.4$), and $Re=40$ (for $n\geq0.4$); for other $n$, it decreases. The maximum increase (73\%) is recorded at $Re=40$, $n=1$ (panel C4 in \fig\ref{fig:cl_g}), while the largest reduction (24\%) occurs at $Re=1$, $n=1.2$ (panel C1 in \fig\ref{fig:cl_g}).
Overall, the variation of $C_L$ with $G$ reflects the intricate coupling of $Re$, $n$, and $\alpha$. Changes in $G$ modify vortex-shedding structures and inter-cylinder flow, which in turn alter lift generation. These findings highlight the complex interactions governing the aerodynamic characteristics of rotating cylinder pairs in non-Newtonian fluids.
\begin{figure}[!bt]
	\centering
	\subfigure[Drag ratio ($C_{DR}$)]{\includegraphics[width=0.8\linewidth]{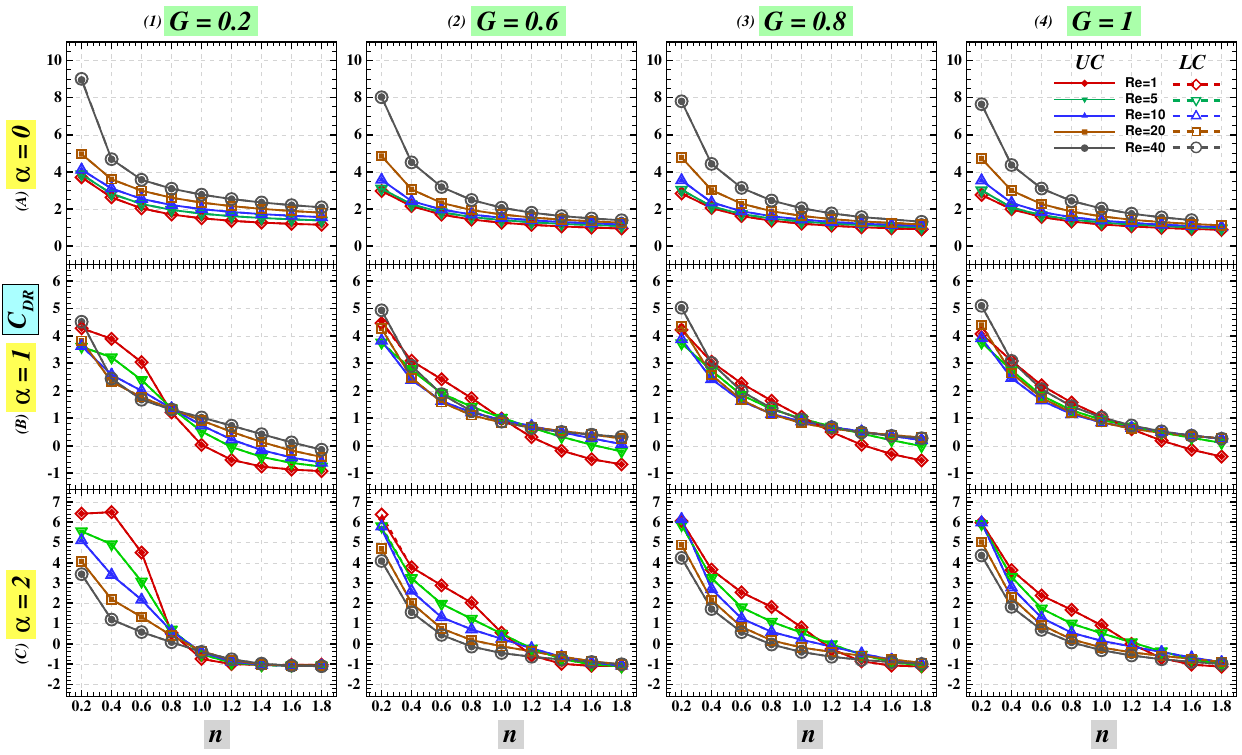}\label{fig:cdr}}
	\subfigure[Lift ratio ($C_{LR}$)]{\includegraphics[width=0.8\linewidth]{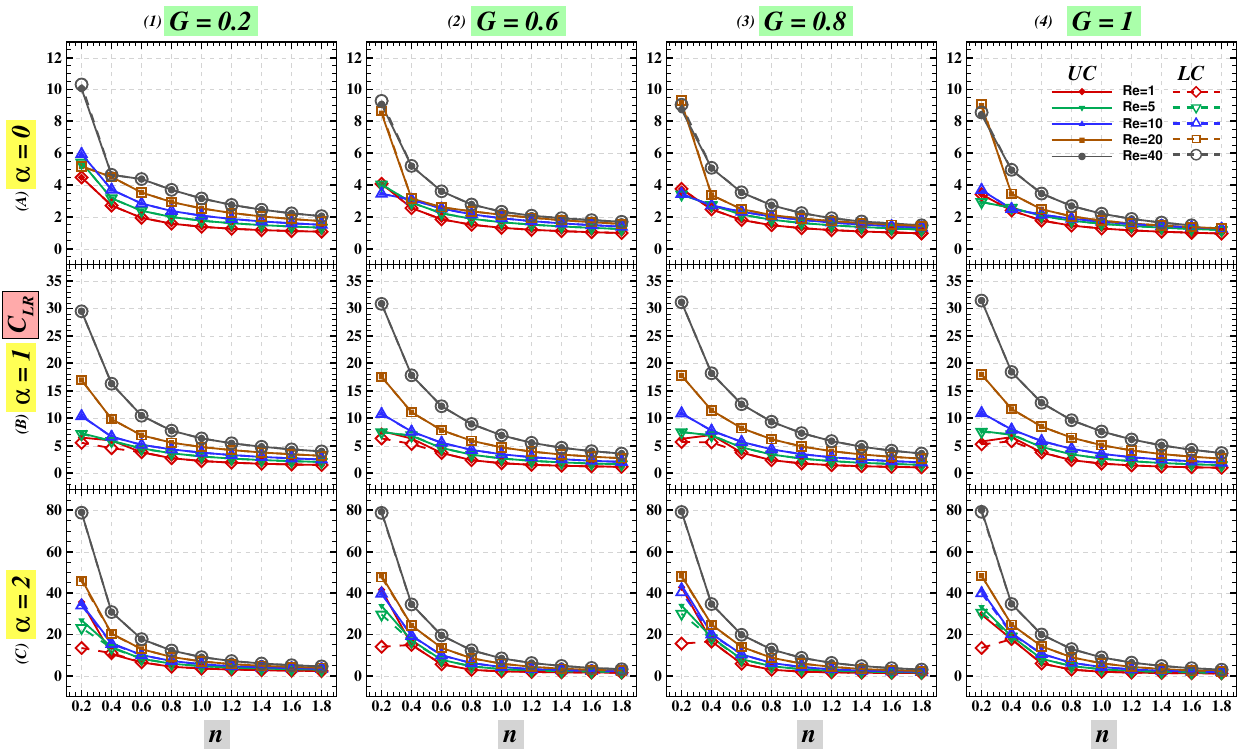}\label{fig:clr}}
	\caption{Variation of drag and lift ratio ($C_{DR}$ and $C_{LR}$) for $Re$ with $n$ at different $\alpha$ (in rows) and $G$ (in columns).}
	\label{fig:cdr-clr-g}
\end{figure}
%
\subsubsection{Drag and lift ratio ($C_{DR}$ and $C_{LR}$)}
%
\noindent
To emphasize the relative contributions of the individual drag and lift components, the drag ratio ($C_{DR} = C_{DP}/C_{DF}$) and lift ratio ($C_{LR} = C_{LP}/C_{LF}$) are shown over a range of Reynolds numbers ($Re$),  power-law indices ($n$), rotational speeds ($\alpha$) and gap ratios ($G$), as illustrated in \fig\ref{fig:cdr-clr-g} (see also \ref{appendix:ratiodragnlift}). In these figures, the solid and dashed lines represent the upper and lower cylinders, respectively. The near overlap of these lines reflects the mirror-symmetric flow behavior, making their responses effectively indistinguishable. Both the drag ratio ($C_{DR}$) and lift ratio ($C_{LR}$) show a decreasing trend with increasing power-law index ($n$) for all gap ratios ($G$) and rotation rates ($\alpha$). For stationary cylinders ($\alpha=0$), both ratios increase with Reynolds number ($Re$) across all $G$ and $n$. Higher values of $C_{DR}$ and $C_{LR}$ occur at lower $n$ (shear-thinning fluids), while lower values correspond to higher $n$ (shear-thickening fluids).  For $\alpha=0$, as $n$ increases from 0.2 to 1.8, $C_{DR}$ decreases by 68\% (for $Re=1$) and 81\% (for $Re=40$) across all $G$, refer panel A in \fig\ref{fig:cdr}; similarly, $C_{LR}$ decreases by 75\% (for $Re=1$) and 81\% (for $Re=40$), refer panel A in \fig\ref{fig:clr}. At $\alpha=1$, for shear-thickening fluids ($n \ge 1.4$), a negative $C_{DR}$ is observed at $Re \le 10$ and $G \le 0.4$ (panel B1 in \fig\ref{fig:cdr}), particularly for $Re=1$, $G=1$ (panel B4 in \fig\ref{fig:cdr}). This negative trend arises from increased flow resistance, leading to reversed pressure contributions.  For example, for $\alpha=1$, $C_{DR}$ and $C_{LR}$ decrease by 115\% and 81\% at $Re=1$, and by 94\% and 88\% at $Re=40$, respectively (panel B in \figs\ref{fig:cdr} and \ref{fig:clr}). These reductions demonstrate the strong influence of cylinder rotation on hydrodynamic forces and system sensitivity to flow conditions. Further,  at $\alpha=2$, a similar trend is noted for $n \ge 1$ across all $Re$ and $G$ (panel C1 in \fig\ref{fig:cdr}), while for shear-thinning fluids ($n \le 1$), $C_{DR}$ increases with $Re$. For $\alpha=2$, $C_{DR}$ and $C_{LR}$ decrease by 118\% and 96\% at $Re=1$, and by 128\% and 96\% at $Re=40$ (panel C1 in \fig\ref{fig:cdr}, and panel C4 in \ref{fig:clr}). Furthermore, $C_{LR}$ increases with $\alpha$ for all $G$, $Re$, and $n$ (panel 2 in \fig\ref{fig:cdr-clr-g}), reflecting the Magnus effect, where rotation induces circulation and enhances lift. These findings underscore the interplay between fluid rheology ($n$), flow parameters ($Re$), and geometric configuration ($G$) in shaping the hydrodynamic response of rotating cylinders.
\begin{figure}[!bt]
	\centering
	\centering
	\includegraphics[width=1\linewidth]{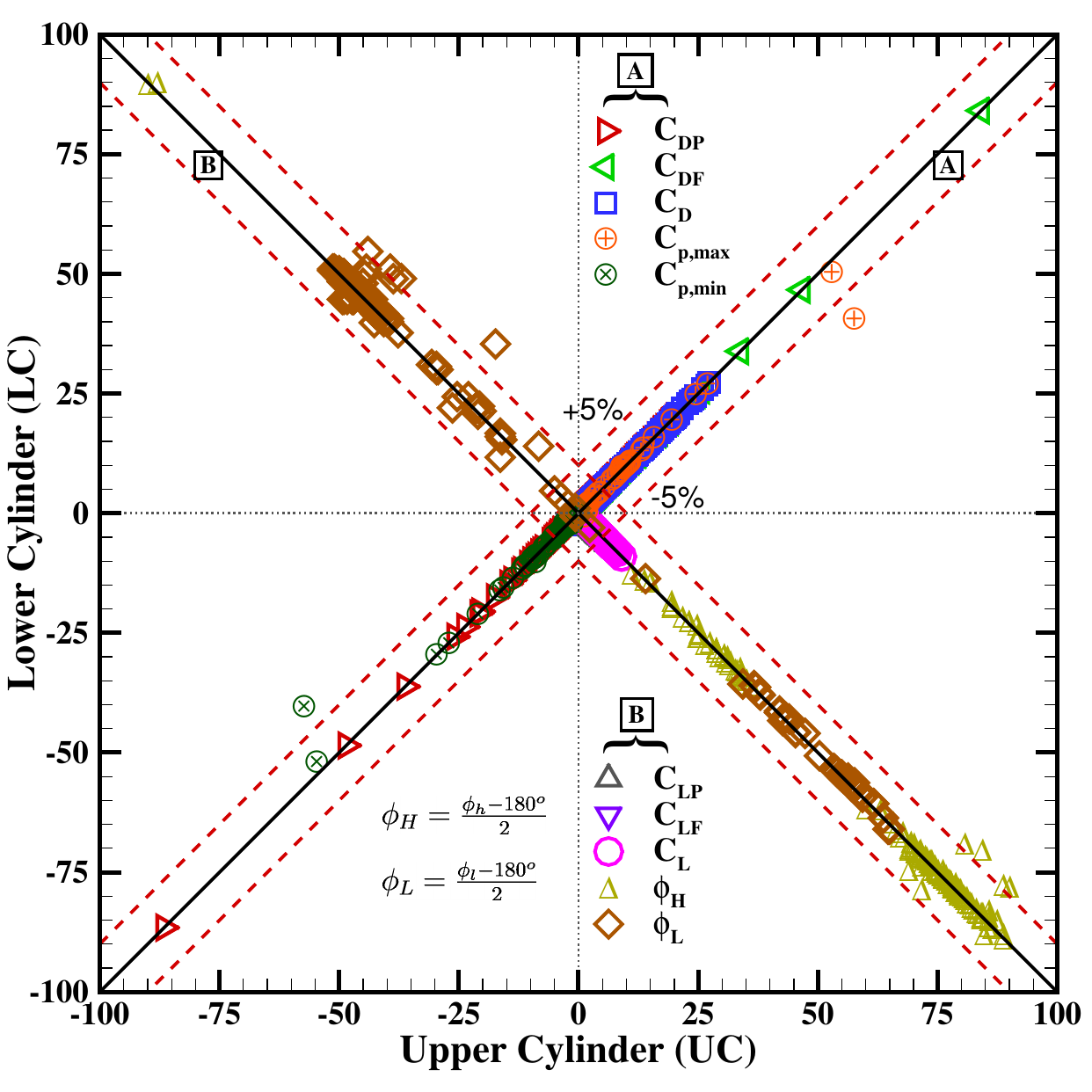}
	\caption{Comparative hydrodynamic features ($\Phi=C_{DP}$, $C_{DF}$, $C_D$, $C_{LP}$, $C_{LF}$, $C_{L}$, $C_{p,max}$, $C_{p,min}$, $\phi_H$, $\phi_L$) for upper and lower cylinders  (UC and LC) for the considered ranges of conditions ($n$, $Re$, $\alpha$, $G$).}
	\label{fig:UC_LC}
\end{figure}
\subsection{Comparative hydrodynamic features for upper and lower cylinders}
\noindent
\fig\ref{fig:UC_LC} illustrates the comparative hydrodynamic characteristics ($\Phi$) between the upper and lower cylinders (UC and LC), such as pressure drag ($C_{DP}$), friction drag ($C_{DF}$), total drag ($C_D$), pressure lift ($C_{LP}$), friction lift ($C_{LF}$), total lift ($C_L$), and minimum and maximum values of pressure coefficients ($C_{p,max}$, $C_{p,min}$) and their corresponding angular positions  ($\phi_h=2\phi_H+180^{\circ}$, $\phi_l=2\phi_L+180^{\circ}$) for the ranges of conditions ($n$, $Re$, $\alpha$, $G$) covered in this work. 
The solid diagonal line (labeled A) in \fig\ref{fig:UC_LC} represents the line of perfect symmetry ($\Phi_{\text{LC}} = \Phi_{\text{UC}}$), corresponding primarily to the drag coefficients ($C_{DP}$, $C_{DF}$, $C_D$) and pressure coefficients  ($C_{p,max}$, $C_{p,min}$). In contrast, the line labeled B, which is inclined at $-45^{\circ}$ to line A, depict the anti-symmetric relationship ($\Phi_{\text{LC}} = - \Phi_{\text{UC}}$) between the lift-related quantities ($C_{LP}$, $C_{LF}$, $C_L$) and the angular positions ($\phi_H$, $\phi_L$). Further, the dashed sub- and super-diagonal lines denote $\pm5\%$ deviation bands to main diagonal lines A and B.
While line A reflects equal magnitude and direction of hydrodynamic quantities ($\Phi_{\text{LC}} = \Phi_{\text{UC}}$), line B corresponds to equal magnitude but opposite direction ($\Phi_{\text{LC}} = - \Phi_{\text{UC}}$). This anti-symmetric distribution is a direct manifestation of the counter-rotating flow configuration, where the induced circulations around each cylinder produce lift forces of equal strength but opposite sense. The data points aligned along line B confirm that the lift components and angles are mirror images between the upper and lower cylinders. Minor deviations from this line (within $\pm5\%$) indicate the influence of small asymmetries in gap flow.

\noindent
Most data points cluster closely around the diagonal line, confirming a high degree of symmetry between the UC and LC for both drag and lift components. The nearly linear correlation between drag coefficients ($C_{DF}$, $C_{DP}$, $C_D$) for both cylinders signifies balanced drag forces, implying that flow parameters ($n$, $Re$, $\alpha$, $G$) do not significantly disturb the stream-wise force equilibrium. In contrast, the lift-related coefficients ($C_{LP}$, $C_{LF}$, $C_L$) exhibit a mirror-symmetric relationship (reflected across the origin, line B), indicating opposite lift directions acting on the cylinders. This anti-symmetric lift distribution arises from the counter-rotating flow patterns generated by the cylinders, which induce upward and downward pressure gradients, respectively.

\noindent
Further, the pressure coefficients ($C_{p,\max}$, $C_{p,\min}$) also follow symmetric distributions but with slight asymmetry in magnitude, where the UC tends to experience lower $C_{p,\min}$ values than the LC. This difference results from flow acceleration near the UC due to rotation direction and gap-induced flow interactions. The angular positions ($\phi_H$, $\phi_L$) corresponding to pressure coefficients ($C_{p,\max}$ and $C_{p,\min}$) for both cylinders further validate this symmetry, following nearly identical trends with a phase shift of $180^\circ$, consistent with the anti-symmetric pressure field between the cylinders.

\noindent
Overall, the results in \fig\ref{fig:UC_LC} confirm that (a) the drag forces are symmetric, indicating balanced stream-wise resistance, (b) the lift forces are anti-symmetric, confirming opposite transverse forces, (c) the pressure distribution and corresponding angular positions variation between the cylinders are responsible for the observed lift reversal. Furthermore, the deviations ($\pm 5\%$) suggest minor influence of flow conditions, such as gap ratio ($G$), Reynolds number ($Re$), rotational rate ($\alpha$), and power-law index ($n$) on overall force symmetry around both cylinders (UC and LC). These comparative features emphasize the inherent hydrodynamic mirror symmetry in the counter-rotating cylinder pair system and validate the numerical consistency of the computed results.
%

\noindent
In summary, the Reynolds number ($Re$), gap ratio ($G$), rotation rate ($\alpha$), and fluid behavior index ($n$) collectively govern the local and overall hydrodynamic characteristics of the counter-rotating cylinders. As $\alpha$ increases from 0 to 2, the pressure coefficient ($C_p$) generally increases with increases $n$ but decreases with increasing $Re$ and $G$. At $\alpha=2$, $C_{p,\text{max}}$ decreases with $G$ for shear-thickening fluids ($n> 1$) but increases for shear-thinning fluids ($n< 1$), showing strong rheological dependence. The stagnation points ($\phi_h$ and $\phi_l$) shift with both $G$ and $\alpha$, reflecting viscosity variations with shear rate and the influence of rotation-induced circulation. The friction coefficient ($C_{DF}$) decreases with $Re$ for all conditions. For stationary cylinders ($\alpha=0$) and low $Re$ ($\leq1$), $C_{DF}$ decreases with $n$, whereas at higher $Re$ and $\alpha>0$, it increases with $n$, indicating enhanced viscous resistance in shear-thickening fluids. Rotation initially reduces $C_{DF}$ by stabilizing the boundary layer but later increases it as the shear intensifies. The pressure drag coefficient ($C_{DP}$) decreases exponentially with $n$ and $Re$ for all $\alpha$ and $G$. For $\alpha>0$ at $Re=1$, $C_{DP}$ transitions from positive to negative in shear-thinning fluids, caused by viscosity reduction and reversed pressure gradients near the surface. Increasing $G$ further decreases $C_{DP}$ for all conditions. The friction lift coefficient ($C_{LF}$) increases with $\alpha$ and $n$ but decreases with $G$ and $Re$. The pressure lift coefficient ($C_{LP}$), though enhanced by rotation, exhibits non-monotonic variation with $n$, decreasing with $G$ and $Re$ but increasing with $\alpha$. Overall, the total drag ($C_D$) decreases with $Re$ and increases with $G$, while the total lift ($C_L$) shows non-monotonic dependence on $n$, $Re$, $\alpha$, and $G$. The drag and lift ratios ($C_{DR}$, $C_{LR}$) generally decrease with increasing $n$ and vary non-linearly with $Re$ and $\alpha$. These complex inter-dependencies underscore the strong coupling between non-Newtonian fluid rheology, flow and geometrical parameters, and rotational effects in determining the hydrodynamic performance of the system.
\section{Concluding remarks}

\noindent
The steady unconfined flow of power-law non-Newtonian fluids past a pair of side-by-side rotating circular cylinders was investigated numerically  using the finite element method over a range of rotational rates ($0 \leq \alpha \leq 2$), gap ratios ($0.2 \leq G \leq 1$), Reynolds numbers ($0 \leq Re \leq 40$), and power-law indices ($0.2 \leq n \leq 1.8$). The analysis focused on macroscopic hydrodynamic quantities, including drag and lift coefficients, and on detailed flow features such as streamline patterns, centerline velocity, and surface pressure distributions. The results reveal that increasing the rotational rate suppresses vortex shedding and promotes the formation of a thin fluid layer along the cylinder surfaces. In shear-thickening fluids, the wake structure is strongly altered by rotation, exhibiting enhanced asymmetry and reduced flow separation. The centerline velocity increases with both $Re$ and $\alpha$, while the pressure coefficient ($C_p$) attains its maximum and minimum at the front and rear stagnation points, respectively. For shear-thickening fluids, the location of maximum pressure shifts toward the direction of rotation, reflecting the influence of rheology on surface pressure distribution.
The total drag coefficient ($C_D$) decreases with increasing $Re$, $n$, and $\alpha$, indicating reduced form and frictional resistance, whereas the total lift coefficient ($C_L$) exhibits a complex dependence on these parameters due to competing effects of rotation and viscous dissipation. The interplay between gap ratio, rotational speed, and non-Newtonian rheology governs the balance between pressure- and shear-induced forces, determining the overall hydrodynamic response of the system. These findings provide a unified understanding of the coupled influence of rotation, rheology, and geometric confinement on non-Newtonian flow around bluff bodies. The results are expected to inform the design of fluid-structure systems in applications such as heat exchangers, particulate transport, and mixing processes involving complex fluids.
%
%
%
\section*{Additional Supplementary Information}
\noindent
Readers may refer \ref{appendix:dragtable} to \ref{appendix:ratiodragnlift} for the detailed data sets and  enlarged figures.  
%
%
%
%
\begin{spacing}{1.1}
\renewcommand{\bibfont}{\small}	
\bibliography{references}
\end{spacing}	
%
%
%
%
%
%
%
%
\appendix
\renewcommand\thesection{Appendix~\Alph{section}}
\renewcommand\thesubsection{\Alph{section}.\arabic{subsection}}
\renewcommand{\thetable}{\Alph{section}.\arabic{table}} \setcounter{table}{0}
\renewcommand{\thefigure}{\Alph{section}.\arabic{figure}} \setcounter{figure}{0} 
\clearpage%
\section{Selection of domain and grid size}\label{appendix:domain-grid}
%
\noindent
The accuracy of the computational results inherently depends on the size of the computational domain and mesh. In the case of an unconfined flow problem, as considered in the present study, the physical domain is characterized by the upstream length ($L_u$), downstream length ($L_d$), and height  ($H$) of the computational domain (refer \fig 1).
\subsection{Domain independence test} 
%
\begin{table}[!b]
	\centering
	\caption{The upstream length ($L_u$)  independence test for extreme gap ratio ($G$), Reynolds numbers ($Re$) and power-law index ($n$) for the maximum rotational rate ($\alpha=2$) of the cylinders.}\label{Tab:dom_Lu-Ld}
	\resizebox{\columnwidth}{!}{%
		\renewcommand{\arraystretch}{1.3}
		\begin{tabular}{@{}|c|c|c|rr|rr|rr|rr|rr|rr|rr|rr|@{}}
			\hline
			\multicolumn{3}{|c|}{$L_d=200$, } &\multicolumn{8}{c|}{$Re = 1$} & \multicolumn{8}{c|}{$Re = 40$} \\ \cline{4-19}
			\multicolumn{3}{|c|}{$H = 100$} &\multicolumn{4}{c|}{Upper cylinder (UC)}&\multicolumn{4}{c|}{Lower cylinder (LC)}  &\multicolumn{4}{c|}{Upper cylinder (UC)}&\multicolumn{4}{c|}{Lower cylinder (LC)} \\
			\hline
			{$G$}&{$n$}&$L_u$&$C_{DP}$&$C_{D}$&$C_{LP}$&$C_{L}$&$C_{DP}$&$C_{D}$&$C_{LP}$&$C_{L}$&$C_{DP}$&$C_{D}$&$C_{LP}$&$C_{L}$&$C_{DP}$&$C_{D}$&$C_{LP}$&$C_{L}$\\
			\hline		
			{0.2}	&0.2	&	40	&	15.208	&	17.577	&	5.511	&	5.7432	&	15.254	&	17.534	&	-5.4128	&	-5.8419	&		0.4193	&	0.5409	&	4.3213	&	4.3760	&	0.4190	&	0.5406	&	-4.3219	&	-4.3768	\\
			&	&	80	&		15.208	&	17.577	&	5.5105	&	5.7427	&	15.253	&	17.534	&	-5.4123	&	-5.8414	&		0.4188	&	0.5408	&	4.3208	&	4.3755	&	0.4190	&	0.5406	&	-4.3212	&	-4.3760	\\
			&	&	120	&		15.208	&	17.576	&	5.5117	&	5.7405	&	15.253	&	17.534	&	-5.4123	&	-5.8413	&		0.4192	&	0.5408	&	4.3206	&	4.3753	&	0.4192	&	0.5407	&	-4.3211	&	-4.3760	\\
			&	&	160	&		15.208	&	17.576	&	5.5117	&	5.7405	&	15.253	&	17.534	&	-5.4123	&	-5.8413	&		0.4192	&	0.5408	&	4.3206	&	4.3753	&	0.4193	&	0.5406	&	-4.3211	&	-4.3759	\\
			\hline																													%
			0.2	&	1 &	40	&	-4.0539	&	1.4674	&	3.3093	&	4.2509	&	-4.0556	&	1.4652	&	-3.3115	&	-4.251	&	-0.1236	&	0.1932	&	2.7932	&	3.1015	&	-0.1238	&	0.1927	&	-2.7932	&	-3.1015	\\
			& &	80	&		-4.0670	&	1.4453	&	3.2999	&	4.2368	&	-4.0687	&	1.443	&	-3.3021	&	-4.2369	&		-0.1242	&	0.1917	&	2.7862	&	3.0936	&	-0.1244	&	0.1913	&	-2.7861	&	-3.0936	\\
			&	&120	&		-4.0679	&	1.4438	&	3.2993	&	4.2358	&	-4.0696	&	1.4416	&	-3.3015	&	-4.2359	&		-0.1243	&	0.1916	&	2.7857	&	3.0930	&	-0.1245	&	0.1912	&	-2.7856	&	-3.0930	\\
			&	& 160	&		-4.0680	&	1.4437	&	3.2992	&	4.2358	&	-4.0697	&	1.4415	&	-3.3014	&	-4.2359	&		-0.1243	&	0.1916	&	2.7856	&	3.0930	&	-0.1245	&	0.1912	&	-2.7856	&	-3.0930	\\
			\hline																																				
			0.2	&1.8	&	40	&	-86.587	&	-2.5274	&	6.3149	&	9.004	&	-86.595	&	-2.5074	&	-6.3379	&	-9.0148	&		-2.3026	&	-0.2193	&	3.8602	&	4.7026	&	-2.3031	&	-0.2197	&	-3.8609	&	-4.7032	\\
			&	&	80	&		-86.579	&	-2.5045	&	6.3238	&	9.0236	&	-86.587	&	-2.4845	&	-6.3469	&	-9.0343	&		-2.3032	&	-0.2203	&	3.8582	&	4.7000	&	-2.3036	&	-0.2207	&	-3.8589	&	-4.7005	\\
			&	&	120	&		-86.578	&	-2.5026	&	6.3245	&	9.0251	&	-86.587	&	-2.4827	&	-6.3476	&	-9.0359	&		-2.3032	&	-0.2204	&	3.8580	&	4.6998	&	-2.3037	&	-0.2207	&	-3.8588	&	-4.7004	\\
			&	&	160	&		-86.578	&	-2.5026	&	6.3245	&	9.0251	&	-86.587	&	-2.4827	&	-6.3476	&	-9.0359	&		-2.3032	&	-0.2204	&	3.8580	&	4.6998	& 	-2.3034	&	-0.2198	&	-3.8592	&	-4.7016	\\
			\hline
			1&		0.2&	40	&	17.4280	&	20.3510	&	6.4142	&	6.5914	&	17.4760	&	20.3100	&	-6.2918	&	-6.7128	&	0.5449	&	0.6700	&	4.2303	&	4.2835	&	0.5443	&	0.6692	&	-4.2225	&	-4.2758	\\
			&	&	80	&		17.4270	&	20.3500	&	6.4139	&	6.5910	&	17.4760	&	20.3090	&	-6.2914	&	-6.7122	&		0.5447	&	0.6698	&	4.2293	&	4.2825	&	0.5441	&	0.6690	&	-4.2216	&	-4.2748	\\
			&	&	120	&		17.4270	&	20.3500	&	6.4138	&	6.5910	&	17.4760	&	20.3090	&	-6.2913	&	-6.7122	&		0.5447	&	0.6698	&	4.2293	&	4.2825	&	0.5441	&	0.6689	&	-4.2216	&	-4.2748	\\
			&	&	160	&		17.4290	&	20.3490	&	6.4104	&	6.5933	&	17.4760	&	20.3090	&	-6.2913	&	-6.7122	&		0.5446	&	0.6695	&	4.2311	&	4.2845	&	0.5443	&	0.6689	&	-4.2211	&	-4.2745	\\
			\hline
			1	&1	&	40	&	2.0571	&	4.2942	&	2.4009	&	3.6005	&	2.0568	&	4.2927	&	-2.4036	&	-3.6061	&		-0.1665	&	0.3765	&	4.8098	&	5.3472	&	-0.1665	&	0.3763	&	-4.8094	&	-5.3469	\\
			&	&	80	&		2.0176	&	4.2178	&	2.3635	&	3.5443	&	2.0173	&	4.2163	&	-2.3662	&	-3.5499	&		-0.1688	&	0.3727	&	4.8002	&	5.3370	&	-0.1688	&	0.3724	&	-4.7998	&	-5.3366	\\
			&	&	120	&		2.0150	&	4.2127	&	2.3610	&	3.5405	&	2.0147	&	4.2112	&	-2.3637	&	-3.5461	&		-0.1690	&	0.3724	&	4.7995	&	5.3363	&	-0.1690	&	0.3721	&	-4.7991	&	-5.3359	\\
			&	&	160	&		2.0148	&	4.2124	&	2.3608	&	3.5403	&	2.0145	&	4.2108	&	-2.3635	&	-3.5458	&		-0.1690	&	0.3724	&	4.7995	&	5.3362	&	-0.1690	&	0.3721	&	-4.7991	&	-5.3358	\\
			\hline
			1	&	1.8	&	40	&	-9.9801	&	-1.1282	&	3.9427	&	7.2487	&	-9.9781	&	-1.1252	&	-3.9465	&	-7.2514	&	-0.7161	&	0.0089	&	3.9224	&	5.2347	&	-0.7161	&	0.0083	&	-3.9222	&	-5.2359	\\
			&	&	80	&		-9.9813	&	-1.1322	&	3.9409	&	7.2447	&	-9.9792	&	-1.1291	&	-3.9447	&	-7.2474	&		-0.7161	&	0.0048	&	3.9032	&	5.2101	&	-0.7161	&	0.0041	&	-3.9031	&	-5.2114	\\
			&	&	120	&		-9.9814	&	-1.1324	&	3.9408	&	7.2445	&	-9.9793	&	-1.1293	&	-3.9446	&	-7.2472	&		-0.7161	&	0.0044	&	3.9019	&	5.2085	&	-0.7161	&	0.0038	&	-3.9017	&	-5.2096	\\
			&		& 160	&		-9.9814	&	-1.1324	&	3.9408	&	7.2445	&	-9.9793	&	-1.1294	&	-3.9446	&	-7.2472	&		-0.7161	&	0.0043	&	3.9018	&	5.2083	&	-0.7161	&	0.0038	&	-3.9016	&	-5.2095	\\
			\hline
		\end{tabular}
	}
\end{table}
%
\begin{table}[!tb]
	\centering
	\caption{The downstream length ($L_d$) independence test for extreme gap ratio ($G$), Reynolds numbers ($Re$) and power-law index ($n$) for the maximum rotational rate ($\alpha=2$)  of the cylinders.}\label{Tab:dom_Ld}
	\resizebox{\columnwidth}{!}{%
		\renewcommand{\arraystretch}{1.3}
		\begin{tabular}{@{}|c|c|c|rr|rr|rr|rr|rr|rr|rr|rr|@{}}
			\hline
			\multicolumn{3}{|c|}{$L_u=120$,} &\multicolumn{8}{c|}{$Re = 1$} & \multicolumn{8}{c|}{$Re = 40$} \\ \cline{4-19}
			\multicolumn{3}{|c|}{$H = 100$} &\multicolumn{4}{c|}{Upper cylinder (UC)}&\multicolumn{4}{c|}{Lower cylinder (LC)}  &\multicolumn{4}{c|}{Upper cylinder (UC)}&\multicolumn{4}{c|}{Lower cylinder (LC)} \\
			\hline
			{$G$}&{$n$}&$L_d$&$C_{DP}$&$C_{D}$&$C_{LP}$&$C_{L}$&$C_{DP}$&$C_{D}$&$C_{LP}$&$C_{L}$&$C_{DP}$&$C_{D}$&$C_{LP}$&$C_{L}$&$C_{DP}$&$C_{D}$&$C_{LP}$&$C_{L}$\\
			\hline
			0.2	&	0.2	&	100	&15.2050	&	17.5830	&	5.5282	&	5.7177	&	15.2350	&	17.5570	&	-5.4442	&	-5.8069	&	 0.4185	&	0.5406	&	4.3196	&	4.3741	&	0.4190	&	0.5404	&	-4.3236	&	-4.3786	\\
			&		&	200	&	15.2080	&	17.5760	&	5.5117	&	5.7405	&	15.2530	&	17.5340	&	-5.4123	&	-5.8413	&	0.4192	&	0.5408	&	4.3206	&	4.3753	&	0.4192	&	0.5407	&	-4.3211	&	-4.3760	\\
			&		&	300	&	15.2050	&	17.5820	&	5.5534	&	5.7009	&	15.2420	&	17.5540	&	-5.4306	&	-5.8279	&	0.4190	&	0.5408	&	4.3197	&	4.3743	&	0.4191	&	0.5404	&	-4.3232	&	-4.3783	\\
			&		&	400	&	15.2070	&	17.5790	&	5.5272	&	5.7239	&	15.2470	&	17.5500	&	-5.4224	&	-5.8318	&	0.4186	&	0.5406	&	4.3224	&	4.3771	&	0.4190	&	0.5406	&	-4.3214	&	-4.3763	\\
			\hline
			0.2	&	1	&100	&	-4.0539	&	1.4674	&	3.3093	&	4.2509	&	-4.0556	&	1.4652	&	-3.3115	&	-4.2510	&	-0.1243	&	0.1913	&	2.7856	&	3.0925	&	-0.1244	&	0.1914	&	-2.7858	&	-3.0932	\\
			&		&200	&	-4.0670	&	1.4453	&	3.2999	&	4.2368	&	-4.0687	&	1.4430	&	-3.3021	&	-4.2369	&		-0.1243	&	0.1916	&	2.7857	&	3.0930	&	-0.1245	&	0.1912	&	-2.7856	&	-3.0930	\\
			&		&300	&	-4.0679	&	1.4438	&	3.2993	&	4.2358	&	-4.0696	&	1.4416	&	-3.3015	&	-4.2359	&		-0.1243	&	0.1915	&	2.7857	&	3.0930	&	-0.1244	&	0.1910	&	-2.7859	&	-3.0933	\\
			&		&400	&	-4.0677	&	1.4452	&	3.2956	&	4.2265	&	-4.0686	&	1.4423	&	-3.2991	&	-4.2306	&		-0.1243	&	0.1913	&	2.7857	&	3.0930	&	-0.1244	&	0.1912	&	-2.7859	&	-3.0932	\\
			\hline
			0.2 &	1.8	&100	 &	-86.5680	&	-2.4648	&	6.2902	&	8.9183	&	-86.5790	&	-2.4891	&	-6.3434	&	-9.0403		 &	-2.3030	&	-0.2199	&	3.8573	&	4.6969	&	-2.3036	&	-0.2209	&	-3.8585	&	-4.7000	\\
			&		&200	&	-86.5780	&	-2.5026	&	6.3245	&	9.0251	&	-86.5870	&	-2.4827	&	-6.3476	&	-9.0359		&	-2.3032	&	-0.2204	&	3.8580	&	4.6998	&	-2.3037	&	-0.2207	&	-3.8588	&	-4.7004	\\
			&		&300		&	-86.5760	&	-2.4780	&	6.3077	&	8.9869	&	-86.5780	&	-2.5071	&	-6.3265	&	-9.0105		&	-2.3031	&	-0.2202	&	3.8579	&	4.6992	&	-2.3034	&	-0.2215	&	-3.8584	&	-4.6998	\\
			&		&400	&	-86.5700	&	-2.4812	&	6.2809	&	8.9307	&	-86.5790	&	-2.4951	&	-6.3304	&	-9.0152	 &	-2.3030	&	-0.2205	&	3.8573	&	4.6979	&	-2.3035	&	-0.2211	&	-3.8584	&	-4.6998	\\
			\hline
			1	&	0.2	&100	&	17.4640	&	20.3300	&	6.3696	&	6.6119	&	17.4700	&	20.3140	&	-6.2826	&	-6.7150	&	0.5443	&	0.6694	&	4.2307	&	4.2837	&	0.5448	&	0.6695	&	-4.2226	&	-4.2759	\\
			&		&200	&		17.4270	&	20.3500	&	6.4138	&	6.5910	&	17.4760	&	20.3090	&	-6.2913	&	-6.7122	&	0.5447	&	0.6698	&	4.2293	&	4.2825	&	0.5441	&	0.6689	&	-4.2216	&	-4.2748	\\
			&		&300	&		17.4310	&	20.3490	&	6.4061	&	6.5968	&	17.4900	&	20.2930	&	-6.2618	&	-6.7380	&	0.5447	&	0.6698	&	4.2312	&	4.2843	&	0.5445	&	0.6690	&	-4.2224	&	-4.2757	\\
			&		&400	&		17.4230	&	20.3490	&	6.4236	&	6.5713	&	17.4710	&	20.3120	&	-6.2801	&	-6.7244	&	0.5446	&	0.6692	&	4.2327	&	4.2859	&	0.5444	&	0.6690	&	-4.2225	&	-4.2760	\\
			\hline
			1	&	1	&100	&	2.0155	&	4.2130	&	2.3601	&	3.5389	&	2.0141	&	4.2084	&	-2.3629	&	-3.5440	&	-0.1689	&	0.3725	&	4.7996	&	5.3362	&	-0.1690	&	0.3722	&	-4.7990	&	-5.3358	\\
			&		&200	&		2.0150	&	4.2127	&	2.3610	&	3.5405	&	2.0147	&	4.2112	&	-2.3637	&	-3.5461	&		-0.1690	&	0.3724	&	4.7995	&	5.3363	&	-0.1690	&	0.3721	&	-4.7991	&	-5.3359	\\
			&		&300	&		2.0151	&	4.2131	&	2.3606	&	3.5384	&	2.0157	&	4.2135	&	-2.3608	&	-3.5383	&		-0.1690	&	0.3725	&	4.7995	&	5.3362	&	-0.1690	&	0.3721	&	-4.7990	&	-5.3357	\\
			&		&400	&		2.0150	&	4.2128	&	2.3619	&	3.5434	&	2.0156	&	4.2125	&	-2.3627	&	-3.5443	&		-0.1690	&	0.3725	&	4.7996	&	5.3364	&	-0.1690	&	0.3722	&	-4.7991	&	-5.3359	\\
			\hline
			1	&	1.8	&100	&	-9.9802	&	-1.1305	&	3.9443	&	7.2521	&	-9.9813	&	-1.1357	&	-3.9466	&	-7.2530	&	-0.7157	&	0.0052	&	3.9014	&	5.2086	&	-0.7161	&	0.0041	&	-3.9017	&	-5.2094	\\
			&		&200	&		-9.9814	&	-1.1324	&	3.9408	&	7.2445	&	-9.9793	&	-1.1293	&	-3.9446	&	-7.2472	&		-0.7161	&	0.0044	&	3.9019	&	5.2085	&	-0.7161	&	0.0038	&	-3.9017	&	-5.2096	\\
			&		&300	&		-9.9792	&	-1.1334	&	3.9406	&	7.2265	&	-9.9845	&	-1.1397	&	-3.9464	&	-7.2548	&		-0.7161	&	0.0049	&	3.9017	&	5.2082	&	-0.7159	&	0.0038	&	-3.9014	&	-5.2093	\\
			&		&400	&		-9.9820	&	-1.1345	&	3.9442	&	7.2561	&	-9.9806	&	-1.1343	&	-3.9477	&	-7.2604	&		-0.7161	&	0.0048	&	3.9020	&	5.2088	&	-0.7160	&	0.0046	&	-3.9017	&	-5.2098	\\
			\hline
		\end{tabular}
	}
\end{table}
\begin{table}[!tb]
	\centering
	\caption{The domain height ($H$) independence test for extreme gap ratio ($G$), Reynolds numbers ($Re$) and power-law index ($n$) for the maximum rotational rate ($\alpha=2$)  of the cylinders.}\label{Tab:dom_H}
	\resizebox{\columnwidth}{!}{%
		\renewcommand{\arraystretch}{1.3}
		\begin{tabular}{@{}|c|c|c|rr|rr|rr|rr|rr|rr|rr|rr|@{}}
			\hline
			\multicolumn{3}{|c|}{$L_u=120$,} &\multicolumn{8}{c|}{$Re = 1$} & \multicolumn{8}{c|}{$Re = 40$} \\ \cline{4-19}
			\multicolumn{3}{|c|}{$L_d=300$} &\multicolumn{4}{c|}{Upper cylinder (UC)}&\multicolumn{4}{c|}{Lower cylinder (LC)}  &\multicolumn{4}{c|}{Upper cylinder (UC)}&\multicolumn{4}{c|}{Lower cylinder (LC)} \\
			\hline
			{$G$}&{$n$}&$H$&$C_{DP}$&$C_{D}$&$C_{LP}$&$C_{L}$&$C_{DP}$&$C_{D}$&$C_{LP}$&$C_{L}$&$C_{DP}$&$C_{D}$&$C_{LP}$&$C_{L}$&$C_{DP}$&$C_{D}$&$C_{LP}$&$C_{L}$\\
			\hline
			0.2	&	0.2 &100	&	15.2050	&	17.5820	&	5.5534	&	5.7009	&	15.2420	&	17.5540	&	-5.4306	&	-5.8279	&	0.4190	&	0.5408	&	4.3197	&	4.3743	&	0.4191	&	0.5404	&	-4.3232	&	-4.3783	\\
			&	&	200	&	15.2060	&	17.5830	&	5.5459	&	5.7072	&	15.2410	&	17.5530	&	-5.4295	&	-5.8267	&	0.4190	&	0.5407	&	4.3190	&	4.3735	&	0.4190	&	0.5403	&	-4.3226	&	-4.3775	\\
			&	&	300	&	15.2040	&	17.5750	&	5.5547	&	5.6964	&	15.2410	&	17.5560	&	-5.4257	&	-5.8282	&	0.4188	&	0.5405	&	4.3187	&	4.3733	&	0.4188	&	0.5402	&	-4.3223	&	-4.3773	\\
			&	&	400	&	15.2050	&	17.5760	&	5.5495	&	5.7029	&	15.2430	&	17.5530	&	-5.4259	&	-5.8290	&	0.4186	&	0.5405	&	4.3187	&	4.3732	&	0.4186	&	0.5402	&	-4.3223	&	-4.3772	\\
			\hline
			0.2	&	1&	100	&	-4.0679	&	1.4438	&	3.2993	&	4.2358	&	-4.0696	&	1.4416	&	-3.3015	&	-4.2359	&	-0.1243	&	0.1915	&	2.7857	&	3.0930	&	-0.1244	&	0.1910	&	-2.7859	&	-3.0933	\\
			&	&	200	&	-4.0856	&	1.4173	&	3.2865	&	4.2159	&	-4.0863	&	1.4124	&	-3.2861	&	-4.2126	&	-0.1247	&	0.1906	&	2.7813	&	3.0881	&	-0.1249	&	0.1901	&	-2.7815	&	-3.0884	\\
			&	&	300	&	-4.0893	&	1.4112	&	3.2841	&	4.2123	&	-4.0900	&	1.4062	&	-3.2837	&	-4.2089	&	-0.1248	&	0.1904	&	2.7806	&	3.0873	&	-0.1249	&	0.1899	&	-2.7808	&	-3.0876	\\
			&	&	400	&	-4.0903	&	1.4094	&	3.2834	&	4.2113	&	-4.0910	&	1.4045	&	-3.2830	&	-4.2079	&	-0.1248	&	0.1904	&	2.7805	&	3.0871	&	-0.1250	&	0.1899	&	-2.7806	&	-3.0874	\\
			\hline
			0.2	& 1.8	&	100	&	-86.5760	&	-2.4780	&	6.3077	&	8.9869	&	-86.5780	&	-2.5071	&	-6.3265	&	-9.0105	&	-2.3031	&	-0.2202	&	3.8579	&	4.6992	&	-2.3034	&	-0.2215	&	-3.8584	&	-4.6998	\\
			&	&	200	&	-86.5630	&	-2.4386	&	6.3221	&	9.0180	&	-86.5650	&	-2.4677	&	-6.3409	&	-9.0415	&	-2.3028	&	-0.2195	&	3.8590	&	4.7006	&	-2.3031	&	-0.2208	&	-3.8595	&	-4.7012	\\
			&	&	300	&	-86.5600	&	-2.4299	&	6.3254	&	9.0250	&	-86.5620	&	-2.4590	&	-6.3442	&	-9.0486	&	-2.3027	&	-0.2193	&	3.8594	&	4.7011	&	-2.3030	&	-0.2206	&	-3.8599	&	-4.7018	\\
			&	&	400	&	-86.5590	&	-2.4273	&	6.3263	&	9.0271	&	-86.5610	&	-2.4564	&	-6.3451	&	-9.0507	&	-2.3026	&	-0.2192	&	3.8596	&	4.7013	&	-2.3029	&	-0.2205	&	-3.8600	&	-4.7020	\\
			\hline
			1	&	0.2 &	100	&	17.4320	&	20.3470	&	6.4002	&	6.6037	&	17.4870	&	20.2960	&	-6.2653	&	-6.7254	&	0.5443	&	0.6694	&	4.2307	&	4.2837	&	0.5448	&	0.6695	&	-4.2226	&	-4.2759	\\
			&	&	200	&	17.4300	&	20.3460	&	6.4002	&	6.5999	&	17.4850	&	20.2940	&	-6.2641	&	-6.7241	&	0.5447	&	0.6699	&	4.2316	&	4.2848	&	0.5445	&	0.6690	&	-4.2227	&	-4.2760	\\
			&	&	300	&	17.4290	&	20.3450	&	6.3998	&	6.5993	&	17.4850	&	20.2930	&	-6.2637	&	-6.7236	&	0.5444	&	0.6693	&	4.2241	&	4.2771	&	0.5445	&	0.6692	&	-4.2232	&	-4.2765	\\
			&	&	400	&	17.4290	&	20.3450	&	6.3996	&	6.5991	&	17.4850	&	20.2930	&	-6.2635	&	-6.7234	&	0.5445	&	0.6693	&	4.2331	&	4.2863	&	0.5444	&	0.6689	&	-4.2219	&	-4.2751	\\
			\hline
			1	& 1	&	100	&	2.0151	&	4.2131	&	2.3606	&	3.5384	&	2.0157	&	4.2135	&	-2.3608	&	-3.5383	&	-0.1689	&	0.3725	&	4.7996	&	5.3362	&	-0.1690	&	0.3722	&	-4.7990	&	-5.3358	\\
			&	&	200	&	1.9550	&	4.0981	&	2.3092	&	3.4627	&	1.9556	&	4.0985	&	-2.3094	&	-3.4626	&	-0.1707	&	0.3698	&	4.7930	&	5.3292	&	-0.1707	&	0.3693	&	-4.7926	&	-5.3288	\\
			&	&	300	&	1.9428	&	4.0748	&	2.2986	&	3.4472	&	1.9434	&	4.0752	&	-2.2988	&	-3.4470	&	-0.1708	&	0.3693	&	4.7915	&	5.3276	&	-0.1710	&	0.3688	&	-4.7917	&	-5.3279	\\
			&	&	400	&	1.9394	&	4.0682	&	2.2957	&	3.4428	&	1.9400	&	4.0687	&	-2.2959	&	-3.4427	&	-0.1710	&	0.3692	&	4.7917	&	5.3278	&	-0.1710	&	0.3688	&	-4.7913	&	-5.3274	\\
			\hline
			1	&	1.8 &	100	&	-9.9819	&	-1.1338	&	3.9408	&	7.2401	&	-9.9805	&	-1.1312	&	-3.9422	&	-7.2413	&	-0.7157	&	0.0052	&	3.9014	&	5.2086	&	-0.7161	&	0.0041	&	-3.9017	&	-5.2094	\\
			&	&	200	&	-9.9768	&	-1.1177	&	3.9473	&	7.2543	&	-9.9754	&	-1.1151	&	-3.9487	&	-7.2555	&	-0.7161	&	0.0036	&	3.8956	&	5.2003	&	-0.7159	&	0.0025	&	-3.8953	&	-5.2014	\\
			&	&	300	&	-9.9754	&	-1.1134	&	3.9491	&	7.2583	&	-9.9741	&	-1.1108	&	-3.9506	&	-7.2595	&	-0.7160	&	0.0044	&	3.8953	&	5.2004	&	-0.7162	&	0.0019	&	-3.8951	&	-5.2008	\\
			&	&	400	&	-9.9750	&	-1.1120	&	3.9497	&	7.2596	&	-9.9737	&	-1.1095	&	-3.9511	&	-7.2607	&	-0.7161	&	0.0035	&	3.8953	&	5.2000	&	-0.7159	&	0.0023	&	-3.8950	&	-5.2010	\\
			\hline
		\end{tabular}
	}
\end{table}
%
\noindent
To represent the unconfined flow condition, the computational domain must be sufficiently large to minimize boundary effects on the flow characteristics. Although a larger domain provides a more accurate physical representation of unconfined flow, it also increases computational cost. Therefore, a domain independence test is performed to evaluate the trade-off between accuracy and computational efficiency by comparing the flow characteristics across different combinations of domain dimensions ($L_u$, $L_d$, $H$).
In this study, various domain dimensions are considered to perform the domain independence test, with upstream length ($40 \le L_u \le 160$), downstream length ($100 \le L_d \le 400$), and height ($100 \le H \le 400$). \tabs\ref{Tab:dom_Lu-Ld} - \ref{Tab:dom_H} summarize the influence of domain size parameters ($L_u$, $L_d$, $H$) on the force coefficients ($C_{DP}$, $C_D$, $C_{LP}$, $C_L$) for both cylinders (UC and LC) at the extreme values of gap ratio ($G = 0.2, 1$), Reynolds number ($Re = 1, 40$), and power-law index ($n = 0.2, 1, 1.8$) for the maximum rotational rate ($\alpha = 2$) using grid G1 (see \tab\ref{tab:grid} for grid specifications).

\noindent
The effect of upstream length ($L_u$) on the force coefficients ($C_{DP}$, $C_D$, $C_{LP}$, $C_L$) is examined in \tab\ref{Tab:dom_Lu-Ld} by fixing the downstream length at $L_d = 200$ and the unconfined domain height at $H = 100$. At $G = 0.2$ and $Re = 1$, the absolute percentage variations in the individual and total drag ($C_{DP}$, $C_D$) and lift ($C_{LP}$, $C_L$) coefficients for $L_u = 40-80$ are below $0.01\%$, $1.5\%$, and $0.9\%$, respectively, for $n = 0.2$, $1$, and $1.8$ for both cylinders. The corresponding variations between $L_u = 80$ and $120$ are below $0.02\%$, $0.10\%$, and $0.07\%$, while further increasing $L_u$ from $120$ to $160$ results in negligible differences ($0.00\%$, $0.006\%$, and $0.00\%$ for $n = 0.2$, $1$, and $1.8$, respectively). At $G = 0.2$ and $Re = 40$, the relative variations for $n = 0.2$, $1$, and $1.8$ are below $0.11\%$, $0.75\%$, and $0.46\%$ when comparing values for $L_u = 40$ and $80$; below $0.09\%$, $0.05\%$, and $0.02\%$ for $L_u = 80-120$; and below $0.01\%$ for $L_u = 120-160$, respectively, for both cylinders.
Furthermore, for both cylinders at a gap ratio of $G = 1$ and Reynolds number $Re = 1$, the absolute percentage variations in the drag and lift coefficients are below $0.01\%$, $1.92\%$, and $0.36\%$ for $L_u = 40$ compared to $80$; below $0.002\%$, $0.13\%$, and $0.02\%$ for $L_u = 80$ compared to $120$; and below $0.05\%$, $0.01\%$, and $0.00\%$ for $L_u = 120$ compared to $160$, for $n = 0.2$, $1$, and $1.8$, respectively. Similarly, at $G = 1$ and $Re = 40$, the corresponding variations are below $0.03\%$, $1.40\%$, and $46.23\%$ for $L_u = 40$ compared to $80$; below $0.002\%$, $0.10\%$, and $7.66\%$ for $L_u = 80$ compared to $120$; and below $0.05\%$, $0.006\%$, and $1.22\%$ for $L_u = 120$ compared to $160$, respectively, for both cylinders. Overall, the relative percentage variations exhibit notable changes when the upstream length increases from $L_u/D = 40$ to $80$, indicating a significant influence on the results. However, the variations become minimal between $L_u/D = 80$ and $120$ and further diminish between $L_u/D = 120$ and $160$. Therefore, $L_u/D = 120$ is chosen as the optimal upstream distance, providing a balance between computational accuracy and efficiency.

\noindent
The effect of downstream length ($L_d$) on the force coefficients ($C_{DP}$, $C_D$, $C_{LP}$, $C_L$) is examined in \tab\ref{Tab:dom_Ld} by fixing the upstream length at $L_u = 120$ and the unconfined domain height at $H = 100$. For the smallest gap ratio ($G = 0.2$) and $Re = 1$, the absolute percentage variations in force coefficients for $n = 0.2$, $1$, and $1.8$ are below $0.40\%$, $1.52\%$, and $1.53\%$ for $L_d = 100-200$; below $0.76\%$, $0.10\%$, and $0.98\%$ for $L_d = 200-300$; and below $0.45\%$, $0.22\%$, and $0.48\%$ for $L_d = 300-400$, respectively, for both cylinders. At $G = 1$ and $Re = 1$, the corresponding variations are below $0.69\%$, $0.07\%$, and $0.56\%$ for $L_d = 100-200$; below $0.47\%$, $0.22\%$, and $0.99\%$ for $L_d = 200-300$; and below $0.39\%$, $0.17\%$, and $0.47\%$ for $L_d = 300-400$, respectively. For $G = 1$ and $Re = 40$, the relative variations are below $0.12\%$, $0.05\%$, and $15.49\%$ for $L_d = 100-200$; below $0.06\%$, $0.03\%$, and $11.14\%$ for $L_d = 200-300$; and below $0.10\%$, $0.02\%$, and $4.85\%$ for $L_d = 300-400$ at $n = 0.2$, $1$, and $1.8$, respectively. Since the force coefficients exhibit negligible changes beyond $L_d = 300$, a downstream length of $L_d = 300$ is adopted to ensure accurate flow representation while maintaining computational efficiency.

\noindent
In unconfined flow analysis, the computational domain is bounded vertically ($L_u$, $L_d$), and minimizing wall effects is essential for accurate flow characterization. The influence of the domain height ($H$) is examined for $H = 100$, $200$, $300$, and $400$, while maintaining fixed upstream and downstream lengths of $L_u = 120$ and $L_d = 300$, respectively, with $\alpha = 2$. For $G = 0.2$, $Re = 1$, and $n = 0.2$, $1$, and $1.8$, the absolute percentage variations in force coefficients are below $0.14\%$, $1.84\%$, and $1.59\%$ when comparing $H = 100-200$; below $0.19\%$, $0.43\%$, and $0.36\%$ for $H = 200$–$300$; and below $0.11\%$, $0.13\%$, and $0.11\%$ for $H = 300-400$, respectively. At $Re = 40$ and the same gap ratio, these variations further reduce to below $0.04\%$, $0.50\%$, and $0.30\%$; $0.04\%$, $0.08\%$, and $0.09\%$; and $0.05\%$, $0.02\%$, and $0.03\%$, respectively, for the same height intervals. For a wider gap ratio ($G = 1$) at $Re = 1$, the corresponding variations are below $0.06\%$, $2.98\%$, and $1.42\%$ for $H = 100-200$; below $0.01\%$, $0.62\%$, and $0.39\%$ for $H = 200-300$; and below $0.00\%$, $0.17\%$, and $0.12\%$ for $H = 300-400$. At higher Reynolds number ($Re = 40$), the respective variations are below $0.08\%$, $1.05\%$, and $39.90\%$; $0.18\%$, $0.21\%$, and $24.19\%$; and $0.21\%$, $0.13\%$, and $21.48\%$. It is evident that domain heights of $H = 100$ and $200$ are insufficient due to noticeable deviations from the results obtained at $H = 300$. However, only marginal differences are observed between $H = 300$ and $400$. Therefore, $H = 300$ is selected as the optimal domain height, ensuring accurate results while maintaining computational efficiency.

\noindent
In summary, a computational domain of size $L_u = 120$, $L_d = 300$, and $H = 300$ is found to be sufficiently large to accurately capture the hydrodynamic characteristics of unconfined non-Newtonian flow past a pair of rotating circular cylinders.
\begin{figure}[!b]
	\centering
	\includegraphics[width=0.7\linewidth]{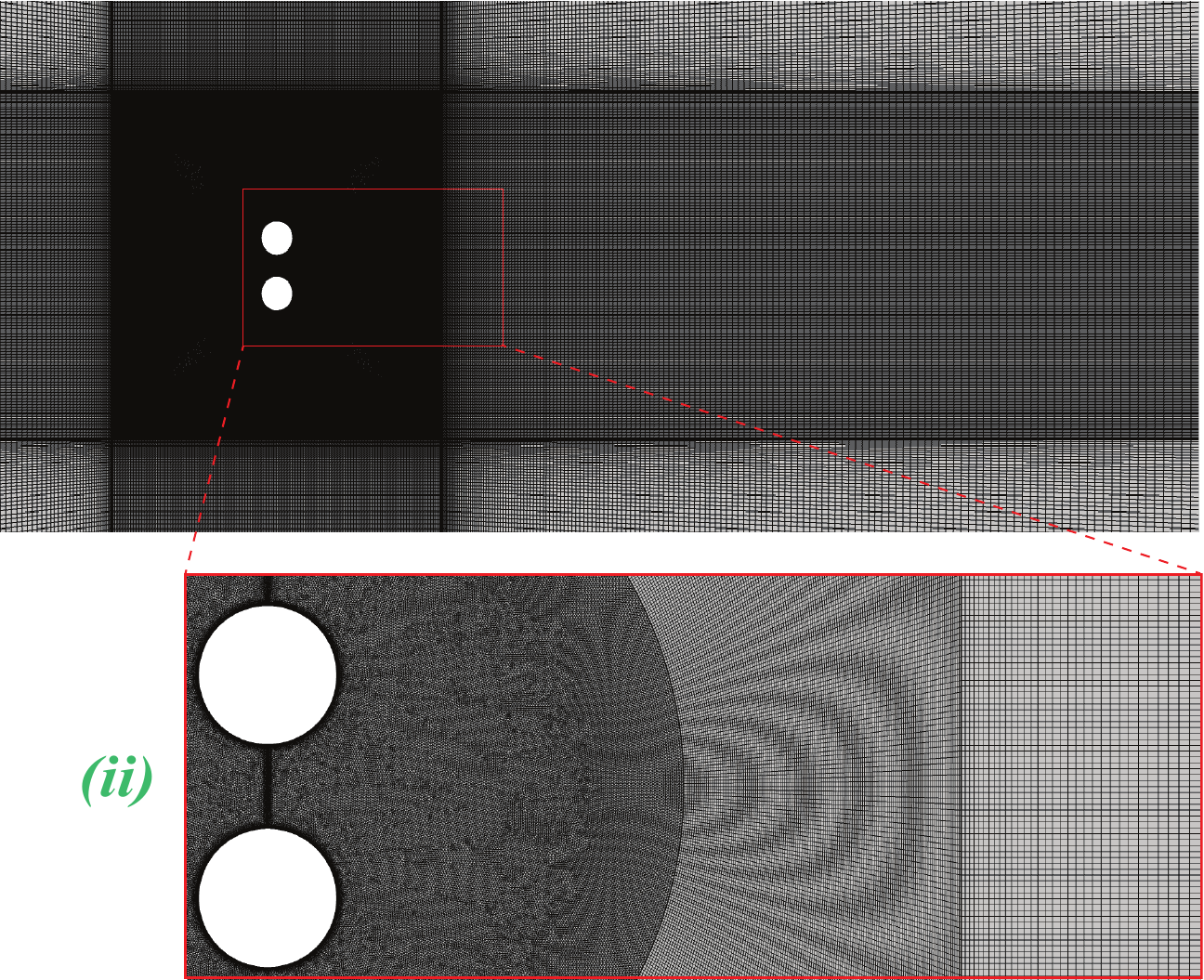} 
	\caption{Computational mesh structure for the considered flow geometry.}
	\label{fig:1g}
\end{figure}
%
%
\begin{table}[!tb]
	\centering
	\caption{Grid characteristics. Here, $N_c$ is the mesh points over the surface of a single cylinder, $\delta_r$ is the minimum mesh spacing, $N_{te}$ is the triangular mesh elements near the cylinders, and $N_e$ is the total number of mesh elements in the selected computational domain ($L_u=120$, $L_d=300$, $H=300$).} \label{tab:grid}
	\renewcommand{\arraystretch}{1.3}
	{\footnotesize
		\begin{tabular}{|c|c|c|c|c|c|c|}
			\hline
			\multirow{2}{*}{{Grid}}& \multirow{2}{*}{$N_c$ }& \multirow{2}{*}{$\delta_r - \delta - \Delta$}& \multicolumn{2}{c|}{$G =0.2$}&\multicolumn{2}{c|}{$G =3$}\\			\cline{4-7}
			& & & {$N_{te}$} & {$N_e$}& {$N_e$}& {$N_e$}\\
			\hline
			$G_1$ & 90		&0.005 - 0.1 - 0.5	&244018	&1320376	 &247506	&1323796  \\
			$G_2$ & 135		&0.005 - 0.1 - 0.5	&249400	&1388060	&252236	&1390896 \\
			$G_3$ & 180		&0.005 - 0.1 - 0.5	&254652	&1455682	&256784	&1457814	\\
			\hline
		\end{tabular}
	}
\end{table}
\begin{table}[!tb]
	\caption{The grid independence test for extreme gap ratio ($G$), Reynolds numbers ($Re$) and power-law index ($n$) for the maximum rotational rate ($\alpha=2$) of the cylinders.}\label{tab:grid1}
	\resizebox{\columnwidth}{!}{%
		\renewcommand{\arraystretch}{1.3}
		\begin{tabular}{@{}|c|c|c|rr|rr|rr|rr|rr|rr|rr|rr|@{}}
			\hline
			\multicolumn{3}{|c|}{$L_u=120, L_d=300$} &\multicolumn{8}{c|}{$Re = 1$} & \multicolumn{8}{c|}{$Re = 40$} \\ \cline{4-19}
			\multicolumn{3}{|c|}{$H=300$} &\multicolumn{4}{c|}{Upper cylinder (UC)}&\multicolumn{4}{c|}{Lower cylinder (LC)}  &\multicolumn{4}{c|}{Upper cylinder (UC)}&\multicolumn{4}{c|}{Lower cylinder (LC)} \\
			\hline
			{$G$}&{$n$}&Grid&$C_{DP}$&$C_{D}$&$C_{LP}$&$C_{L}$&$C_{DP}$&$C_{D}$&$C_{LP}$&$C_{L}$&$C_{DP}$&$C_{D}$&$C_{LP}$&$C_{L}$&$C_{DP}$&$C_{D}$&$C_{LP}$&$C_{L}$\\
			\hline			
			0.2	&	0.2 &	$G_1$	&	15.1950	&	17.5900	&	5.5378	&	5.7083	&	15.2250	&	17.5610	&	-5.4305	&	-5.8144	&	0.4179	&	0.5370	&	4.3275	&	4.3772	&	0.4181	&	0.5369	&	-4.3209	&	-4.3706	\\
			&	&	$G_2$	&	15.2060	&	17.5770	&	5.5491	&	5.7045	&	15.2430	&	17.5530	&	-5.4261	&	-5.8293	&	0.4188	&	0.5405	&	4.3187	&	4.3733	&	0.4188	&	0.5402	&	-4.3223	&	-4.3773	\\
			&	&		$G_3$	&	15.2070	&	17.5820	&	5.5382	&	5.7280	&	15.2360	&	17.5480	&	-5.4421	&	-5.8057	&	0.4192	&	0.5418	&	4.3244	&	4.3825	&	0.4188	&	0.5414	&	-4.3210	&	-4.3792	\\
			\hline			
			0.2	& 1	&		$G_1$	&	-4.0890	&	1.4250	&	3.2841	&	4.2011	&	-4.0912	&	1.4166	&	-3.2864	&	-4.2053	&	-0.1250	&	0.1902	&	2.7816	&	3.0816	&	-0.1252	&	0.1897	&	-2.7813	&	-3.0816	\\
			&	&		$G_2$	&	-4.0893	&	1.4112	&	3.2841	&	4.2123	&	-4.0900	&	1.4062	&	-3.2837	&	-4.2089	&	-0.1248	&	0.1904	&	2.7806	&	3.0873	&	-0.1249	&	0.1899	&	-2.7808	&	-3.0876	\\
			&	&		$G_3$	&	-4.0907	&	1.4026	&	3.2802	&	4.2085	&	-4.0885	&	1.4077	&	-3.2872	&	-4.2168	&	-0.1248	&	0.1903	&	2.7804	&	3.0904	&	-0.1248	&	0.1899	&	-2.7800	&	-3.0898	\\
			\hline			
			0.2	& 1.8	&		$G_1$	&	-86.5050	&	-2.2142	&	6.3250	&	9.0061	&	-86.5110	&	-2.2662	&	-6.3463	&	-9.0240	&		-2.3013	&	-0.2144	&	3.8594	&	4.6912	&	-2.3020	&	-0.2163	&	-3.8604	&	-4.6927	\\
			&	&		$G_2$	&	-86.5600	&	-2.4299	&	6.3254	&	9.0250	&	-86.5620	&	-2.4590	&	-6.3442	&	-9.0486	&	-2.3027	&	-0.2193	&	3.8594	&	4.7010	&	-2.3030	&	-0.2207	&	-3.8599	&	-4.7018	\\
			&	&		$G_3$	&	-86.5860	&	-2.5520	&	6.3076	&	9.0163	&	-86.5770	&	-2.5273	&	-6.3517	&	-9.0443	&	-2.3031	&	-0.2223	&	3.8591	&	4.7063	&	-2.3033	&	-0.2239	&	-3.8595	&	-4.7046	\\
			\hline			
			1	& 0.2	&		$G_1$	&	17.4340	&	20.3350	&	6.3912	&	6.5756	&	17.4550	&	20.3150	&	-6.2978	&	-6.7003	&	0.5437	&	0.6661	&	4.2304	&	4.2786	&	0.5434	&	0.6657	&	-4.2220	&	-4.2700	\\
			&	&		$G_2$	&	17.4260	&	20.3450	&	6.3935	&	6.6095	&	17.4800	&	20.2990	&	-6.2645	&	-6.7295	&	0.5446	&	0.6697	&	4.2221	&	4.2748	&	0.5446	&	0.6692	&	-4.2236	&	-4.2769	\\
			&	&		$G_3$	&	17.4250	&	20.3470	&	6.4295	&	6.5842	&	17.4710	&	20.3090	&	-6.2955	&	-6.7048	&	0.5445	&	0.6704	&	4.2323	&	4.2887	&	0.5448	&	0.6704	&	-4.2262	&	-4.2826	\\
			\hline			
			1	&	1&		$G_1$	&	1.9442	&	4.0749	&	2.2977	&	3.4389	&	1.9399	&	4.0625	&	-2.3007	&	-3.4459	&	-0.1713	&	0.3698	&	4.7922	&	5.3169	&	-0.1714	&	0.3693	&	-4.7919	&	-5.3169	\\
			&	&		$G_2$	&	1.9438	&	4.0746	&	2.2987	&	3.4495	&	1.9420	&	4.0713	&	-2.2988	&	-3.4469	&	-0.1708	&	0.3693	&	4.7915	&	5.3276	&	-0.1710	&	0.3688	&	-4.7917	&	-5.3279	\\
			&	&		$G_3$	&	1.9426	&	4.0734	&	2.2990	&	3.4517	&	1.9430	&	4.0744	&	-2.3009	&	-3.4561	&	-0.1708	&	0.3687	&	4.7920	&	5.3340	&	-0.1709	&	0.3686	&	-4.7914	&	-5.3334	\\
			\hline			
			1	&	1.8 &		$G_1$	&	-9.9659	&	-1.0668	&	3.9483	&	7.2445	&	-9.9805	&	-1.1105	&	-3.9562	&	-7.2643	&	-0.7164	&	0.0049	&	3.8958	&	5.1905	&	-0.7169	&	0.0033	&	-3.8960	&	-5.1919	\\
			&	&		$G_2$	&	-9.9732	&	-1.1135	&	3.9493	&	7.2648	&	-9.9783	&	-1.1241	&	-3.9510	&	-7.2594	&	-0.7160	&	0.0043	&	3.8953	&	5.2004	&	-0.7162	&	0.0025	&	-3.8950	&	-5.2014	\\
			&	&		$G_3$	&	-9.9781	&	-1.1370	&	3.9496	&	7.2681	&	-9.9773	&	-1.1349	&	-3.9551	&	-7.2812	&	-0.7159	&	0.0019	&	3.8952	&	5.2044	&	-0.7157	&	0.0014	&	-3.8949	&	-5.2055	\\
			\hline			
		\end{tabular}
	}
\end{table}
%
\subsection{Grid independence test}
\noindent
The grid sensitivity analysis is performed using three levels of non-uniform unstructured meshes, denoted as $G_1$, $G_2$, and $G_3$ (refer to \tab\ref{tab:grid} for detailed specifications), with the optimized computational domain ($L_u = 120$, $L_d = 300$, $H = 300$). \fig\ref{fig:1g} shows the computational mesh structure used in this study. The influence of grid resolution on the computed flow characteristics is assessed by varying the number of mesh elements on each cylinder surface ($90 \le N_c \le 180$). 
\tab\ref{tab:grid1} presents the influence of mesh refinement on the absolute percentage variations in the force coefficients ($C_{DP}$, $C_D$, $C_{LP}$, and $C_L$) for both cylinders under different combinations of governing parameters: $n = 0.2$, $1$, and $1.8$; $Re = 1$ and $40$; $G = 0.2$ and $1$; and $\alpha = 2$ (maximum rotational rate). For $Re = 1$, the variations between grids $G_2$ and $G_1$ are found to be  under $0.256\%$, $0.968\%$, and $9.742\%$, whereas the corresponding differences between $G_3$ and $G_2$ are below $0.412\%$, $0.609\%$, and $5.025\%$. Similarly, for $Re = 40$ and the same set of parameters, the variations between $G_2$ and $G_1$ were below $0.659\%$, $0.208\%$, and $2.280\%$, while those between $G_3$ and $G_2$ are below $0.231\%$, $0.1\%$, and $1.487\%$.
Furthermore, at $G=1$ and $Re=1$, the variations in the drag and lift components are below $0.529\%$, $0.308\%$, and $4.378\%$ between $G_2$ and $G_1$, and below $0.563\%$, $0.267\%$, and $2.11\%$ between $G_3$ and $G_2$. Similarly, at $G=1$ and $Re=40$, the corresponding variations are below $0.543\%$, $0.263\%$, and $0.191\%$ between $G_2$ and $G_1$, and below $0.325\%$, $0.162\%$, and $0.079\%$ between $G_3$ and $G_2$. Based on these results, the grid configuration $G_2$ is selected for the present study, as it provides an optimal balance between computational accuracy and efficiency.
\clearpage%
\clearpage
\section{Pressure and friction drag coefficients ($C_{DP}$ and $C_{DF}$)}\label{appendix:dragtable}
\setcounter{table}{0}\setcounter{figure}{0} 
%
\begin{table}[ht!]
	\centering
	\caption{Dependence of the pressure ($C_{DP}$) and friction ($C_{DF}$) drag coefficients  on the Reynolds number ($Re=1$), rotational velocity ($\alpha$) and power-law index ($n$) for upper ({UC}) and lower ({LC}) cylinders.}\label{tab:drag1}
	\resizebox{\columnwidth}{!}{%
	\renewcommand{\arraystretch}{1.3}
\begin{tabular}{|c|c|r|r|r|r|r|r|r|r|r|r|r|r|r|r|r|r|r|r|}	 %
		\hline
			\multicolumn{2}{|c|}{$Re=1$}  & \multicolumn{9}{c|}{$C_{DP}$ (for UC)}&\multicolumn{9}{c|}{$C_{DP}$ (for LC)} \\ \hline	
			$\alpha$ &	$G$ &$n\rightarrow$ 0.2&0.4&0.6&0.8&1&1.2&1.4&1.6&1.8&0.2&0.4&0.6&0.8&1&1.2&1.4&1.6&1.8 \\ \hline
0&	0.2	&	19.255	&	13.149	&	8.2116	&	5.5667	&	4.1126	&	3.2396	&	2.6811	&	2.3058	&	2.0409	&	19.255	&	13.149	&	8.2113	&	5.5663	&	4.1123	&	3.2393	&	2.6808	&	2.3054	&	2.0406	\\	\cline{2-20}
&	0.4	&	19.763	&	13.302	&	8.2119	&	5.5284	&	4.0639	&	3.1896	&	2.6327	&	2.2597	&	1.9965	&	19.762	&	13.301	&	8.2112	&	5.5278	&	4.0633	&	3.1891	&	2.6322	&	2.2592	&	1.996	\\	\cline{2-20}
&	0.6	&	20.056	&	13.504	&	8.2746	&	5.5444	&	4.063	&	3.183	&	2.6243	&	2.2504	&	1.9864	&	20.056	&	13.503	&	8.2741	&	5.5439	&	4.0626	&	3.1826	&	2.6239	&	2.25	&	1.986	\\	\cline{2-20}
&	0.8	&	20.132	&	13.699	&	8.3615	&	5.5862	&	4.0872	&	3.2006	&	2.6394	&	2.2639	&	1.9983	&	20.131	&	13.699	&	8.3611	&	5.5857	&	4.0867	&	3.2001	&	2.6389	&	2.2634	&	1.9979	\\	\cline{2-20}
&	1	&	20.078	&	13.8680	&	8.4556	&	5.6403	&	4.1246	&	3.2317	&	2.6679	&	2.2908	&	2.0235	&	20.078	&	13.868	&	8.4552	&	5.6398	&	4.1241	&	3.2311	&	2.6674	&	2.2903	&	2.023	\\	\hline
1&	0.2	&	16.739	&	11.505	&	6.7605	&	3.2949	&	0.1225	&	-3.5203	&	-8.2432	&	-14.706	&	-23.744	&	16.719	&	11.511	&	6.7602	&	3.2945	&	0.1220	&	-3.5209	&	-8.2439	&	-14.706	&	-23.745	\\
\end{tabular}
%
			\multicolumn{2}{|c|}{$Re=1$}  & \multicolumn{9}{c|}{$C_{DF}$ (for UC)}&\multicolumn{9}{c|}{$C_{DF}$ (for LC)} \\ \hline	
$\alpha$ &	$G$ &$n\rightarrow$ 0.2&0.4&0.6&0.8&1&1.2&1.4&1.6&1.8&0.2&0.4&0.6&0.8&1&1.2&1.4&1.6&1.8 \\ \hline
0&	0.2	&	5.2057	&	5.0228	&	4.0525	&	3.2814	&	2.7519	&	2.3854	&	2.1237	&	1.9297	&	1.7807	&	5.2049	&	5.022	&	4.0521	&	3.2811	&	2.7516	&	2.3852	&	2.1236	&	1.9296	&	1.7806	\\	\cline{2-20}
&	0.4	&	6.1607	&	5.7449	&	4.5309	&	3.6273	&	3.0254	&	2.6170	&	2.3297	&	2.1191	&	1.9584	&	6.1592	&	5.7443	&	4.5305	&	3.6269	&	3.025	&	2.6165	&	2.3293	&	2.1186	&	1.9579	\\	\cline{2-20}
&	0.6	&	6.7665	&	6.2778	&	4.8925	&	3.8897	&	3.2328	&	2.7930	&	2.4875	&	2.2655	&	2.0971	&	6.7661	&	6.2775	&	4.8922	&	3.8894	&	3.2325	&	2.7928	&	2.4872	&	2.2653	&	2.0969	\\	\cline{2-20}
&	0.8	&	7.1134	&	6.6834	&	5.1820	&	4.1025	&	3.4018	&	2.9370	&	2.6170	&	2.3863	&	2.2118	&	7.1133	&	6.6831	&	5.1816	&	4.1020	&	3.4012	&	2.9365	&	2.6164	&	2.3857	&	2.2112	\\	\cline{2-20}
&	1	&	7.2992	&	6.9928	&	5.4182	&	4.2799	&	3.544	&	3.0593	&	2.7278	&	2.4901	&	2.3107	&	7.2987	&	6.9927	&	5.4179	&	4.2795	&	3.5435	&	3.0587	&	2.7272	&	2.4895	&	2.3101	\\ \hline
1&	0.2	&	3.9106	&	2.955	&	2.2211	&	2.7216	&	4.2015	&	6.8094	&	10.8700	&	16.9000	&	25.661	&	3.9973	&	2.9335	&	2.2200	&	2.7198	&	4.1993	&	6.8064	&	10.866	&	16.894	&	25.654	\\
\hline
		\end{tabular}
	}
\end{table}
\begin{table}[h]
	\centering
	\caption{Dependence of the pressure ($C_{DP}$) and friction ($C_{DF}$) drag coefficients  on the Reynolds number ($Re=5$), rotational velocity ($\alpha$) and power-law index ($n$) for upper ({UC}) and lower ({LC}) cylinders.}\label{tab:drag3}
	\resizebox{\columnwidth}{!}{%
		\renewcommand{\arraystretch}{1.3}
		\begin{tabular}{|c|c|r|r|r|r|r|r|r|r|r|r|r|r|r|r|r|r|r|r|}	 %
			\hline
			\multicolumn{2}{|c|}{$Re=5$}  & \multicolumn{9}{c|}{$C_{DP}$ (for UC)}&\multicolumn{9}{c|}{$C_{DP}$ (for LC)} \\ \hline	
			$\alpha$ &	$G$ &$n\rightarrow$ 0.2&0.4&0.6&0.8&1&1.2&1.4&1.6&1.8&0.2&0.4&0.6&0.8&1&1.2&1.4&1.6&1.8 \\ \hline
			0	&	0.2	&	4.302	&	3.1987	&	2.5232	&	2.113	&	1.8451	&	1.6599	&	1.5232	&	1.4162	&	1.3284	&	4.3019	&	3.1985	&	2.523	&	2.1128	&	1.8449	&	1.6597	&	1.523	&	1.4159	&	1.3282	\\	\cline{2-20}
			&	0.4	&	4.3636	&	3.2396	&	2.5376	&	2.1134	&	1.8373	&	1.6468	&	1.506	&	1.3953	&	1.3039	&	4.3632	&	3.2392	&	2.5372	&	2.1131	&	1.837	&	1.6465	&	1.5057	&	1.395	&	1.3036	\\	\cline{2-20}
			&	0.6	&	4.3932	&	3.2881	&	2.572	&	2.1374	&	1.8546	&	1.6584	&	1.5129	&	1.3982	&	1.3032	&	4.3931	&	3.2879	&	2.5718	&	2.1372	&	1.8543	&	1.6581	&	1.5126	&	1.398	&	1.303	\\	\cline{2-20}
			&	0.8	&	4.4051	&	3.3302	&	2.613	&	2.1731	&	1.8857	&	1.6851	&	1.5354	&	1.4169	&	1.3184	&	4.405	&	3.33	&	2.6127	&	2.1728	&	1.8854	&	1.6848	&	1.535	&	1.4165	&	1.3181	\\	\cline{2-20}
			&	1	&	4.4072	&	3.363	&	2.6539	&	2.2135	&	1.924	&	1.7205	&	1.5674	&	1.4457	&	1.3444	&	4.4071	&	3.3627	&	2.6536	&	2.2132	&	1.9237	&	1.7201	&	1.5671	&	1.4454	&	1.344	\\	\hline
			1	&	0.2	&	3.5608	&	2.536	&	1.8363	&	1.2508	&	0.641	&	-0.091	&	-1.0466	&	-2.3513	&	-4.1705	&	3.5596	&	2.5364	&	1.8368	&	1.2507	&	0.6407	&	-0.0913	&	-1.047	&	-2.3517	&	-4.1709	\\
\end{tabular}
%
			\multicolumn{2}{|c|}{$Re=5$}  & \multicolumn{9}{c|}{$C_{DF}$ (for UC)}&\multicolumn{9}{c|}{$C_{DF}$ (for LC)} \\ \hline	
			$\alpha$ &	$G$ &$n\rightarrow$ 0.2&0.4&0.6&0.8&1&1.2&1.4&1.6&1.8&0.2&0.4&0.6&0.8&1&1.2&1.4&1.6&1.8 \\ \hline
		0&	0.2	&	1.1142	&	1.1361	&	1.1138	&	1.0861	&	1.0582	&	1.0324	&	1.0086	&	0.9859	&	0.9635	&	1.1138	&	1.1358	&	1.1136	&	1.0859	&	1.0581	&	1.0323	&	1.0085	&	0.9858	&	0.9634	\\	\cline{2-20}
		&	0.4	&	1.3296	&	1.3308	&	1.2855	&	1.244	&	1.2082	&	1.1773	&	1.1494	&	1.1232	&	1.0971	&	1.3294	&	1.3306	&	1.2854	&	1.2438	&	1.208	&	1.1771	&	1.1492	&	1.1229	&	1.0968	\\	\cline{2-20}
		&	0.6	&	1.4204	&	1.4677	&	1.4158	&	1.3675	&	1.328	&	1.2951	&	1.2655	&	1.2372	&	1.2087	&	1.4207	&	1.4676	&	1.4157	&	1.3674	&	1.3279	&	1.2949	&	1.2654	&	1.2371	&	1.2085	\\	\cline{2-20}
		&	0.8	&	1.4548	&	1.5626	&	1.5182	&	1.4696	&	1.4295	&	1.3964	&	1.3666	&	1.3375	&	1.3074	&	1.4547	&	1.5624	&	1.5179	&	1.4692	&	1.4291	&	1.396	&	1.3662	&	1.3371	&	1.307	\\	\cline{2-20}
		&	1	&	1.4669	&	1.6259	&	1.5983	&	1.5547	&	1.5171	&	1.4856	&	1.4568	&	1.4281	&	1.3976	&	1.4669	&	1.6257	&	1.5981	&	1.5544	&	1.5167	&	1.4853	&	1.4565	&	1.4277	&	1.3973	\\	\hline			
		1&	0.2	&	0.987	&	0.788	&	0.7673	&	0.9162	&	1.2308	&	1.7524	&	2.5522	&	3.7384	&	5.4666	&	0.9869	&	0.7765	&	0.7654	&	0.9155	&	1.2302	&	1.7517	&	2.5512	&	3.7372	&	5.4647	\\
\hline		
		\end{tabular}
	}
\end{table}
\begin{table}[ht!]
	\centering
	\caption{Dependence of the pressure ($C_{DP}$) and friction ($C_{DF}$) drag coefficients  on the Reynolds number ($Re=10$), rotational velocity ($\alpha$) and power-law index ($n$) for upper (UC) and lower (LC) cylinders.}\label{tab:drag5}
	\resizebox{\columnwidth}{!}{%
		\renewcommand{\arraystretch}{1.3}
		\begin{tabular}{|c|c|r|r|r|r|r|r|r|r|r|r|r|r|r|r|r|r|r|r|}	 %
			\hline
			\multicolumn{2}{|c|}{$Re=10$}  & \multicolumn{9}{c|}{$C_{DP}$ (for UC)}&\multicolumn{9}{c|}{$C_{DP}$ (for LC)} \\ \hline	
			$\alpha$ &	$G$ &$n\rightarrow$ 0.2&0.4&0.6&0.8&1&1.2&1.4&1.6&1.8&0.2&0.4&0.6&0.8&1&1.2&1.4&1.6&1.8 \\ \hline
		0&	0.2	&	2.4644	&	2.0012	&	1.7318	&	1.559	&	1.4371	&	1.3429	&	1.2654	&	1.1984	&	1.1385	&	2.4642	&	2.001	&	1.7316	&	1.5588	&	1.4369	&	1.3428	&	1.2652	&	1.1982	&	1.1383	\\	\cline{2-20}
		&	0.4	&	2.5182	&	2.0354	&	1.7529	&	1.5678	&	1.4363	&	1.3352	&	1.252	&	1.18	&	1.1149	&	2.518	&	2.0351	&	1.7527	&	1.5675	&	1.436	&	1.3349	&	1.2517	&	1.1797	&	1.1147	\\	\cline{2-20}
		&	0.6	&	2.5459	&	2.0812	&	1.7948	&	1.6019	&	1.462	&	1.3535	&	1.264	&	1.1866	&	1.1165	&	2.5458	&	2.0811	&	1.7947	&	1.6017	&	1.4618	&	1.3533	&	1.2638	&	1.1864	&	1.1163	\\	\cline{2-20}
		&	0.8	&	2.5452	&	2.1175	&	1.8405	&	1.6463	&	1.5018	&	1.388	&	1.2935	&	1.2116	&	1.1374	&	2.5452	&	2.1174	&	1.8403	&	1.6461	&	1.5016	&	1.3877	&	1.2933	&	1.2113	&	1.1371	\\	\cline{2-20}
		&	1	&	2.5402	&	2.136	&	1.8768	&	1.6887	&	1.5449	&	1.4295	&	1.3328	&	1.2489	&	1.173	&	2.5401	&	2.1359	&	1.8766	&	1.6885	&	1.5447	&	1.4292	&	1.3326	&	1.2486	&	1.1728	\\	\hline	
		1&	0.2	&	1.9555	&	1.4332	&	1.1262	&	0.8638	&	0.5813	&	0.2333	&	-0.23	&	-0.8703	&	-1.7701	&	1.9552	&	1.4334	&	1.1262	&	0.8636	&	0.5811	&	0.2331	&	-0.23	&	-0.8703	&	-1.7701	\\
\end{tabular}
%
			\multicolumn{2}{|c|}{$Re=10$}  & \multicolumn{9}{c|}{$C_{DF}$ (for UC)}&\multicolumn{9}{c|}{$C_{DF}$ (for LC)} \\ \hline	
			$\alpha$ &	$G$ &$n\rightarrow$ 0.2&0.4&0.6&0.8&1&1.2&1.4&1.6&1.8&0.2&0.4&0.6&0.8&1&1.2&1.4&1.6&1.8 \\ \hline
		0	&	0.2	&	0.5994	&	0.6461	&	0.6818	&	0.7065	&	0.7225	&	0.732	&	0.7363	&	0.7359	&	0.7311	&	0.5992	&	0.6459	&	0.6817	&	0.7063	&	0.7223	&	0.7319	&	0.7362	&	0.7358	&	0.731	\\	\cline{2-20}
		&	0.4	&	0.696	&	0.7784	&	0.814	&	0.8389	&	0.8548	&	0.8634	&	0.8658	&	0.8627	&	0.8544	&	0.696	&	0.7783	&	0.8139	&	0.8388	&	0.8547	&	0.8632	&	0.8656	&	0.8625	&	0.8541	\\	\cline{2-20}
		&	0.6	&	0.7143	&	0.8595	&	0.9126	&	0.9436	&	0.9629	&	0.9734	&	0.9762	&	0.9722	&	0.9617	&	0.7143	&	0.8595	&	0.9126	&	0.9435	&	0.9628	&	0.9733	&	0.9761	&	0.9721	&	0.9616	\\	\cline{2-20}
		&	0.8	&	0.7189	&	0.9009	&	0.9823	&	1.0265	&	1.0533	&	1.0683	&	1.0739	&	1.0711	&	1.0604	&	0.7188	&	0.9008	&	0.9821	&	1.0263	&	1.053	&	1.068	&	1.0735	&	1.0708	&	1.0601	\\	\cline{2-20}
		&	1	&	0.7193	&	0.92	&	1.026	&	1.0873	&	1.1251	&	1.1478	&	1.1588	&	1.16	&	1.1523	&	0.7192	&	0.9199	&	1.0258	&	1.0871	&	1.1249	&	1.1475	&	1.1585	&	1.1597	&	1.152	\\	\hline			
		1	&	0.2	&	0.5387	&	0.5544	&	0.5637	&	0.6353	&	0.7848	&	1.035	&	1.4255	&	2.0104	&	2.8626	&	0.5381	&	0.5536	&	0.5632	&	0.635	&	0.7845	&	1.0345	&	1.4247	&	2.0091	&	2.8613	\\
\hline
		\end{tabular}
	}
\end{table}
\begin{table}[ht!]
	\centering
	\caption{Dependence of the pressure ($C_{DP}$) and friction ($C_{DF}$) drag coefficients  on the Reynolds number ($Re=20$), rotational velocity ($\alpha$) and power-law index ($n$) for upper ({UC}) and lower ({LC}) cylinders.}\label{tab:drag7}
	\resizebox{\columnwidth}{!}{%
		\renewcommand{\arraystretch}{1.3}
		\begin{tabular}{|c|c|r|r|r|r|r|r|r|r|r|r|r|r|r|r|r|r|r|r|}	 %
			\hline
			\multicolumn{2}{|c|}{$Re=20$}  & \multicolumn{9}{c|}{$C_{DP}$ (for UC)}&\multicolumn{9}{c|}{$C_{DP}$ (for LC)} \\ \hline	
			$\alpha$ &	$G$ &$n\rightarrow$ 0.2&0.4&0.6&0.8&1&1.2&1.4&1.6&1.8&0.2&0.4&0.6&0.8&1&1.2&1.4&1.6&1.8 \\ \hline
		0	&	0.2	&	1.5988	&	1.3928	&	1.2941	&	1.2254	&	1.1681	&	1.1165	&	1.0677	&	1.0202	&	0.9722	&	1.5987	&	1.3927	&	1.294	&	1.2252	&	1.168	&	1.1163	&	1.0676	&	1.02	&	0.972	\\	\cline{2-20}
		&	0.4	&	1.6445	&	1.4675	&	1.3421	&	1.2519	&	1.18	&	1.1178	&	1.0606	&	1.0056	&	0.9499	&	1.6445	&	1.4673	&	1.3419	&	1.2517	&	1.1798	&	1.1176	&	1.0604	&	1.0054	&	0.9497	\\	\cline{2-20}
		&	0.6	&	1.6258	&	1.4943	&	1.398	&	1.3065	&	1.2254	&	1.1536	&	1.088	&	1.0256	&	0.9628	&	1.6257	&	1.4943	&	1.3979	&	1.3064	&	1.2253	&	1.1535	&	1.0879	&	1.0254	&	0.9627	\\	\cline{2-20}
		&	0.8	&	1.6081	&	1.4923	&	1.4184	&	1.3463	&	1.2741	&	1.204	&	1.137	&	1.0723	&	1.0076	&	1.6081	&	1.4923	&	1.4183	&	1.3463	&	1.274	&	1.2038	&	1.1368	&	1.0721	&	1.0074	\\	\cline{2-20}
		&	1	&	1.5929	&	1.4851	&	1.4219	&	1.3635	&	1.3046	&	1.2452	&	1.1861	&	1.1278	&	1.0699	&	1.5929	&	1.485	&	1.4218	&	1.3633	&	1.3044	&	1.245	&	1.1859	&	1.1276	&	1.0697	\\	\hline			
		1	&	0.2	&	1.1296	&	0.8725	&	0.7256	&	0.6093	&	0.4842	&	0.326	&	0.1099	&	-0.1951	&	-0.6299	&	1.1295	&	0.8724	&	0.7254	&	0.6091	&	0.484	&	0.3259	&	0.1096	&	-0.1953	&	-0.6298	\\
\end{tabular}
%
			\multicolumn{2}{|c|}{$Re=20$}  & \multicolumn{9}{c|}{$C_{DF}$ (for UC)}&\multicolumn{9}{c|}{$C_{DF}$ (for LC)} \\ \hline	
			$\alpha$ &	$G$ &$n\rightarrow$ 0.2&0.4&0.6&0.8&1&1.2&1.4&1.6&1.8&0.2&0.4&0.6&0.8&1&1.2&1.4&1.6&1.8 \\ \hline
			0	&	0.2	&	0.3225	&	0.3874	&	0.4341	&	0.471	&	0.4992	&	0.5198	&	0.5334	&	0.5407	&	0.5416	&	0.3224	&	0.3873	&	0.434	&	0.471	&	0.4992	&	0.5197	&	0.5334	&	0.5406	&	0.5415	\\	\cline{2-20}
			&	0.4	&	0.3293	&	0.4709	&	0.5459	&	0.5916	&	0.6233	&	0.6445	&	0.6568	&	0.6611	&	0.6569	&	0.3293	&	0.4709	&	0.5459	&	0.5916	&	0.6232	&	0.6444	&	0.6567	&	0.6609	&	0.6567	\\	\cline{2-20}
			&	0.6	&	0.3353	&	0.4895	&	0.6024	&	0.6755	&	0.7213	&	0.7498	&	0.7655	&	0.7704	&	0.7644	&	0.3352	&	0.4895	&	0.6023	&	0.6754	&	0.7212	&	0.7497	&	0.7654	&	0.7703	&	0.7644	\\	\cline{2-20}
			&	0.8	&	0.3373	&	0.4954	&	0.622	&	0.7179	&	0.7853	&	0.8299	&	0.8569	&	0.8696	&	0.8691	&	0.3372	&	0.4954	&	0.6219	&	0.7177	&	0.7851	&	0.8297	&	0.8566	&	0.8693	&	0.8688	\\	\cline{2-20}
			&	1	&	0.3383	&	0.4974	&	0.6293	&	0.7356	&	0.8178	&	0.8787	&	0.9214	&	0.9487	&	0.9626	&	0.3383	&	0.4973	&	0.6291	&	0.7354	&	0.8177	&	0.8785	&	0.9212	&	0.9485	&	0.9624	\\	\hline			
			1	&	0.2	&	0.2952	&	0.3745	&	0.4112	&	0.4547	&	0.5275	&	0.6458	&	0.8304	&	1.1104	&	1.528	&	0.2949	&	0.3742	&	0.411	&	0.4545	&	0.5272	&	0.6455	&	0.8301	&	1.1101	&	1.527	\\
\hline
		\end{tabular}
	}
\end{table}
\begin{table}[ht!]
	\centering
	\caption{Dependence of the pressure ($C_{DP}$) and friction ($C_{DF}$) drag coefficients  on the Reynolds number ($Re=40$), rotational velocity ($\alpha$) and power-law index ($n$) for upper ({UC}) and lower ({LC}) cylinders.}\label{tab:drag9}
	\resizebox{\columnwidth}{!}{%
		\renewcommand{\arraystretch}{1.3}
		\begin{tabular}{|c|c|r|r|r|r|r|r|r|r|r|r|r|r|r|r|r|r|r|r|}	 %
			\hline
			\multicolumn{2}{|c|}{$Re=40$}  & \multicolumn{9}{c|}{$C_{DP}$ (for UC)}&\multicolumn{9}{c|}{$C_{DP}$ (for LC)} \\ \hline	
			$\alpha$ &	$G$ &$n\rightarrow$ 0.2&0.4&0.6&0.8&1&1.2&1.4&1.6&1.8&0.2&0.4&0.6&0.8&1&1.2&1.4&1.6&1.8 \\ \hline
			0	&	0.2	&	1.1836	&	1.115	&	1.0474	&	1.0075	&	0.9729	&	0.9374	&	0.8989	&	0.85541	&	0.80074	&	1.1797	&	1.1148	&	1.0473	&	1.0074	&	0.9728	&	0.9373	&	0.8988	&	0.8553	&	0.8006	\\	\cline{2-20}
			&	0.4	&	1.1674	&	1.1462	&	1.1238	&	1.0763	&	1.0196	&	0.9656	&	0.9129	&	0.8573	&		&	1.1593	&	1.1457	&	1.1236	&	1.0762	&	1.0195	&	0.9655	&	0.9127	&	0.8572	&		\\	\cline{2-20}
			&	0.6	&	1.1417	&	1.1394	&	1.1374	&	1.119	&	1.0836	&	1.0352	&	0.9793	&	0.9182	&	0.8391	&	1.1388	&	1.139	&	1.1373	&	1.1189	&	1.0835	&	1.0351	&	0.9792	&	0.9181	&	0.839	\\	\cline{2-20}
			&	0.8	&	1.1194	&	1.1267	&	1.1348	&	1.1297	&	1.1115	&	1.0824	&	1.0443	&		&	0.9465	&	1.1182	&	1.1265	&	1.1347	&	1.1296	&	1.1114	&	1.0823	&	1.0442	&		&	0.9464	\\	\cline{2-20}
			&	1	&	1.101	&	1.1136	&	1.1268	&	1.1285	&	1.1192	&	1.1011	&	1.0765	&	1.047	&		&	1.1007	&	1.1133	&	1.1267	&	1.1284	&	1.1191	&	1.101	&	1.0764	&	1.0469	&		\\	\hline
			1	&	0.2	&	0.7132	&	0.5837	&	0.493	&	0.4364	&	0.3835	&	0.3155	&	0.219	&	0.0789	&	-0.1251	&	0.7127	&	0.5835	&	0.4928	&	0.4362	&	0.3834	&	0.3153	&	0.2188	&	0.0788	&	-0.1252	\\
\end{tabular}
%
			\multicolumn{2}{|c|}{$Re=40$}  & \multicolumn{9}{c|}{$C_{DF}$ (for UC)}&\multicolumn{9}{c|}{$C_{DF}$ (for LC)} \\ \hline	
			$\alpha$ &	$G$ &$n\rightarrow$ 0.2&0.4&0.6&0.8&1&1.2&1.4&1.6&1.8&0.2&0.4&0.6&0.8&1&1.2&1.4&1.6&1.8 \\ \hline
			0	&	0.2	&	0.1323	&	0.2384	&	0.2935	&	0.3261	&	0.3514	&	0.3704	&	0.3829	&	0.3884	&	0.3838	&	0.131	&	0.2383	&	0.2934	&	0.3261	&	0.3514	&	0.3703	&	0.3829	&	0.3884	&	0.3838	\\	\cline{2-20}
			&	0.4	&	0.1395	&	0.2475	&	0.3468	&	0.4217	&	0.4658	&	0.4905	&	0.5031	&	0.505	&		&	0.1389	&	0.2475	&	0.3468	&	0.4217	&	0.4658	&	0.4905	&	0.5031	&	0.5049	&		\\	\cline{2-20}
			&	0.6	&	0.1422	&	0.2528	&	0.3581	&	0.4509	&	0.5249	&	0.5765	&	0.6066	&	0.6182	&	0.6046	&	0.1421	&	0.2528	&	0.358	&	0.4508	&	0.5249	&	0.5764	&	0.6066	&	0.6182	&	0.6046	\\	\cline{2-20}
			&	0.8	&	0.1436	&	0.2547	&	0.3625	&	0.4609	&	0.5461	&	0.616	&	0.6695	&		&	0.725	&	0.1434	&	0.2547	&	0.3625	&	0.4608	&	0.5459	&	0.6158	&	0.6693	&		&	0.7248	\\	\cline{2-20}
			&	1	&	0.1442	&	0.2552	&	0.364	&	0.4648	&	0.5545	&	0.6319	&	0.6968	&	0.7495	&		&	0.144	&	0.2551	&	0.3639	&	0.4647	&	0.5543	&	0.6317	&	0.6966	&	0.7494	&		\\	\hline
			1	&	0.2	&	0.1579	&	0.2387	&	0.2949	&	0.3314	&	0.373	&	0.4326	&	0.522	&	0.6562	&	0.856	&	0.1578	&	0.2385	&	0.2947	&	0.3312	&	0.3728	&	0.4324	&	0.5218	&	0.6559	&	0.8555	\\
\hline
		\end{tabular}
	}
\end{table}
%
\clearpage
\section{Pressure and friction lift coefficients ($C_{LP}$ and $C_{LF}$)}\label{appendix:lifttable}
\setcounter{table}{0}\setcounter{figure}{0} 
%
\begin{table}[ht!]
	\centering
	\caption{Dependence of the pressure ($C_{LP}$) and friction ($C_{LF}$) lift coefficients  on the Reynolds number ($Re=1$), rotational velocity ($\alpha$) and power-law index ($n$) for upper ({UC}) and lower ({LC}) cylinders.}\label{tab:lift1}
	\resizebox{\columnwidth}{!}{%
		\renewcommand{\arraystretch}{1.3}
\begin{tabular}{|c|c|r|r|r|r|r|r|r|r|r|r|r|r|r|r|r|r|r|r|}	 %
			\hline
			\multicolumn{2}{|c|}{$Re=1$}  & \multicolumn{9}{c|}{$C_{LP}$ (for UC)}&\multicolumn{9}{c|}{$C_{LP}$ (for LC)} \\ \hline	
			$\alpha$ &	$G$ &$n\rightarrow$ 0.2&0.4&0.6&0.8&1&1.2&1.4&1.6&1.8&0.2&0.4&0.6&0.8&1&1.2&1.4&1.6&1.8 \\ \hline
			0	&	0.2	&	1.9803	&	1.1548	&	1.0302	&	1.1129	&	1.1445	&	1.1362	&	1.112	&	1.0815	&	1.0474	&	-1.9818	&	-1.1565	&	-1.0316	&	-1.1141	&	-1.1455	&	-1.137	&	-1.1128	&	-1.0822	&	-1.048	\\	\cline{2-20}
			&	0.4	&	2.069	&	1.2165	&	1.0858	&	1.1586	&	1.1751	&	1.1524	&	1.1157	&	1.0751	&	1.0342	&	-2.0716	&	-1.2185	&	-1.0874	&	-1.1598	&	-1.1761	&	-1.1533	&	-1.1165	&	-1.0758	&	-1.0349	\\	\cline{2-20}
			&	0.6	&	1.9164	&	1.2535	&	1.1322	&	1.1996	&	1.2052	&	1.1718	&	1.1252	&	1.0756	&	1.0273	&	-1.9189	&	-1.2557	&	-1.1339	&	-1.201	&	-1.2064	&	-1.1728	&	-1.1261	&	-1.0764	&	-1.0281	\\	\cline{2-20}
			&	0.8	&	1.5728	&	1.2623	&	1.1689	&	1.2354	&	1.2335	&	1.192	&	1.1377	&	1.0809	&	1.0264	&	-1.5756	&	-1.2646	&	-1.1707	&	-1.2368	&	-1.2346	&	-1.193	&	-1.1386	&	-1.0816	&	-1.0272	\\	\cline{2-20}
			&	1	&	1.1846	&	1.2453	&	1.1967	&	1.2662	&	1.2593	&	1.2117	&	1.1511	&	1.0884	&	1.029	&	-1.1869	&	-1.2477	&	-1.1985	&	-1.2676	&	-1.2605	&	-1.2127	&	-1.1521	&	-1.0893	&	-1.0298	\\	\hline
			1	&	0.2	&	3.6681	&	1.9769	&	1.5602	&	1.7404	&	1.9824	&	2.1783	&	2.3196	&	2.4168	&	2.4843	&	-3.5751	&	-1.935	&	-1.5607	&	-1.7404	&	-1.9827	&	-2.1793	&	-2.3216	&	-2.4209	&	-2.4892	\\
\end{tabular}
%
			\multicolumn{2}{|c|}{$Re=1$}  & \multicolumn{9}{c|}{$C_{LF}$ (for UC)}&\multicolumn{9}{c|}{$C_{LF}$ (for LC)} \\ \hline	
			$\alpha$ &	$G$ &$n\rightarrow$ 0.2&0.4&0.6&0.8&1&1.2&1.4&1.6&1.8&0.2&0.4&0.6&0.8&1&1.2&1.4&1.6&1.8 \\ \hline
			0	&	0.2	&	0.442	&	0.4258	&	0.5311	&	0.7071	&	0.8319	&	0.9077	&	0.9516	&	0.9748	&	0.9842	&	-0.4427	&	-0.4268	&	-0.5321	&	-0.7081	&	-0.8329	&	-0.9085	&	-0.9525	&	-0.9756	&	-0.9849	\\	\cline{2-20}
			&	0.4	&	0.4757	&	0.4622	&	0.5738	&	0.7537	&	0.8769	&	0.949	&	0.989	&	1.0093	&	1.0169	&	-0.477	&	-0.4634	&	-0.5748	&	-0.7546	&	-0.8777	&	-0.9497	&	-0.9897	&	-1.0099	&	-1.0175	\\	\cline{2-20}
			&	0.6	&	0.4696	&	0.4908	&	0.6128	&	0.7974	&	0.9191	&	0.9873	&	1.0229	&	1.0388	&	1.043	&	-0.4707	&	-0.4925	&	-0.6143	&	-0.7987	&	-0.9201	&	-0.9883	&	-1.0237	&	-1.0395	&	-1.0436	\\	\cline{2-20}
			&	0.8	&	0.4197	&	0.5105	&	0.6476	&	0.8377	&	0.9582	&	1.0227	&	1.0534	&	1.0644	&	1.0643	&	-0.4194	&	-0.5113	&	-0.6483	&	-0.8382	&	-0.9586	&	-1.023	&	-1.0537	&	-1.0646	&	-1.0645	\\	\cline{2-20}
			&	1	&	0.3477	&	0.5193	&	0.677	&	0.8738	&	0.9937	&	1.0546	&	1.0805	&	1.0863	&	1.0814	&	-0.3482	&	-0.5204	&	-0.678	&	-0.8747	&	-0.9945	&	-1.0554	&	-1.0812	&	-1.087	&	-1.0821	\\	\hline
			1	&	0.2	&	0.5664	&	0.3324	&	0.4123	&	0.6399	&	0.9013	&	1.1449	&	1.3559	&	1.5321	&	1.6777	&	-0.646	&	-0.4254	&	-0.4096	&	-0.6386	&	-0.9003	&	-1.1444	&	-1.3561	&	-1.5335	&	-1.6797	\\
\hline
		\end{tabular}
	}
\end{table}
\begin{table}[h]
	\centering
	\caption{Dependence of the pressure ($C_{LP}$) and friction ($C_{LF}$) lift coefficients on the Reynolds number ($Re=5$), rotational velocity ($\alpha$) and power-law index ($n$) for upper ({UC}) and lower ({LC}) cylinders.}\label{tab:lift3}
	\resizebox{\columnwidth}{!}{%
		\renewcommand{\arraystretch}{1.3}
		\begin{tabular}{|c|c|r|r|r|r|r|r|r|r|r|r|r|r|r|r|r|r|r|r|}	 %
			\hline
			\multicolumn{2}{|c|}{$Re=5$}  & \multicolumn{9}{c|}{$C_{LP}$ (for UC)}&\multicolumn{9}{c|}{$C_{LP}$ (for LC)} \\ \hline	
			$\alpha$ &	$G$ &$n\rightarrow$ 0.2&0.4&0.6&0.8&1&1.2&1.4&1.6&1.8&0.2&0.4&0.6&0.8&1&1.2&1.4&1.6&1.8 \\ \hline
			0	&	0.2	&	1.8243	&	1.2511	&	1.1551	&	1.1212	&	1.0877	&	1.0515	&	1.0144	&	0.9786	&	0.9445	&	-1.8248	&	-1.2515	&	-1.1555	&	-1.1216	&	-1.0881	&	-1.0519	&	-1.0148	&	-0.979	&	-0.9449	\\	\cline{2-20}
			&	0.4	&	1.7022	&	1.3002	&	1.1974	&	1.1502	&	1.1035	&	1.0562	&	1.0108	&	0.9683	&	0.9285	&	-1.7027	&	-1.3006	&	-1.1978	&	-1.1506	&	-1.1039	&	-1.0565	&	-1.0112	&	-0.9687	&	-0.9288	\\	\cline{2-20}
			&	0.6	&	1.1998	&	1.2745	&	1.204	&	1.1543	&	1.1008	&	1.0465	&	0.995	&	0.947	&	0.9022	&	-1.2004	&	-1.275	&	-1.2045	&	-1.1548	&	-1.1013	&	-1.047	&	-0.9954	&	-0.9475	&	-0.9026	\\	\cline{2-20}
			&	0.8	&	0.7624	&	1.1741	&	1.1711	&	1.1304	&	1.0758	&	1.0186	&	0.9639	&	0.9127	&	0.8648	&	-0.7629	&	-1.1746	&	-1.1716	&	-1.1309	&	-1.0763	&	-1.0191	&	-0.9643	&	-0.9132	&	-0.8652	\\	\cline{2-20}
			&	1	&	0.4706	&	1.0297	&	1.1034	&	1.0794	&	1.0282	&	0.9711	&	0.9155	&	0.8633	&	0.8144	&	-0.4714	&	-1.0303	&	-1.1039	&	-1.0799	&	-1.0287	&	-0.9716	&	-0.916	&	-0.8638	&	-0.8149	\\	\hline
			1	&	0.2	&	3.1114	&	2.1408	&	1.9338	&	1.9731	&	2.0537	&	2.1401	&	2.2205	&	2.292	&	2.3542	&	-3.1081	&	-2.137	&	-1.9314	&	-1.973	&	-2.0539	&	-2.1404	&	-2.221	&	-2.2928	&	-2.3561	\\
\end{tabular}
%
			\multicolumn{2}{|c|}{$Re=5$}  & \multicolumn{9}{c|}{$C_{LF}$ (for UC)}&\multicolumn{9}{c|}{$C_{LF}$ (for LC)} \\ \hline	
			$\alpha$ &	$G$ &$n\rightarrow$ 0.2&0.4&0.6&0.8&1&1.2&1.4&1.6&1.8&0.2&0.4&0.6&0.8&1&1.2&1.4&1.6&1.8 \\ \hline
			0	&	0.2	&	0.3371	&	0.3951	&	0.4852	&	0.5626	&	0.619	&	0.6579	&	0.6843	&	0.7023	&	0.7142	&	-0.3373	&	-0.3954	&	-0.4855	&	-0.563	&	-0.6194	&	-0.6583	&	-0.6847	&	-0.7027	&	-0.7147	\\	\cline{2-20}
			&	0.4	&	0.3476	&	0.4271	&	0.5194	&	0.5943	&	0.6457	&	0.6801	&	0.7033	&	0.7188	&	0.7285	&	-0.348	&	-0.4274	&	-0.5198	&	-0.5947	&	-0.6461	&	-0.6805	&	-0.7037	&	-0.7192	&	-0.7289	\\	\cline{2-20}
			&	0.6	&	0.2977	&	0.4398	&	0.541	&	0.6151	&	0.6628	&	0.6928	&	0.7119	&	0.7239	&	0.7302	&	-0.2979	&	-0.4403	&	-0.5415	&	-0.6156	&	-0.6632	&	-0.6932	&	-0.7123	&	-0.7243	&	-0.7305	\\	\cline{2-20}
			&	0.8	&	0.224	&	0.4294	&	0.5465	&	0.6222	&	0.6672	&	0.6935	&	0.7088	&	0.7169	&	0.7194	&	-0.2239	&	-0.4295	&	-0.5466	&	-0.6223	&	-0.6674	&	-0.6937	&	-0.7089	&	-0.717	&	-0.7195	\\	\cline{2-20}
			&	1	&	0.161	&	0.4003	&	0.5355	&	0.6142	&	0.6577	&	0.6809	&	0.6927	&	0.6972	&	0.6962	&	-0.1611	&	-0.4005	&	-0.5358	&	-0.6145	&	-0.658	&	-0.6813	&	-0.6931	&	-0.6976	&	-0.6966	\\	\hline
			1	&	0.2	&	0.4282	&	0.3646	&	0.4293	&	0.5458	&	0.6761	&	0.8075	&	0.9333	&	1.0515	&	1.1619	&	-0.4356	&	-0.3647	&	-0.4336	&	-0.5464	&	-0.6761	&	-0.8076	&	-0.9335	&	-1.0519	&	-1.1625	\\
\hline
		\end{tabular}
	}
\end{table}
\begin{table}[ht!]
	\centering
	\caption{Dependence of the pressure ($C_{LP}$) and friction ($C_{LF}$) lift coefficients  on the Reynolds number ($Re=10$), rotational velocity ($\alpha$) and power-law index ($n$) for upper ({UC}) and lower ({LC}) cylinders.}\label{tab:lift5}
	\resizebox{\columnwidth}{!}{%
		\renewcommand{\arraystretch}{1.3}
		\begin{tabular}{|c|c|r|r|r|r|r|r|r|r|r|r|r|r|r|r|r|r|r|r|}	 %
			\hline
			\multicolumn{2}{|c|}{$Re=10$}  & \multicolumn{9}{c|}{$C_{LP}$ (for UC)}&\multicolumn{9}{c|}{$C_{LP}$ (for LC)} \\ \hline	
			$\alpha$ &	$G$ &$n\rightarrow$ 0.2&0.4&0.6&0.8&1&1.2&1.4&1.6&1.8&0.2&0.4&0.6&0.8&1&1.2&1.4&1.6&1.8 \\ \hline
			0	&	0.2	&	1.665	&	1.2825	&	1.1663	&	1.0995	&	1.0477	&	1.0025	&	0.9618	&	0.924	&	0.8882	&	-1.6653	&	-1.2828	&	-1.1665	&	-1.0998	&	-1.0479	&	-1.0028	&	-0.962	&	-0.9243	&	-0.8885	\\	\cline{2-20}
			&	0.4	&	0.9853	&	1.2226	&	1.1507	&	1.087	&	1.0316	&	0.9823	&	0.9374	&	0.8954	&	0.8554	&	-0.9856	&	-1.2229	&	-1.151	&	-1.0873	&	-1.0319	&	-0.9826	&	-0.9376	&	-0.8957	&	-0.8556	\\	\cline{2-20}
			&	0.6	&	0.4897	&	0.9601	&	1.0286	&	1.0053	&	0.9656	&	0.9233	&	0.8813	&	0.8404	&	0.8002	&	-0.49	&	-0.9604	&	-1.0289	&	-1.0056	&	-0.9659	&	-0.9236	&	-0.8817	&	-0.8407	&	-0.8005	\\	\cline{2-20}
			&	0.8	&	0.3053	&	0.671	&	0.8308	&	0.8613	&	0.8495	&	0.8233	&	0.7918	&	0.7581	&	0.7233	&	-0.3052	&	-0.6712	&	-0.8311	&	-0.8617	&	-0.8499	&	-0.8237	&	-0.7921	&	-0.7584	&	-0.7237	\\	\cline{2-20}
			&	1	&	0.2215	&	0.4766	&	0.6352	&	0.6938	&	0.7052	&	0.6959	&	0.6773	&	0.6543	&	0.6288	&	-0.2222	&	-0.4769	&	-0.6356	&	-0.6942	&	-0.7056	&	-0.6963	&	-0.6777	&	-0.6547	&	-0.6292	\\	\hline
			1	&	0.2	&	2.9064	&	2.257	&	2.0635	&	2.0603	&	2.1015	&	2.1551	&	2.2141	&	2.2735	&	2.3295	&	-2.9061	&	-2.257	&	-2.0632	&	-2.0605	&	-2.1017	&	-2.1554	&	-2.2141	&	-2.2735	&	-2.33	\\
\end{tabular}
%
			\multicolumn{2}{|c|}{$Re=10$}  & \multicolumn{9}{c|}{$C_{LF}$ (for UC)}&\multicolumn{9}{c|}{$C_{LF}$ (for LC)} \\ \hline	
			$\alpha$ &	$G$ &$n\rightarrow$ 0.2&0.4&0.6&0.8&1&1.2&1.4&1.6&1.8&0.2&0.4&0.6&0.8&1&1.2&1.4&1.6&1.8 \\ \hline
			0	&	0.2	&	0.2797	&	0.3458	&	0.4109	&	0.4639	&	0.5052	&	0.537	&	0.5613	&	0.5795	&	0.5923	&	-0.2798	&	-0.346	&	-0.4111	&	-0.4642	&	-0.5055	&	-0.5373	&	-0.5616	&	-0.5798	&	-0.5926	\\	\cline{2-20}
			&	0.4	&	0.23	&	0.3531	&	0.4263	&	0.4776	&	0.5155	&	0.5444	&	0.5663	&	0.5821	&	0.5925	&	-0.2301	&	-0.3533	&	-0.4265	&	-0.4779	&	-0.5158	&	-0.5447	&	-0.5665	&	-0.5824	&	-0.5928	\\	\cline{2-20}
			&	0.6	&	0.1419	&	0.3121	&	0.4077	&	0.4659	&	0.5054	&	0.5337	&	0.5542	&	0.5683	&	0.5766	&	-0.142	&	-0.3124	&	-0.408	&	-0.4662	&	-0.5057	&	-0.534	&	-0.5545	&	-0.5686	&	-0.5769	\\	\cline{2-20}
			&	0.8	&	0.0887	&	0.2473	&	0.358	&	0.4264	&	0.471	&	0.5016	&	0.523	&	0.5372	&	0.5454	&	-0.0887	&	-0.2473	&	-0.3581	&	-0.4265	&	-0.4711	&	-0.5017	&	-0.523	&	-0.5373	&	-0.5454	\\	\cline{2-20}
			&	1	&	0.0602	&	0.1911	&	0.2968	&	0.3688	&	0.4174	&	0.4509	&	0.4745	&	0.4907	&	0.5009	&	-0.0601	&	-0.1912	&	-0.297	&	-0.3691	&	-0.4176	&	-0.4512	&	-0.4748	&	-0.491	&	-0.5012	\\	\hline
			1	&	0.2	&	0.2787	&	0.3398	&	0.3997	&	0.4767	&	0.5637	&	0.6551	&	0.7494	&	0.8452	&	0.9381	&	-0.2798	&	-0.3407	&	-0.4004	&	-0.4768	&	-0.5638	&	-0.6553	&	-0.75	&	-0.8467	&	-0.9395	\\
\hline
		\end{tabular}
	}
\end{table}
\begin{table}[ht!]
	\centering
	\caption{Dependence of the pressure ($C_{LP}$) and friction ($C_{LF}$) lift coefficients on the Reynolds number ($Re=20$), rotational velocity ($\alpha$) and power-law index ($n$) for upper ({UC}) and lower ({LC}) cylinders.}\label{tab:lift7}
	\resizebox{\columnwidth}{!}{%
		\renewcommand{\arraystretch}{1.3}
		\begin{tabular}{|c|c|r|r|r|r|r|r|r|r|r|r|r|r|r|r|r|r|r|r|}	 %
			\hline
			\multicolumn{2}{|c|}{$Re=20$}  & \multicolumn{9}{c|}{$C_{LP}$ (for UC)}&\multicolumn{9}{c|}{$C_{LP}$ (for LC)} \\ \hline	
			$\alpha$ &	$G$ &$n\rightarrow$ 0.2&0.4&0.6&0.8&1&1.2&1.4&1.6&1.8&0.2&0.4&0.6&0.8&1&1.2&1.4&1.6&1.8 \\ \hline
		0	&	0.2	&	0.9002	&	1.1523	&	1.0757	&	1.0141	&	0.9643	&	0.9208	&	0.8808	&	0.8425	&	0.8042	&	-0.9004	&	-1.1525	&	-1.0759	&	-1.0143	&	-0.9645	&	-0.921	&	-0.881	&	-0.8426	&	-0.8044	\\	\cline{2-20}
		&	0.4	&	0.3528	&	0.6282	&	0.8394	&	0.8817	&	0.8726	&	0.8475	&	0.8163	&	0.7821	&	0.7451	&	-0.3526	&	-0.6285	&	-0.8396	&	-0.8819	&	-0.8728	&	-0.8477	&	-0.8165	&	-0.7823	&	-0.7453	\\	\cline{2-20}
		&	0.6	&	0.2729	&	0.3622	&	0.5002	&	0.6148	&	0.6735	&	0.693	&	0.6905	&	0.6756	&	0.6519	&	-0.2732	&	-0.3623	&	-0.5004	&	-0.6151	&	-0.6738	&	-0.6932	&	-0.6908	&	-0.6758	&	-0.6521	\\	\cline{2-20}
		&	0.8	&	0.2303	&	0.277	&	0.3473	&	0.4152	&	0.4723	&	0.5115	&	0.5329	&	0.5401	&	0.5358	&	-0.2304	&	-0.2771	&	-0.3475	&	-0.4154	&	-0.4726	&	-0.5118	&	-0.5332	&	-0.5404	&	-0.5361	\\	\cline{2-20}
		&	1	&	0.1942	&	0.2323	&	0.2792	&	0.3193	&	0.3529	&	0.3802	&	0.4006	&	0.4139	&	0.4202	&	-0.1941	&	-0.2325	&	-0.2794	&	-0.3196	&	-0.3532	&	-0.3805	&	-0.4009	&	-0.4142	&	-0.4205	\\	\hline
		1	&	0.2	&	2.7848	&	2.4275	&	2.2136	&	2.1521	&	2.1576	&	2.185	&	2.2254	&	2.2739	&	2.3261	&	-2.7854	&	-2.4277	&	-2.2138	&	-2.1523	&	-2.1578	&	-2.1851	&	-2.2256	&	-2.2741	&	-2.3261	\\
\end{tabular}
%
			\multicolumn{2}{|c|}{$Re=20$}  & \multicolumn{9}{c|}{$C_{LF}$ (for UC)}&\multicolumn{9}{c|}{$C_{LF}$ (for LC)} \\ \hline	
			$\alpha$ &	$G$ &$n\rightarrow$ 0.2&0.4&0.6&0.8&1&1.2&1.4&1.6&1.8&0.2&0.4&0.6&0.8&1&1.2&1.4&1.6&1.8 \\ \hline
			0	&	0.2	&	0.174	&	0.2556	&	0.3046	&	0.3459	&	0.3813	&	0.4106	&	0.4341	&	0.4519	&	0.4637	&	-0.1741	&	-0.2557	&	-0.3047	&	-0.3461	&	-0.3815	&	-0.4108	&	-0.4344	&	-0.4521	&	-0.464	\\	\cline{2-20}
			&	0.4	&	0.0702	&	0.1852	&	0.271	&	0.3271	&	0.3678	&	0.3989	&	0.4225	&	0.4396	&	0.4498	&	-0.0697	&	-0.1853	&	-0.2712	&	-0.3273	&	-0.368	&	-0.3991	&	-0.4228	&	-0.4398	&	-0.4501	\\	\cline{2-20}
			&	0.6	&	0.0321	&	0.1149	&	0.1922	&	0.2619	&	0.3161	&	0.3564	&	0.386	&	0.407	&	0.4197	&	-0.0316	&	-0.115	&	-0.1923	&	-0.2621	&	-0.3163	&	-0.3566	&	-0.3862	&	-0.4072	&	-0.4199	\\	\cline{2-20}
			&	0.8	&	0.0251	&	0.0823	&	0.1399	&	0.1966	&	0.2493	&	0.2944	&	0.3306	&	0.3581	&	0.3771	&	-0.0247	&	-0.0822	&	-0.1399	&	-0.1966	&	-0.2493	&	-0.2944	&	-0.3306	&	-0.3582	&	-0.3772	\\	\cline{2-20}
			&	1	&	0.0216	&	0.0674	&	0.1128	&	0.1573	&	0.2003	&	0.2402	&	0.2754	&	0.3049	&	0.3283	&	-0.0213	&	-0.0674	&	-0.1129	&	-0.1574	&	-0.2005	&	-0.2404	&	-0.2756	&	-0.3052	&	-0.3287	\\	\hline
			1	&	0.2	&	0.1642	&	0.2466	&	0.3214	&	0.3877	&	0.4535	&	0.5209	&	0.5903	&	0.6621	&	0.7372	&	-0.1645	&	-0.2468	&	-0.3215	&	-0.3878	&	-0.4536	&	-0.5211	&	-0.5905	&	-0.6626	&	-0.7382	\\
\hline
		\end{tabular}
	}
\end{table}
\begin{table}[ht!]
	\centering
	\caption{Dependence of the pressure ($C_{LP}$) and friction ($C_{LF}$)lift coefficients  on the Reynolds number ($Re=40$), rotational velocity ($\alpha$) and power-law index ($n$) for upper ({UC}) and lower ({LC}) cylinders.}\label{tab:lift9}
	\resizebox{\columnwidth}{!}{%
		\renewcommand{\arraystretch}{1.3}
		\begin{tabular}{|c|c|r|r|r|r|r|r|r|r|r|r|r|r|r|r|r|r|r|r|}	 %
			\hline
			\multicolumn{2}{|c|}{$Re=40$}  & \multicolumn{9}{c|}{$C_{LP}$ (for UC)}&\multicolumn{9}{c|}{$C_{LP}$ (for LC)} \\ \hline	
			$\alpha$ &	$G$ &$n\rightarrow$ 0.2&0.4&0.6&0.8&1&1.2&1.4&1.6&1.8&0.2&0.4&0.6&0.8&1&1.2&1.4&1.6&1.8 \\ \hline
			0	&	0.2	&	0.4234	&	0.561	&	0.8201	&	0.8518	&	0.8327	&	0.8021	&	0.7674	&	0.7288	&	0.6817	&	-0.4301	&	-0.5613	&	-0.8203	&	-0.8519	&	-0.8328	&	-0.8023	&	-0.7675	&	-0.729	&	-0.6819	\\	\cline{2-20}
			&	0.4	&	0.3214	&	0.3474	&	0.3863	&	0.485	&	0.5839	&	0.6307	&	0.6402	&	0.6275	&		&	-0.3287	&	-0.3484	&	-0.3866	&	-0.4851	&	-0.584	&	-0.6308	&	-0.6404	&	-0.6276	&		\\	\cline{2-20}
			&	0.6	&	0.2505	&	0.2759	&	0.3041	&	0.3337	&	0.3717	&	0.4169	&	0.4555	&	0.4773	&	0.4741	&	-0.2576	&	-0.2769	&	-0.3043	&	-0.3339	&	-0.3719	&	-0.417	&	-0.4557	&	-0.4775	&	-0.4743	\\	\cline{2-20}
			&	0.8	&	0.2019	&	0.2257	&	0.25	&	0.2731	&	0.293	&	0.3116	&	0.3307	&		&	0.3628	&	-0.2076	&	-0.2262	&	-0.2503	&	-0.2734	&	-0.2932	&	-0.3118	&	-0.3309	&		&	-0.3631	\\	\cline{2-20}
			&	1	&	0.1635	&	0.1873	&	0.2097	&	0.23	&	0.2468	&	0.2598	&	0.2699	&	0.2782	&		&	-0.1657	&	-0.1885	&	-0.21	&	-0.2303	&	-0.2471	&	-0.26	&	-0.2702	&	-0.2785	&		\\	\hline
			1	&	0.2	&	2.5323	&	2.492	&	2.3862	&	2.2631	&	2.2176	&	2.2184	&	2.237	&	2.2686	&	2.3098	&	-2.5339	&	-2.4926	&	-2.3865	&	-2.2633	&	-2.2178	&	-2.2185	&	-2.2372	&	-2.2688	&	-2.3099	\\
\end{tabular}
%
			\multicolumn{2}{|c|}{$Re=40$}  & \multicolumn{9}{c|}{$C_{LF}$ (for UC)}&\multicolumn{9}{c|}{$C_{LF}$ (for LC)} \\ \hline	
			$\alpha$ &	$G$ &$n\rightarrow$ 0.2&0.4&0.6&0.8&1&1.2&1.4&1.6&1.8&0.2&0.4&0.6&0.8&1&1.2&1.4&1.6&1.8 \\ \hline
			0	&	0.2	&	0.0421	&	0.1209	&	0.1874	&	0.2303	&	0.264	&	0.2915	&	0.313	&	0.3279	&	0.3336	&	-0.0417	&	-0.121	&	-0.1875	&	-0.2304	&	-0.2642	&	-0.2916	&	-0.3131	&	-0.3281	&	-0.3338	\\	\cline{2-20}
			&	0.4	&	0.0344	&	0.0653	&	0.1067	&	0.1635	&	0.2173	&	0.2576	&	0.2865	&	0.3054	&		&	-0.0344	&	-0.0653	&	-0.1068	&	-0.1636	&	-0.2174	&	-0.2578	&	-0.2866	&	-0.3055	&		\\	\cline{2-20}
			&	0.6	&	0.0276	&	0.0532	&	0.0843	&	0.1202	&	0.16	&	0.201	&	0.2377	&	0.266	&	0.2805	&	-0.0277	&	-0.0532	&	-0.0844	&	-0.1203	&	-0.1602	&	-0.2012	&	-0.2379	&	-0.2662	&	-0.2807	\\	\cline{2-20}
			&	0.8	&	0.0231	&	0.0446	&	0.0708	&	0.1001	&	0.1311	&	0.1627	&	0.1938	&		&	0.2471	&	-0.0229	&	-0.0445	&	-0.0707	&	-0.1001	&	-0.1311	&	-0.1627	&	-0.1939	&		&	-0.2472	\\	\cline{2-20}
			&	1	&	0.0194	&	0.038	&	0.0605	&	0.0858	&	0.1122	&	0.1388	&	0.1647	&	0.1894	&		&	-0.0194	&	-0.038	&	-0.0606	&	-0.0859	&	-0.1124	&	-0.139	&	-0.1649	&	-0.1897	&		\\	\hline
			1	&	0.2	&	0.0859	&	0.1529	&	0.2281	&	0.2924	&	0.3507	&	0.4078	&	0.4654	&	0.5239	&	0.5837	&	-0.0859	&	-0.153	&	-0.2282	&	-0.2925	&	-0.3508	&	-0.408	&	-0.4657	&	-0.5244	&	-0.5841	\\
\hline
		\end{tabular}
	}
\end{table}
%
%
\clearpage%
\section{Streamline profiles}\label{appendix:streamline}
 \setcounter{table}{0} \setcounter{figure}{0}
%
\begin{figure}[H]
	\centering
	\includegraphics[width=0.85\linewidth]{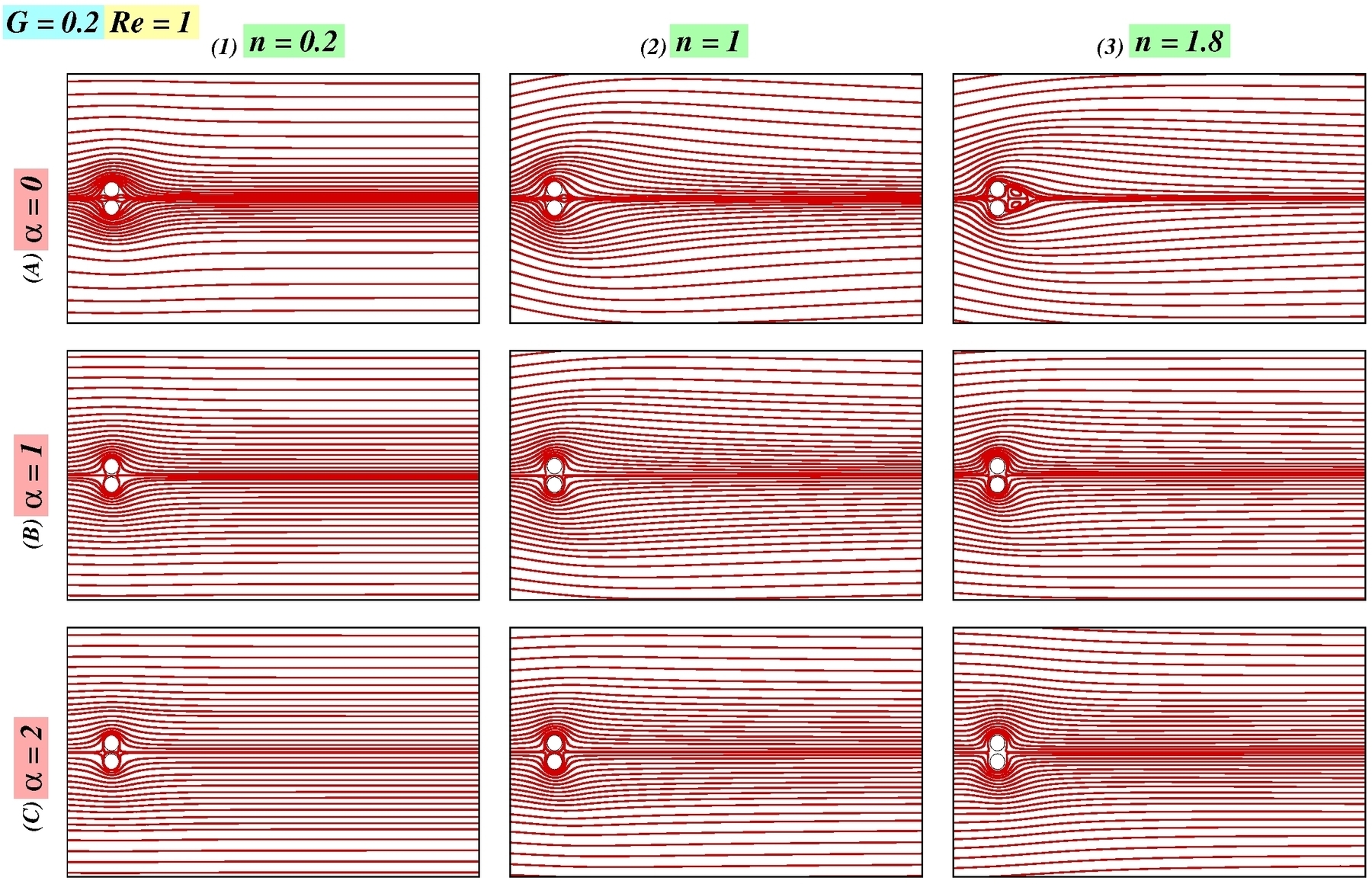}
	\caption{Streamline profiles for $Re=1$ at $G=0.2$.}
	\label{fig:st1}
\end{figure}
\begin{figure}[H]
	\centering
	\includegraphics[width=0.85\linewidth]{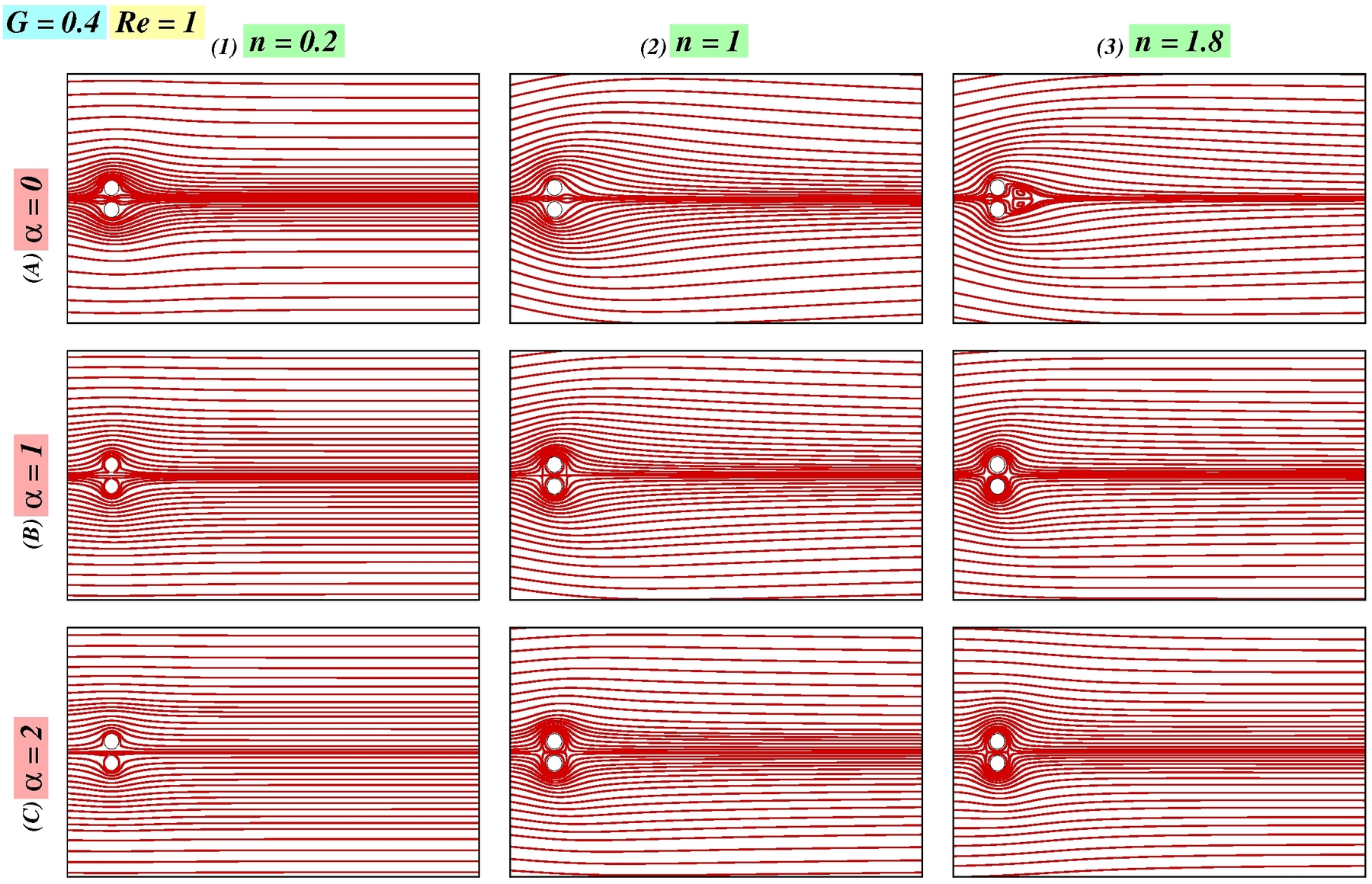}
	\caption{Streamline profiles for $Re=1$ at $G=0.4$.}
	\label{fig:st2}
\end{figure}
\begin{figure}[H]
	\centering
	\includegraphics[width=0.9\linewidth]{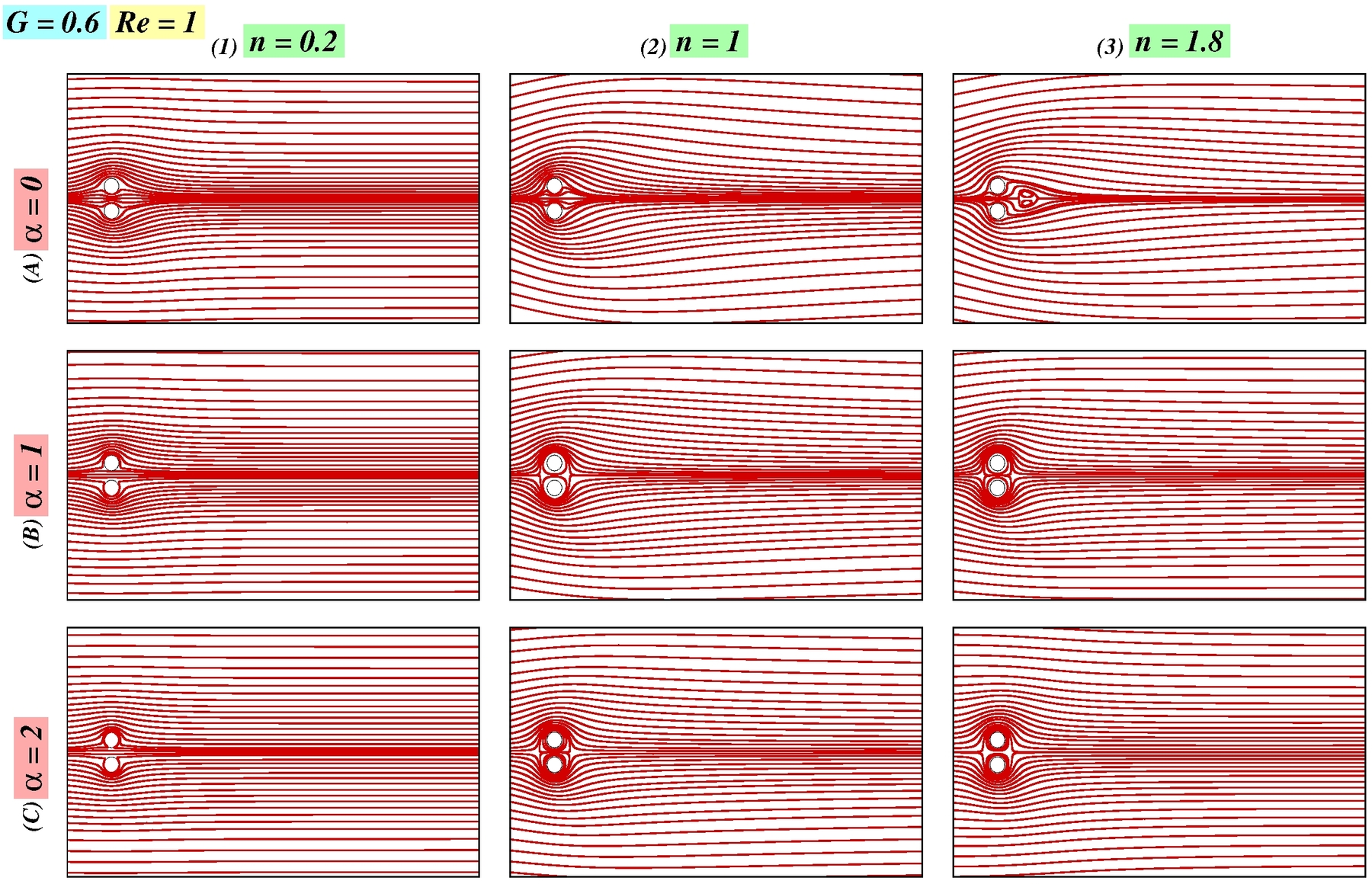}
	\caption{Streamline profiles for $Re=1$ at $G=0.6$.}
	\label{fig:st3}
\end{figure}
\begin{figure}[H]
	\centering
	\includegraphics[width=0.9\linewidth]{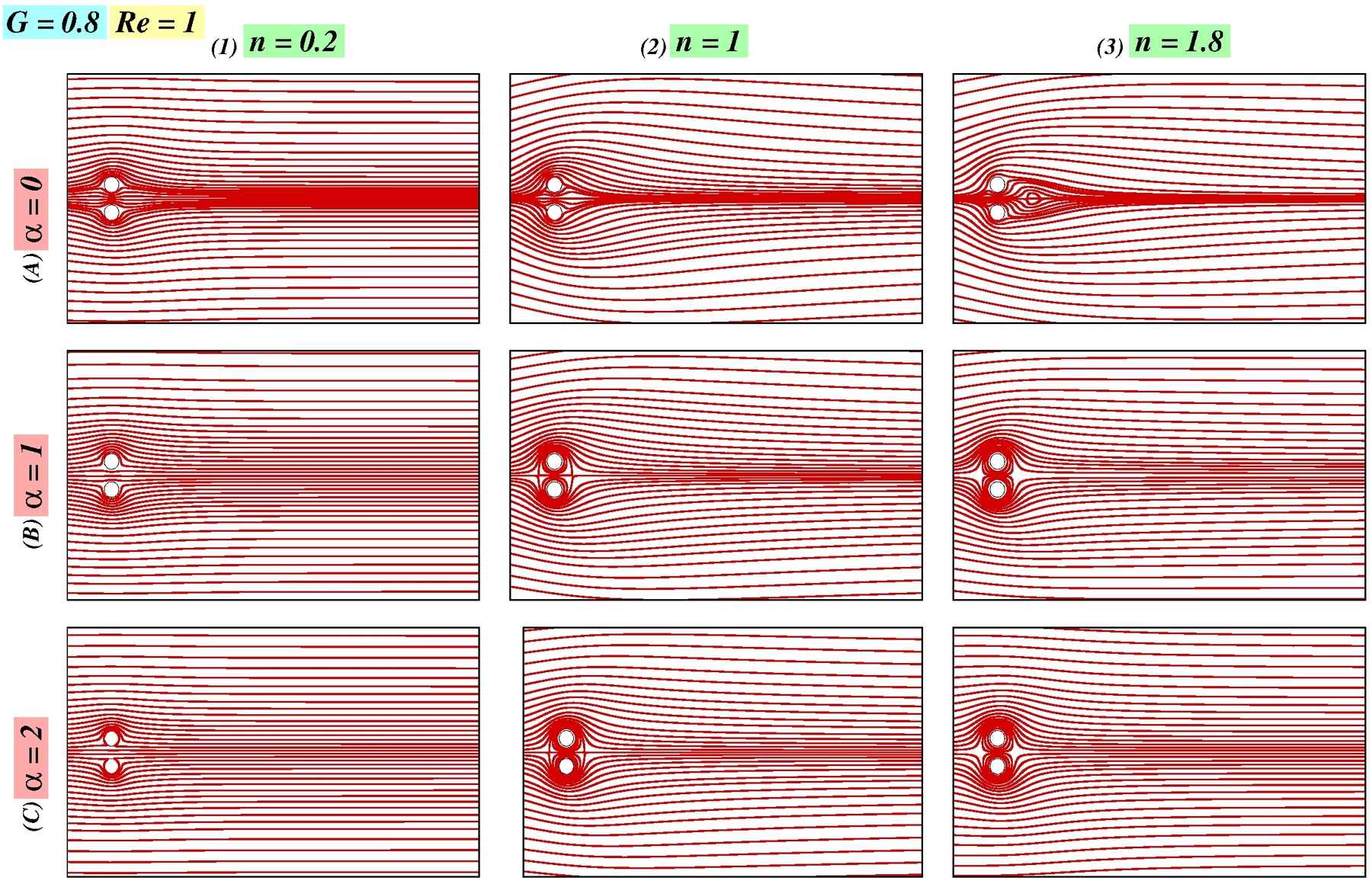}
	\caption{Streamline profiles for $Re=1$ at .$G=0.8$.}
	\label{fig:st4}
\end{figure}
\begin{figure}[H]
	\centering
	\includegraphics[width=0.9\linewidth]{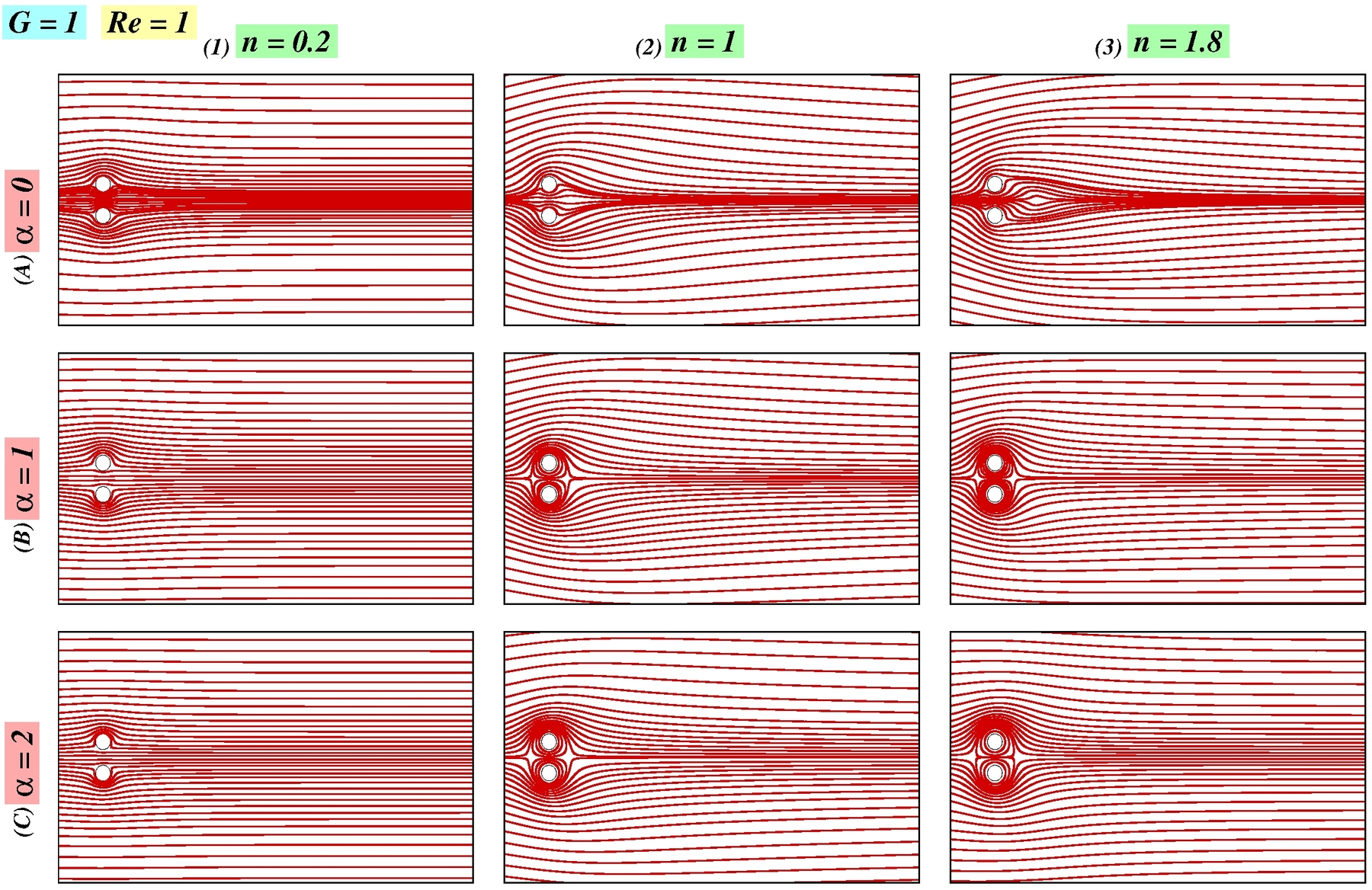}
	\caption{Streamline profiles for $Re=1$ at $G=1$.}
	\label{fig:st5}
\end{figure}
\begin{figure}[H]
	\centering
	\includegraphics[width=0.9\linewidth]{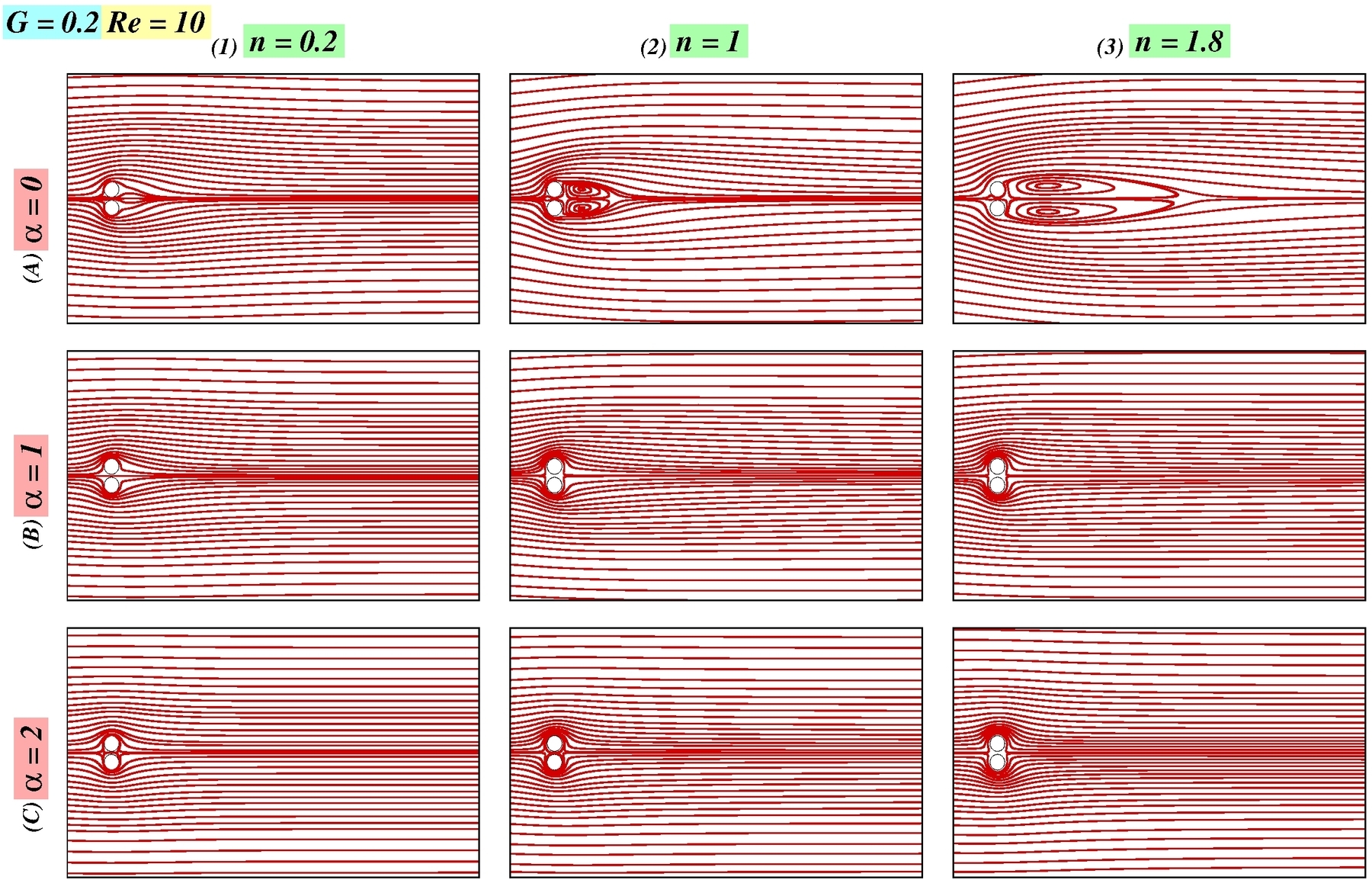}
	\caption{Streamline profiles for $Re=10$ at $G=0.2$.}
	\label{fig:st6}
\end{figure}
\begin{figure}[H]
	\centering
	\includegraphics[width=0.9\linewidth]{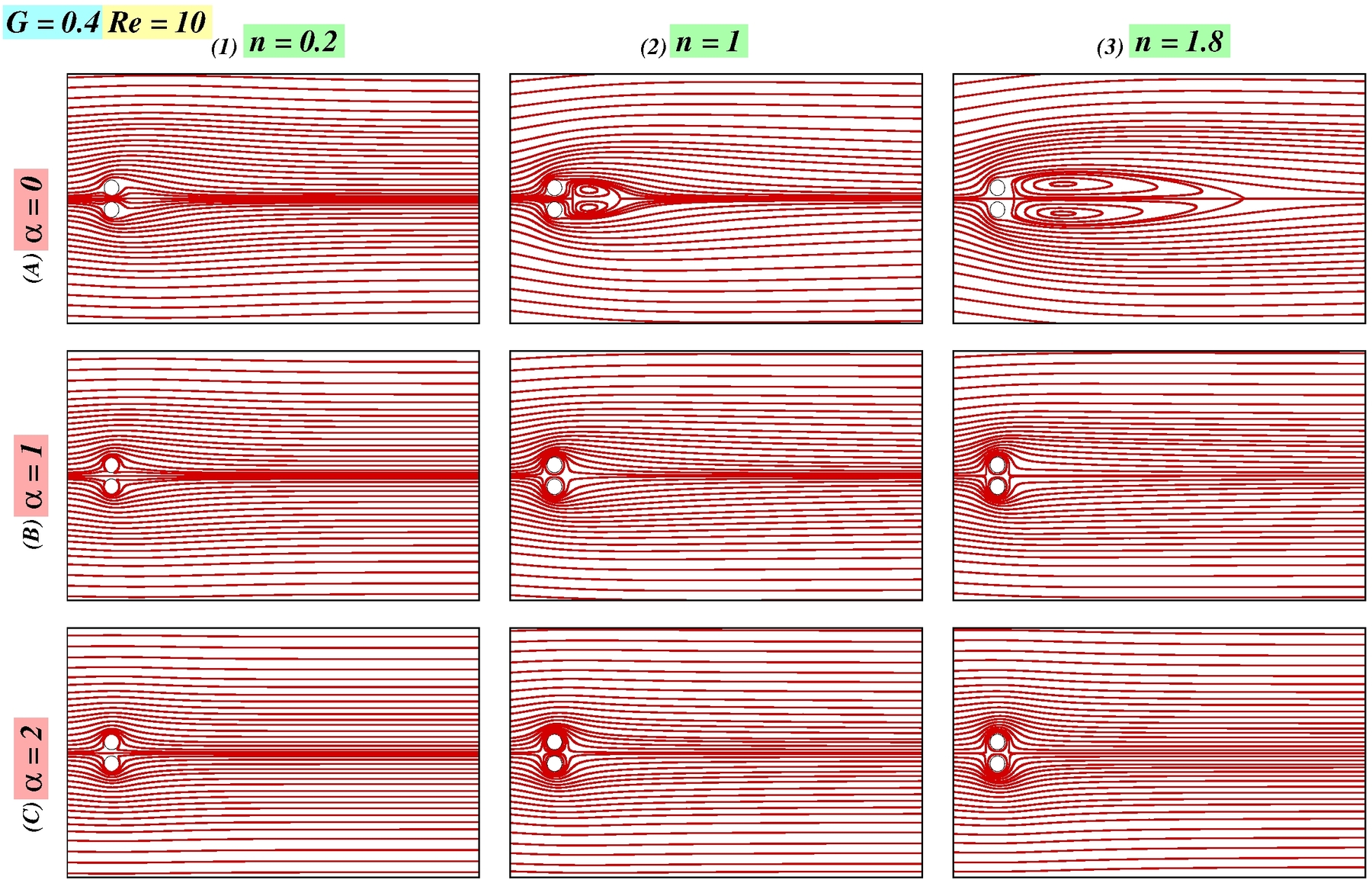}
	\caption{Streamline profiles for $Re=10$ at $G=0.4$.}
	\label{fig:st7}
\end{figure}
\begin{figure}[H]
	\centering
	\includegraphics[width=0.9\linewidth]{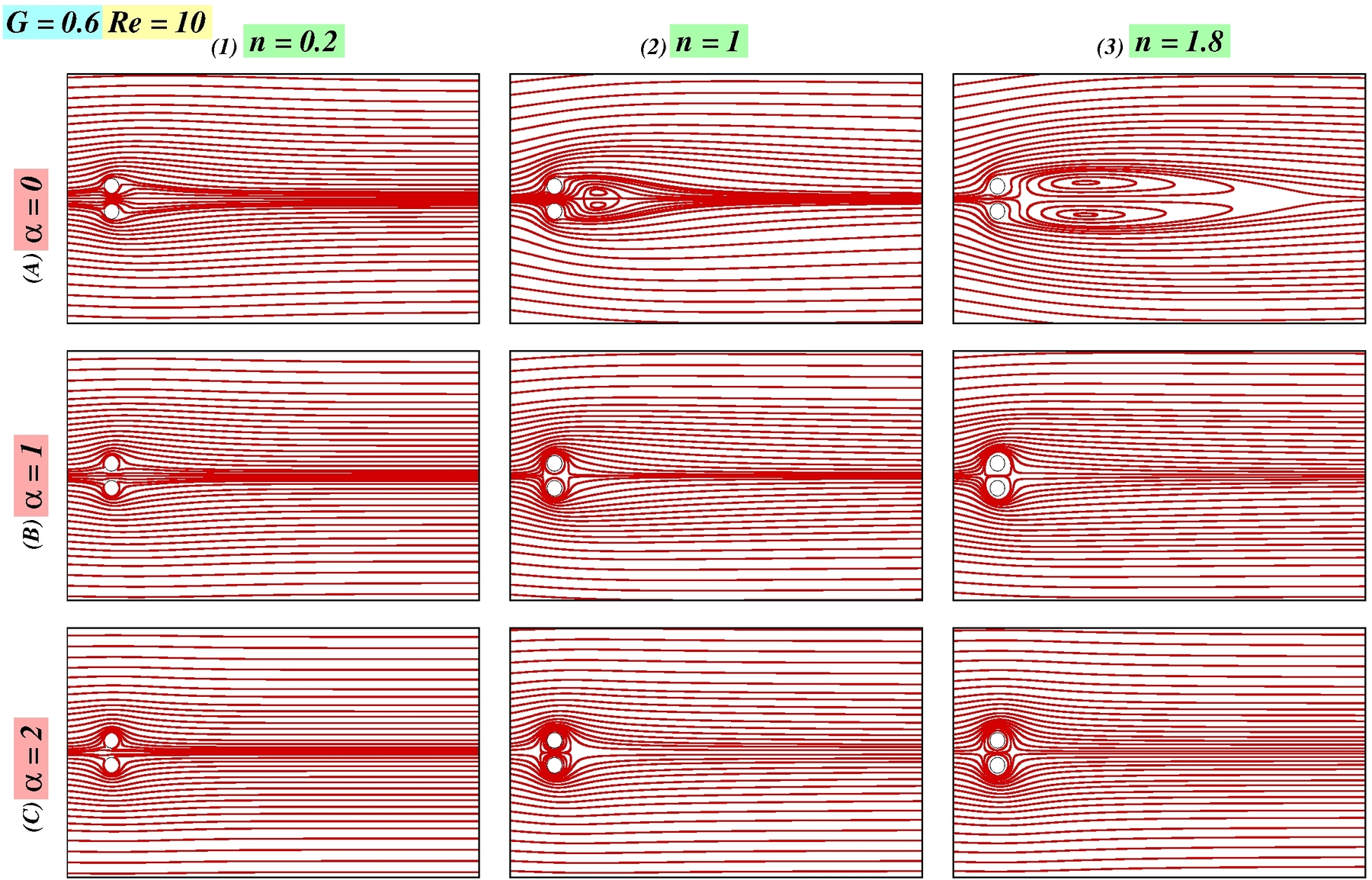}
	\caption{Streamline profiles for $Re=10$ at $G=0.6$.}
	\label{fig:st8}
\end{figure}
\begin{figure}[H]
	\centering
	\includegraphics[width=0.9\linewidth]{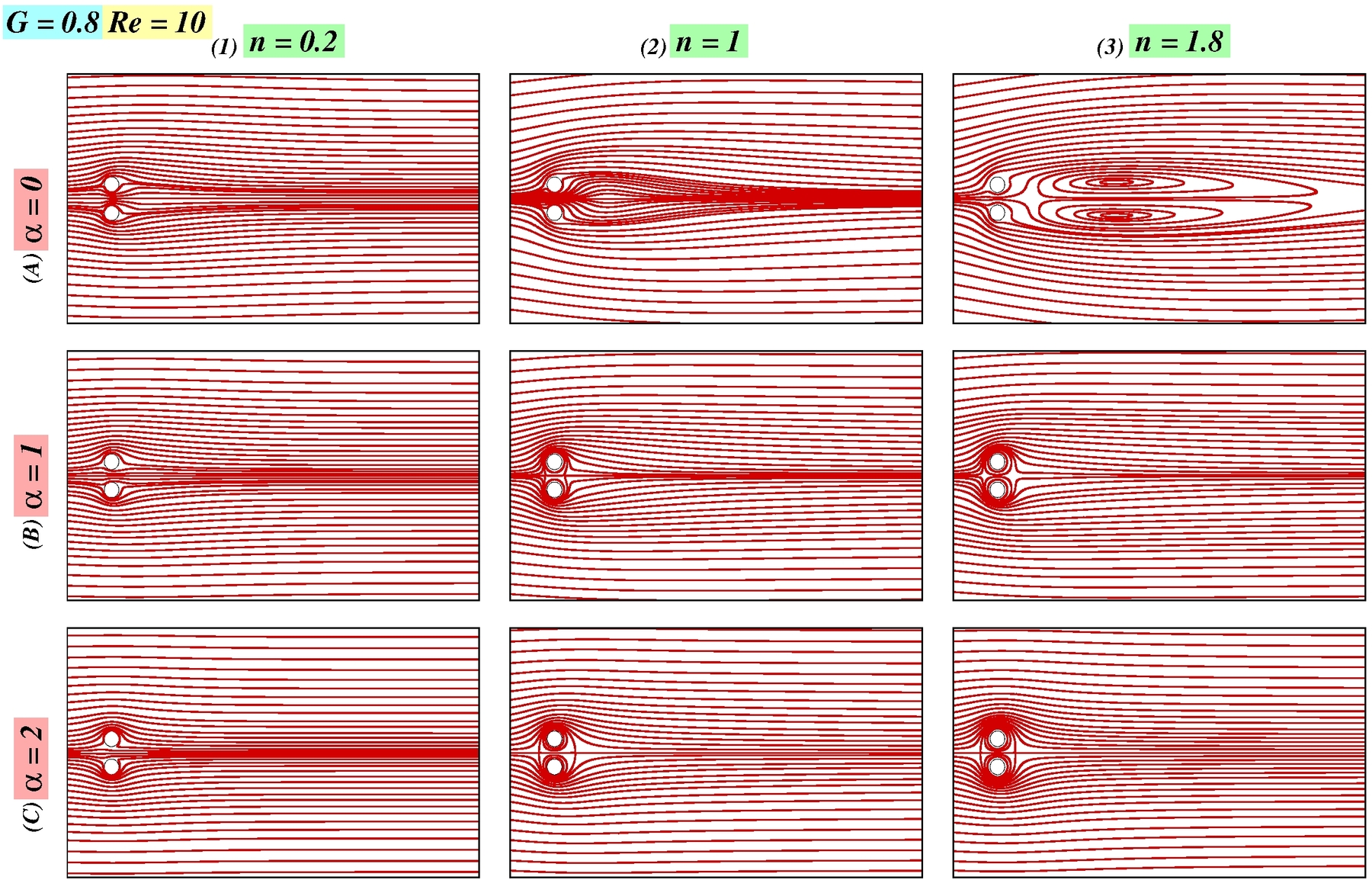}
	\caption{Streamline profiles for $Re=10$ at $G=0.8$.}
	\label{fig:st9}
\end{figure}
\begin{figure}[H]
	\centering
	\includegraphics[width=0.9\linewidth]{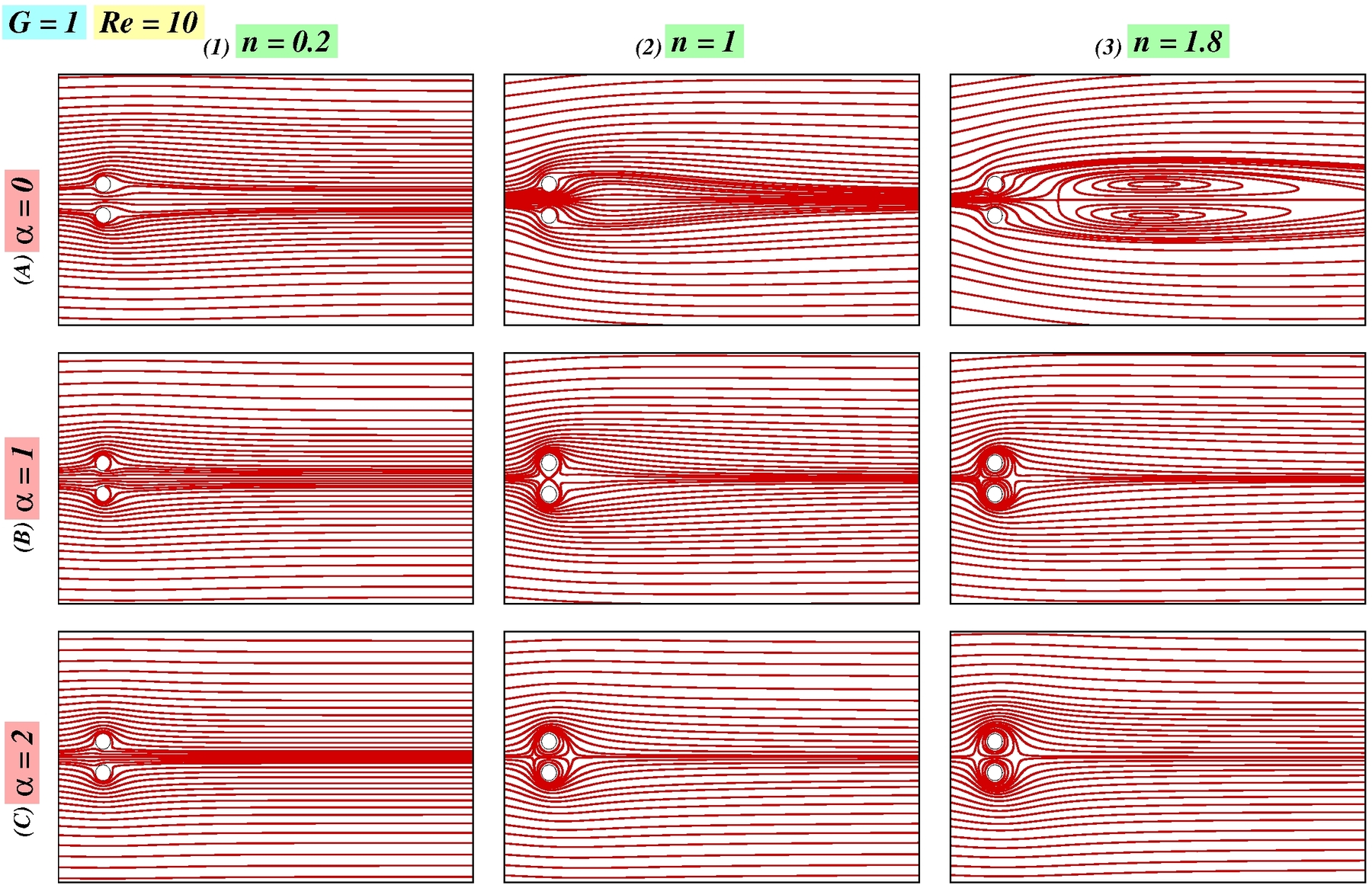}
	\caption{Streamline profiles for $Re=10$ at $G=1$.}
	\label{fig:st10}
\end{figure}
\begin{figure}[H]
	\centering
	\includegraphics[width=0.9\linewidth]{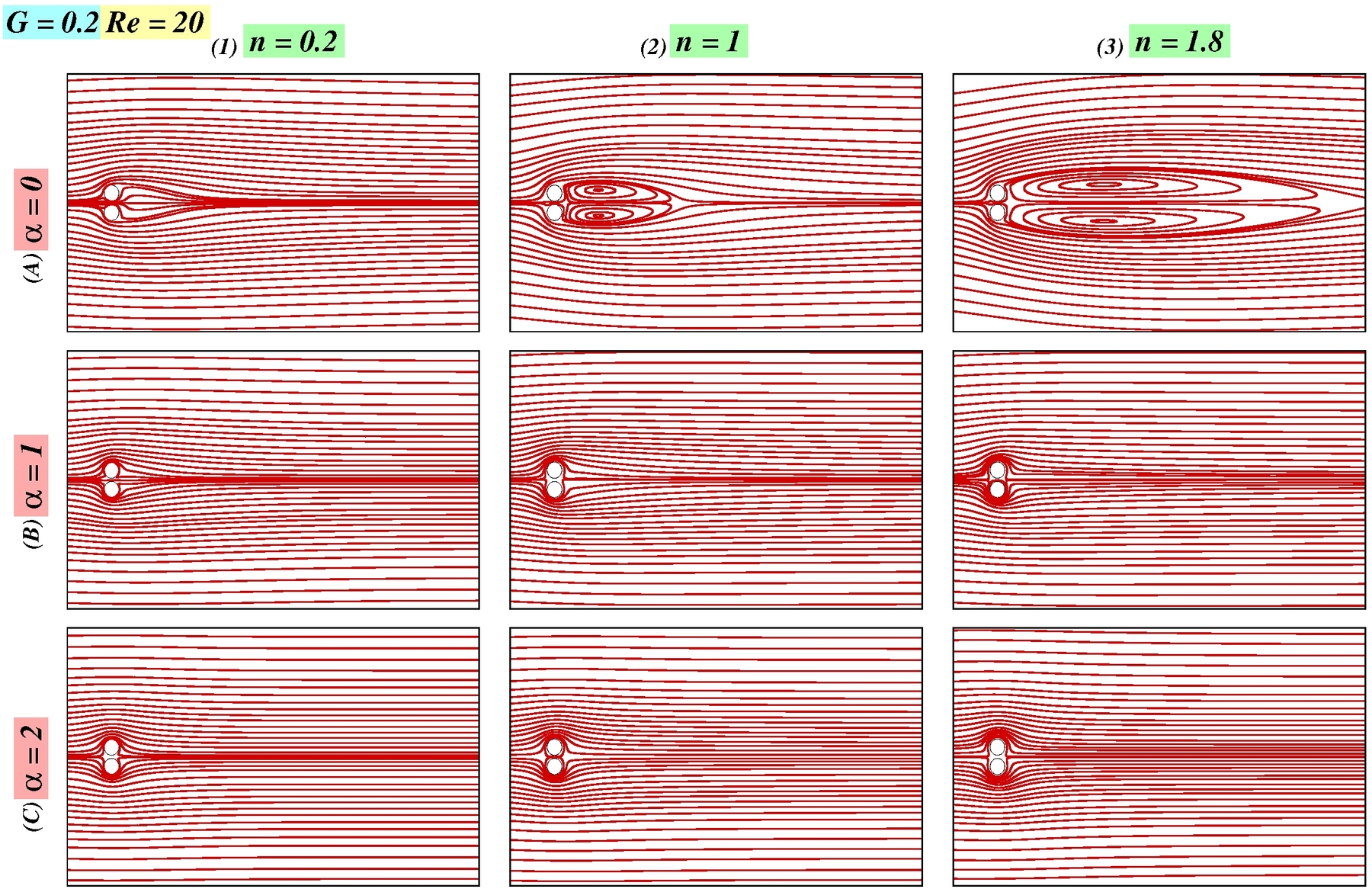}
	\caption{Streamline profiles for $Re=20$ at $G=0.2$.}
	\label{fig:st11}
\end{figure}
\begin{figure}[H]
	\centering
	\includegraphics[width=0.9\linewidth]{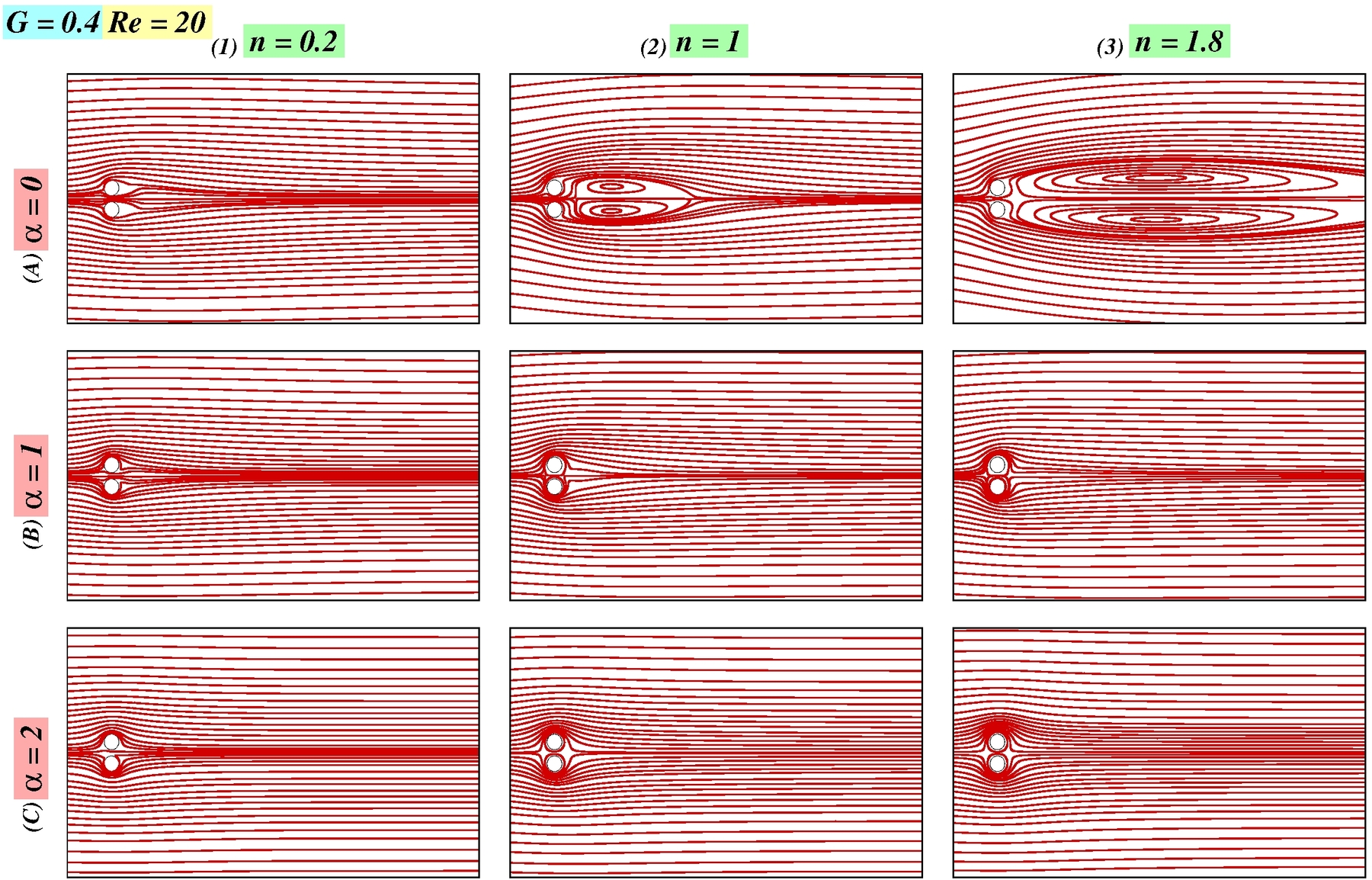}
	\caption{Streamline profiles for $Re=20$ at $G=0.4$.}
	\label{fig:st12}
\end{figure}
\begin{figure}[H]
	\centering
	\includegraphics[width=0.9\linewidth]{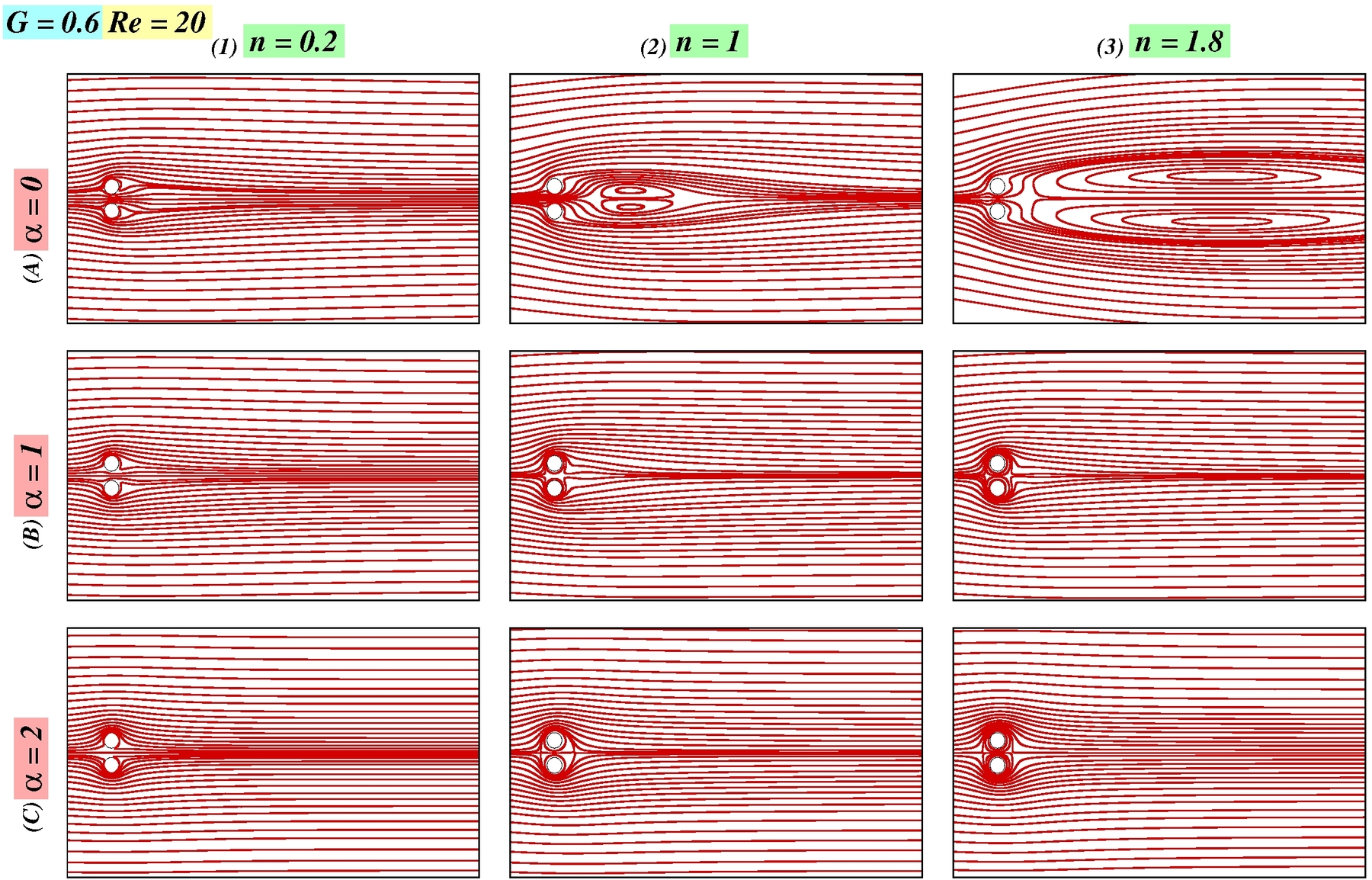}
	\caption{Streamline profiles for $Re=20$ at $G=0.6$.}
	\label{fig:st13}
\end{figure}
\begin{figure}[H]
	\centering
	\includegraphics[width=0.9\linewidth]{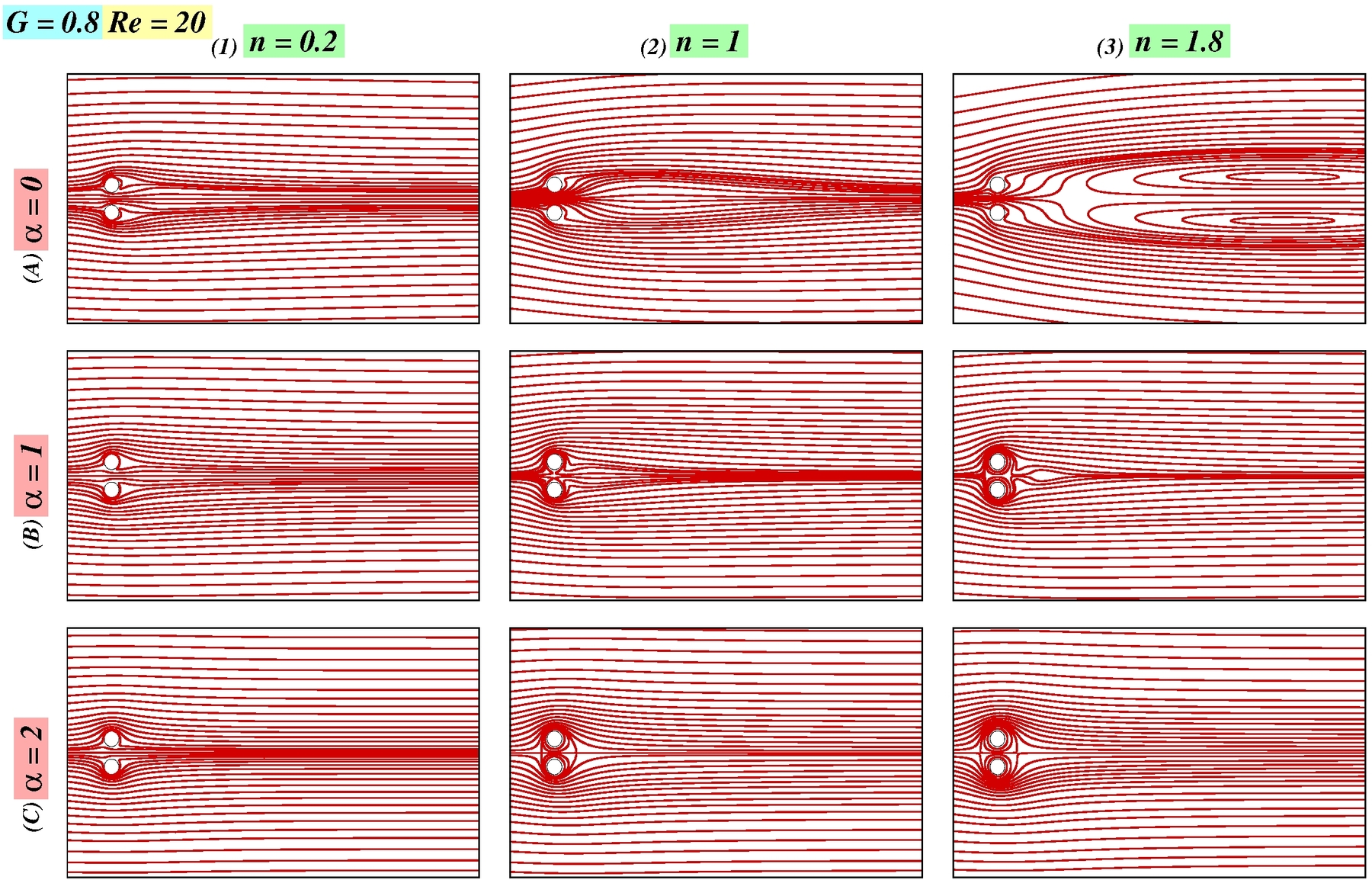}
	\caption{Streamline profiles for $Re=20$ at $G=0.8$.}
	\label{fig:st14}
\end{figure}
\begin{figure}[H]
	\centering
	\includegraphics[width=0.9\linewidth]{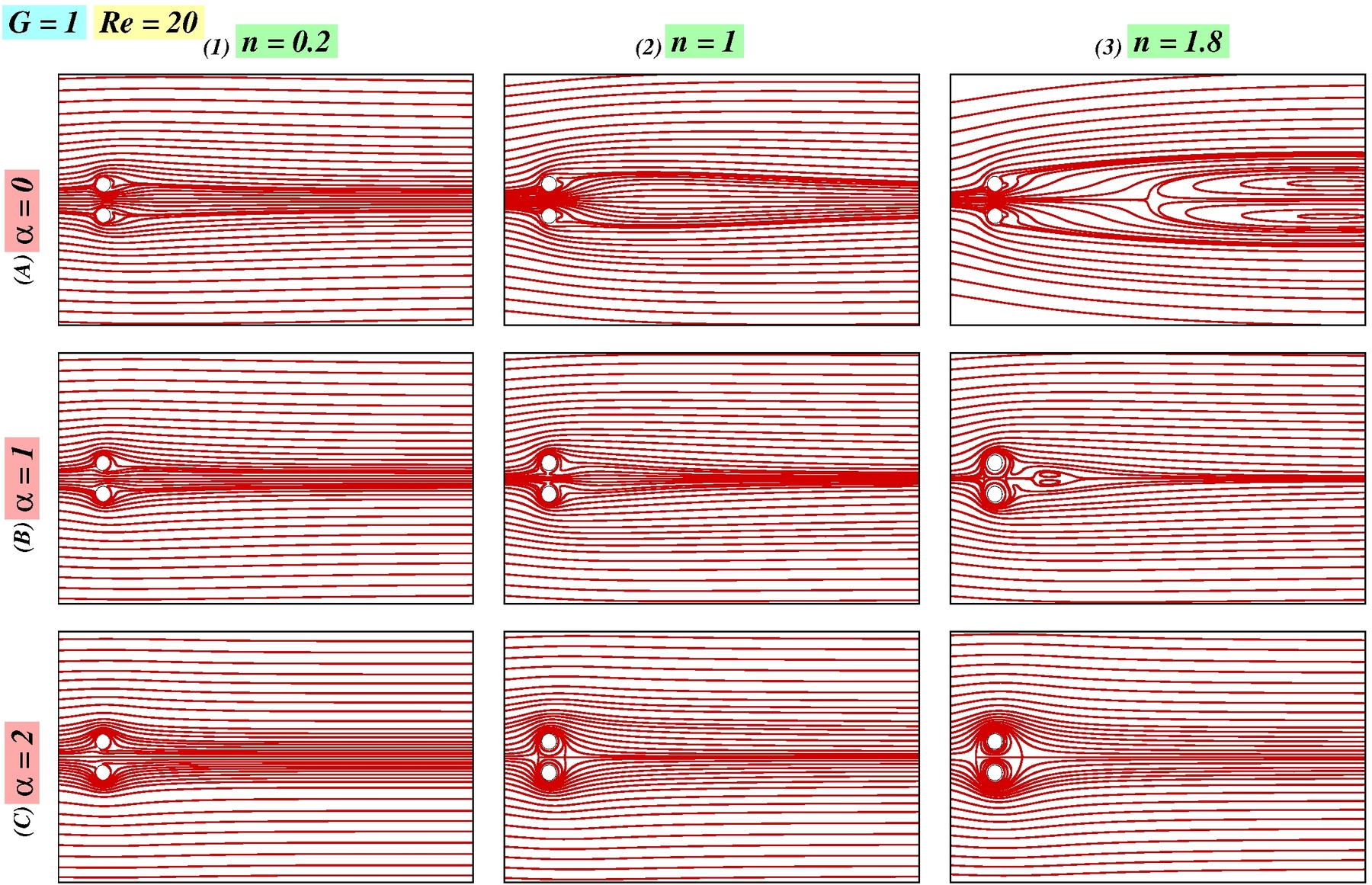}
	\caption{Streamline profiles for $Re=20$ at $G=1$.}
	\label{fig:st15}
\end{figure}
\begin{figure}[H]
	\centering
	\includegraphics[width=0.9\linewidth]{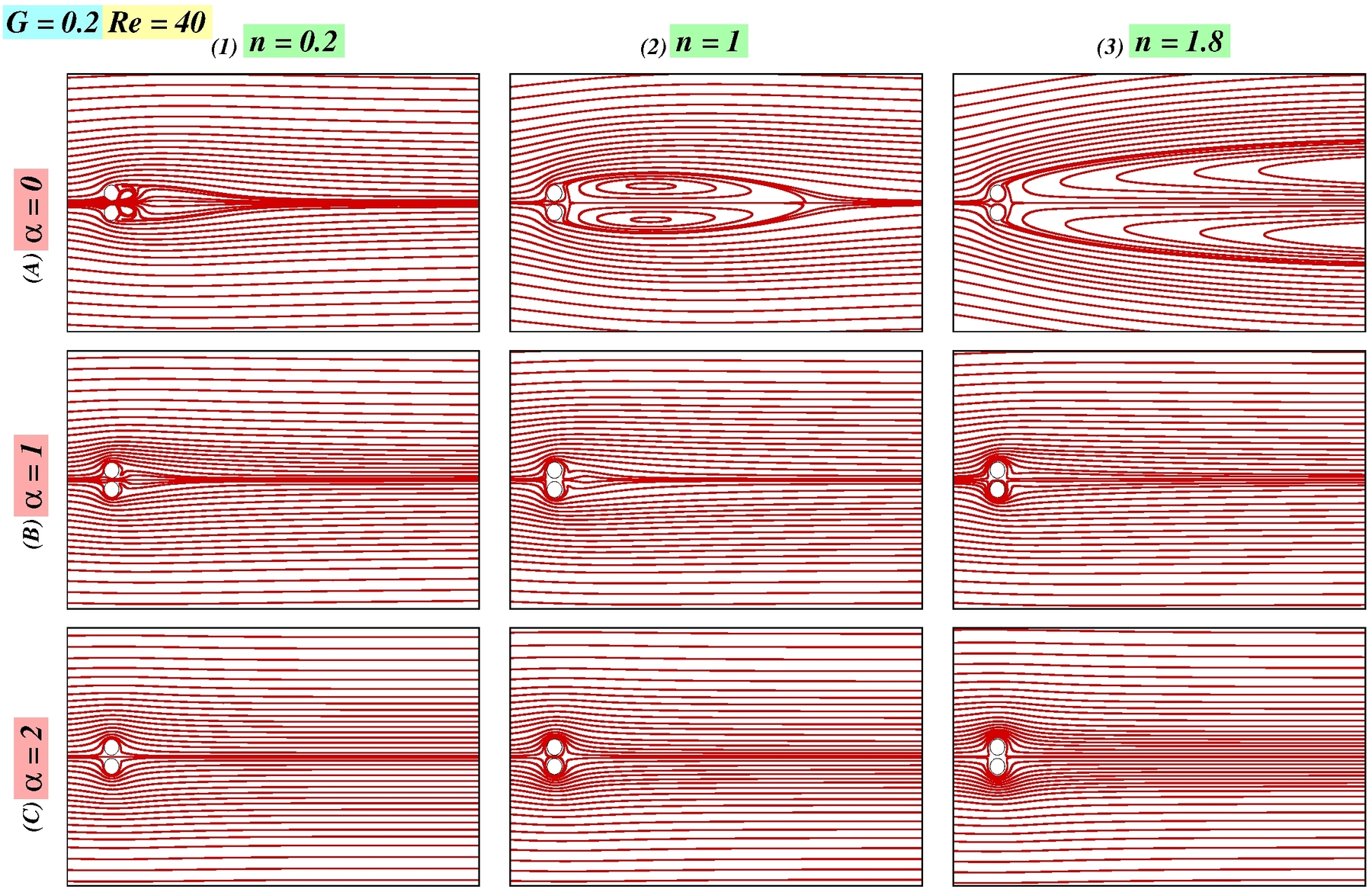}
	\caption{Streamline profiles for $Re=40$ at $G=0.2$.}
	\label{fig:st16}
\end{figure}
\begin{figure}[H]
	\centering
	\includegraphics[width=0.9\linewidth]{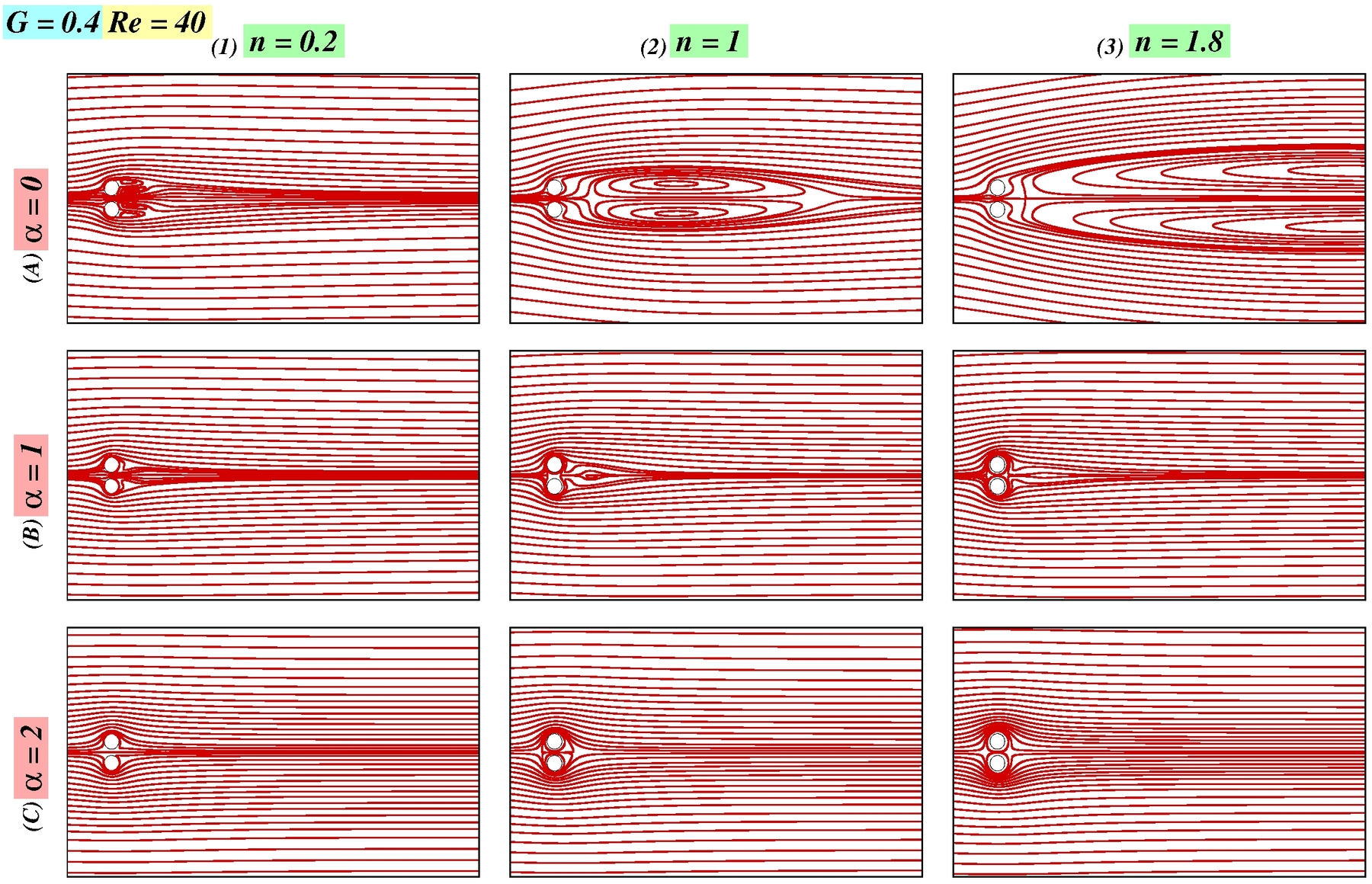}
	\caption{Streamline profiles for $Re=40$ at $G=0.4$.}
	\label{fig:st17}
\end{figure}
\begin{figure}[H]
	\centering
	\includegraphics[width=0.9\linewidth]{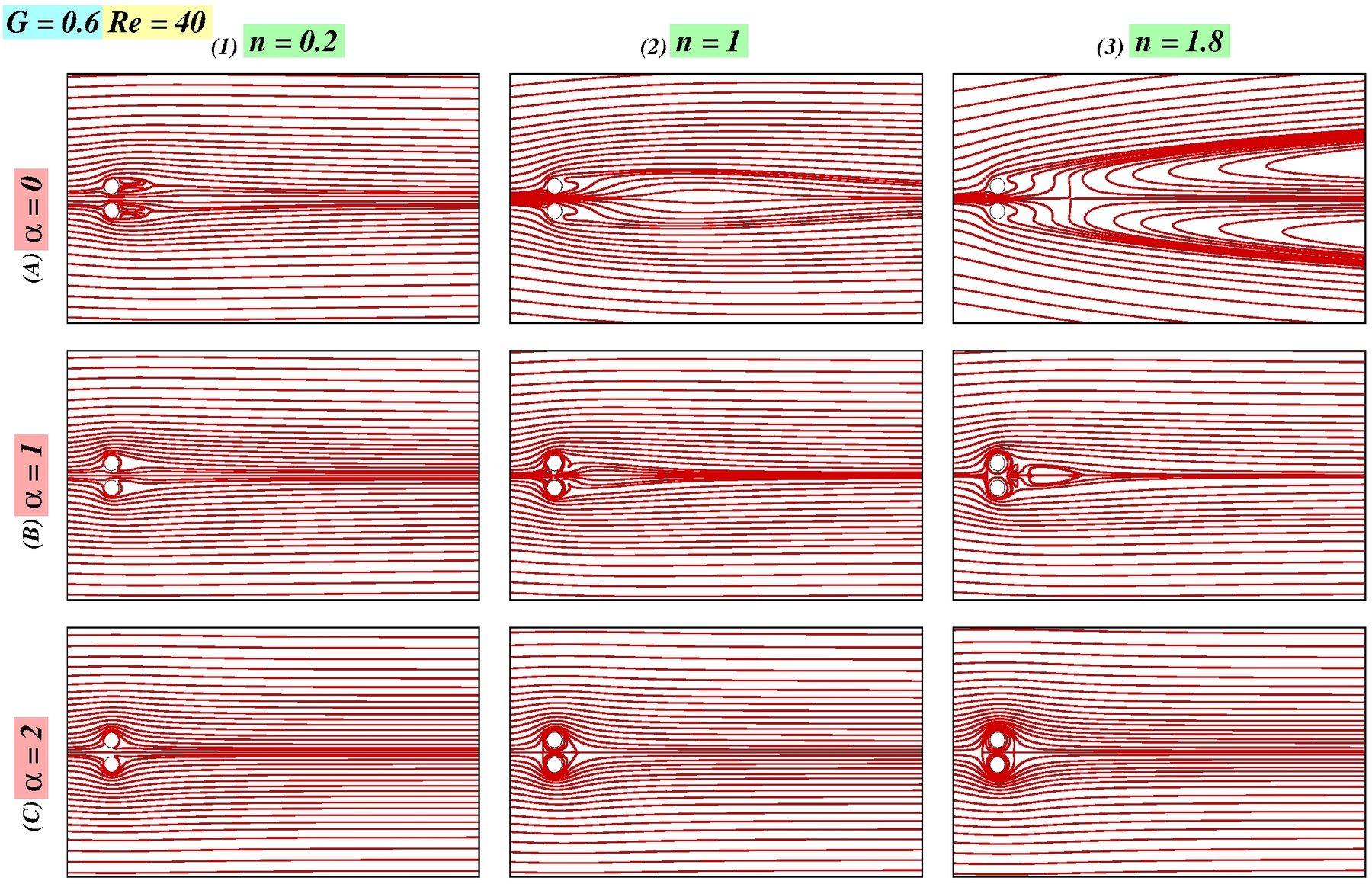}
	\caption{Streamline profiles for $Re=40$ at $G=0.6$.}
	\label{fig:st18}
\end{figure}
\begin{figure}[H]
	\centering
	\includegraphics[width=0.9\linewidth]{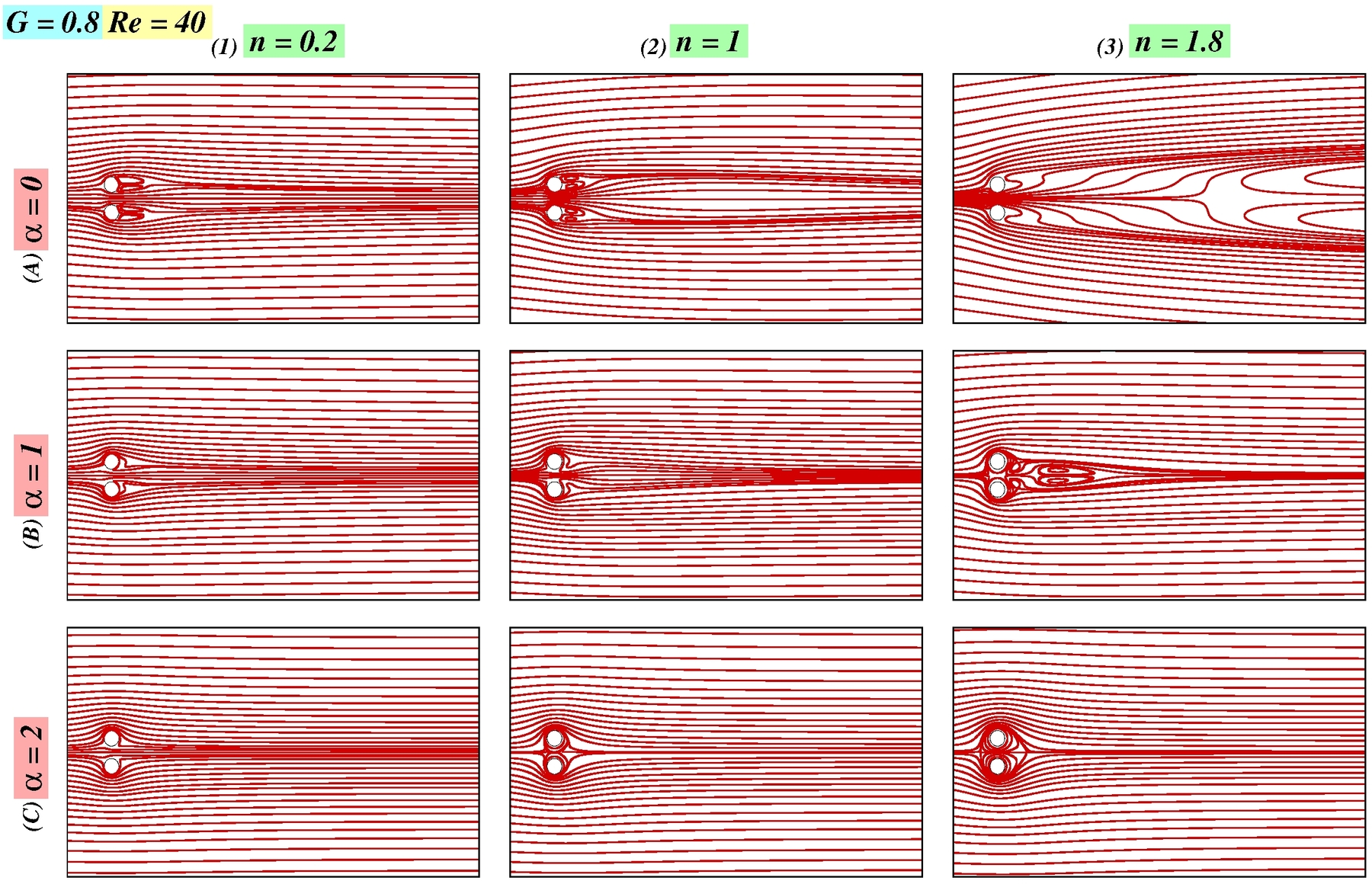}
	\caption{Streamline profiles for $Re=40$ at $G=0.8$.}
	\label{fig:st19}
\end{figure}
\begin{figure}[H]
	\centering
	\includegraphics[width=0.9\linewidth]{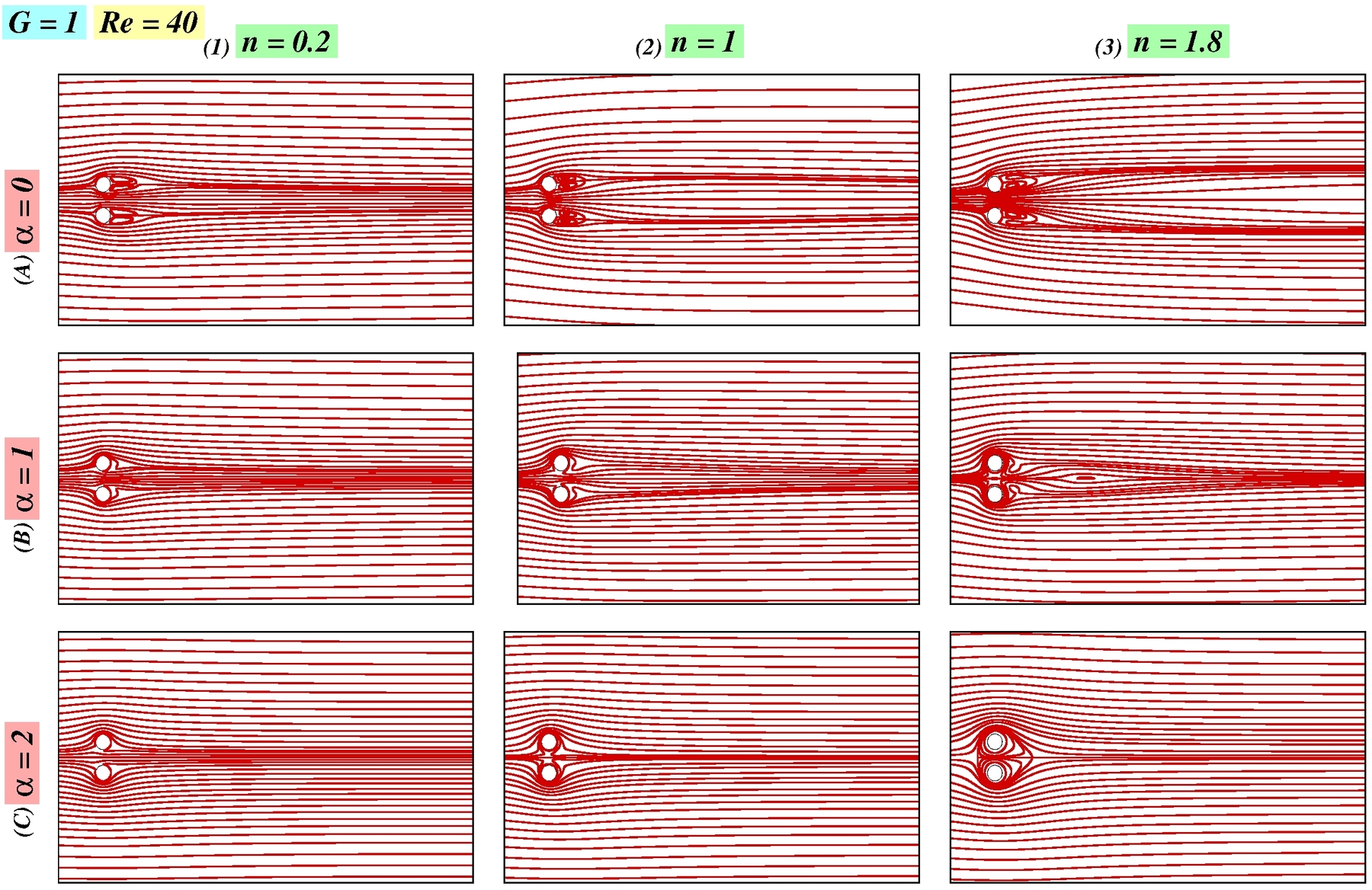}
	\caption{Streamline profiles for $Re=40$ at $G=1$.}
	\label{fig:st20}
\end{figure}
%
\clearpage
\section{Centerline velocity profiles:}\label{appendix:velocity}
 \setcounter{table}{0} \setcounter{figure}{0}
%
\begin{figure}[H]
	\centering
	\includegraphics[width=1\linewidth]{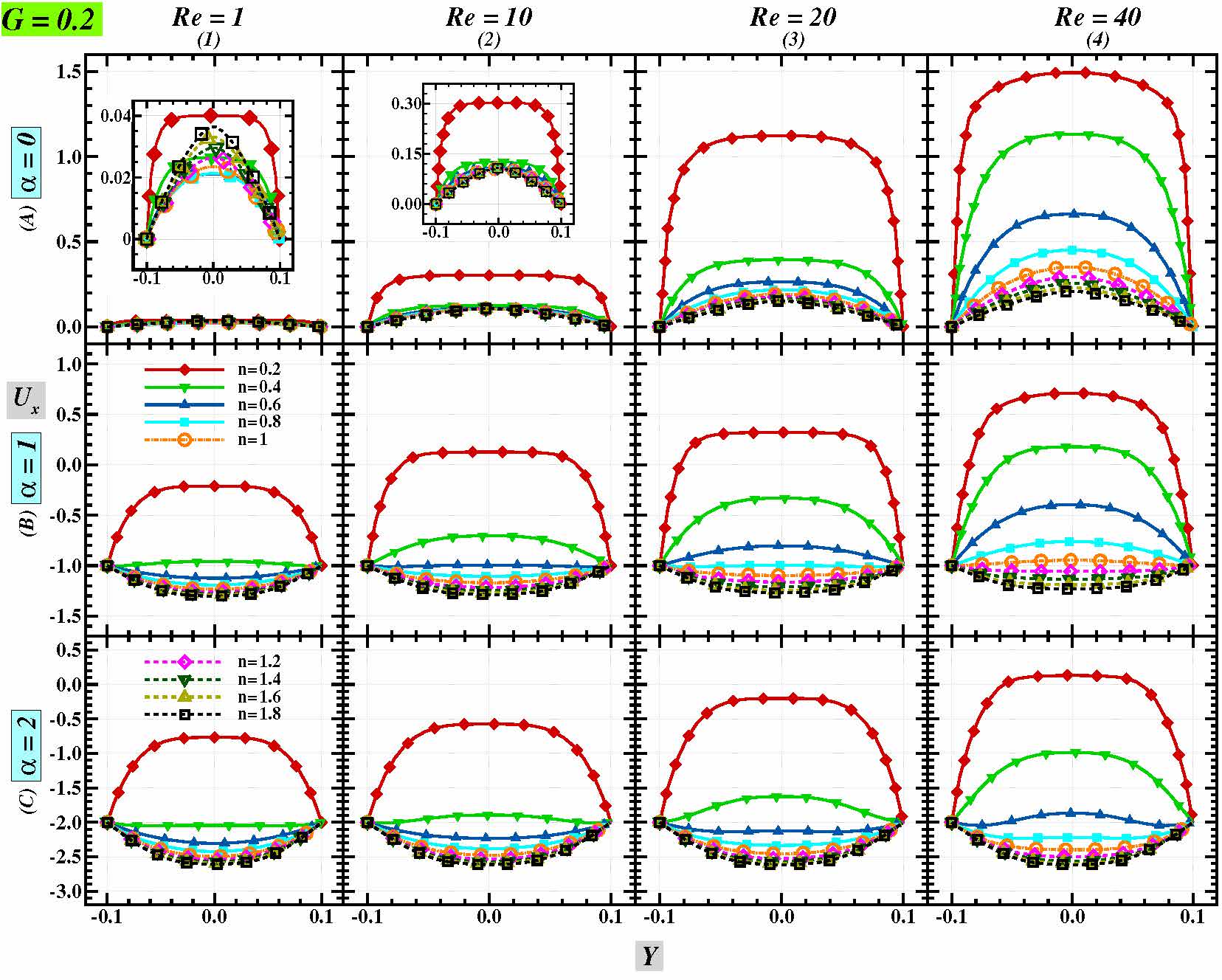}
	\caption{Centerline velocity profiles ($U_x$ at line 6, \fig\ref{fig:1}) for different power-law indices ($n$) at various rotational rates ($\alpha$, rows) and Reynolds numbers ($Re$, columns) for $G=0.2$.}
	\label{fig:velocity-0.2}
\end{figure}
\begin{figure}[H]
	\centering
	\includegraphics[width=1\linewidth]{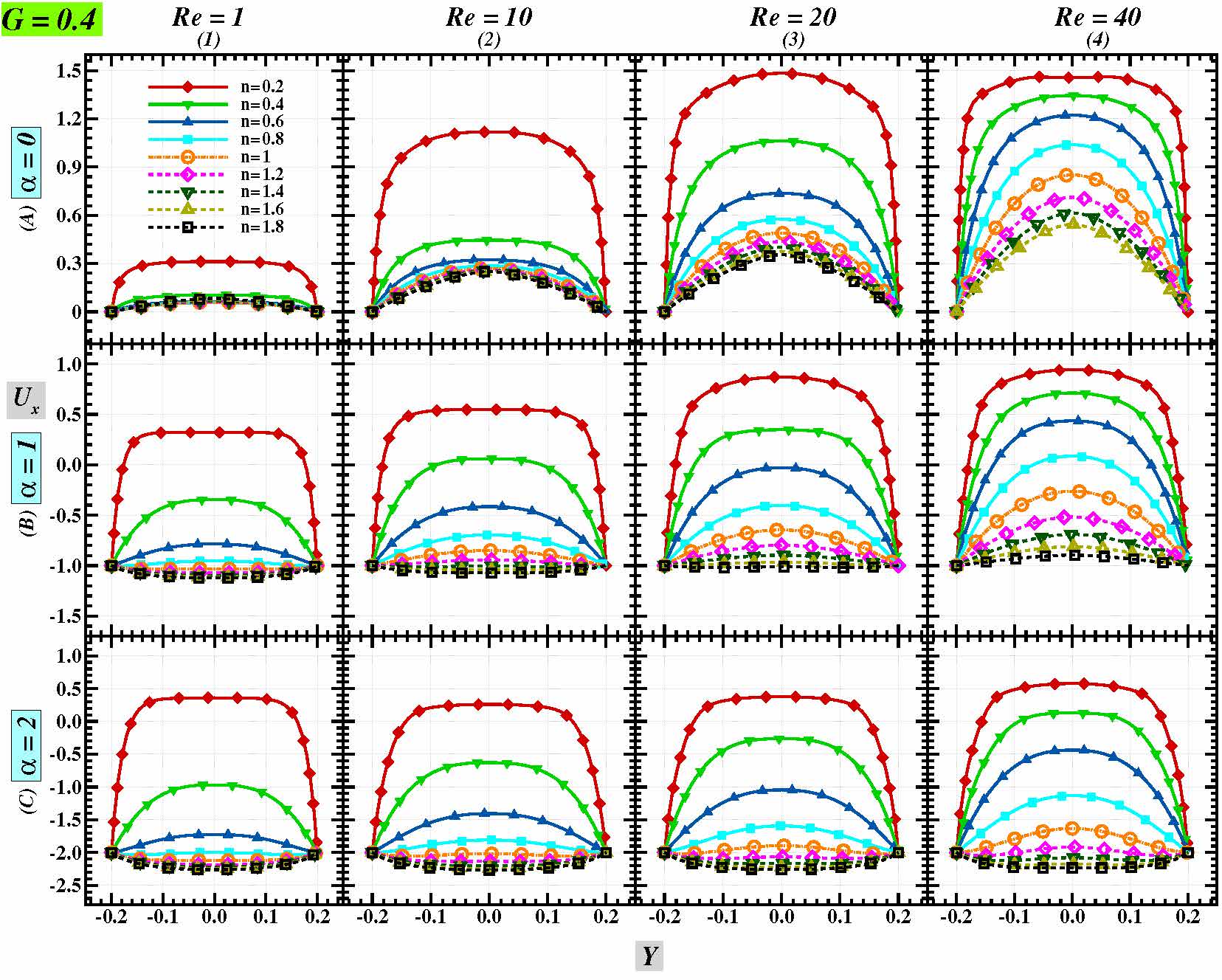}
	\caption{Centerline velocity profiles ($U_x$ at line 6, \fig\ref{fig:1}) for different power-law indices ($n$) at various rotational rates ($\alpha$, rows) and Reynolds numbers ($Re$, columns) for $G=0.4$.}
	\label{fig:velocity-0.4}
\end{figure}
\begin{figure}[H]
	\centering
	\includegraphics[width=1\linewidth]{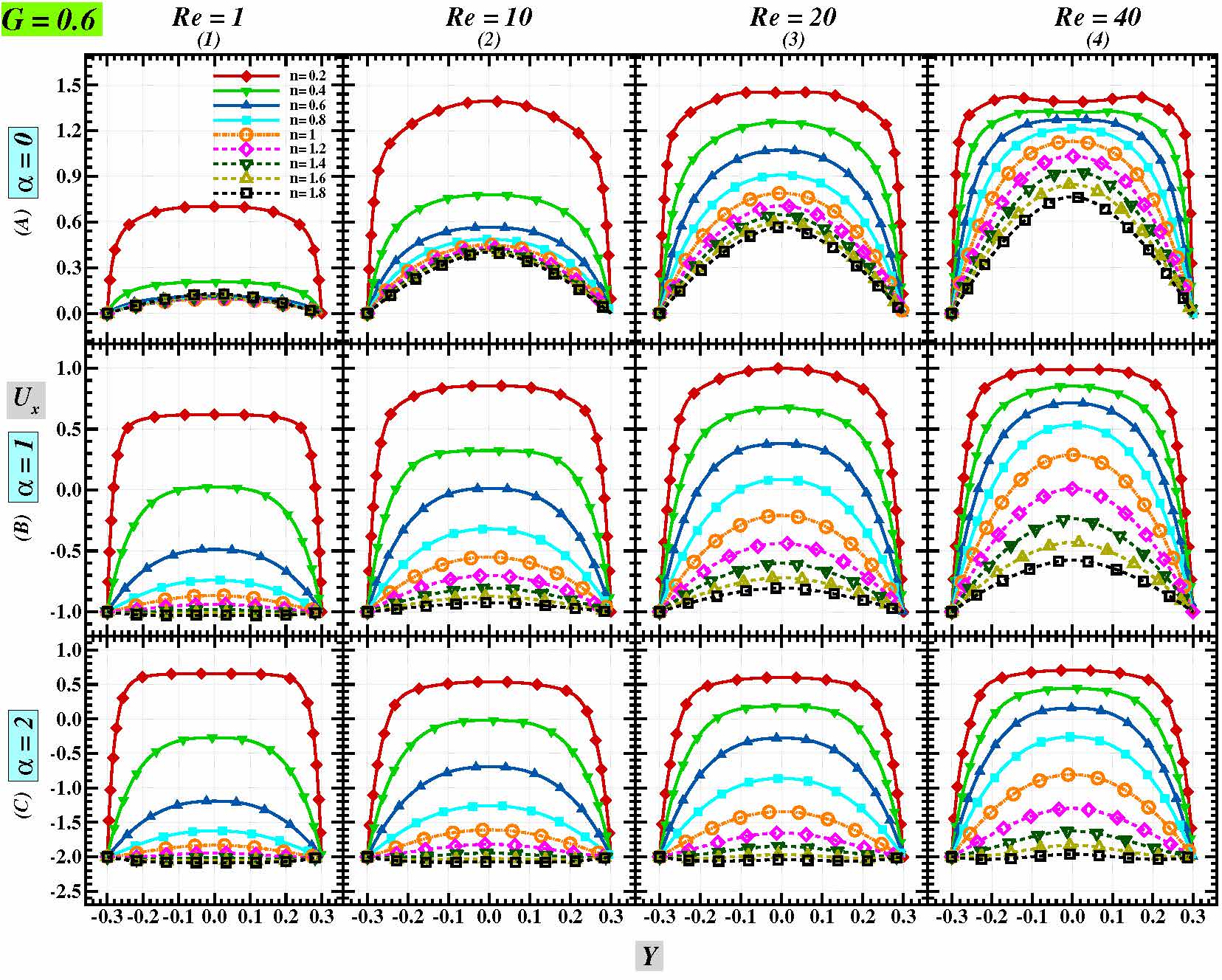}
	\caption{Centerline velocity profiles ($U_x$ at line 6, \fig\ref{fig:1}) for different power-law indices ($n$) at various rotational rates ($\alpha$, rows) and Reynolds numbers ($Re$, columns) for $G=0.6$.}
	\label{fig:velocity-0.6}
\end{figure}
\begin{figure}[H]
	\centering
	\includegraphics[width=1\linewidth]{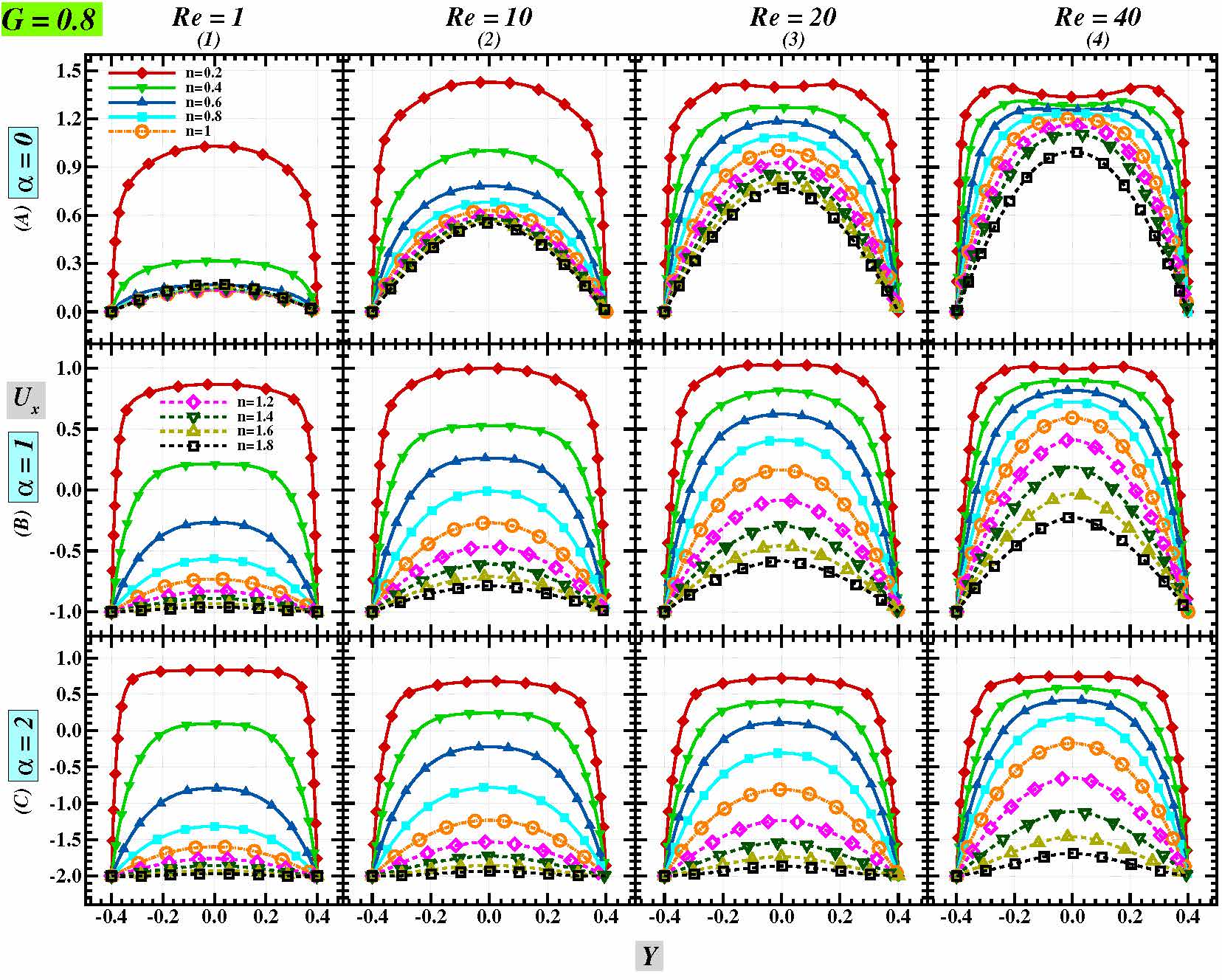}
	\caption{Centerline velocity profiles ($U_x$ at line 6, \fig\ref{fig:1}) for different power-law indices ($n$) at various rotational rates ($\alpha$, rows) and Reynolds numbers ($Re$, columns) for $G=0.8$.}
	\label{fig:velocity-0.8}
\end{figure}
\begin{figure}[H]
	\centering
	\includegraphics[width=1\linewidth]{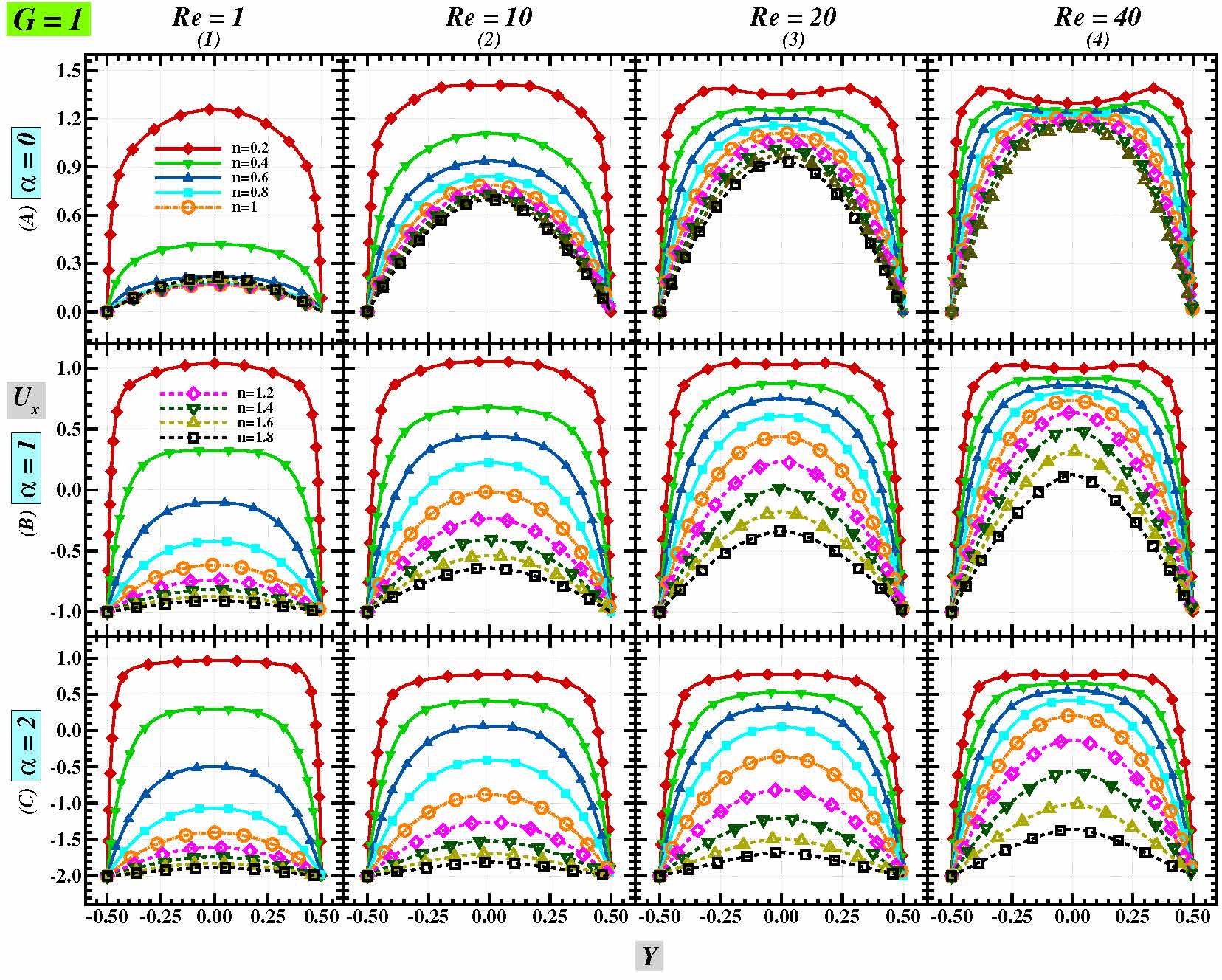}
	\caption{Centerline velocity profiles ($U_x$ at line 6, \fig\ref{fig:1}) for different power-law indices ($n$) at various rotational rates ($\alpha$, rows) and Reynolds numbers ($Re$, columns) for $G=1$.}
	\label{fig:velocity-1}
\end{figure}
%
\clearpage
\section{Pressure coefficient ($C_p$) profiles}\label{appendix:pressure}
 \setcounter{table}{0} \setcounter{figure}{0}
%
\begin{figure}[H]
	\centering
	\subfigure[Upper cylinder (UC)] {\includegraphics[width=0.7\linewidth]{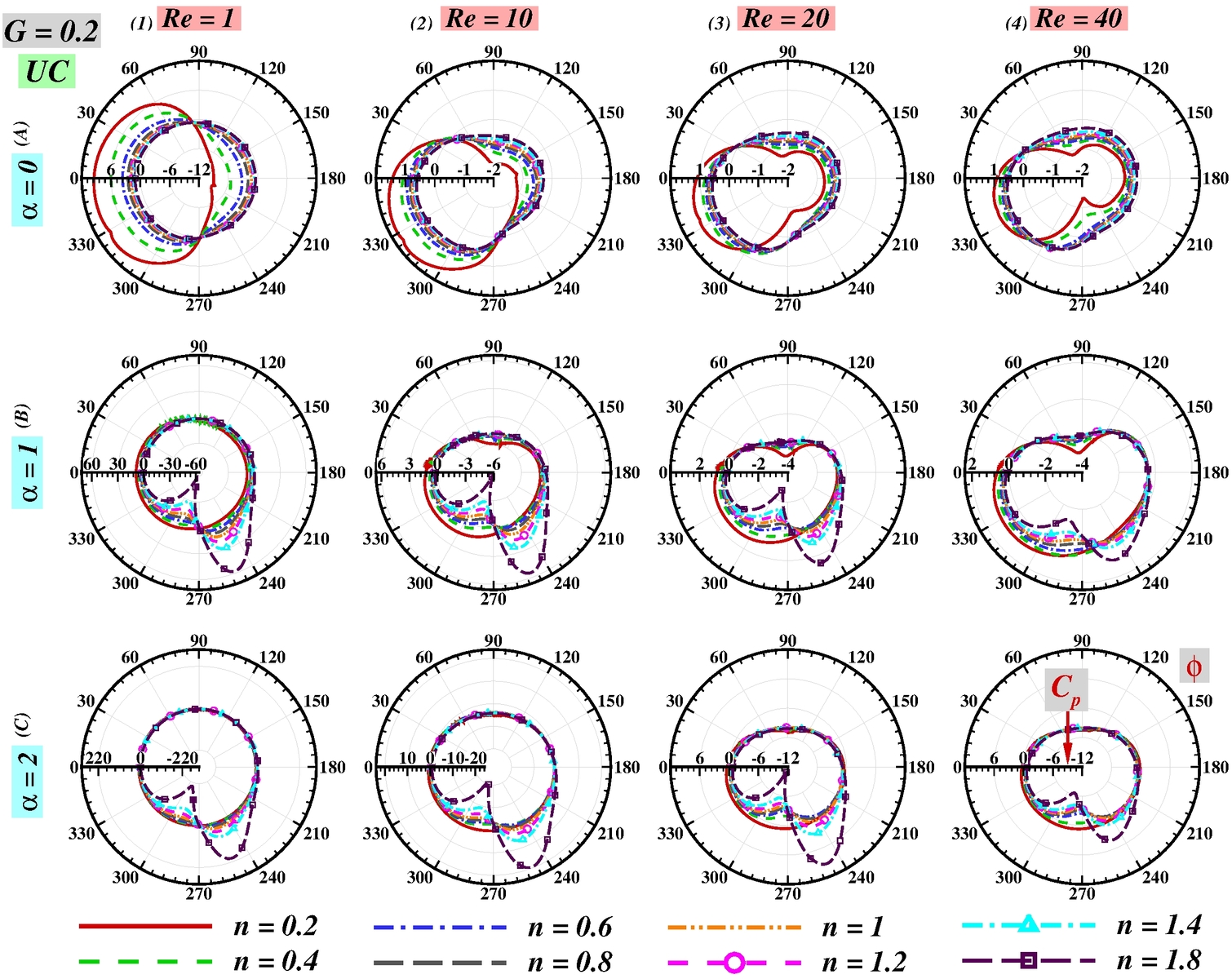}\label{fig:pre-up0.2}}
	\subfigure[Lower cylinder (LC)] {\includegraphics[width=0.7\linewidth]{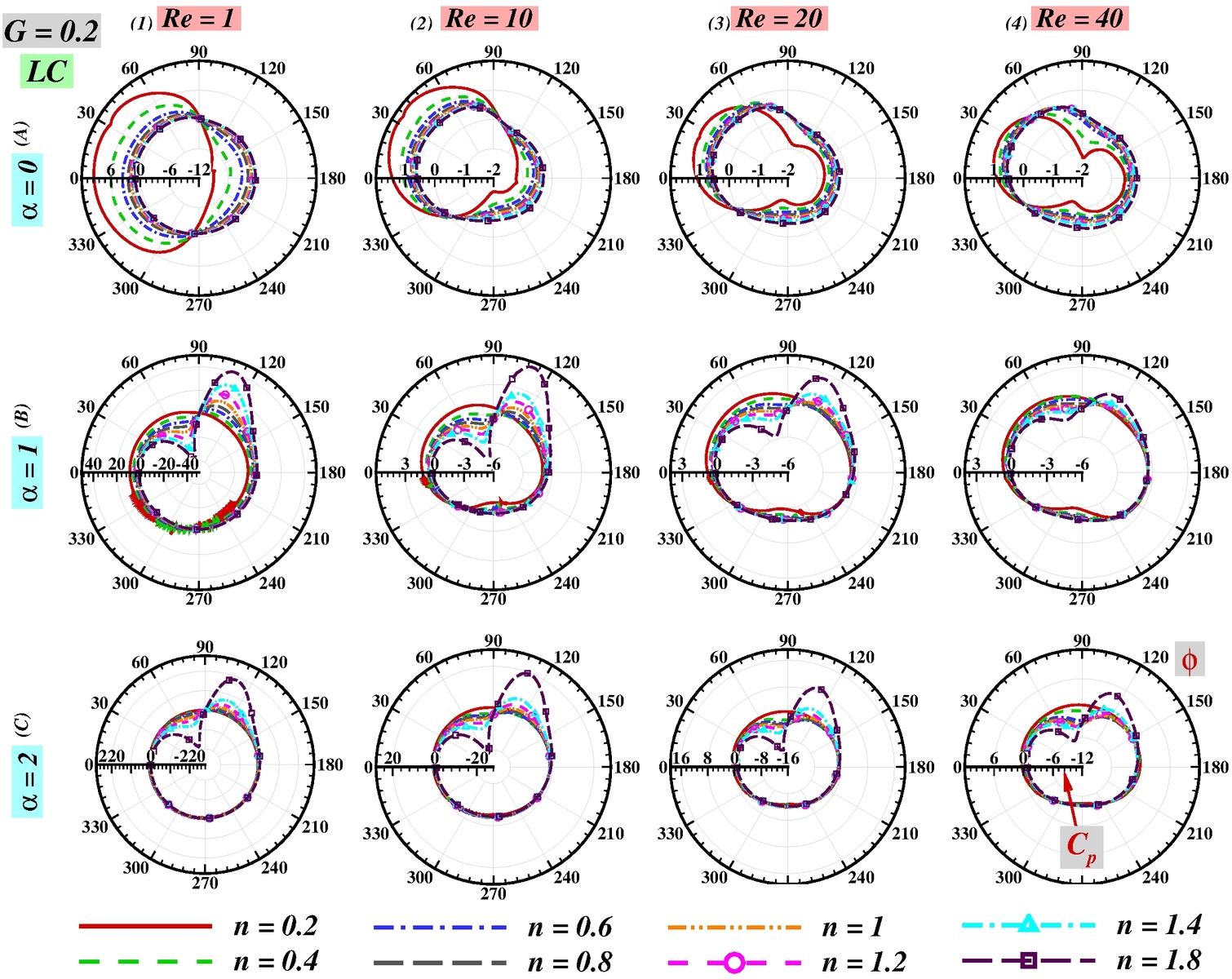}\label{fig:pre-low0.2}}
	\caption{Distribution of pressure coefficient on the surface of cylinders at $G=0.2$.}
	\label{fig:pre-0.2}
\end{figure}
\begin{figure}[H]
	\centering
	\subfigure[Upper cylinder (UC)]
	{\includegraphics[width=0.75\linewidth]{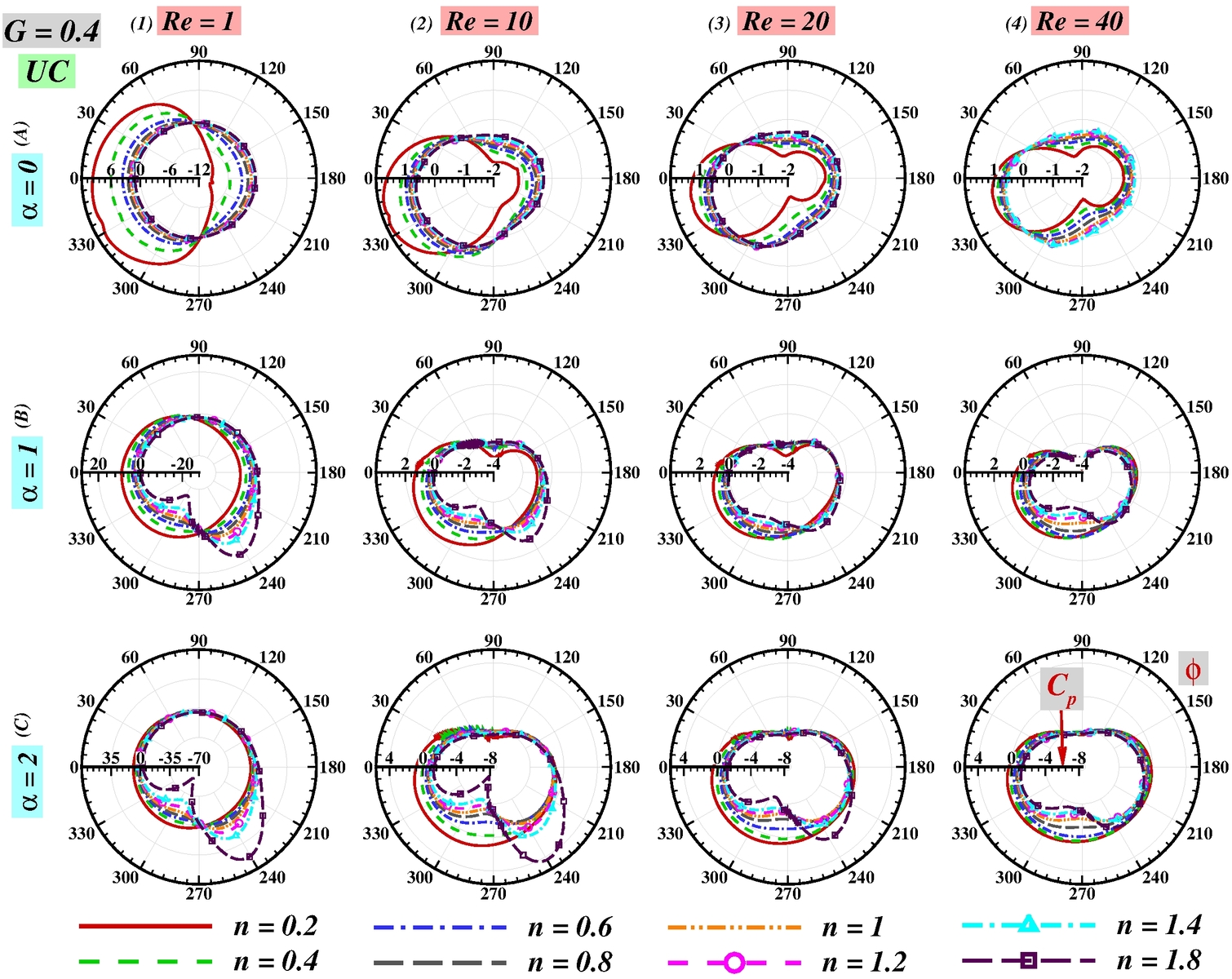}\label{fig:pre-up0.4}}
	\subfigure[Lower cylinder (LC)] {\includegraphics[width=0.75\linewidth]{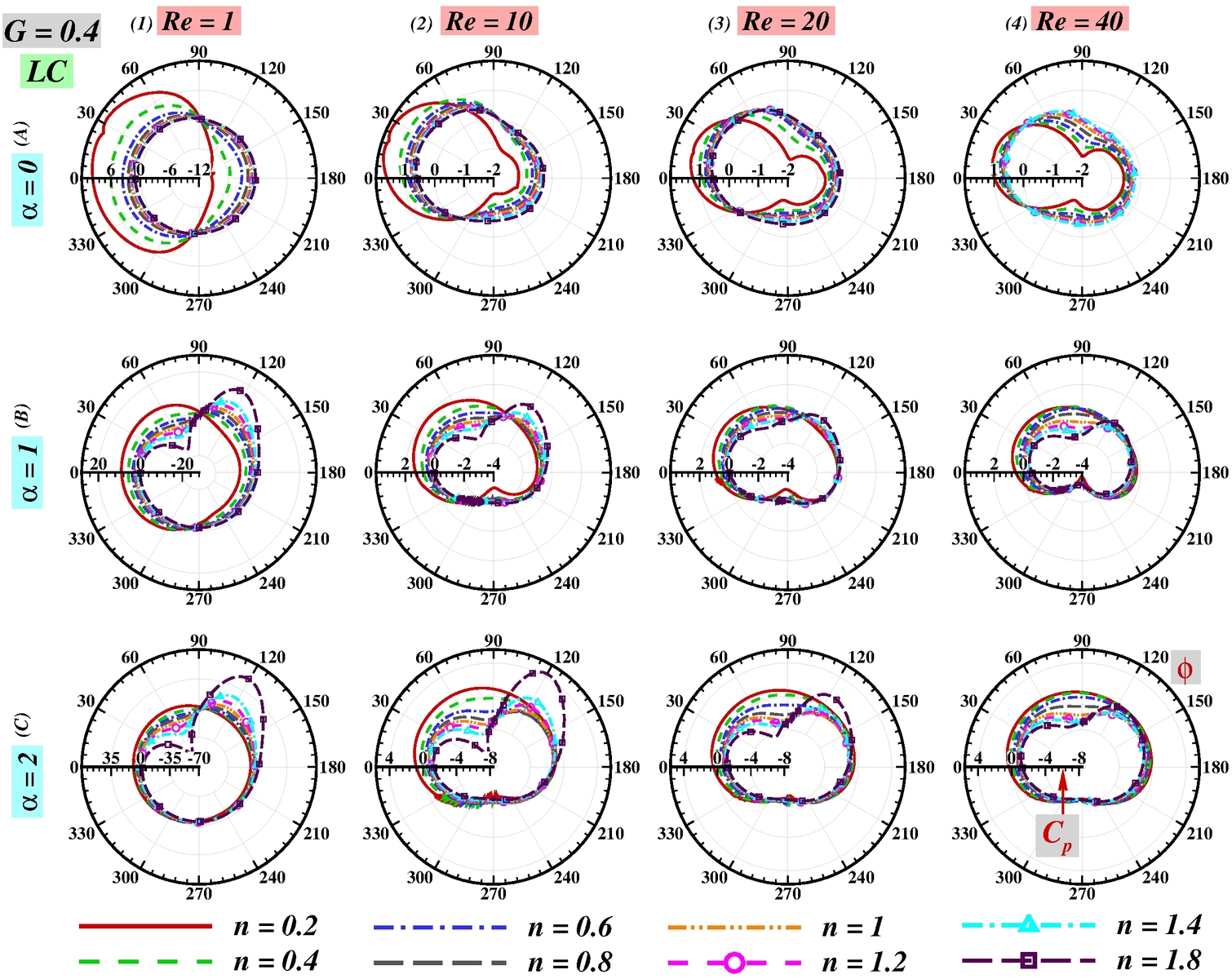}\label{fig:pre-low0.4}}	
	\caption{Distribution of pressure coefficient on the surface of cylinders at $G=0.4$.}
	\label{fig:pre-0.4}
\end{figure}
\begin{figure}[H]
	\centering
	\subfigure[Upper cylinder (UC)]
	{\includegraphics[width=0.75\linewidth]{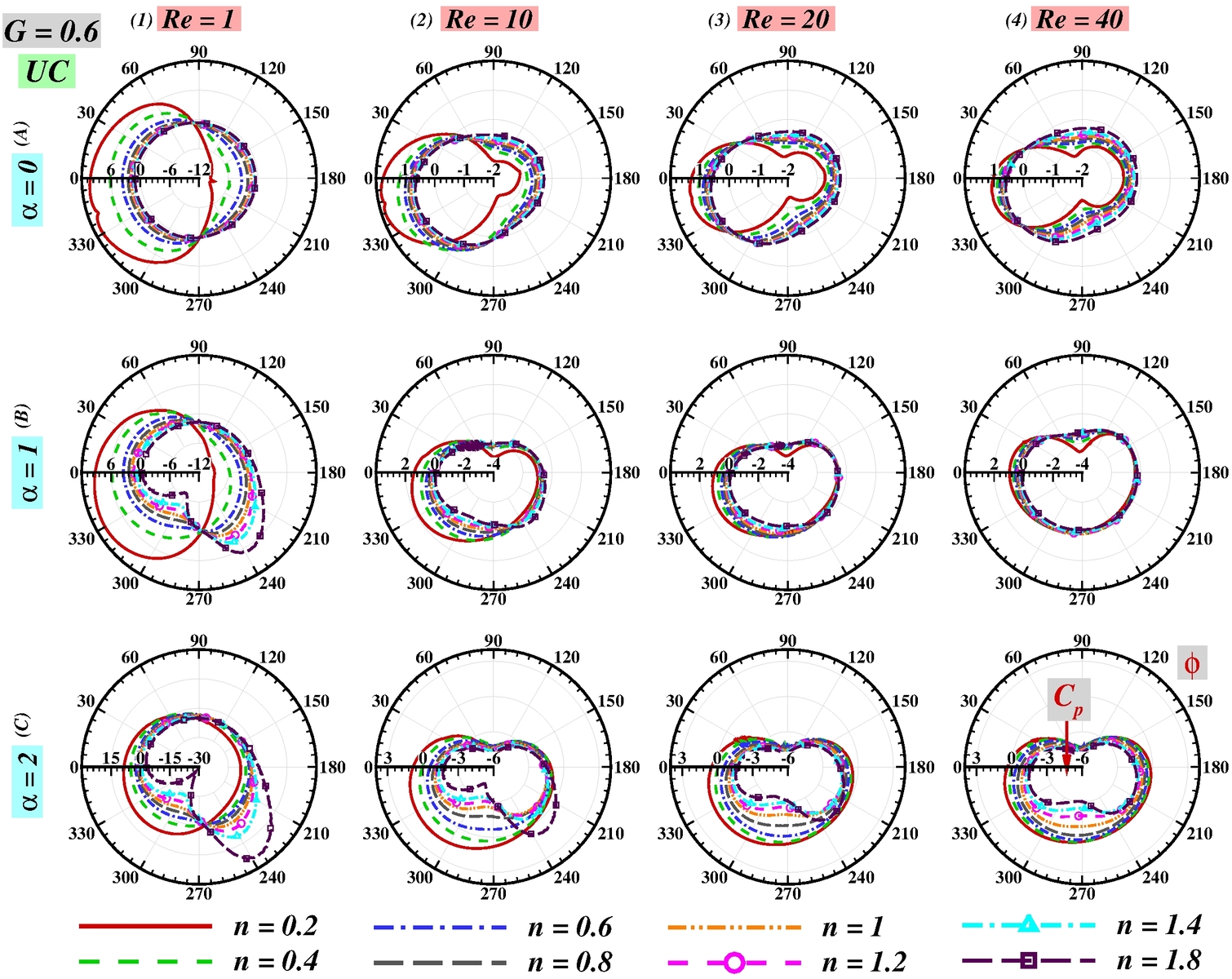}\label{fig:pre-up0.6}}
	\subfigure[Lower cylinder (LC)] {\includegraphics[width=0.75\linewidth]{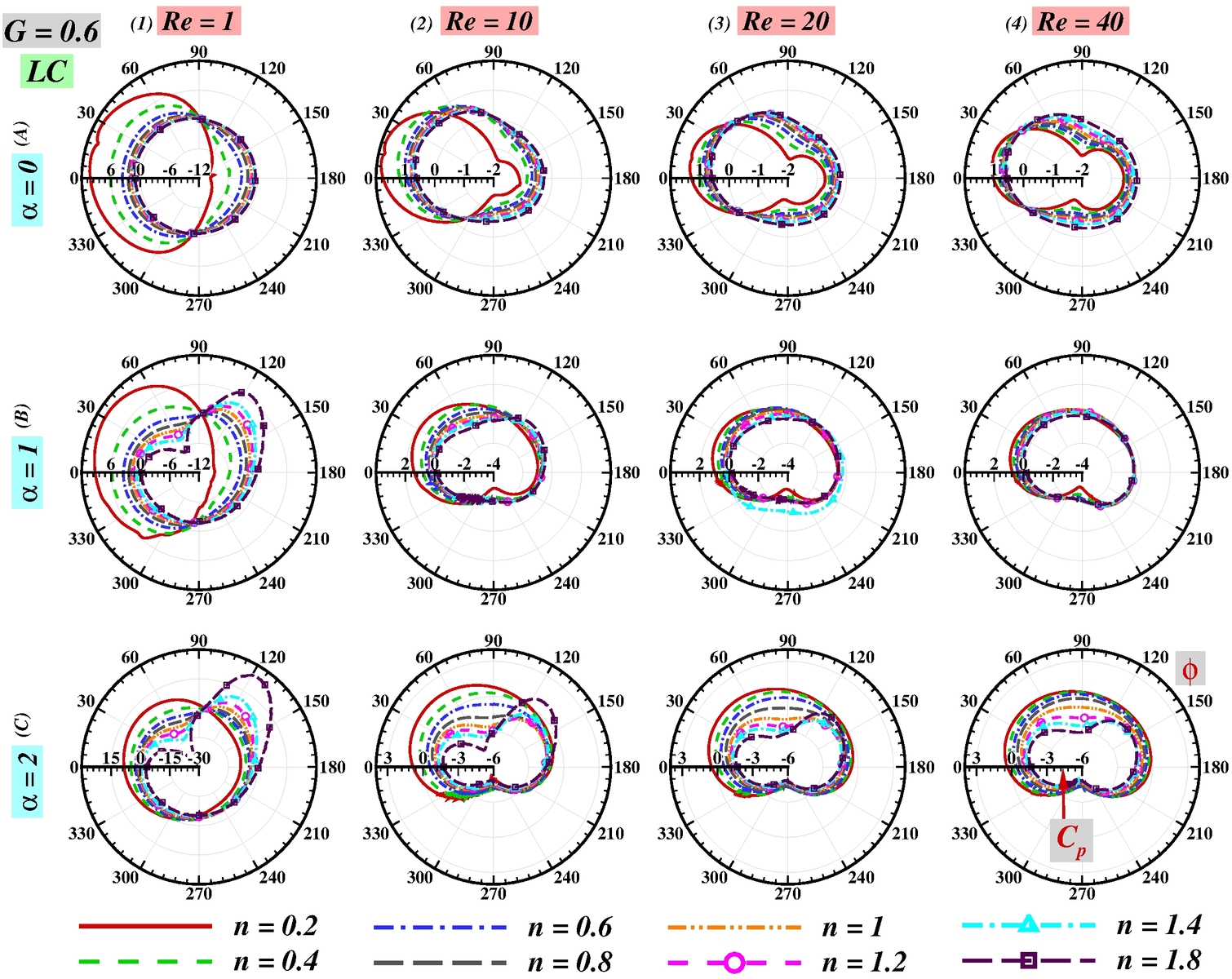}\label{fig:pre-low0.6}}
	\caption{Distribution of pressure coefficient on the surface of cylinders at $G=0.6$.}
	\label{fig:pre-0.6}
\end{figure}
\begin{figure}[H]
	\centering
	\subfigure[Upper cylinder (UC)]
	{\includegraphics[width=0.75\linewidth]{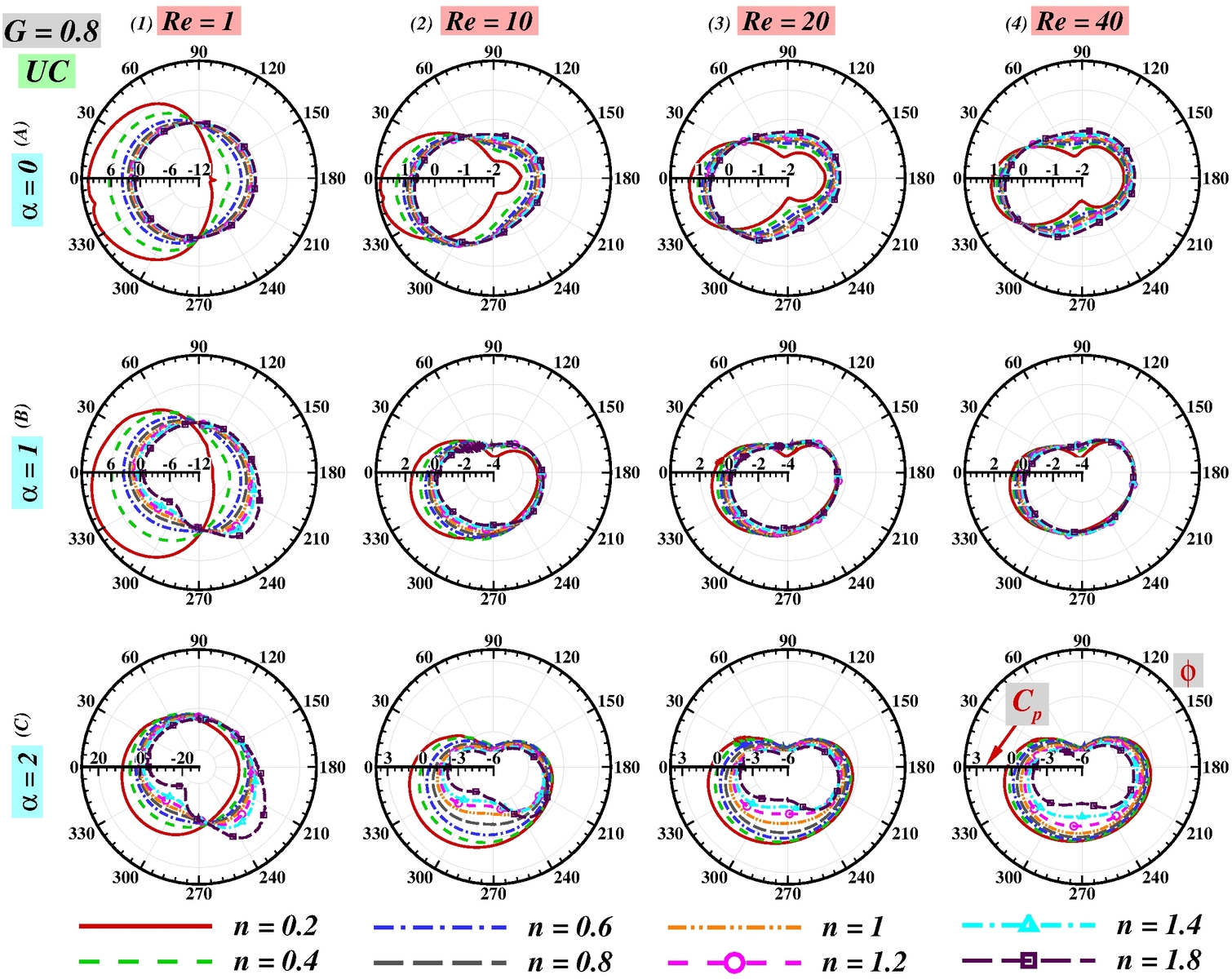}\label{fig:pre-up0.8}}
	\subfigure[Lower cylinder (LC)] 
	{\includegraphics[width=0.75\linewidth]{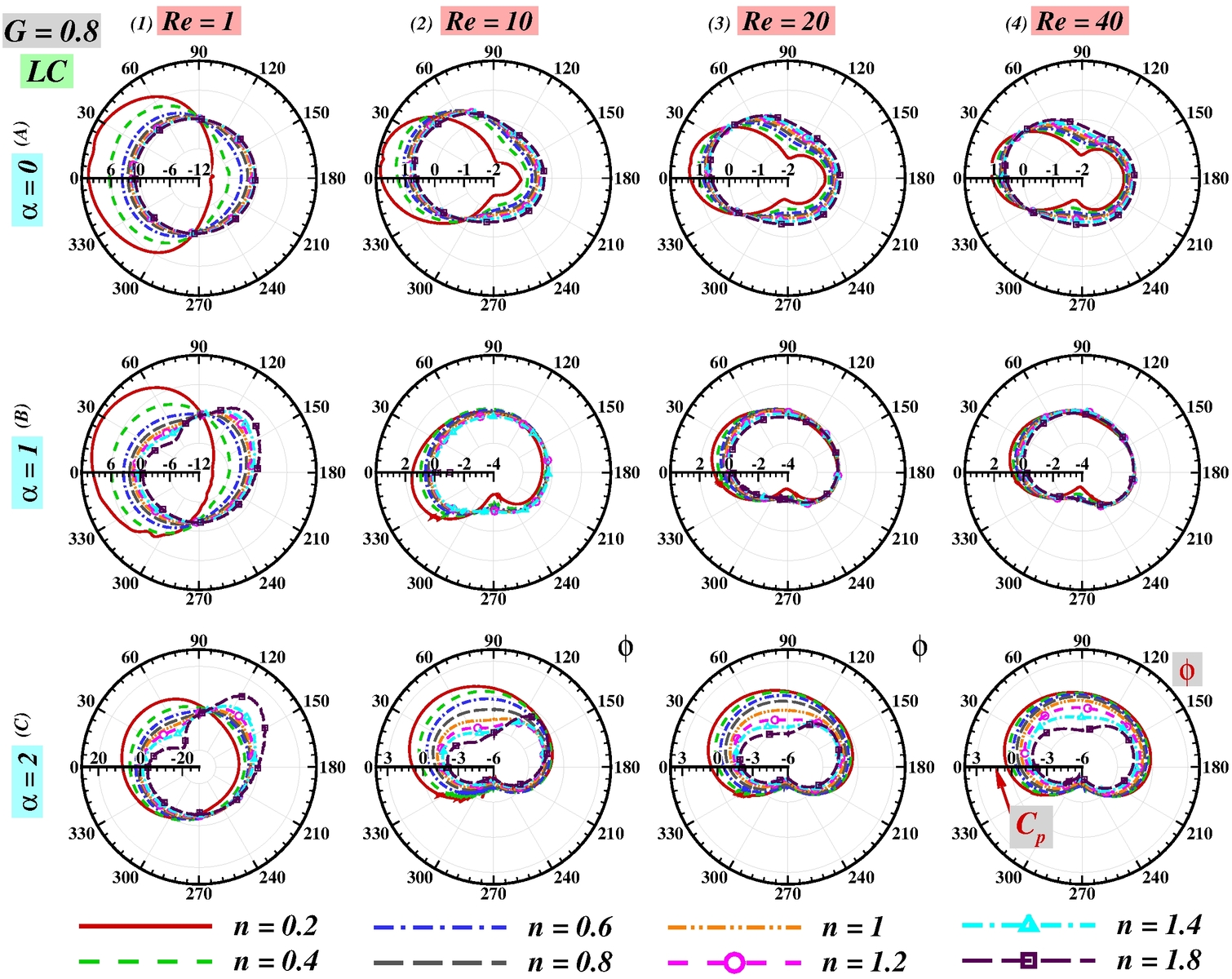}\label{fig:pre-low0.8}}
	\caption{Distribution of pressure coefficient on the surface of cylinders at $G=0.8$.}
	\label{fig:pre-0.8}
\end{figure}
\begin{figure}[H]
	\centering
	\subfigure[Upper cylinder (UC)]
	{\includegraphics[width=0.75\linewidth]{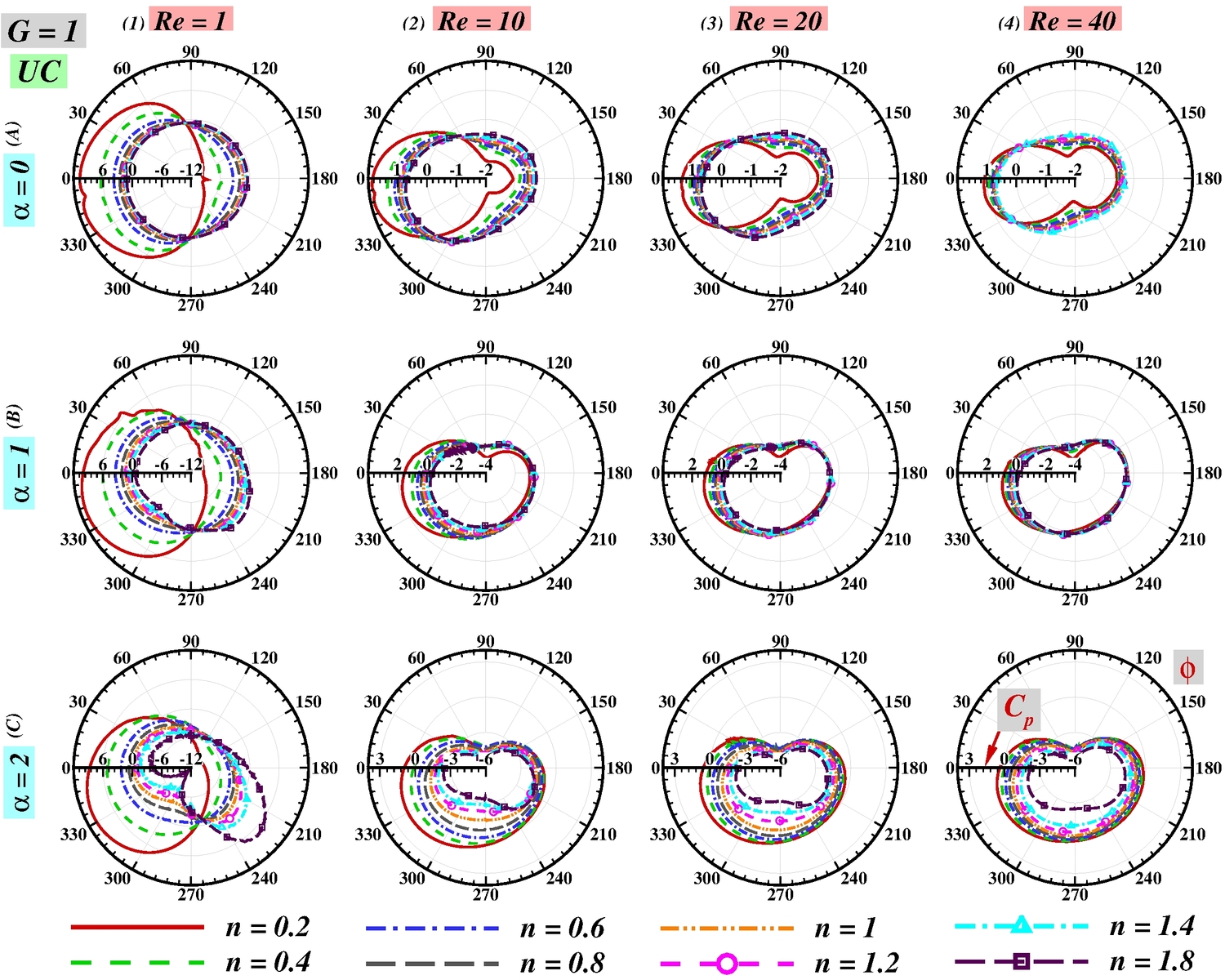}\label{fig:pre-up1}}
	\subfigure[Lower cylinder (LC)] 
	{\includegraphics[width=0.75\linewidth]{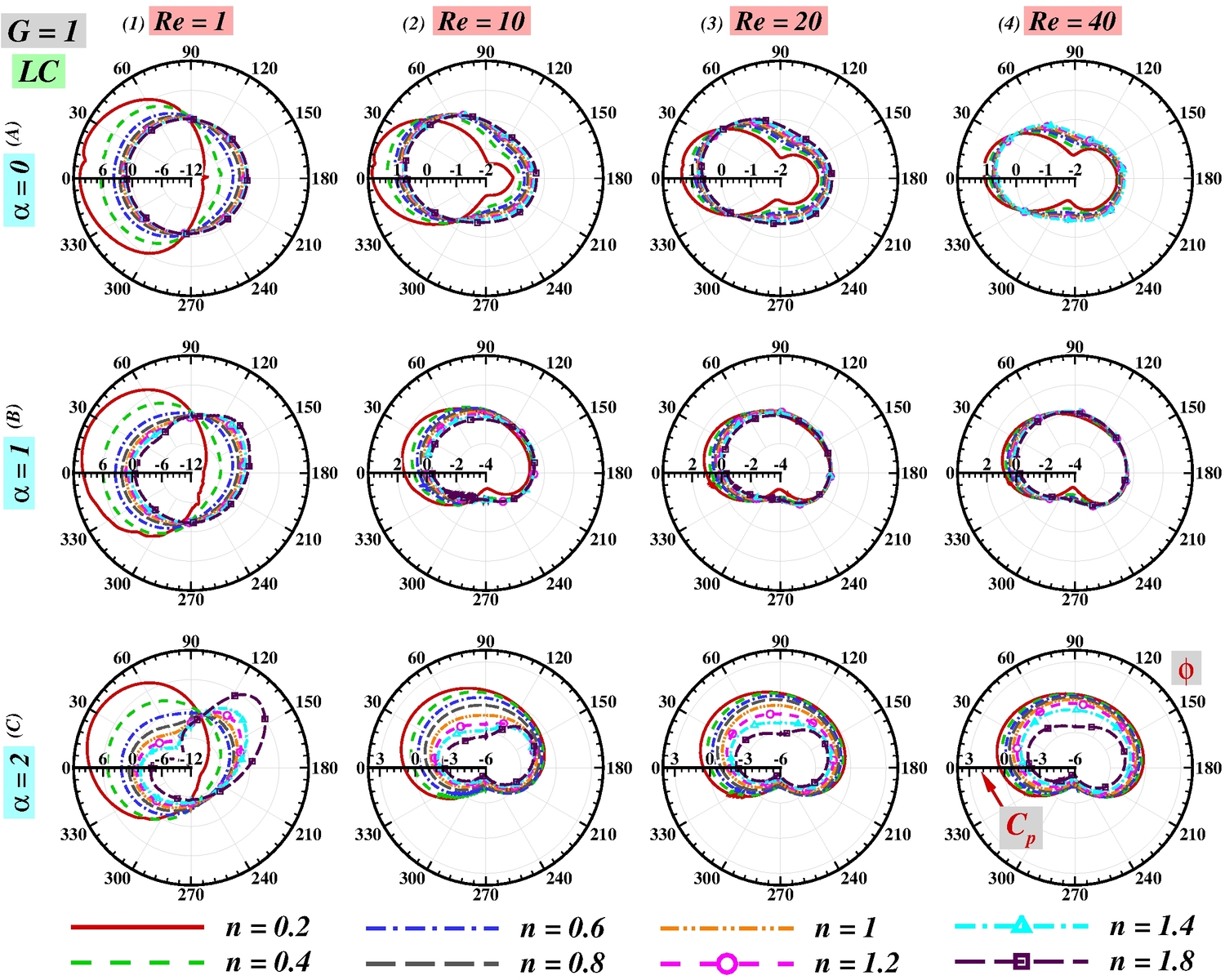}\label{fig:pre-low1}}
	\caption{Distribution of pressure coefficient on the surface of cylinders at $G=1$.}
	\label{fig:pre-1}
\end{figure}
%
\clearpage
\section{Friction and pressure drag coefficient ($C_{DF}$ and $C_{DP}$)}\label{appendix:Idrag}
\setcounter{table}{0} \setcounter{figure}{0}
%
\begin{figure}[H]
	\centering
	\includegraphics[width=1\linewidth]{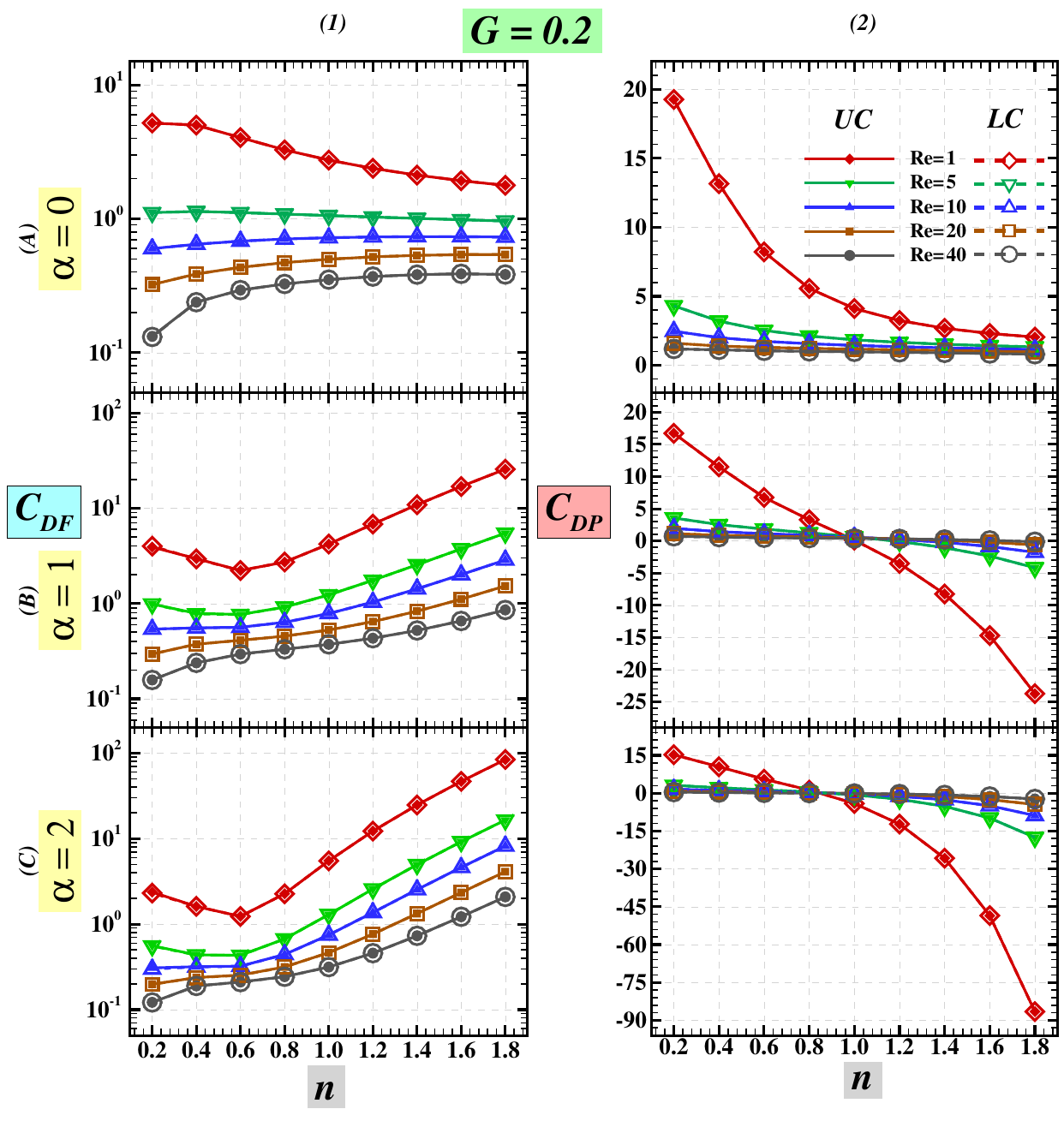}
	\caption{The variation of friction and pressure drag coefficient ($C_{DF}$ and $C_{DP}$) with $n$ and $Re$ for $G=0.2$.}
	\label{fig:cdpcdfg0.2}
\end{figure}
\begin{figure}[H]
	\centering
	\includegraphics[width=1\linewidth]{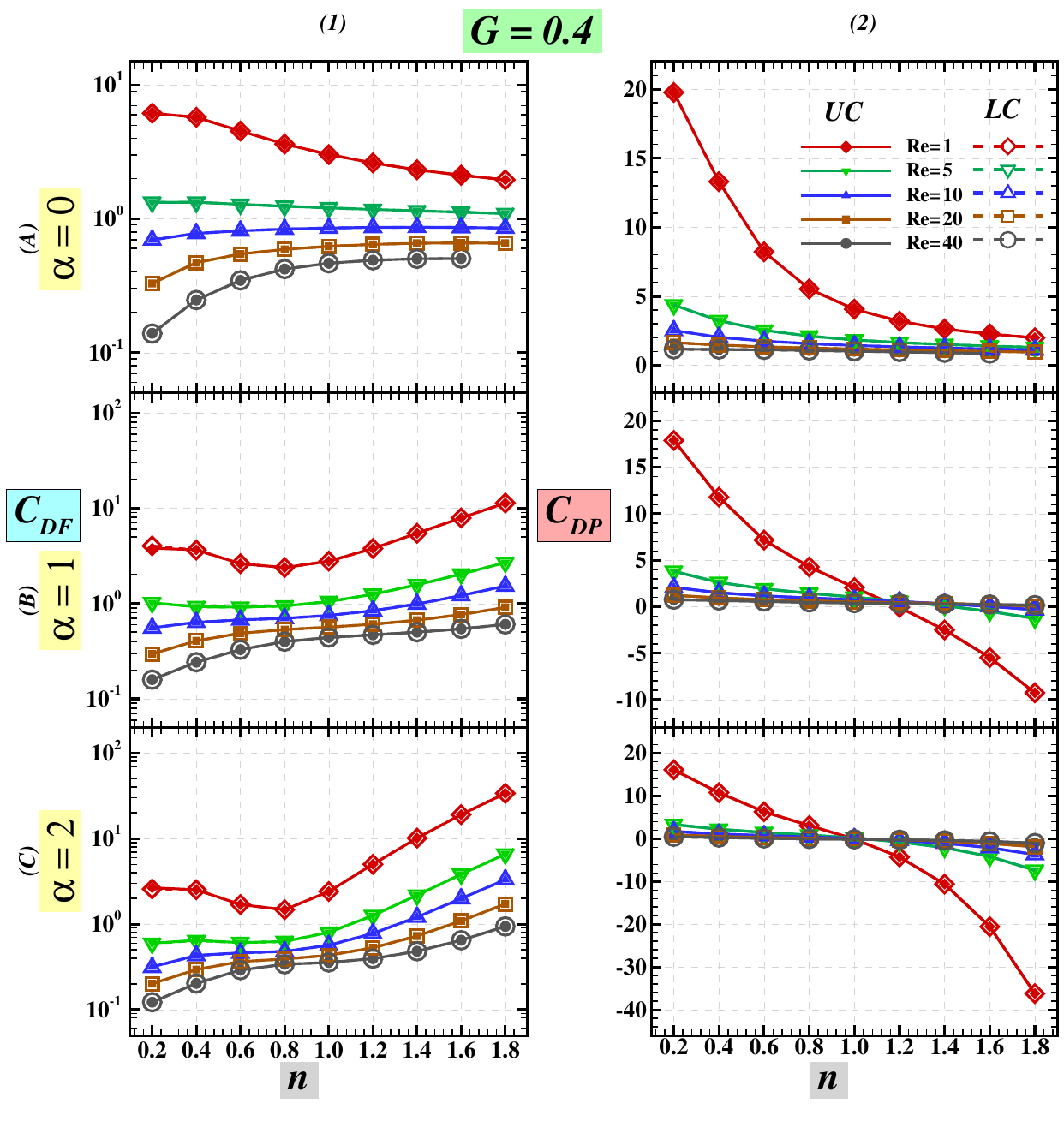}
	\caption{The variation of friction and pressure drag coefficient ($C_{DF}$ and $C_{DP}$) with $n$ and $Re$ for $G=0.4$.}
	\label{fig:cdpcdf-g0.4}
\end{figure}
\begin{figure}[H]
	\centering
	\includegraphics[width=1\linewidth]{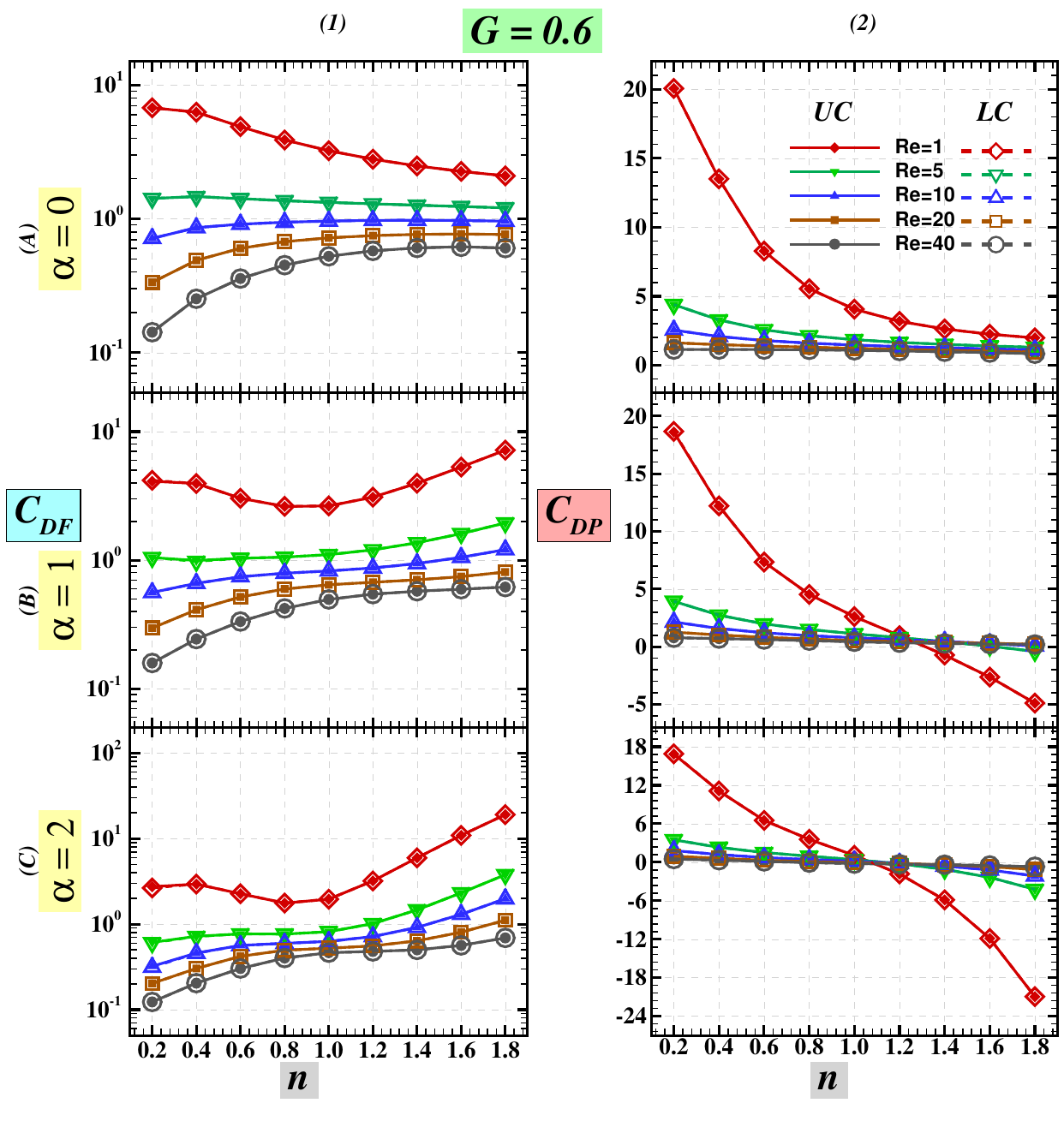}
	\caption{The variation of friction and pressure drag coefficient ($C_{DF}$ and $C_{DP}$) with $n$ and $Re$ for $G=0.6$.}
	\label{fig:cdpcdfg0.6}
\end{figure}
\begin{figure}[H]
	\centering
	\includegraphics[width=1\linewidth]{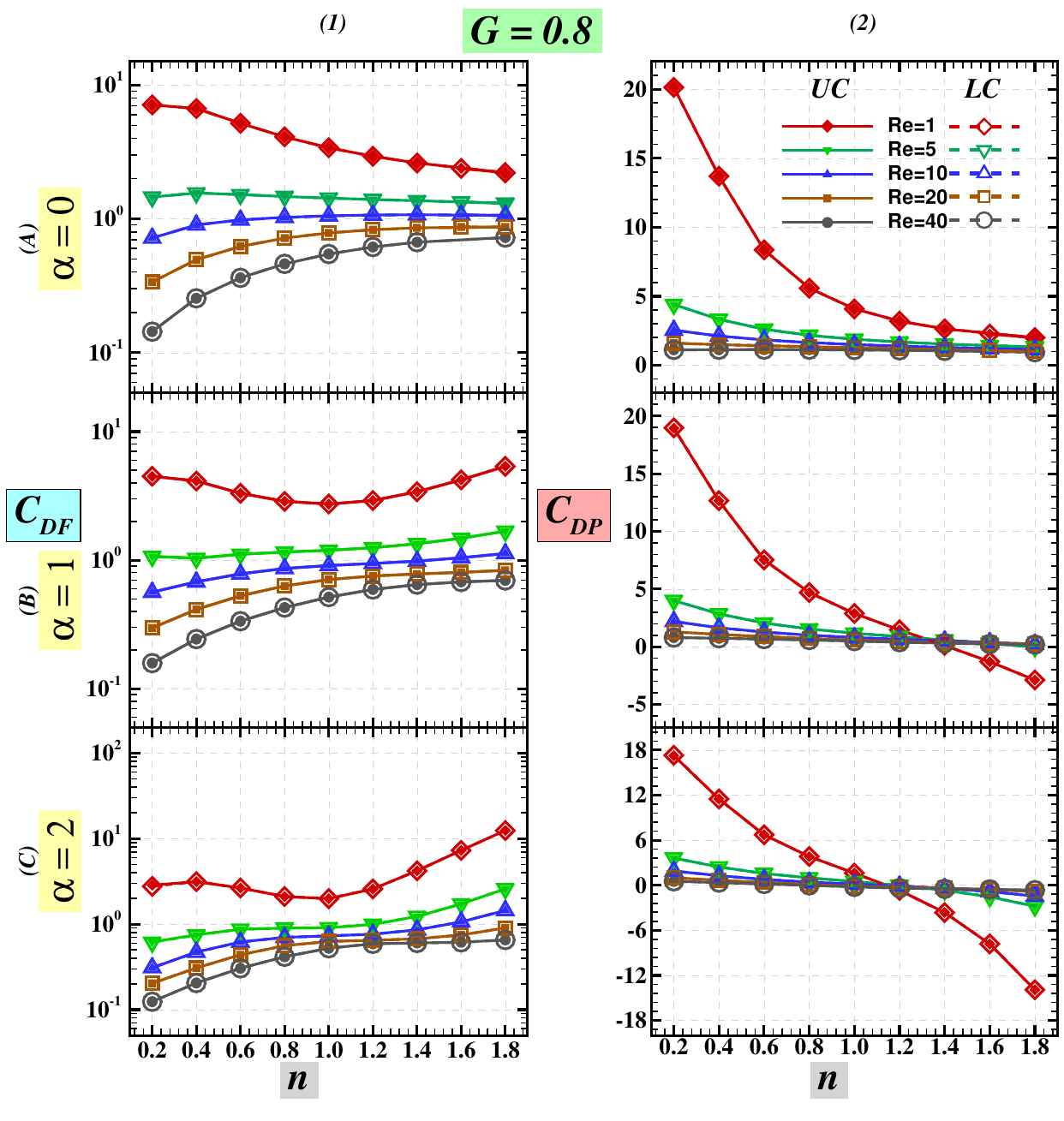}
	\caption{The variation of friction and pressure drag coefficient ($C_{DF}$ and $C_{DP}$) with $n$ and $Re$ for $G=0.8$.}
	\label{fig:cdpcdf-g0.8}
\end{figure}
\begin{figure}[H]
	\centering
	\includegraphics[width=1\linewidth]{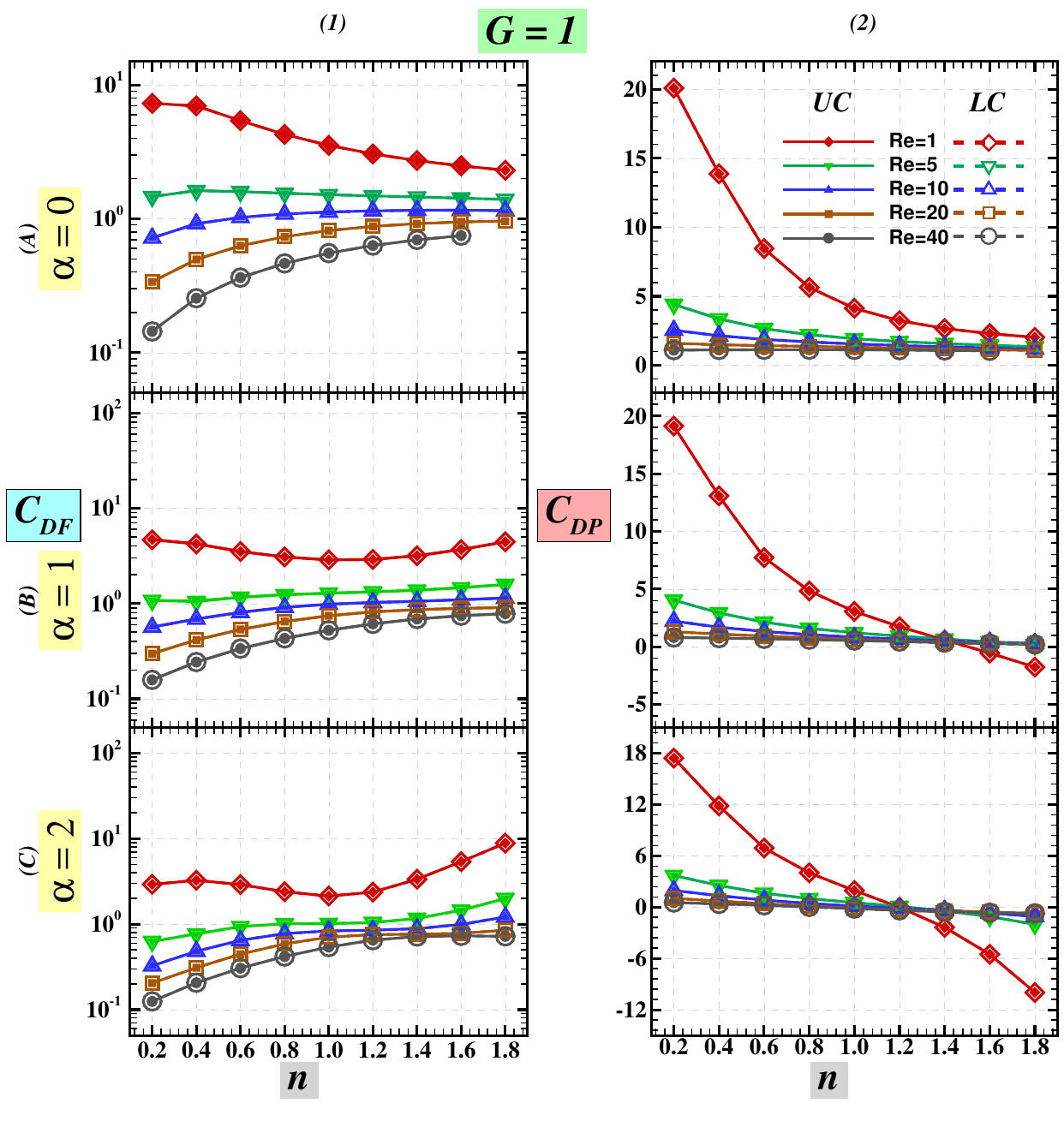}
	\caption{The variation of friction and pressure drag coefficient ($C_{DF}$ and $C_{DP}$) with $n$ and $Re$ for $G=1$.}
	\label{fig:cdpcdfg1}
\end{figure}
%
\clearpage
\section{Friction and pressure lift coefficient ($C_{LF}$ and $C_{LP}$)}\label{appendix:Ilift}
\setcounter{table}{0}\setcounter{figure}{0} 
%
\begin{figure}[H]
	\centering
	\includegraphics[width=1\linewidth]{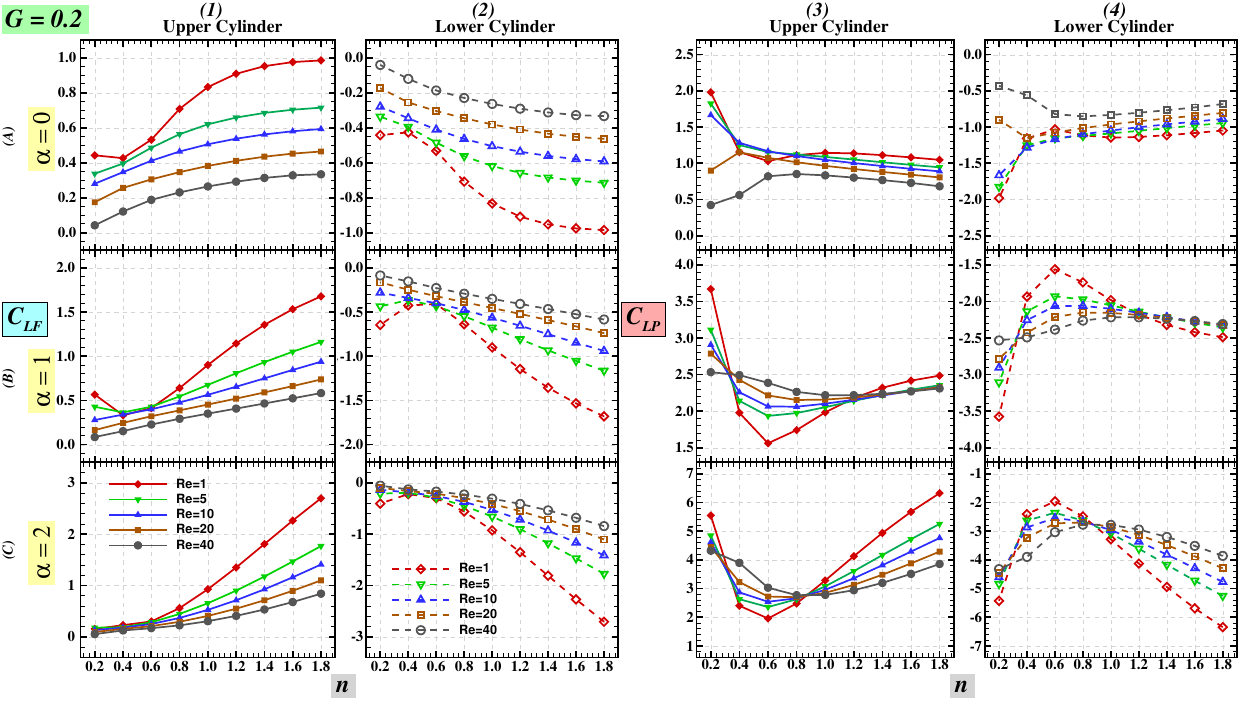}
	\caption{Variation of friction and pressure lift coefficient ($C_{LF}$ and $C_{LP}$) with $n$ and $Re$ for $G=0.2$.}
	\label{fig:clp-clf-g0.2}
\end{figure}
\begin{figure}[H]
	\centering
	\includegraphics[width=1\linewidth]{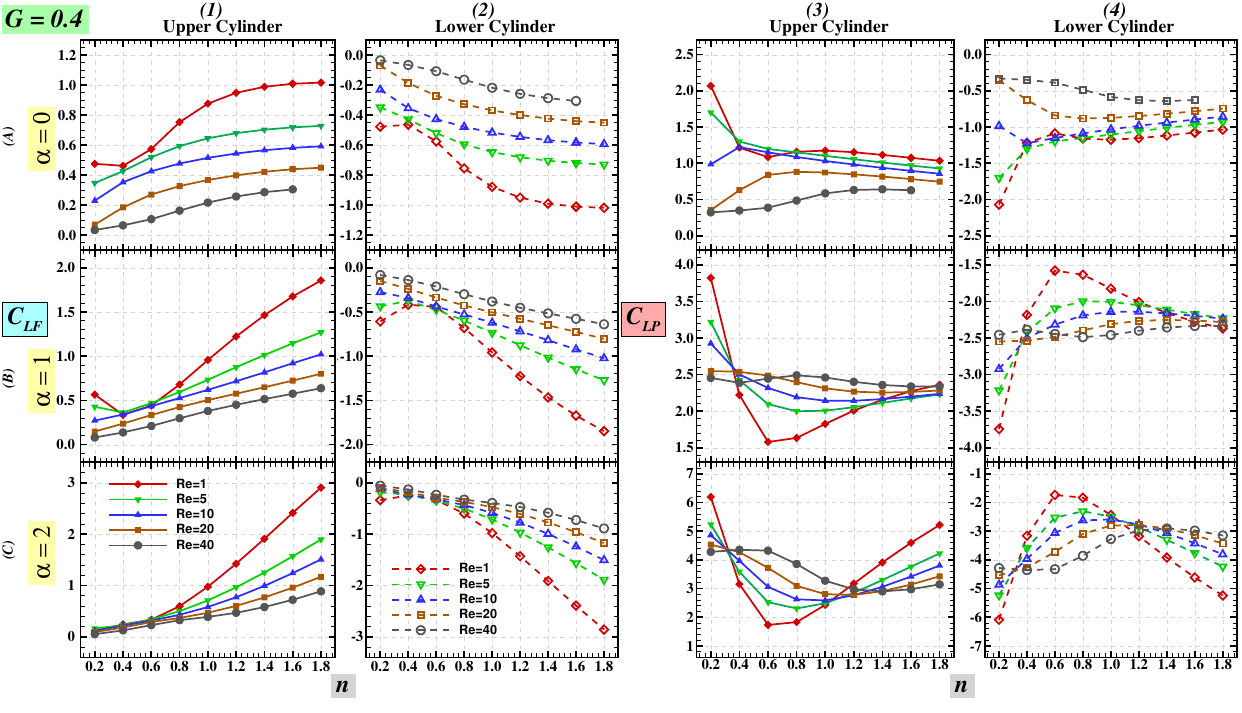}
	\caption{Variation of friction and pressure lift coefficient ($C_{LF}$ and $C_{LP}$) with $n$ and $Re$ for $G=0.4$.}
	\label{fig:clp-clf-g0.4}
\end{figure}
\begin{figure}[H]
	\centering
	\includegraphics[width=1\linewidth]{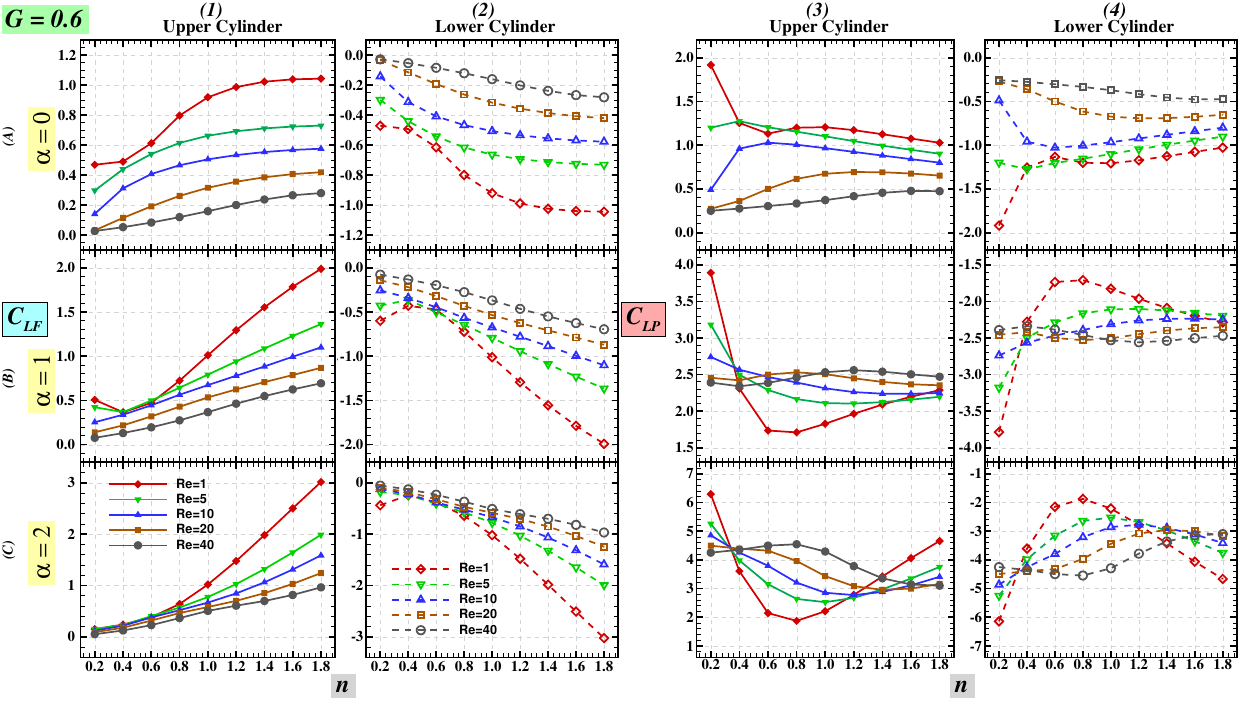}
	\caption{Variation of friction and pressure lift coefficient ($C_{LF}$ and $C_{LP}$) with power-law index ($n$) and Reynolds number ($Re$) for gap ratio $G=0.6$}
	\label{fig:clp-clf-g0.6}
\end{figure}
\begin{figure}[H]
	\centering
	\includegraphics[width=1\linewidth]{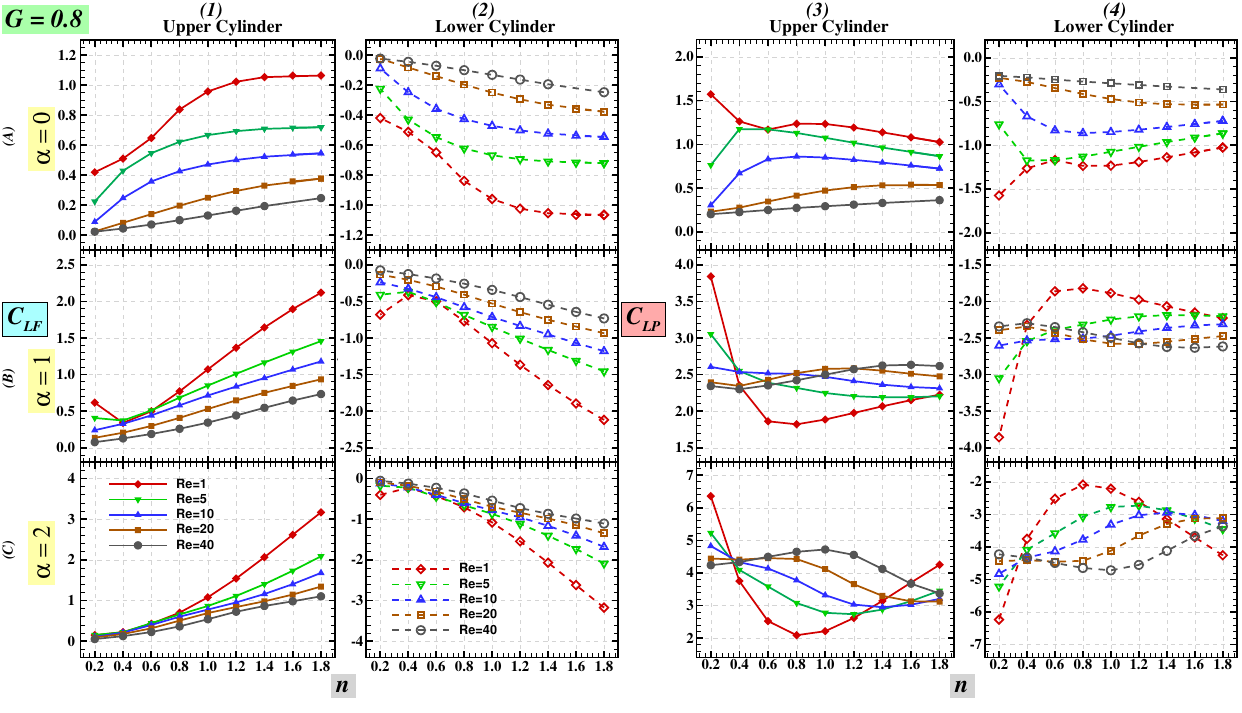}
	\caption{Variation of friction and pressure lift coefficient ($C_{LF}$ and $C_{LP}$) with $n$ and $Re$ for $G=0.8$.}
	\label{fig:clp-clf-g0.8}
\end{figure}
\begin{figure}[H]
	\centering
	\includegraphics[width=1\linewidth]{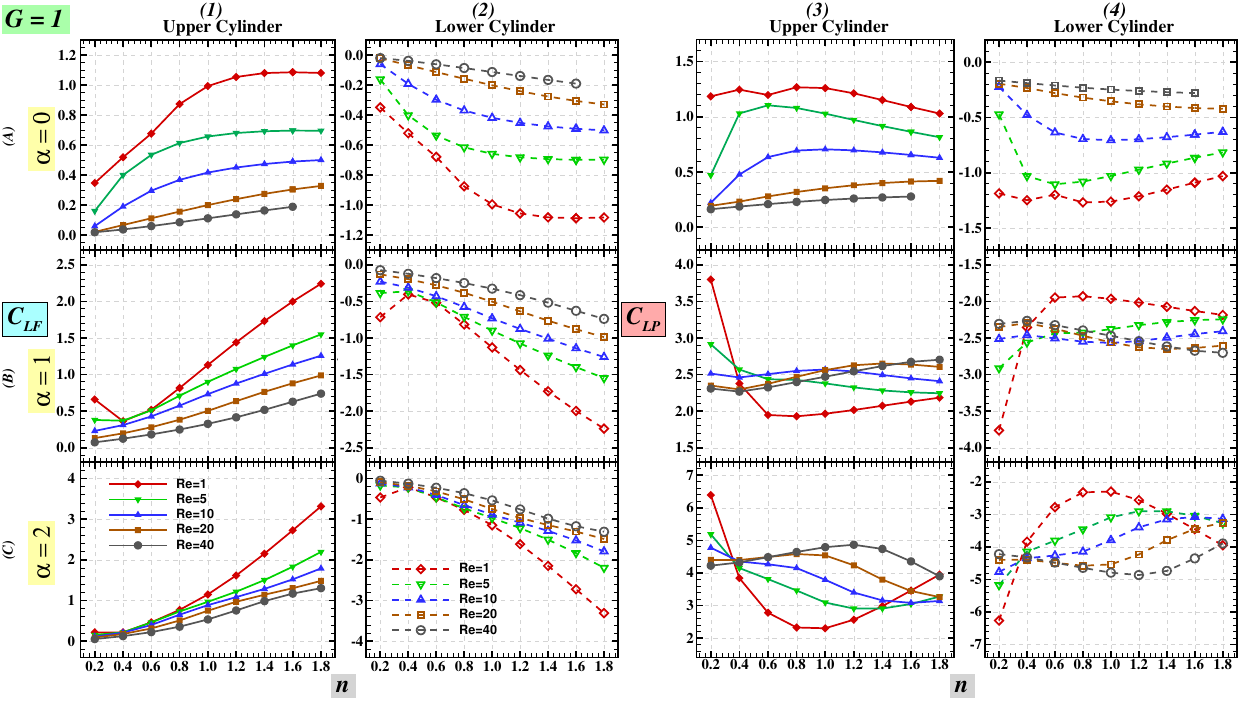}
	\caption{Variation of friction and pressure lift coefficient ($C_{LF}$ and $C_{LP}$) with $n$ and $Re$ for $G=1$.}
	\label{fig:clp-clf-g1}
\end{figure}
\begin{figure}[H]
	\centering
		\subfigure[Friction lift coefficient ($C_{LF}$)]  {\includegraphics[width=\linewidth]{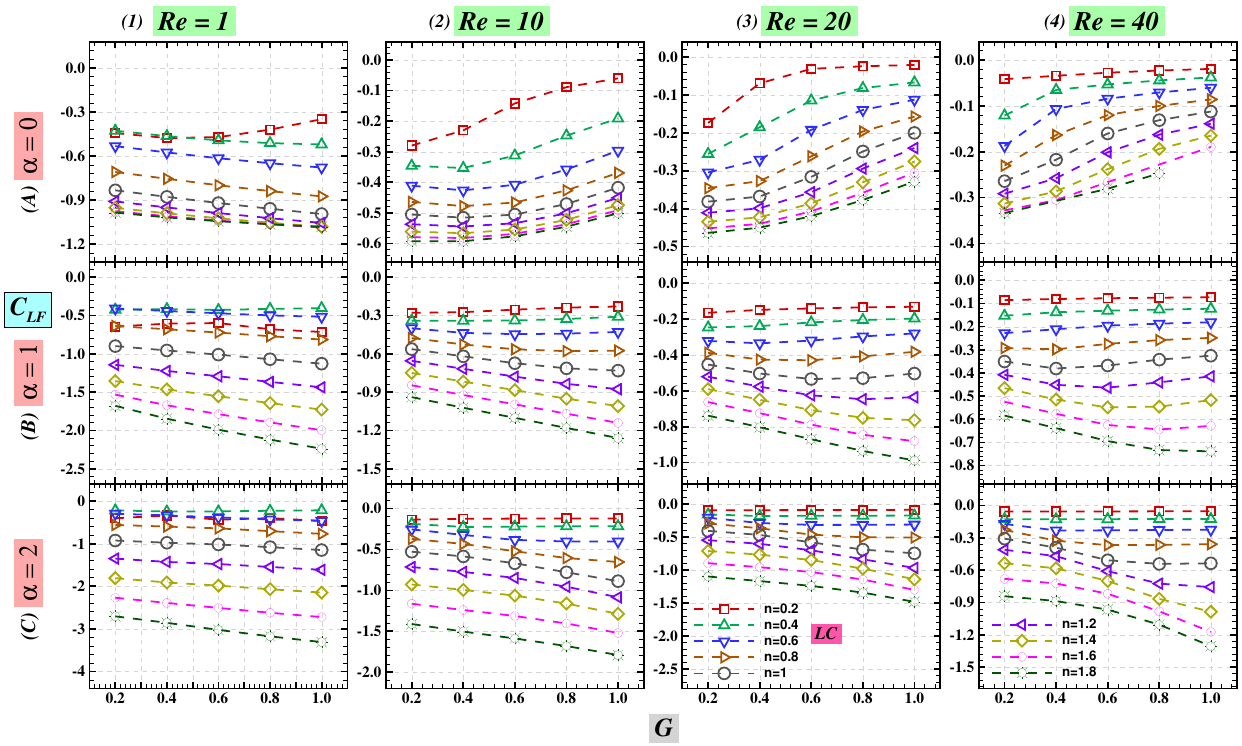}\label{fig:clf_G-LC}}
		\subfigure[Pressure lift coefficient ($C_{LP}$)] {\includegraphics[width=\linewidth]{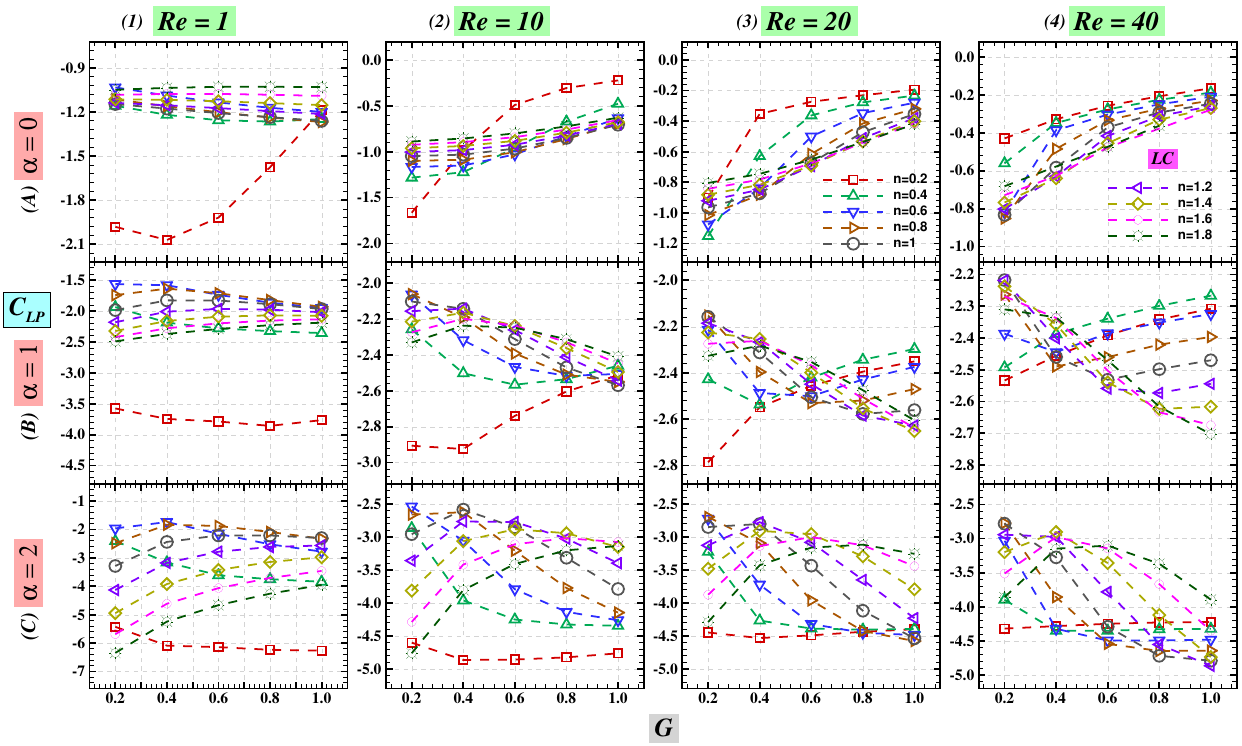}\label{fig:clp_G-LC}}
	\caption{Variation of individual lift coefficients ($C_{LF}$ and $C_{LP}$) for LC with $G$ and $n$ for different $Re$.}
	\label{fig:clpf_G-LC}
\end{figure}
%
\clearpage
\section{Total drag and lift coefficients ($C_{D}$ and $C_{L}$)}\label{appendix:Tdragnlift}
\setcounter{table}{0}\setcounter{figure}{0}  
%
\begin{figure}[H]
	\centering
	\includegraphics[width=1\linewidth]{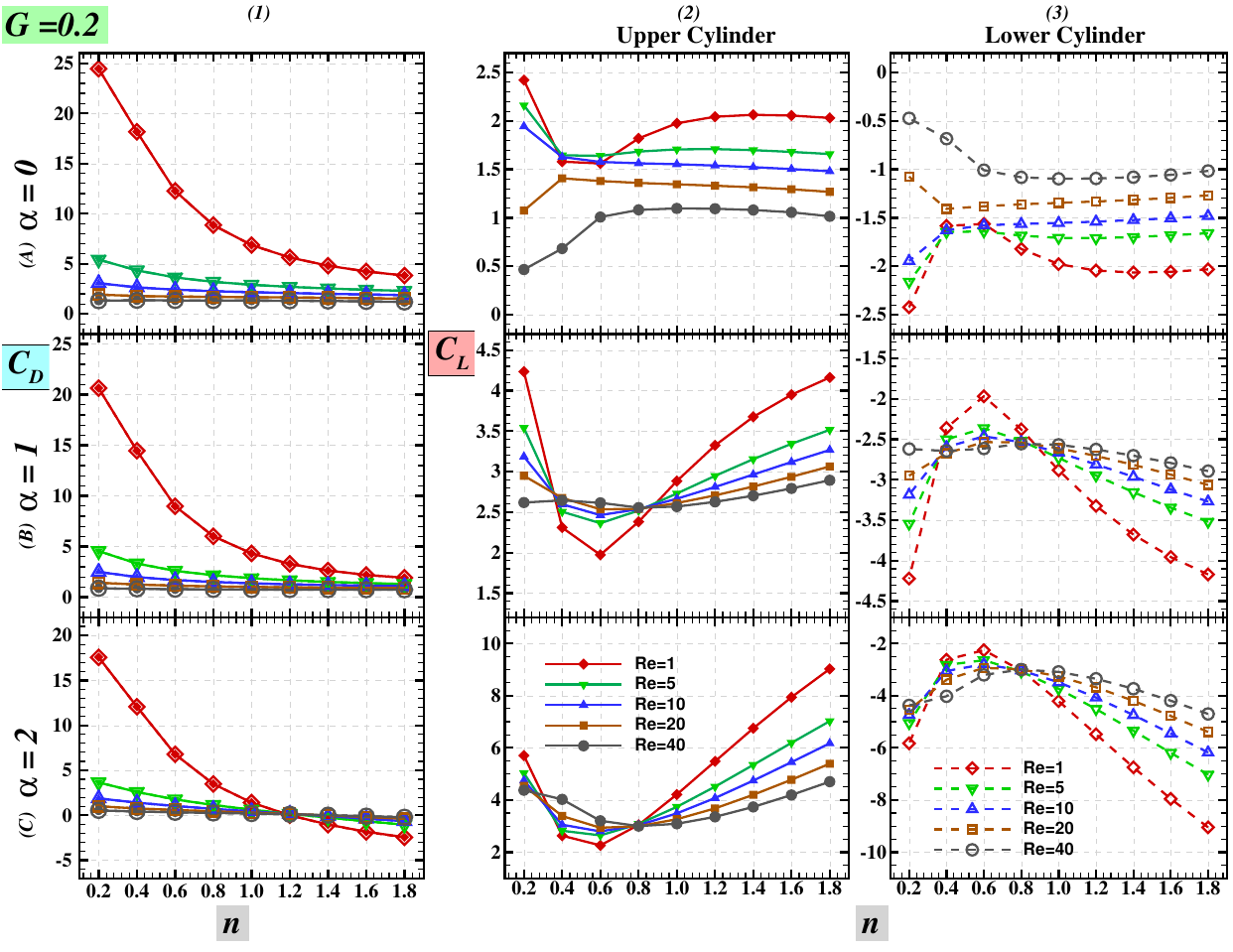}
	\caption{Variation of drag and lift coefficients ($C_{D}$ and $C_{L}$) with $n$ and $Re$ for $G=0.2$.}
	\label{fig:cdcl-g0.2}
\end{figure}
\begin{figure}[H]
	\centering
	\includegraphics[width=1\linewidth]{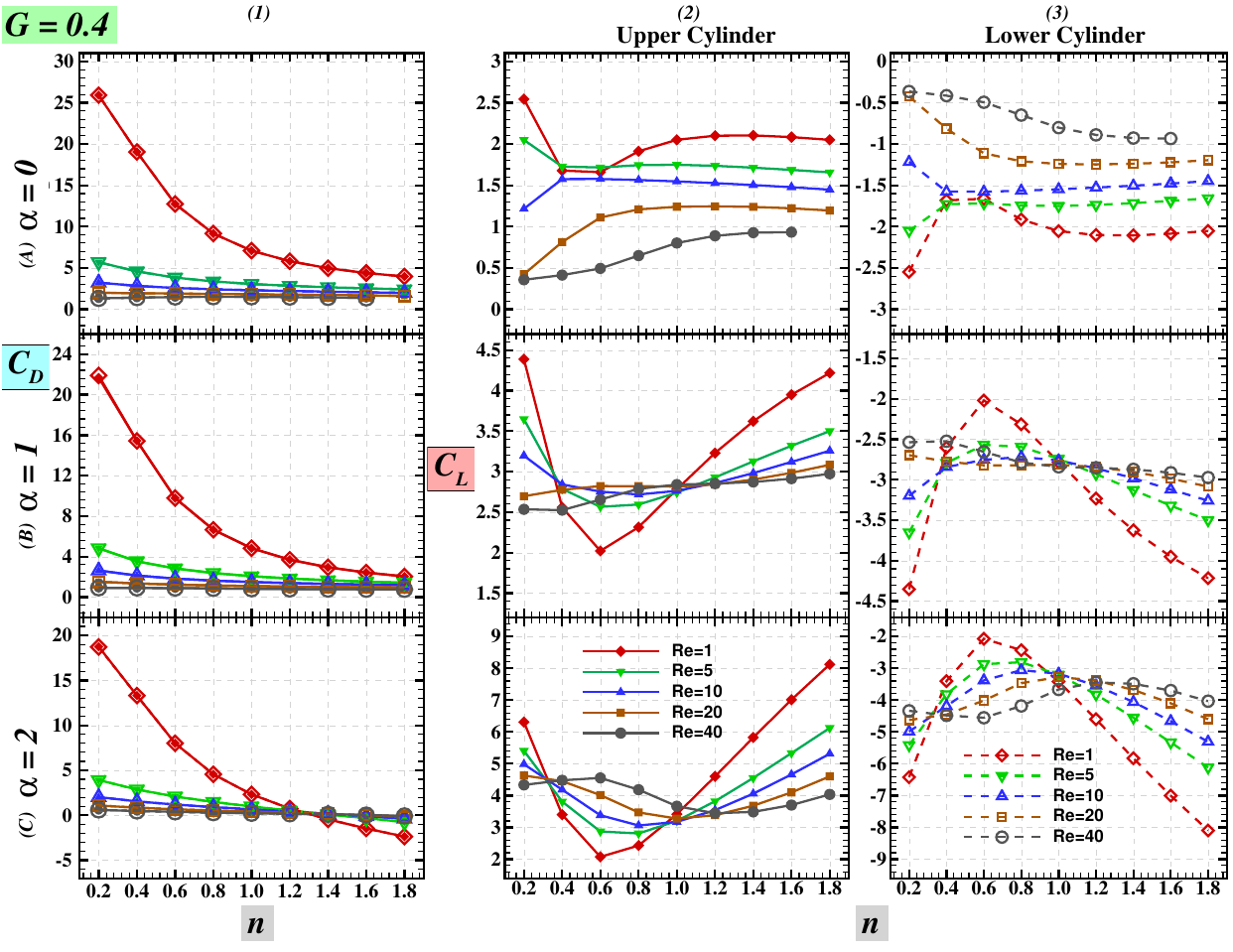}
	\caption{Variation of drag and lift coefficients ($C_{D}$ and $C_{L}$) with $n$ and $Re$ for $G=0.4$.}
	\label{fig:cdcl-g0.4}
\end{figure}
\begin{figure}[H]
	\centering
	\includegraphics[width=1\linewidth]{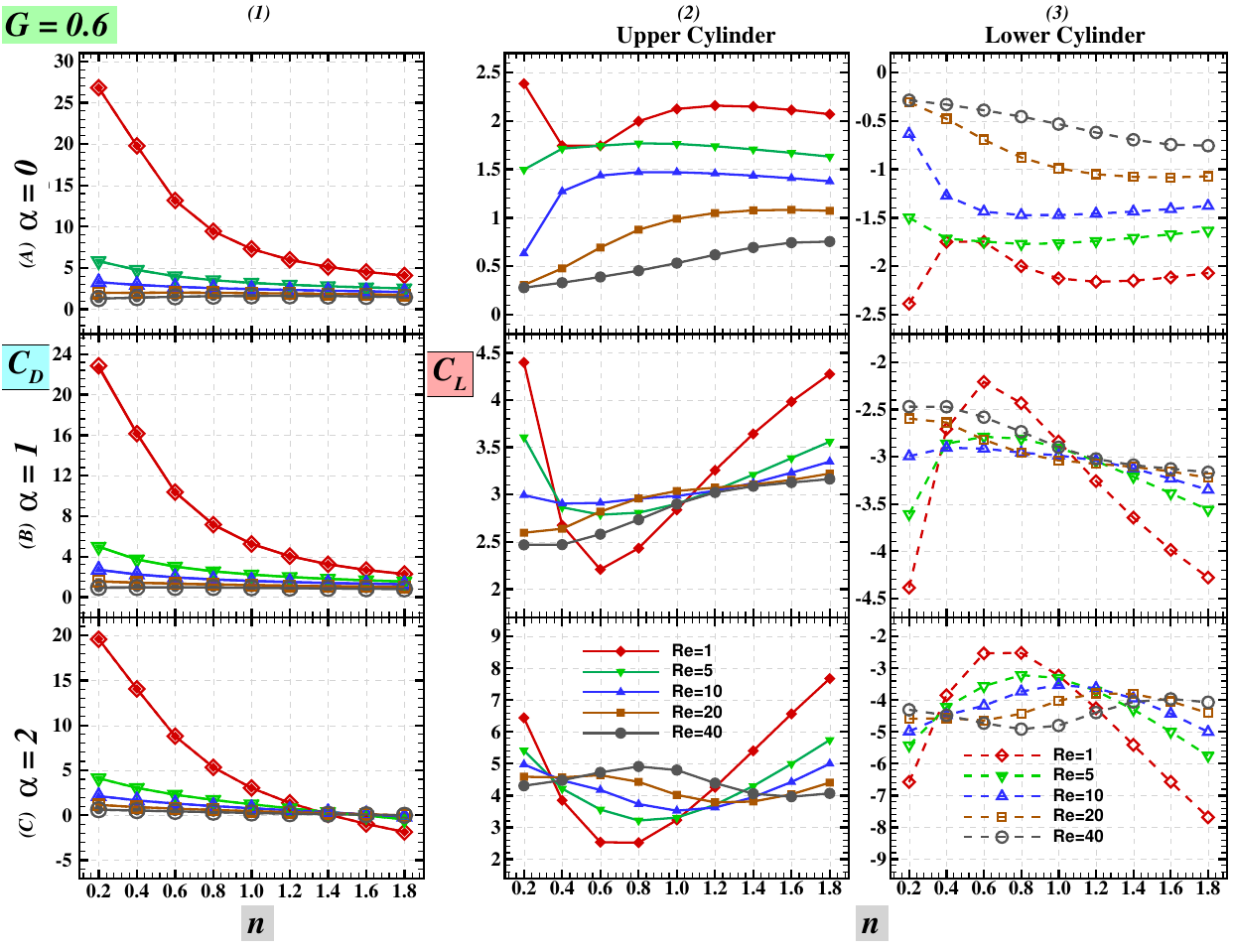}
	\caption{Variation of drag and lift coefficients ($C_{D}$ and $C_{L}$) with $n$ and $Re$ for $G=0.6$.}
	\label{fig:cdcl-g0.6}
\end{figure}
\begin{figure}[H]
	\centering
	\includegraphics[width=1\linewidth]{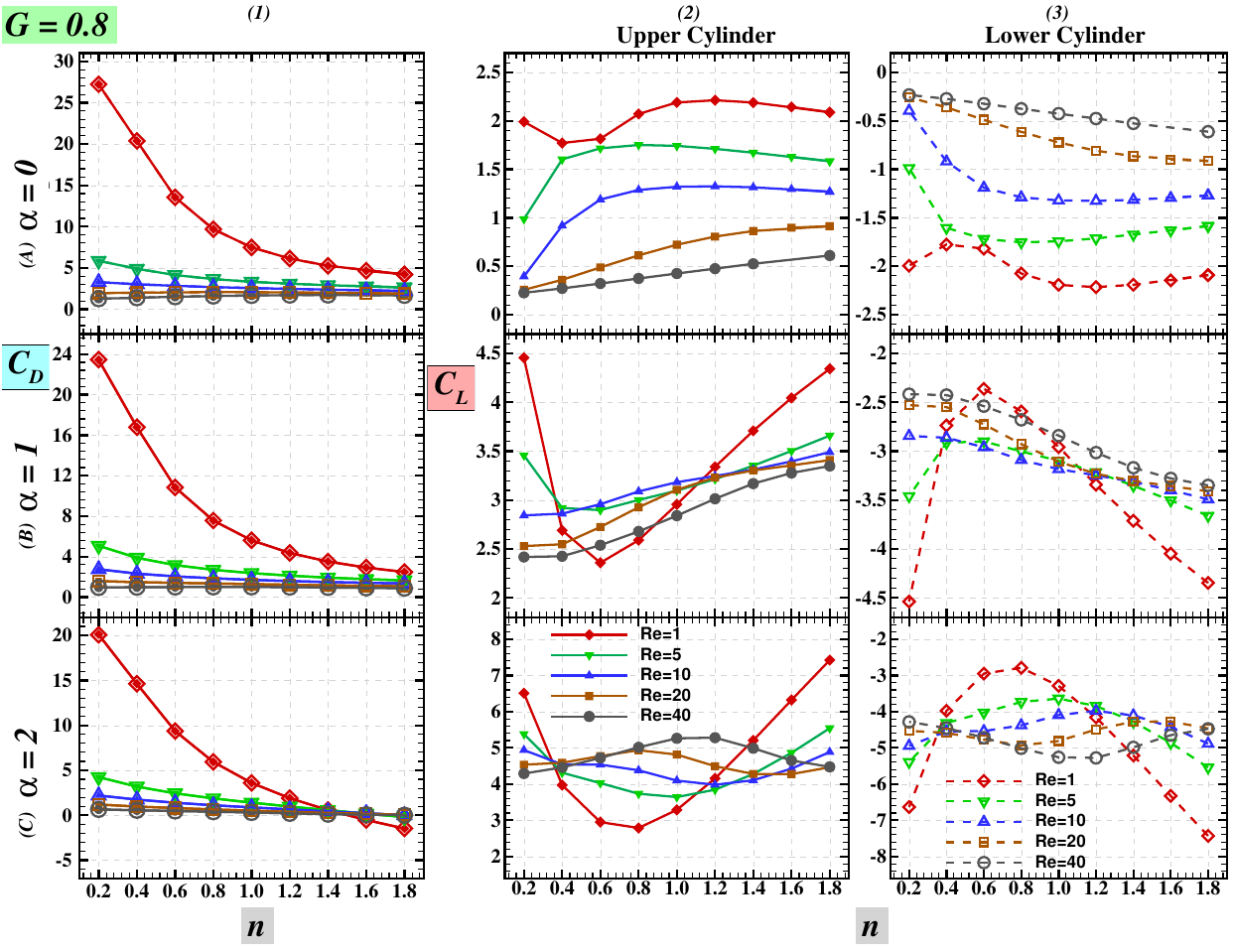}
	\caption{Variation of drag and lift coefficients ($C_{D}$ and $C_{L}$) with $n$ and $Re$ for $G=0.8$.}
	\label{fig:cdcl-g0.8}
\end{figure}
\begin{figure}[H]
	\centering
	\includegraphics[width=1\linewidth]{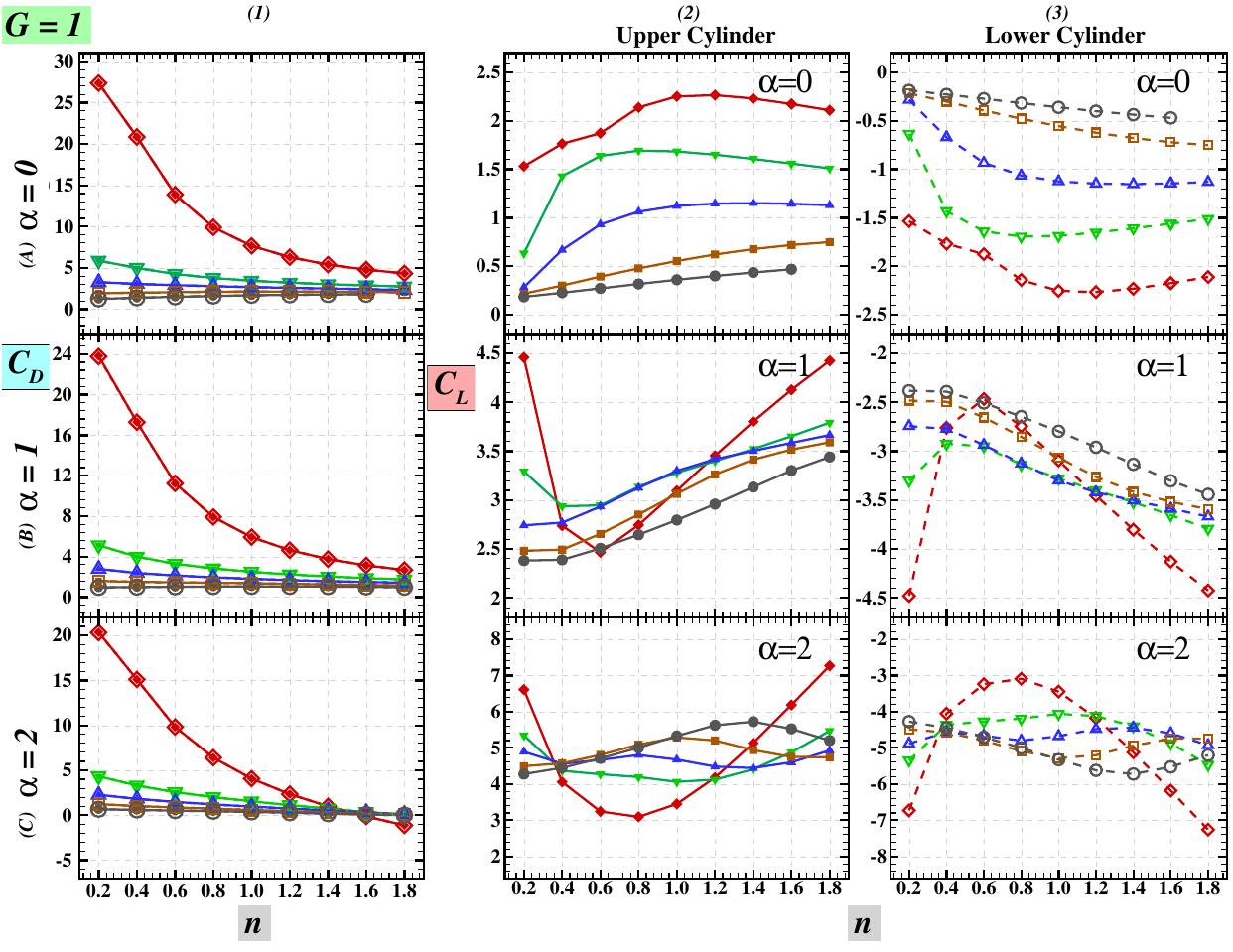}
	\caption{Variation of drag and lift coefficients ($C_{D}$ and $C_{L}$) with $n$ and $Re$ for $G=1$.}
	\label{fig:cdcl-g1}
\end{figure}
\begin{figure}[t!]
	\centering
	\subfigure[Total lift coefficient ($C_L$) at different $n$] {\includegraphics[width=1\linewidth]{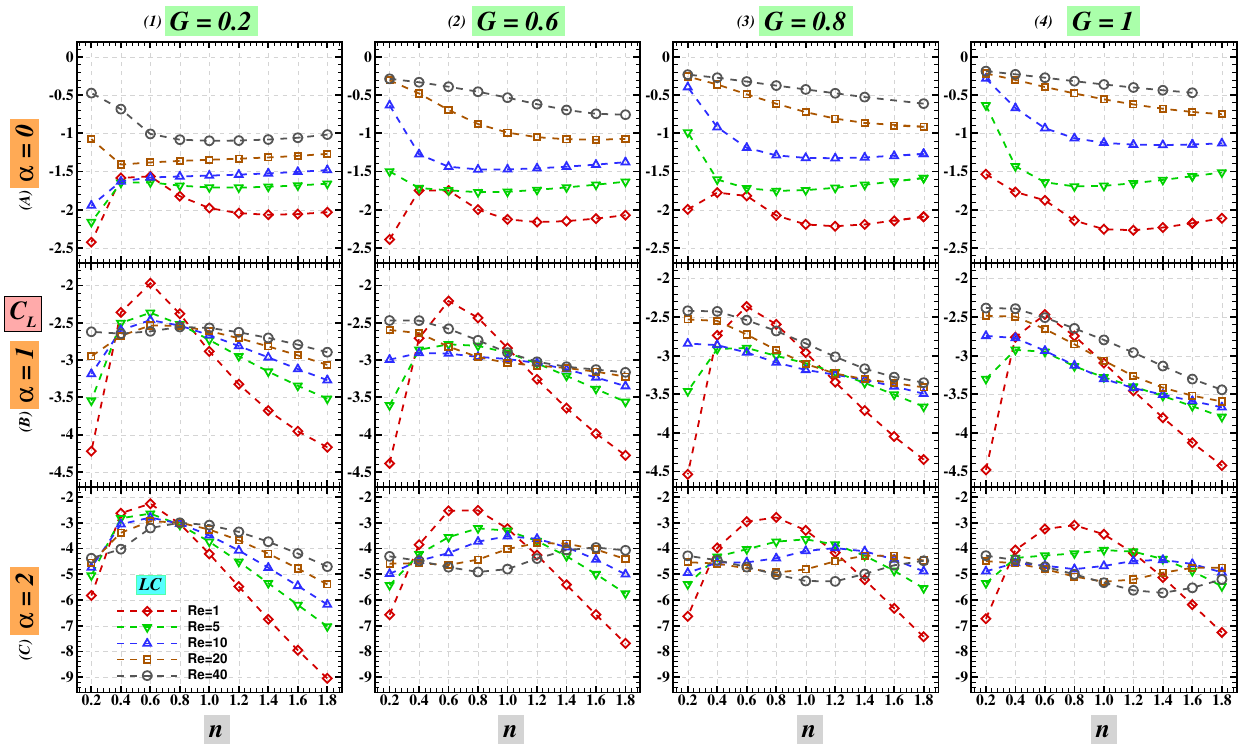}\label{fig:cl-LC}}
	\subfigure[Total drag coefficient ($C_L$) at different $G$] {\includegraphics[width=1\linewidth]{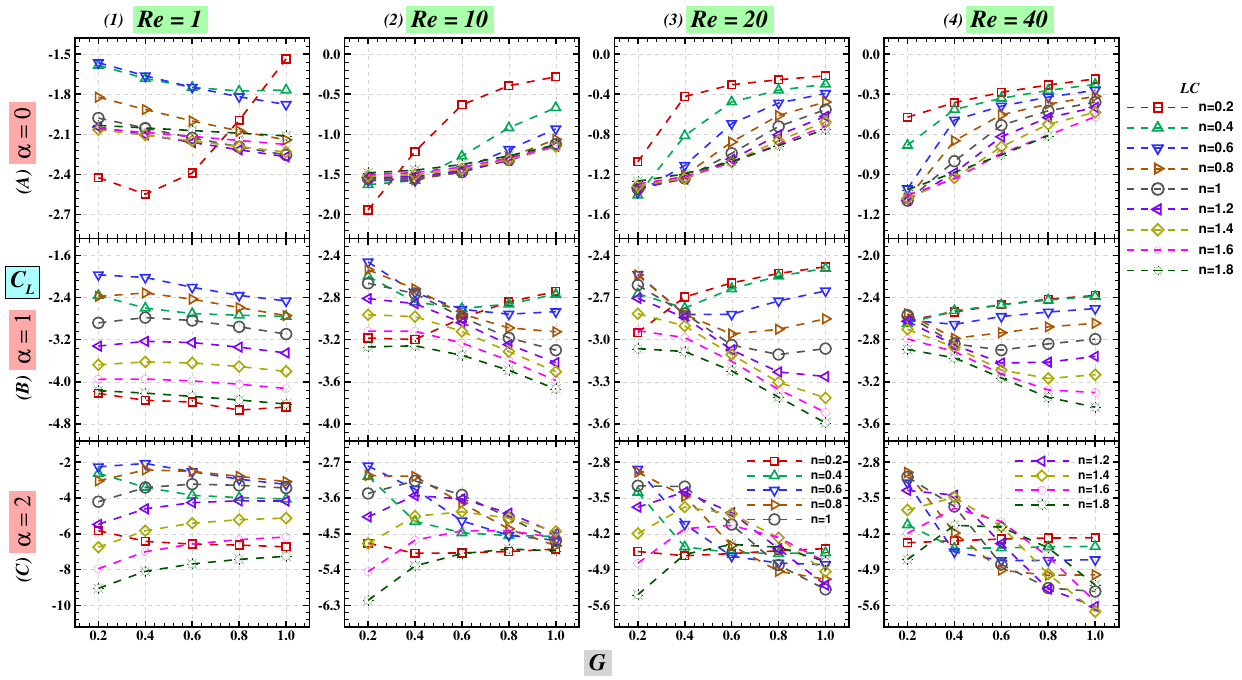}\label{fig:cl_g-LC}}
	\caption{Variation of total lift coefficient ($C_{L}$) for LC with $Re$.}
	\label{fig:cdclg-LC}
\end{figure}
%
\clearpage
\section{Drag and lift ratio ($C_{DR}$ and $C_{LR}$)}\label{appendix:ratiodragnlift}
\setcounter{table}{0}\setcounter{figure}{0} 
%
\begin{figure}[H]
	\centering
	\includegraphics[width=1\linewidth]{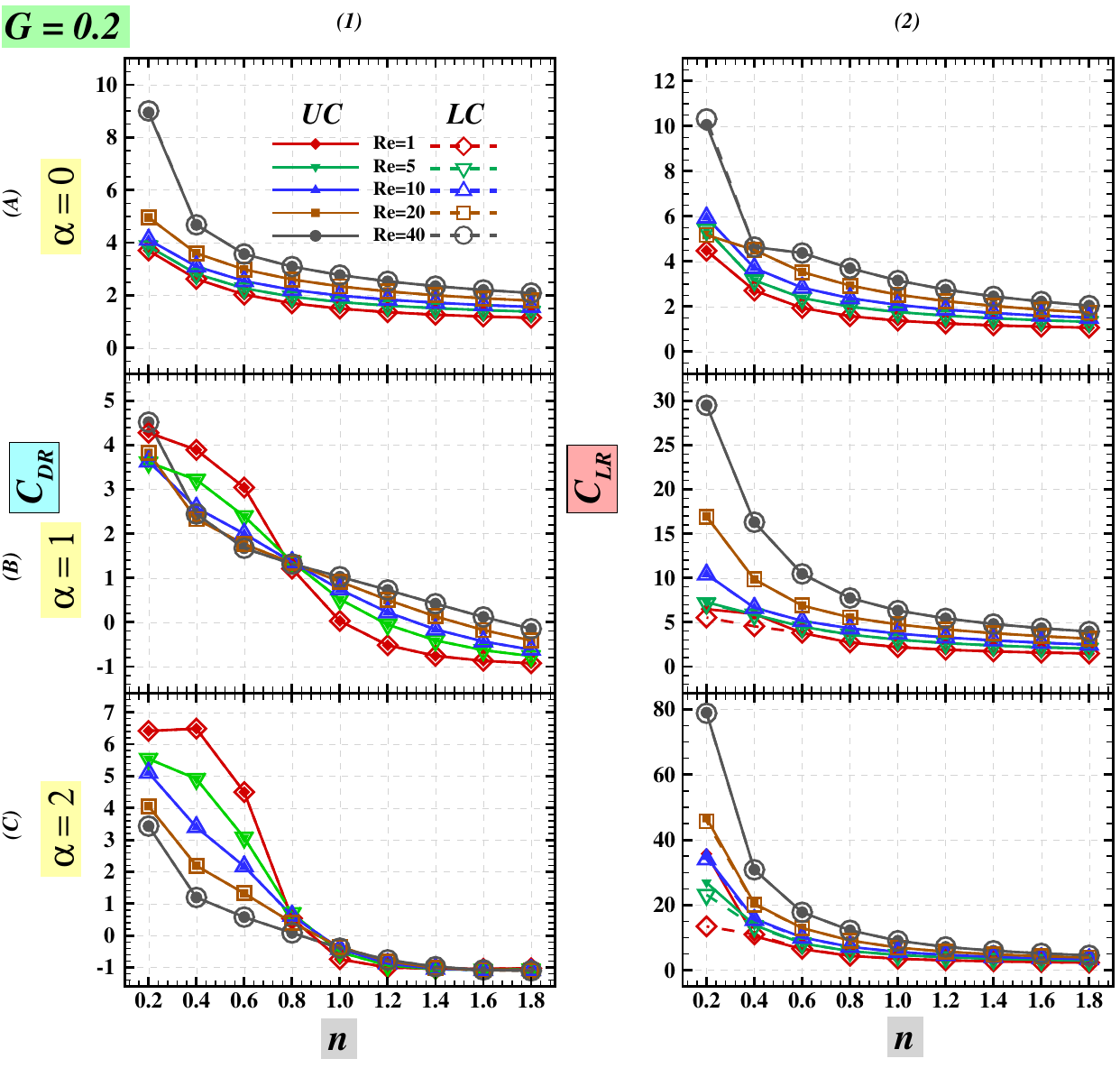}
	\caption{Variation of drag and lift ratio ($C_{DR}$ and $C_{LR}$) with $n$ and $Re$ for $G=0.2$.}
	\label{fig:cdrclr-g0.2}
\end{figure}
\begin{figure}[H]
	\centering
	\includegraphics[width=1\linewidth]{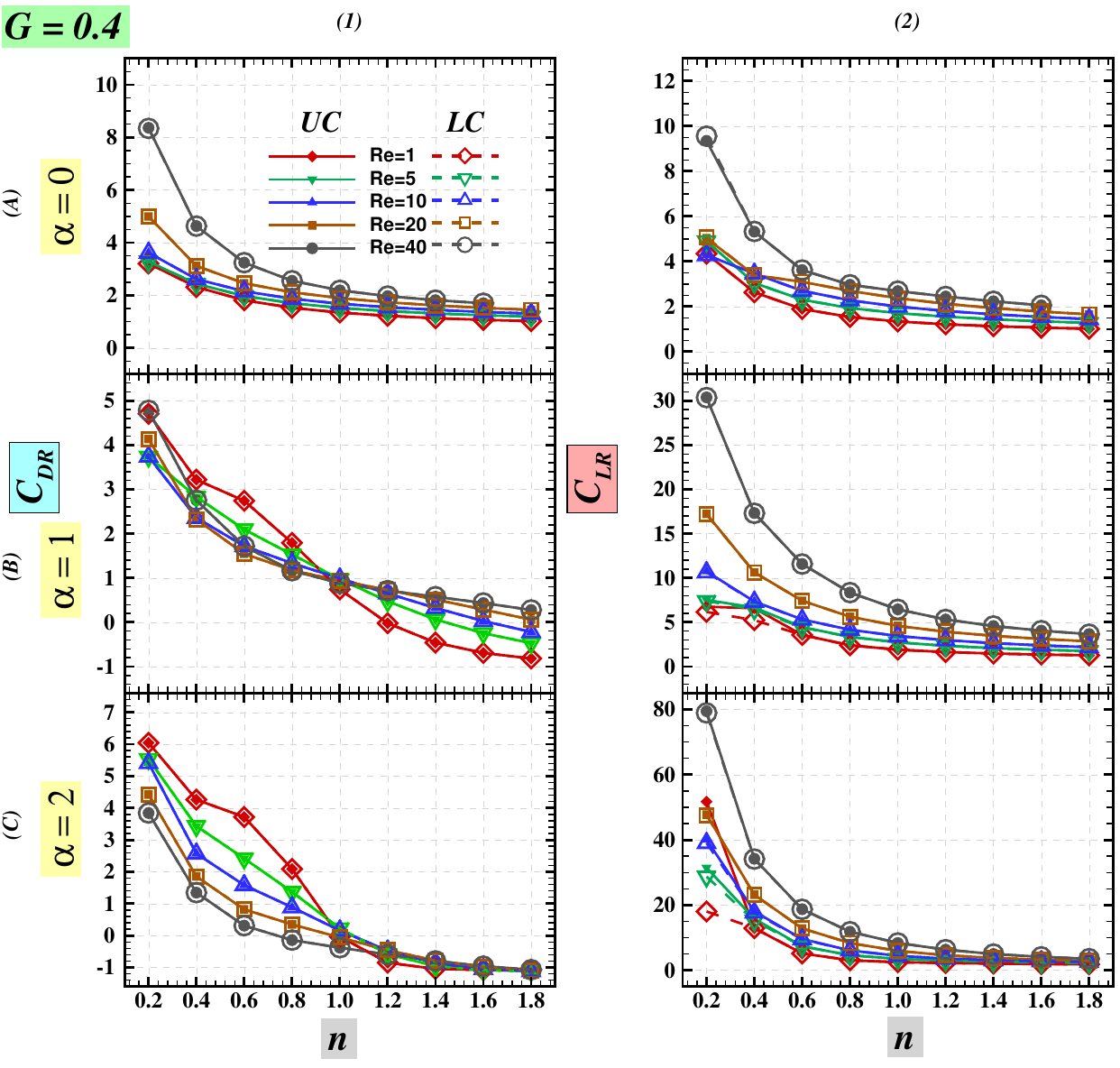}
	\caption{Variation of drag and lift ratio ($C_{DR}$ and $C_{LR}$) with $n$ and $Re$ for $G=0.4$.}
	\label{fig:cdrclr-g0.4}
\end{figure}
\begin{figure}[H]
	\centering
	\includegraphics[width=1\linewidth]{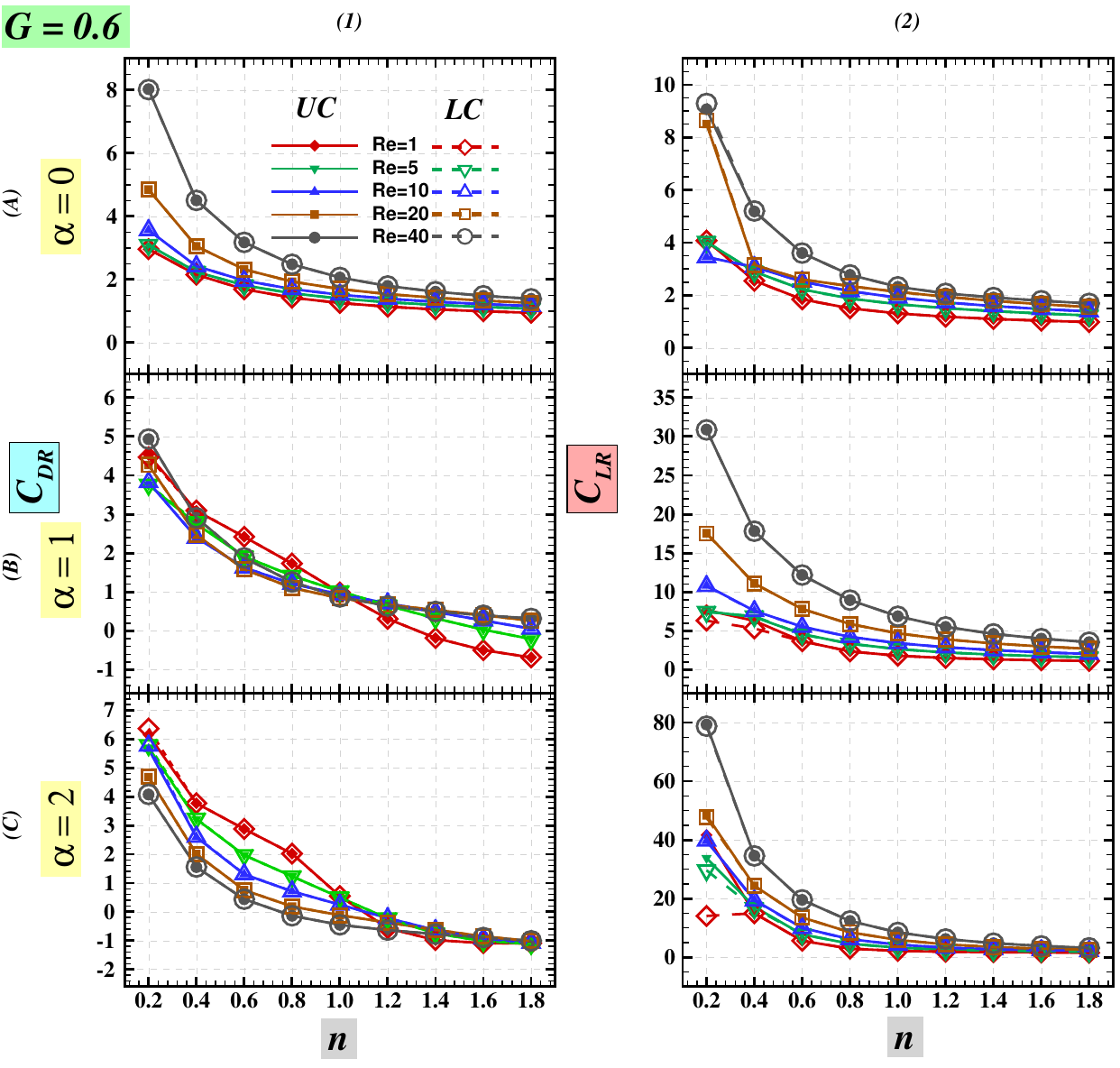}
	\caption{Variation of drag and lift ratio ($C_{DR}$ and $C_{LR}$) with $n$ and $Re$ for $G=0.6$.}
	\label{fig:cdrclr-g0.6}
\end{figure}
\begin{figure}[H]
	\centering
	\includegraphics[width=1\linewidth]{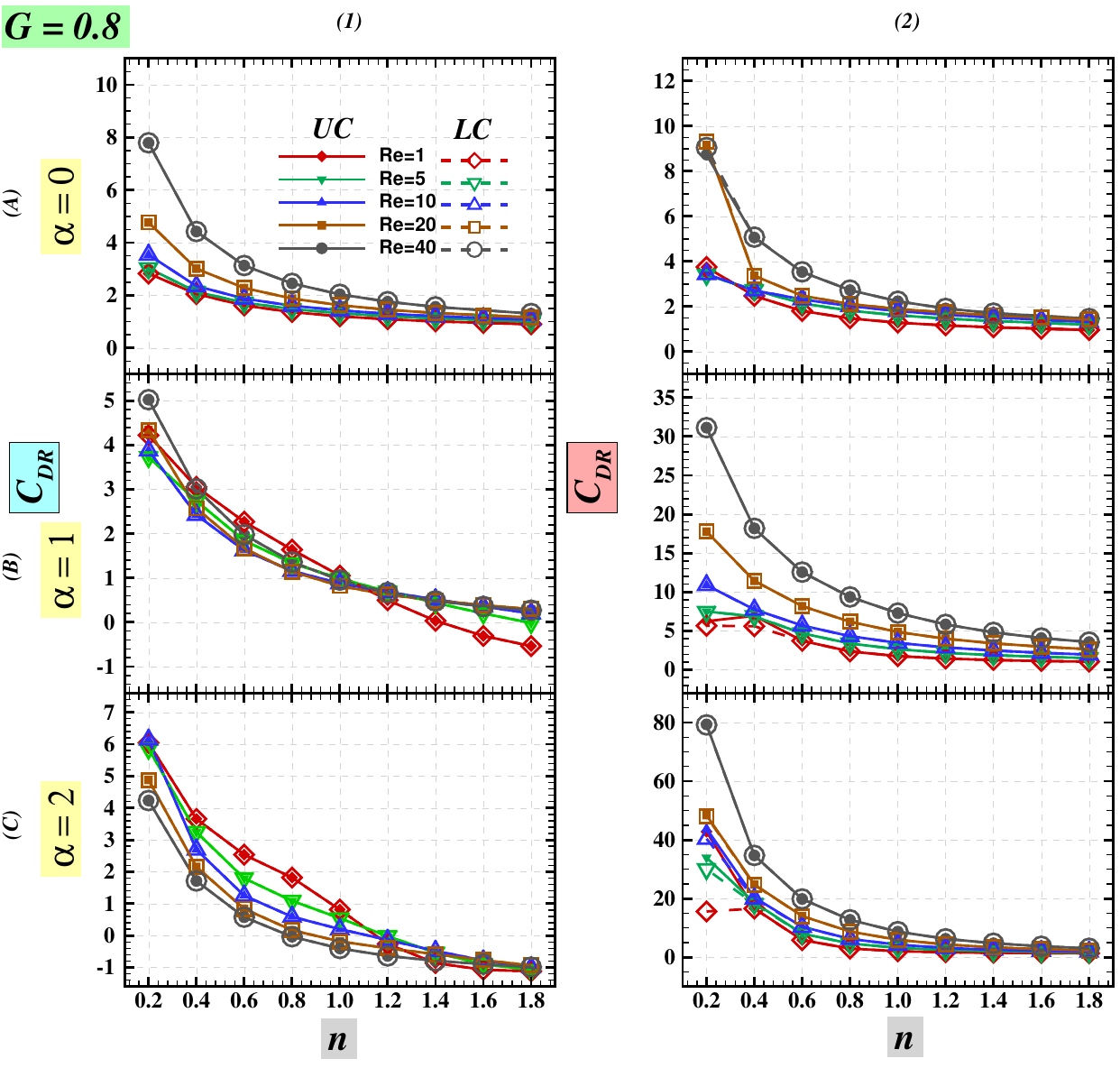}
	\caption{Variation of drag and lift ratio ($C_{DR}$ and $C_{LR}$) with $n$ and $Re$ for $G=0.8$.}
	\label{fig:cdrclr-g0.8}
\end{figure}
\begin{figure}[H]
	\centering
	\includegraphics[width=1\linewidth]{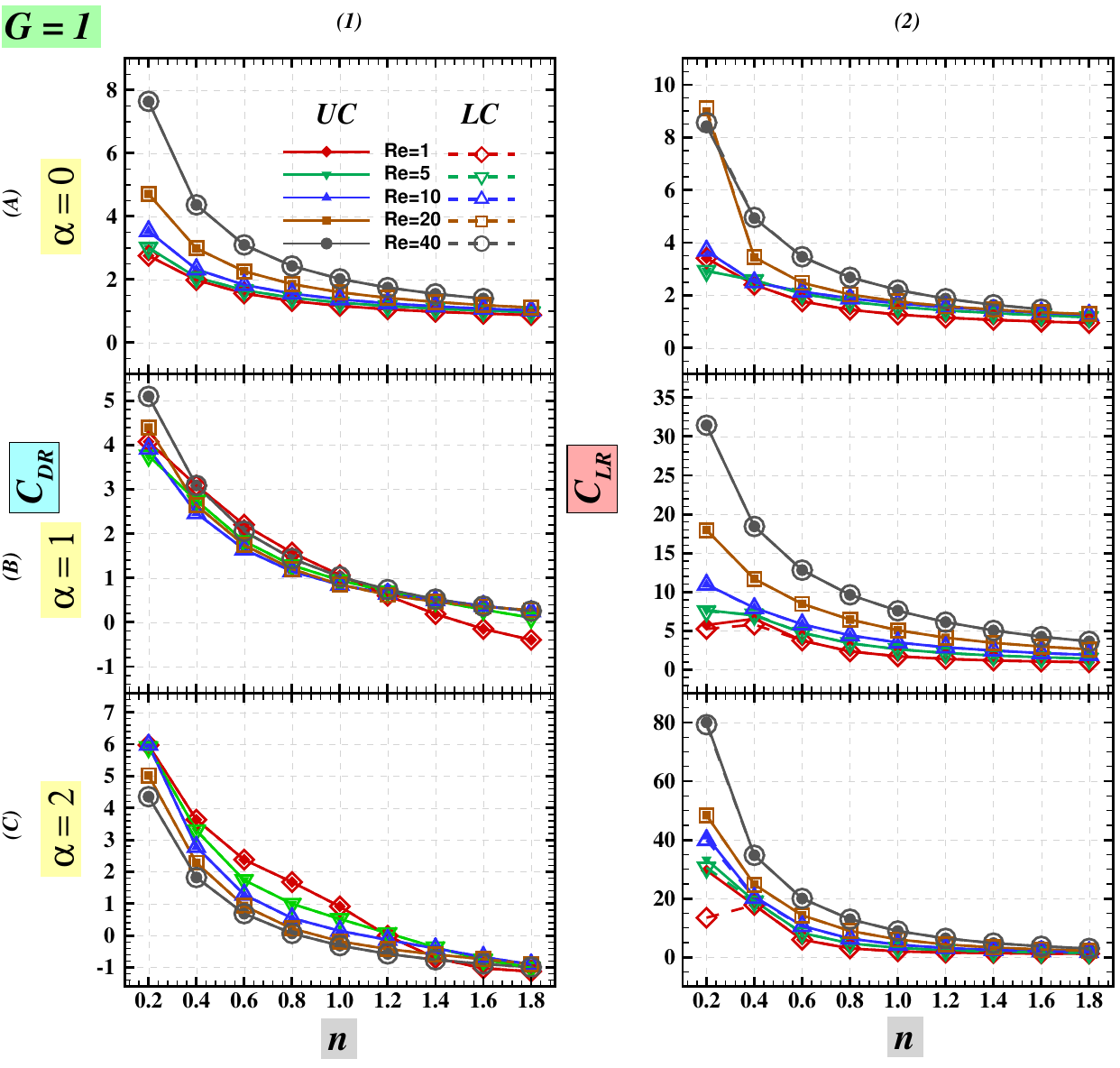}
	\caption{Variation of drag and lift ratio ($C_{DR}$ and $C_{LR}$) with $n$ and $Re$ for $G=1$.}
	\label{fig:cdrclr-g1}
\end{figure}
%
%
%
%
%
%
%
%
\end{document}